\documentclass[10pt]{article}

\usepackage{footmisc}
\usepackage{amsfonts}

\usepackage{amsmath,amsthm,amssymb}
\usepackage{siunitx}

\usepackage{graphicx}
\usepackage{epstopdf}

\usepackage{./style/bxascmac}

\usepackage{./style/slashbox}
\usepackage{./style/fancybox}
\usepackage{./style/framed}
\usepackage{./style/extarrows}
\usepackage{./style/pifont}

\usepackage{makeidx,epsfig,lscape}

\usepackage[colorlinks = false,
           linkcolor = red,
            urlcolor  = magenta,
           citecolor = green]{hyperref}

\newcommand{\bd}[1]{\mbox{\boldmath$#1$}}
\DeclareMathAlphabet{\mathpzc}{OT1}{pzc}{m}{it}

\def\bt{\begin{tabular}}
\def\et{\end{tabular}}
\def\and{\mbox{ and }}

\def\1{{\bf 1}}

\begin{document}

$\mbox{ }$

 \vskip 12mm

{ 
% "Title of the Paper"
{\noindent{\Large\bf
Memristor Circuits for Simulating Nonlinear Dynamics and Their Periodic Forcing}}
%\runtitle{ Memristor Circuits for Simulating Nonlinear Dynamics and Their Periodic Forcing}
\\[6mm]
{\centering
{\large Makoto Itoh\footnote{After retirement from Fukuoka Institute of Technology, he has continued to study the nonlinear dynamics on memristors. }}\\
{\it 1-19-20-203, Arae, Jonan-ku, \\ 
Fukuoka, 814-0101 JAPAN\\
Email: {itoh-makoto@jcom.home.ne.jp}}\\[1mm]
}

{ % \fontfamily{Cambria}\selectfont
{\noindent{\\[3mm]\textup{ 
In this paper, we show that the dynamics of a wide variety of nonlinear systems such as engineering, physical, chemical, biological, and ecological systems, can be simulated or modeled by the dynamics of memristor circuits.  
It has the advantage that we can apply nonlinear circuit theory to analyze the dynamics of memristor circuits. 
Applying an external source to these memristor circuits, they exhibit complex behavior, such as chaos and non-periodic oscillation.   
If the memristor circuits have an integral invariant, they can exhibit quasi-periodic or non-periodic behavior by the sinusoidal forcing.  
Their behavior greatly depends on the initial conditions, the parameters, and the maximum step size of the numerical integration.  
Furthermore, an overflow is likely to occur due to the numerical instability in long-time simulations. 
In order to generate a non-periodic oscillation, we have to choose the initial conditions, the parameters, and the maximum step size, carefully.  
We also show that we can reconstruct chaotic attractors by using the terminal voltage and current of the memristor.   
Furthermore, in many memristor circuits, the active memristor switches between passive and active modes of operation, depending on its terminal voltage. 
We can measure its complexity order by defining the binary coding for the operation modes.  
By using this coding, we show that in the forced memristor Toda lattice equations, the memristor's operation modes exhibit the higher complexity.  
Furthermore, in the memristor Chua circuit, the memristor has the special operation modes.  
 }}}
\\[4mm]
 
%Keywords
{\noindent{Keywords: \
 }

%
%
%%%%%%%%%%%%%%%%%%%%%%%%%%%%%%%%%%%%%%%%%%%%%%%%%%%%%%%%%%%%%%%%%%%%%%%
\section{Introduction}
\label{sec: introduction}
%%%%%%%%%%%%%%%%%%%%%%%%%%%%%%%%%%%%%%%%%%%%%%%%%%%%%%%%%%%%%%%%%%%%%%%
%
%
%
The dynamics of $n$-dimensional autonomous systems can be transformed into the dynamics of two-element \emph{extended memristor} circuits.
The internal state of the  memristors in these two-element circuits have the same dynamics as $n$-dimensional autonomous systems \cite{Itoh(2017)}.   
Thus, the memristors are essential dynamical elements needed in the modeling of complex nonlinear dynamical phenomena.  
In this paper, based on the above research results, we show that the dynamics of a wide variety of nonlinear systems, not only in physical and engineering systems, but also in biological and chemical systems and, even, in ecological systems, can be simulated or modeled by the dynamics of memristor circuits.  
It has the advantage that we can apply nonlinear circuit theory to analyze the dynamics of memristor circuits.

It is known that the dynamics of Chua's circuit and Van der Pol oscillator can be realized by using an ideal active memristor and some linear elements \cite{Itoh(2013)}. 
However, almost nonlinear systems can not satisfy the circuits equations without change.   
Thus, in order to transform their nonlinear equations into the memristor circuit equations, we use two methods, one is the exponential coordinate transformation, and the other is the time-scaling change \cite{Itoh(2017), Itoh(2011), Itoh(2014)}.  
The resulting memristor circuits have the same dynamics as the nonlinear systems.  
Furthermore, by connecting an external periodic forcing to these memristor circuits,  
they can exhibit complex behavior, such as chaos and non-periodic oscillation.  
If the memristor circuits have an integral invariant, then they can exhibit quasi-periodic or non-periodic behavior, which greatly depends on the initial conditions, the circuit parameters, and the maximum step size of the numerical integration.   
Furthermore, an overflow (outside the range of data) is likely to occur due to the numerical instability in long-time simulations.  
Thus, in order to generate a non-periodic oscillation, we have to choose the initial conditions, the parameters, and the maximum step size, carefully.  
Furthermore, noise may considerably affect the behavior of physical circuits.  

We also show that if we plot the terminal voltage against current of the memristor in the circuits, we can get the reconstruction of chaotic attractor on the two-dimensional plane. 
Furthermore, if we plot the instantaneous power $p$ versus the terminal voltage $v$ of the active memristor, then the $v-p$ locus lies in the first and the third quadrants, and it is pinched at the origin in many memristor circuits.   
It looks exactly like the $i-v$ loci of the passive memristor when a periodic source is supplied.  
Thus, the active memristor switches between passive and active modes of operation depending on its terminal voltage. 
However, in the forced memristor Toda lattice equations, the $v-p$ locus exhibits more complicated behavior, that is, it switches between four modes of operation.  
In order to measure the \emph{complexity order}, we define the binary coding for the above memristor's operation modes.  
By using this coding, we show that in the forced memristor Toda lattice equations, the memristor's operation modes exhibit the higher complexity.  
Furthermore, in the memristor Chua circuit, the active memristor exhibits the special operation modes, which is quite different from the other memristor circuits.  

%\clearpage
%
%
%%%%%%%%%%%%%%%%%%%%%%%%%%%%%%%%%%%%%%%%%%%%%%%%%%%%%%%%%%%%%%%%%%%%%%%
\section{Three-element Memristor Circuit}
\label{sec: 3-memristor}
%%%%%%%%%%%%%%%%%%%%%%%%%%%%%%%%%%%%%%%%%%%%%%%%%%%%%%%%%%%%%%%%%%%%%%%
%
%
%
Let us consider the three-element memristor circuit 
in Figure \ref{fig:memristor-inductor-battery}, 
which consists of an inductor $L$, a battery $E$, and a current-controlled extended memristor.    

The terminal voltage $v_{M}$ and the terminal current $i_{M}$ of the current-controlled extended memristor are described
by
\begin{center}
\begin{minipage}{8.7cm}
\begin{shadebox}
\underline{\emph{V-I characteristics of the extended memristor}}
\begin{equation}
\begin{array}{lll}
  v_{M} &=& \hat{R}( \bd{x}, \ i_{M} ) \, i_{M},  \\ 
    && \hat{R}( \bd{x}, \ 0) \ne \infty,
  \vspace{1mm} \\
  \displaystyle \frac{d\bd{x}}{dt} &=& \tilde{\bd{f}}(\bd{x}, \ i_{M}).
\end{array}
\vspace{2mm}
\label{eqn: extended-20}
\end{equation}
\end{shadebox}
\end{minipage}
\end{center}
Here, $\bd{x} = (x_{1}, \, x_{2}, \, \cdots, \, x_{n}) \in \mathbb{R}^{n}$,  
$\hat{R}( \bd{x}, \ i_{M} )$ is a continuous scalar-valued function, \\
and $\tilde{\bd{f}} = (\tilde{f}_{1}, \, \tilde{f}_{2}, \, \cdots, \, \tilde{f}_{n}): \mathbb{R}^{n} \rightarrow \mathbb{R}^{n}$ 
(see Appendix A).  

The dynamics of the above three-element memristor circuit is given by 
\begin{center}
\begin{minipage}{8.7cm}
\begin{shadebox}
\underline{\emph{Three-element memristor circuit equations}}
\begin{equation}
\begin{array}{cll}
  \displaystyle L \frac{di}{dt} &=& - v_{M} + E = - \hat{R}( \bd{x}, \ i ) \, i + E,
  \vspace{1mm} \\
  \displaystyle \frac{d\bd{x}}{dt} &=& \tilde{\bd{f}}(\bd{x}, \ i),
\end{array}
\vspace{2mm}
\label{eqn: dynamics-1}
\end{equation}
where $L$ denotes the inductance of the inductor, $E$ denotes the voltage of the battery, and $i_{M}=i$.
\end{shadebox}
\end{minipage}
\end{center}

Assume that $n=1$ and $L=1$.  
Then Eq. (\ref{eqn: dynamics-1}) can be recast into the form 
\begin{center}
\begin{minipage}{11.0cm}
\begin{shadebox}
\underline{\emph{Three-element memristor circuit equations with $n=1$ and $L=1$}} \vspace{1mm} 
\begin{equation}
\begin{array}{cll}
  \displaystyle \frac{di}{dt} &=& - v_{M} + E = - \hat{R}( x, \ i ) \, i + E,
  \vspace{1mm} \\
  \displaystyle \frac{dx}{dt} &=& \tilde{f}_{1}(x, \ i),
\end{array}
\vspace{2mm}
\label{eqn: dynamics-n-1}
\end{equation}
where $\bd{x} = x$ and $\tilde{\bd{f}} = \tilde{f}_{1}$. 
\end{shadebox}
\end{minipage}
\end{center}
%
%

%---Fig. 1-------%
\begin{figure}[hpbt]
 \centering
  \psfig{file=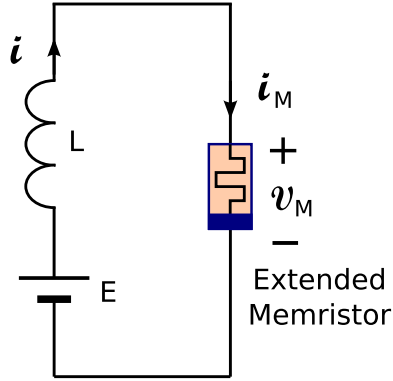, height=5cm} 
  \caption{A three-element memristor circuit which consists of an inductor $L$, a battery $E$, and a current-controlled extended memristor (right side).}
  \label{fig:memristor-inductor-battery} 
\end{figure}
%
%

%---Fig. 2-------%
\begin{figure}[hpbt]
 \centering
 \psfig{file=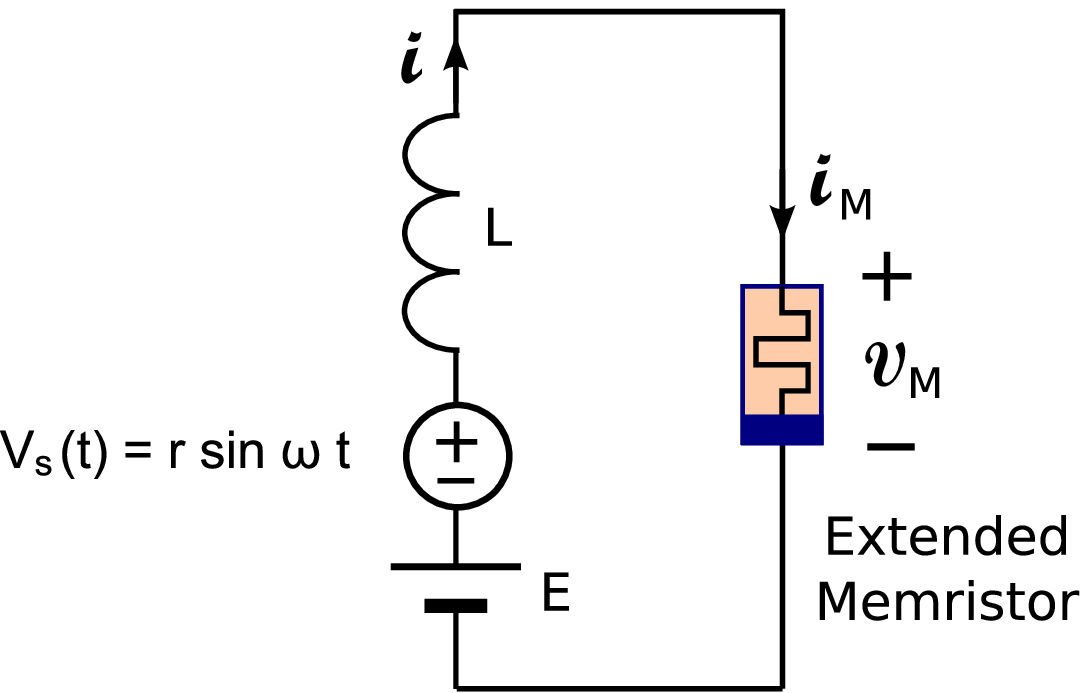, width=8cm} 
  \caption{Four-element memristor circuit driven by a periodic voltage source $v_{s}(t) = r \sin ( \omega t)$, where $r$ and $\omega$ are constants.}
  \label{fig:memristive-inductor-battery-source} 
\end{figure}
%
%

%==============================================================%
\subsection{Brusselator equations}
\label{sec: Brusselator}
%==============================================================%
%
The Brusselator is a theoretical model for a type of autocatalytic reaction. 
The dynamics of the Brusselator is given by 
\begin{center}
\begin{minipage}{8.7cm}
\begin{shadebox}
\underline{\emph{Brusselator equations}}
\begin{equation}
\begin{array}{lll}
 \displaystyle \frac{d u}{dt} &=& A + \bigl \{ uv - (B+1) \bigr \} u, 
 \vspace{2mm} \\
 \displaystyle \frac{d v}{dt} &=& B \, u - u^{2}\,v,
\end{array}
\vspace{2mm}
\label{eqn: Brusselator-1}
\end{equation}
where $A$ and $B$ are constants. 
\end{shadebox}
\end{minipage}
\end{center}
Consider the three-element memristor circuit in Figure \ref{fig:memristor-inductor-battery} with $L=1$.  
Then the dynamics of this circuit is given by Eq. (\ref{eqn: dynamics-n-1}).  
Assume that Eq. (\ref{eqn: dynamics-n-1}) satisfies 
\begin{equation}
 \begin{array}{ccc}
  E &=& A, \vspace{2mm} \\
  \hat{R}(x, \, i) &=& - \bigl \{ i \, x - (B+1) \bigr \}, \vspace{2mm} \\
  \tilde{f}_{1}(x, i) &=& Bi - i^{2}\, x.
 \end{array}
\end{equation}
Then we obtain 
\begin{center}
\begin{minipage}{8.7cm}
\begin{shadebox}
\underline{\emph{Memristor Brusselator equations}}
\begin{equation}
\begin{array}{lll}
 \displaystyle \frac{d i}{dt} &=& A + \bigl \{ ix - (B+1) \bigr \} i, 
 \vspace{2mm} \\
 \displaystyle \frac{d x}{dt} &=& B \, i - i^{2}\, x,
\end{array}
\vspace{2mm}
\label{eqn: Brusselator-2}
\end{equation}
where $A$ and $B$ are constants.
\end{shadebox}
\end{minipage}
\end{center}
Equations (\ref{eqn: Brusselator-1}) and (\ref{eqn: Brusselator-2}) are equivalent if we change the variables  
\begin{equation}
  i = u, \ x = v.  
\end{equation}
The terminal voltage $v_{M}$ and the terminal current $i_{M}$ of the current-controlled extended memristor 
in Figure \ref{fig:memristor-inductor-battery} are given by
\begin{center}
\begin{minipage}{8.9cm}
\begin{shadebox}
\underline{\emph{V-I characteristics of the extended memristor}}
\begin{equation}
\begin{array}{lll}
  v_{M} &=& \hat{R}(x, \, i_{M}) \, i_{M} =  - \bigl \{ i_{M} \, x - (B+1) \bigr \} i_{M},   
  \vspace{3mm} \\
        && \hat{R}(x, \, 0) \ne \infty,
  \vspace{1mm} \\
 \displaystyle \frac{d x}{dt} &=& B \, i_{M} - {i_{M}}^{2}\, x,
\end{array}
\label{eqn: Brusselator-3}
\end{equation}
where $\hat{R}(x, \, i_{M}) =  - \bigl \{ i_{M} \, x - (B+1) \bigr \}$ and $i_{M}=i$.   \vspace{2mm}
\end{shadebox}
\end{minipage}
\end{center}
It follows that the Brusselator equations (\ref{eqn: Brusselator-1}) can be realized by 
the three-element memristor circuit in Figure \ref{fig:memristor-inductor-battery}.  
Equations (\ref{eqn: Brusselator-1}) and (\ref{eqn: Brusselator-2}) exhibit periodic oscillation (limit cycle). 
When an external source is added as shown in Figure \ref{fig:memristive-inductor-battery-source}, 
the forced memristor Brusselator equations can exhibit chaotic oscillation \cite{Bashkirtseva(2005)}.   
The dynamics of this circuit is given by 
\begin{center}
\begin{minipage}{8.7cm}
\begin{shadebox}
\underline{\emph{Forced memristor Brusselator equations}}
\begin{equation}
\begin{array}{lll}
 \displaystyle \frac{d i}{dt} &=& A + \bigl \{ ix - (B+1) \bigr \} i + r \sin ( \omega t), 
 \vspace{2mm} \\
 \displaystyle \frac{d x}{dt} &=& B \, i - i^{2}\, x,
\end{array}
\vspace{2mm}
\label{eqn: Brusselator-4}
\end{equation}
where $r$ and $\omega$ are constants.  
\end{shadebox}
\end{minipage}
\end{center}
We show the chaotic attractor, Poincar\'e map, and $i_{M}-v_{M}$ locus of Eq. (\ref{eqn: Brusselator-4}) in Figures \ref{fig:Brusselator-attractor}, \ref{fig:Brusselator-poincare}, and \ref{fig:Brusselator-pinch}(a), respectively.  
The following parameters are used in our computer simulations:
\begin{equation}
  A = E = 0.4, \ B = 1.2, \ r = 0.05, \ \omega = 0.81.  
\end{equation}
The $i_{M}-v_{M}$ locus moves in the first quadrant, 
that is, it moves in the \emph{passive} region, since the instantaneous power of the extended memristoris positive, that is, 
\begin{equation}
  P_{M}(t) \stackrel{\triangle}{=}i_{M}(t) \, v_{M}(t) > 0.  
\end{equation}
Hence, the instantaneous power $P_{M}(t)$ is dissipated in the extended memristor, which is delivered from the forcing signal and the inductor.  
Furthermore, the $i_{M}-v_{M}$ locus is not pinched at the origin as shown in Figure \ref{fig:Brusselator-pinch}(a), since the trajectory does not tend to the origin.  

We define next the instantaneous power of the two circuit elements, that is, the instantaneous power of the extended memristor and the battery by 
\begin{equation}
  p_{ME}(t) \stackrel{\triangle}{=} i_{M}(t)\, v_{ME}(t), 
\end{equation}
where $v_{ME}(t) = v_{M}(t)-E$, and $E$ denotes the voltage of the battery. 
That is, $v_{ME}(t)$ denotes the voltage across the extended memristor and the battery.  
We show the $v_{ME}-p_{ME}$ locus in Figure \ref{fig:Brusselator-pinch}(b).  
Observe that the locus is pinched at the origin, and it lies in the first and the third quadrants.  
Thus, the instantaneous power $p_{ME}(t)$ delivered from the forced signal and the inductor is dissipated when $v_{M}(t) - E > 0$.   
However, the instantaneous power $p_{ME}(t)$ is \emph{not} dissipated when $v_{M}(t) - E <0 $.
We conclude as follow: \\
\begin{center}
\begin{minipage}{.7\textwidth}
\begin{itembox}[l]{Behavior of the extended memristor}
Assume that Eq. (\ref{eqn: Brusselator-4}) exhibits chaotic oscillation.  
Then, we obtain the following results: 
\begin{enumerate}
\item The extended memristor defined by Eq. (\ref{eqn: Brusselator-3}) is operated as a \emph{passive} element.  
The instantaneous power $P_{M}(t)$ of the memristor is dissipated in this extended memristor, 
which is delivered from the forcing signal and the inductor.  \\
\item 
When $v_{M}(t) - E <0 $, the instantaneous power $p_{ME}(t)$ of the extended memristor and the battery  is \emph{not} dissipated.
However, when $v_{M}(t) - E > 0$, the instantaneous power $p_{ME}(t)$ is dissipated.    
\end{enumerate}
\end{itembox} 
\\
\end{minipage}
\end{center}

Note that $x(t)$ in Eq. (\ref{eqn: Brusselator-3}) is the internal state of the extended memristor. 
Thus, we might not be able to observe it.  
However, we can reconstruct the chaotic attractor into two dimensional Euclidean space (plane) by using 
\begin{equation}
  \bigl (i(t), \ i'(t) \bigr ), 
\end{equation}
where $\displaystyle i'(t) \stackrel{\triangle}{=} \frac{d i(t)}{dt}$ (see \cite{Packard} for more details).  
Furthermore, the $i_{M}-v_{M}$ locus in Figure \ref{fig:Brusselator-pinch}(a) is considered to be the reconstruction of the chaotic attractor on the two-dimensional plane, since 
\begin{equation}
  \Bigl ( i_{M}(t), \,  v_{M}(t) \Bigr )  \equiv \Bigl ( i(t), \ - i'(t) + A + r \sin ( \omega t) \Bigr ), 
\end{equation}
where $i_{M}(t)=i(t)$.
We show their trajectories and Poincar\'e maps in Figures \ref{fig:Brusselator-reconstruction} and \ref{fig:Brusselator-reconstruction-poincare}, respectively.     
We can also reconstruct the chaotic attractor into the three-dimensional Euclidean space by using 

\begin{equation}
  \bigl ( i(t), \ i'(t), \ i''(t) \bigr ), 
\end{equation}
or
\begin{equation}
\scalebox{0.85}{$\displaystyle  \Bigl ( i_{M}(t), \,  v_{M}(t), \, i_{M}''(t) \Bigr )  \equiv \Bigl ( i(t), \ - i'(t) + A + r \sin ( \omega t), \, i''(t) \Bigr ), $}
\end{equation}
where $\displaystyle i''(t) \stackrel{\triangle}{=} \frac{d^{2} i(t)}{dt^{2}}$.  
We show the reconstructed three-dimensional attractors in Figure \ref{fig:Brusselator-reconstruction-2}.  
We can apply the above reconstruction methods to other examples in this paper.

%---Fig. 3-------%
\begin{figure}[hpbt]
 \centering
  \psfig{file=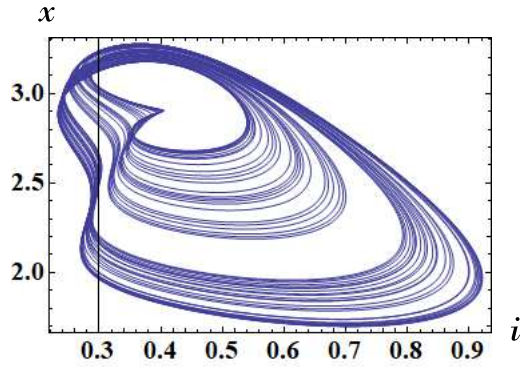, height=4.8cm}
  \caption{Chaotic attractor of the forced memristor Brusselator equations (\ref{eqn: Brusselator-4}).  
  Parameters: $A = 0.4, \ B = 1.2, \ r = 0.05, \ \omega = 0.81$. 
  Initial conditions: $i(0)=1.1, \  x(0)=1.1$.}
  \label{fig:Brusselator-attractor} 
\end{figure}
%
%

%---Fig. 4-------%
\begin{figure}[hpbt]
 \centering
  \psfig{file=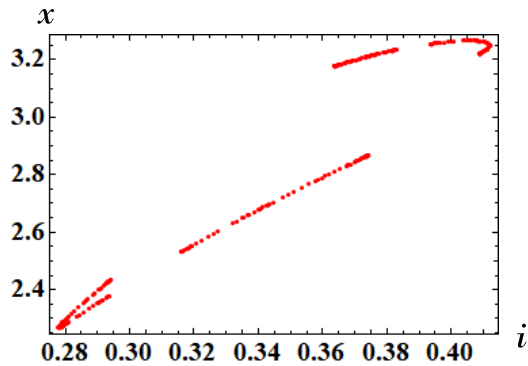, height=4.8cm}
  \caption{Poincar\'e map of the forced memristor Brusselator equations (\ref{eqn: Brusselator-4}). 
   Parameters: $A = 0.4, \ B = 1.2, \ r = 0.05, \ \omega = 0.81$.  
   Initial conditions: $i(0)=1.1, \  x(0)=1.1$.}
  \label{fig:Brusselator-poincare} 
\end{figure}
%
%

%---Fig. 5-------%
\begin{figure}[hpbt]
 \centering
   \begin{tabular}{cc}
   \psfig{file=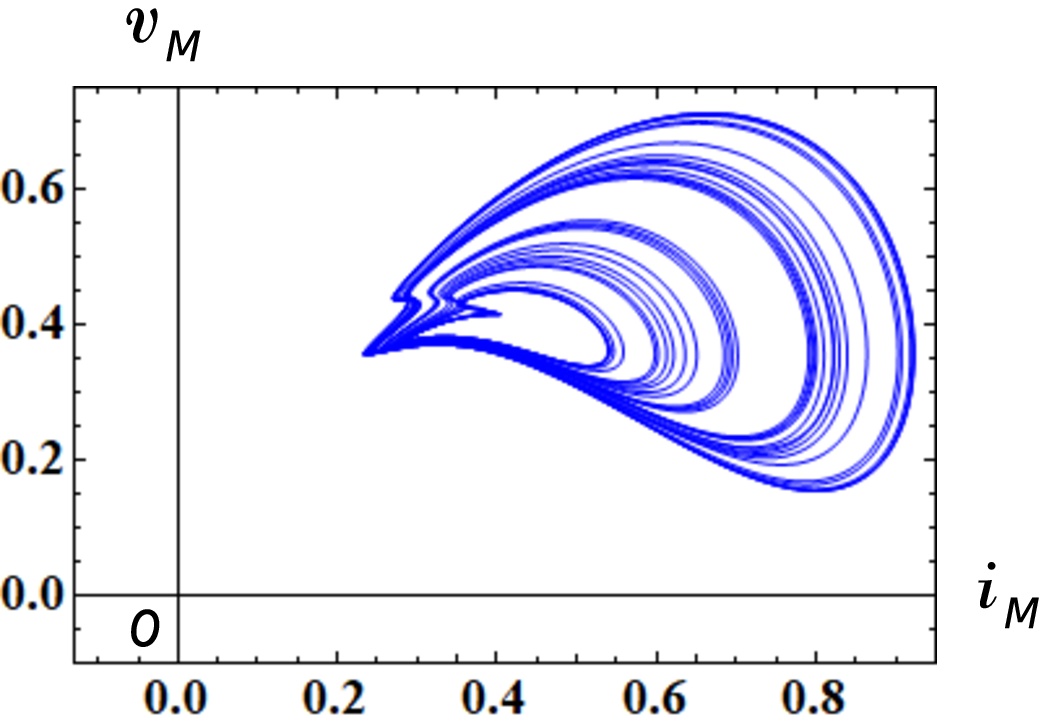, height=4.7cm} & 
   \psfig{file=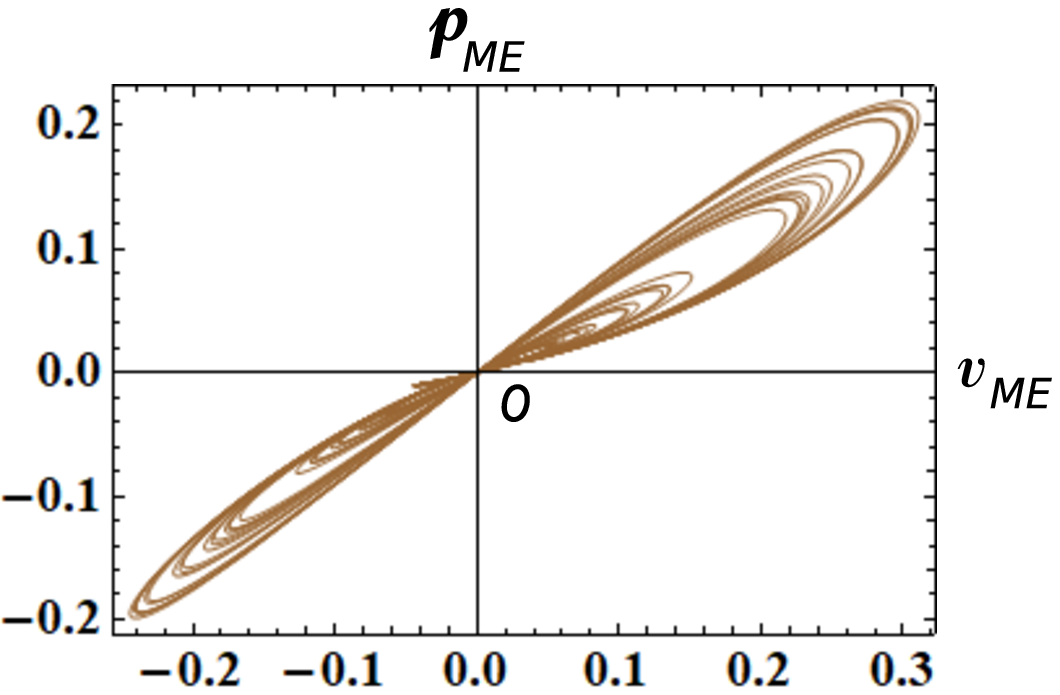, height=4.7cm} \vspace{1mm} \\
   (a) $i_{M}-v_{M}$ locus & (b) $i_{ME}-p_{ME}$ locus \\
   \end{tabular}
 \caption{The $i_{M}-v_{M}$ and $i_{ME}-p_{ME}$ loci of the forced memristor Brusselator equations (\ref{eqn: Brusselator-4}).  
   Here, $v_{M}$ and  $i_{M}$ denote the terminal voltage and the terminal current of the current-controlled extended memristor, 
   and $p_{ME}(t)$ are instantaneous powers defined by $p_{ME}(t)=i_{M}(t)\, v_{ME}(t)$.  
   Observe that the $v_{M}-i_{M}$ locus is \emph{not} pinched at the origin, and the locus lies in the first quadrant only. 
   That is, it moves in the \emph{passive} region. 
   However, the $v_{ME}-p_{ME}$ locus is pinched at the origin, and it lies in the first and the third quadrants.    
   Parameters: $A = 0.4, \ B = 1.2, \ r = 0.05, \ \omega = 0.81, \ d=0.7$.  
   Initial conditions: $i(0)=1.1, \  x(0)=1.1$.}
  \label{fig:Brusselator-pinch} 
\end{figure}
%
%

%---Fig. 6-------%
\begin{figure}[hpbt]
  \centering
   \begin{tabular}{cc}
   \psfig{file=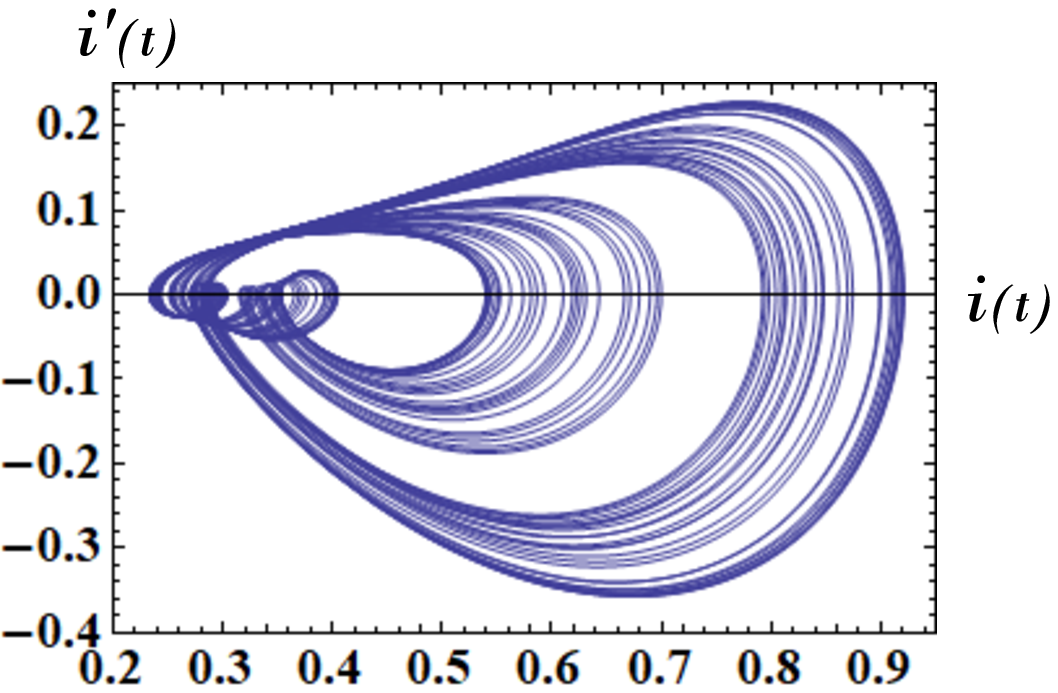, height=4.8cm} & 
   \psfig{file=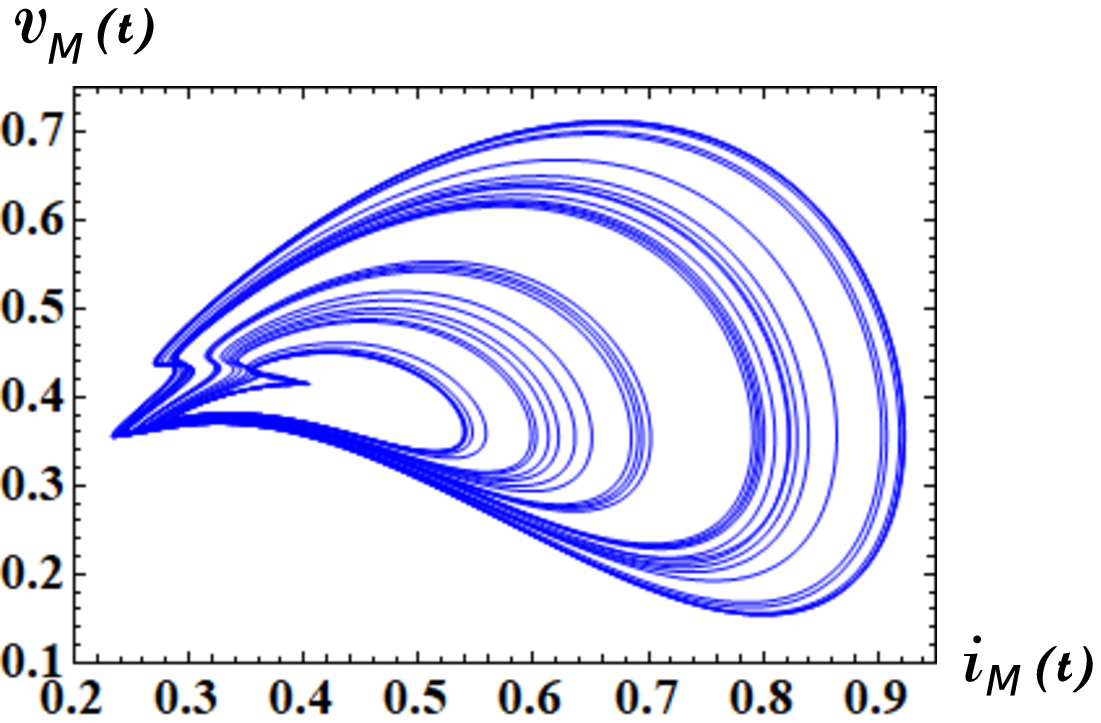,  height=4.8cm}  \\
   (a) $\bigl (i(t), \, i'(t) \bigr )$ reconstruction & (b) $\bigl ( i_{M}(t), \  v_{M}(t) \bigr )$ reconstruction \\
   \end{tabular}
  \caption{Reconstructed chaotic attractors using $\bigl ( i(t), \, i'(t) \bigr )$ and $\bigl ( i_{M}(t), \  v_{M}(t) \bigr )$,      
  where $v_{M}$ and  $i_{M}$ denote the terminal voltage and the terminal current of the current-controlled extended memristor. 
  Parameters: $A = 0.4, \ B = 1.2, \ r = 0.05, \ \omega = 0.81$. 
  Initial conditions: $i(0)=1.1, \  x(0)=1.1$.}
  \label{fig:Brusselator-reconstruction} 
\end{figure}
%
%

%---Fig. 7-------%
\begin{figure}[hpbt]
  \centering
   \begin{tabular}{cc}
   \psfig{file=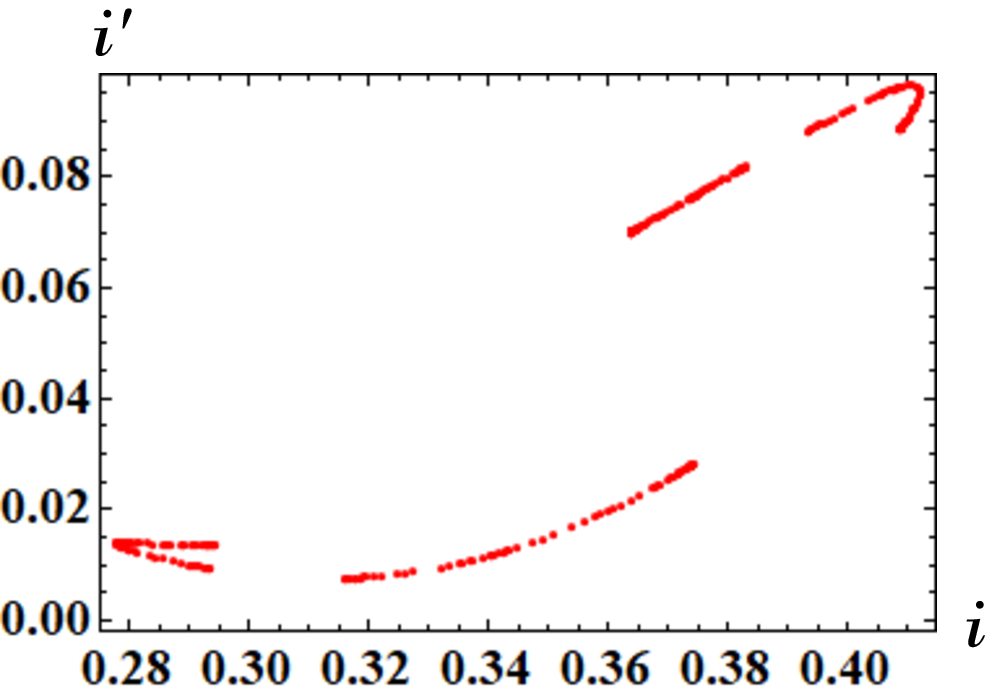, height=4.8cm} & 
   \psfig{file=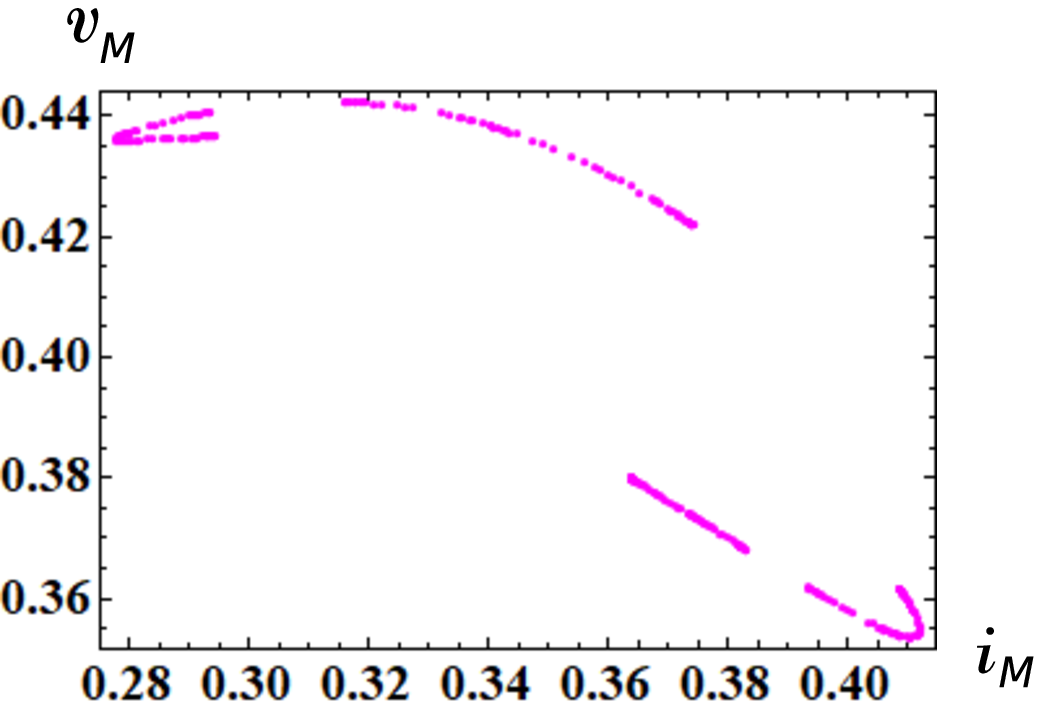, height=4.8cm} \\
   (a) Poincar\'e map for the attractor in Figure \ref{fig:Brusselator-reconstruction}(a) & 
   (b) Poincar\'e map for the attractor in Figure \ref{fig:Brusselator-reconstruction}(b)  \\
   \end{tabular}
  \caption{Poincar\'e maps for the reconstructed chaotic attractors in Figure \ref{fig:Brusselator-reconstruction}.  
  Observe that these two Poincar\'e maps are quite similar.  
  Parameters: $A = 0.4, \ B = 1.2, \ r = 0.05, \ \omega = 0.81$. 
  Initial conditions: $i(0)=1.1, \  x(0)=1.1$.}
  \label{fig:Brusselator-reconstruction-poincare} 
\end{figure}
%
%

%---Fig. 8-------%
\begin{figure}[hpbt]
  \centering
   \begin{tabular}{cc}
   \psfig{file=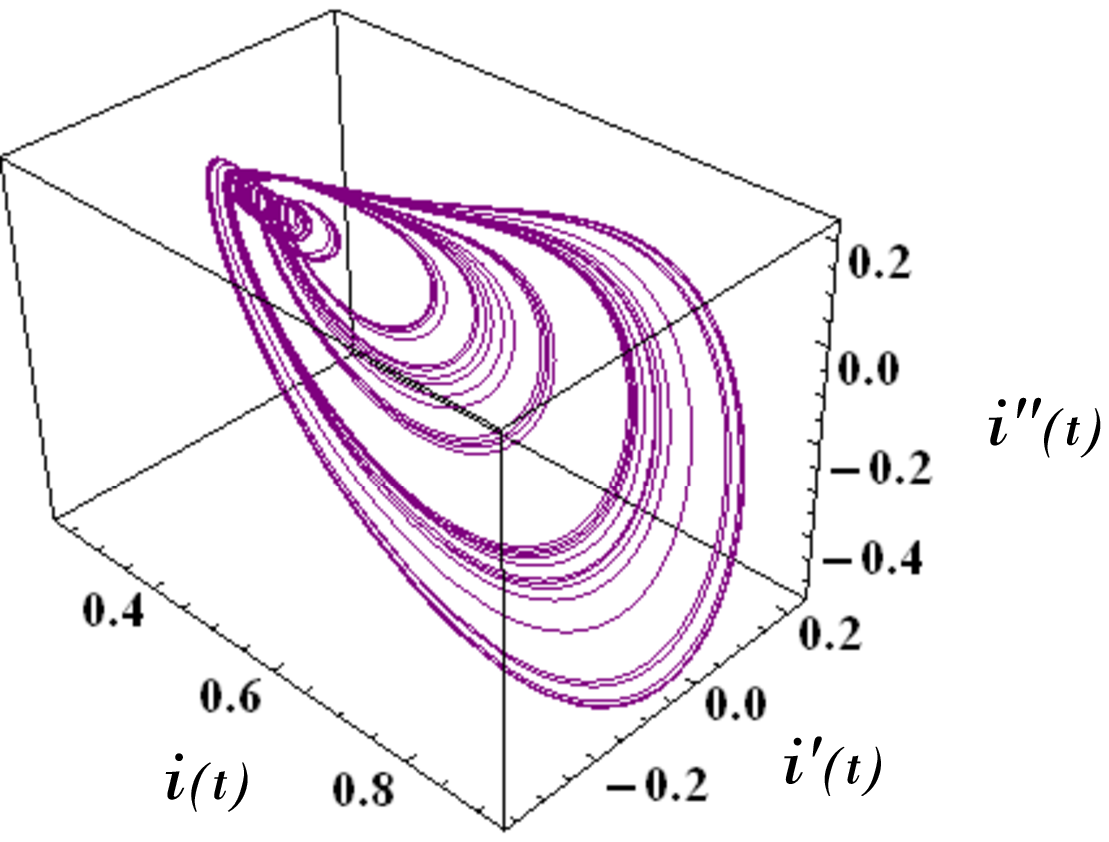,  height=4.8cm} & 
   \psfig{file=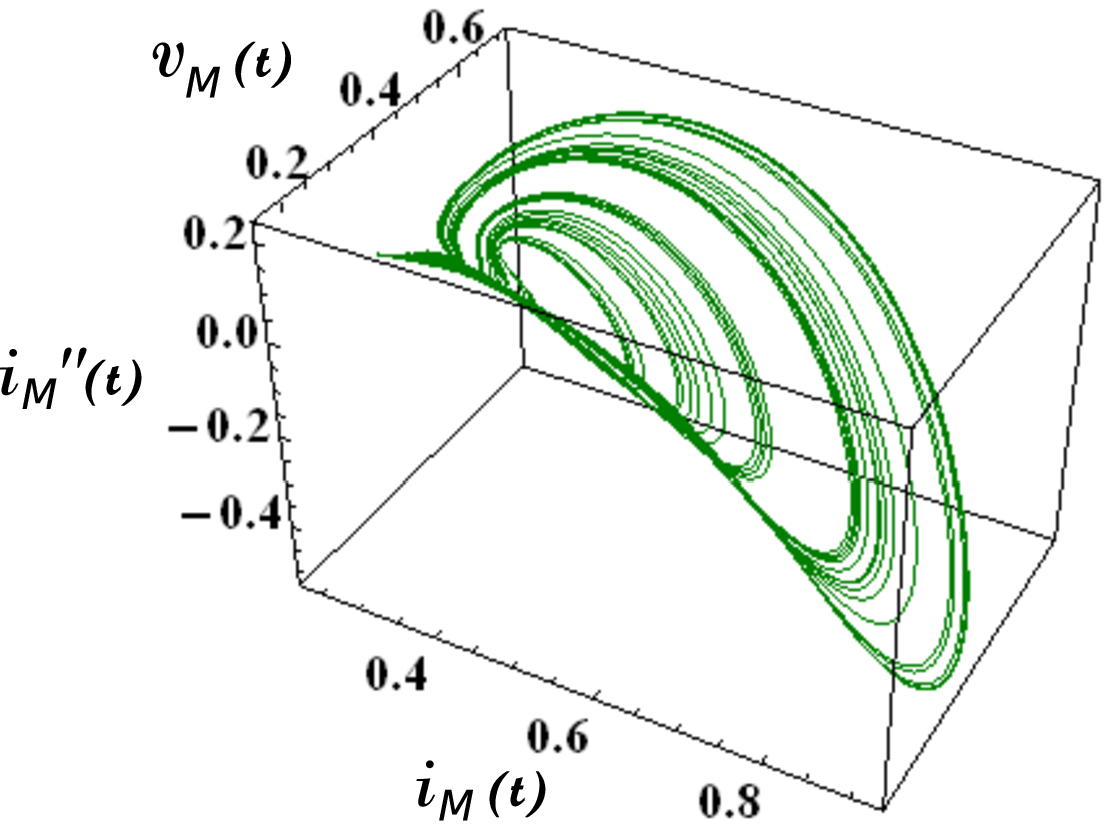,  height=4.8cm} \\
    (a) $\bigl ( i(t), \ i'(t), \ i''(t) \bigr )$ reconstruction &
    (b) $\bigl ( i_{M}(t), \  v_{M}(t), \ {i_{M}}''(t) \bigr )$ reconstruction \\
   \end{tabular}
  \caption{Reconstructed chaotic attractor using  $\bigl ( i(t), \ i'(t) \ i''(t) \bigr )$ and $\bigl ( i_{M}(t), \  v_{M}(t), \ {i_{M}}''(t) \bigr )$, 
  where $v_{M}$ and  $i_{M}$ denote the terminal voltage and the terminal current of the current-controlled extended memristor.   
  Parameters: $A = 0.4, \ B = 1.2, \ r = 0.05, \ \omega = 0.81$. 
  Initial conditions: $i(0)=1.1, \  x(0)=1.1$.}
  \label{fig:Brusselator-reconstruction-2} 
\end{figure}
%
%

%\clearpage
%==============================================================%
\subsection{Diffusion-less Gierer-Meinhardt equations}
%==============================================================%
%

Diffusion-less Gierer-Meinhardt equations \cite{Gierer(1972), Gonpot(2008), Li(2017)} is defined by 
\begin{center}
\begin{minipage}{8.7cm}
\begin{shadebox}
\underline{\emph{Diffusion-less Gierer-Meinhardt equations}}
\begin{equation}
\begin{array}{lll}
 \displaystyle \frac{d u}{dt} &=& \displaystyle \frac{u^{2}}{v}  - b \, u =  \left ( \frac{u}{v}  - b \right ) u, 
 \vspace{2mm} \\
 \displaystyle \frac{d v}{dt} &=& u^{2} - c \, v,
\end{array}
\vspace{2mm}
\label{eqn: Gierer-Meinhardt-1}
\end{equation}
where $b$ and $c$ are constants.
\end{shadebox}
\end{minipage}
\end{center}
Let us consider the three-element memristor circuit in Figure \ref{fig:memristor-inductor-battery}.  
The dynamics of this circuit given by Eq. (\ref{eqn: dynamics-n-1}).  
Assume that Eq. (\ref{eqn: dynamics-n-1}) satisfies 
\begin{equation}
 \begin{array}{ccc}
  E &=& 0, \vspace{2mm} \\
  \hat{R}(x, i) &=& \displaystyle - \left (\frac{i}{x}  - b \right ),        \vspace{2mm} \\
  \tilde{f}_{1}(x, i) &=& i^{2} - c \, x.  
 \end{array}
\end{equation}
Then we obtain 
\begin{center}
\begin{minipage}{9.5cm}
\begin{shadebox}
\underline{\emph{Memristor diffusion-less Gierer-Meinhardt equations}}
\begin{equation}
\begin{array}{lll}
 \displaystyle \frac{d i}{dt} &=& \displaystyle \left ( \frac{i}{x}  - b \right ) i, 
 \vspace{2mm} \\
 \displaystyle \frac{d x}{dt} &=& i^{2} - c \, x,
\end{array}
\vspace{2mm}
\label{eqn: Gierer-Meinhardt-2}
\end{equation}
where $b$ and $c$ are constants.
\end{shadebox}
\end{minipage}
\end{center}
Equations (\ref{eqn: Gierer-Meinhardt-1}) and (\ref{eqn: Gierer-Meinhardt-2}) are equivalent if we change the variables 
\begin{equation}
 i=u, \ x=v. 
\end{equation}
The terminal voltage $v_{M}$ and the terminal current $i_{M}$ of the current-controlled extended memristor in Figure \ref{fig:memristor-inductor-battery} are given by 
\begin{center}
\begin{minipage}{8.7cm}
\begin{shadebox}
\underline{\emph{V-I characteristics of the extended memristor}}
\begin{equation}
\begin{array}{lll}
  v_{M} &=& \displaystyle \hat{R}(x, \, i_{M}) \, i_{M} =  - \left (\frac{i_{M}}{x}  - b \right ) \, i_{M},   
  \vspace{3mm} \\
        && \hat{R}(x, \, 0) \ne \infty,
  \vspace{1mm} \\
 \displaystyle \frac{d x}{dt} &=& {i_{M}}^{2} - c \, x,
\end{array}
\label{eqn: Gierer-Meinhardt-3}
\end{equation}
where $\displaystyle \hat{R}(x, \, i_{M}) =  - \left (\frac{i_{M}}{x}  - b \right )$. \vspace{2mm}
\end{shadebox}
\end{minipage}
\end{center}
The above small-signal memristance $\hat{R}(x, \, i_{M})$ satisfies 
\begin{equation}
 \lim_{x \to 0} |\hat{R}(x, \, i_{M})| =  \left  | - \left (\frac{i_{M}}{x}  - b \right ) \right | = \infty, 
\end{equation}
when $i_{M}  \ne 0$.  
In order to avoid this singularity, we use the different time-scaling \cite{Andronov}.     
That is, after time scaling by $ d\tau = v \, dt$,  
Eqs. (\ref{eqn: Gierer-Meinhardt-1}), (\ref{eqn: Gierer-Meinhardt-2}), and (\ref{eqn: Gierer-Meinhardt-3}) assume the equivalent forms 
\begin{center}
\begin{minipage}{11.7cm}
\begin{shadebox}
\underline{\emph{Diffusion-less Gierer-Meinhardt equations with time scaling}}
\begin{equation}
\begin{array}{lll}
 \displaystyle \frac{d u}{d\tau} &=& \displaystyle (u  - b v) \, u, 
 \vspace{2mm} \\
 \displaystyle \frac{d v}{d\tau} &=& (u^{2} - c \, v) v,
\end{array}
\vspace{2mm}
\label{eqn: Gierer-Meinhardt-50}
\end{equation}
where $b$ and $c$ are constants, 
\end{shadebox}
\end{minipage}
\end{center}
\begin{center}
\begin{minipage}{11.7cm}
\begin{shadebox}
\underline{\emph{Memristor diffusion-less Gierer-Meinhardt equations with time scaling}}
\begin{equation}
\begin{array}{lll}
 \displaystyle \frac{d i}{dt} &=& \displaystyle  ( i   - b \, x ) i, 
 \vspace{2mm} \\
 \displaystyle \frac{d x}{dt} &=& (i^{2} - c \, x) x,
\end{array}
\vspace{2mm}
\label{eqn: Gierer-Meinhardt-51}
\end{equation}
where $b$ and $c$ are constants,
\end{shadebox}
\end{minipage}
\end{center}
and
\begin{center}
\begin{minipage}{11.7cm}
\begin{shadebox}
\underline{\emph{V-I characteristics of the extended memristor}}
\begin{equation}
\begin{array}{lll}
  v_{M} &=& \displaystyle \hat{R}(x, \, i_{M}) \, i_{M} =  - ( i_{M}  - b \, x  ) \, i_{M},   
  \vspace{3mm} \\
        && \hat{R}(x, \, 0) \ne \infty,
  \vspace{1mm} \\
 \displaystyle \frac{d x}{dt} &=& ({i_{M}}^{2} - c \, x) x,
\end{array}
\label{eqn: Gierer-Meinhardt-52}
\end{equation}
where $\displaystyle \hat{R}(x, \, i_{M}) =  - ( i_{M}  - b x  )$, 
\end{shadebox}
\end{minipage}
\end{center}
respectively.  
Similarly, Eq. (\ref{eqn: Gierer-Meinhardt-51}) can be realized by 
the three-element memristor circuit in Figure \ref{fig:memristor-inductor-battery}, where   
\begin{equation}
 \begin{array}{ccc}
  E &=& 0, \vspace{2mm} \\
  \hat{R}(x, i) &=& \displaystyle - ( i  - bx ),        \vspace{2mm} \\
  \tilde{f}_{1}(x, i) &=& (i^{2} - c \, x) \, x,
 \end{array}
\end{equation}
Note that the above time scaling maps orbits between systems (\ref{eqn: Gierer-Meinhardt-1}) and (\ref{eqn: Gierer-Meinhardt-50}) in a one-to-one manner except at the singularity $v = 0$, although it may not preserve the time orientation of orbits. 

Equations (\ref{eqn: Gierer-Meinhardt-1}) and (\ref{eqn: Gierer-Meinhardt-51}) exhibit periodic oscillation (limit cycle).  
When an external source is added as shown in Figure \ref{fig:memristive-inductor-battery-source}, 
the forced memristor circuit can exhibit chaotic oscillation.  
The dynamics of this circuit is given by 
\begin{center}
\begin{minipage}{13.0cm}
\begin{shadebox}
\underline{\emph{Forced memristor diffusion-less Gierer-Meinhardt equations with time scaling}}
\begin{equation}
\begin{array}{lll}
 \displaystyle \frac{d i}{dt} &=& \displaystyle  ( i   - b \, x ) i + r \sin ( \omega t), 
 \vspace{2mm} \\
 \displaystyle \frac{d x}{dt} &=& (i^{2} - c \, x) x,
\end{array}
\vspace{2mm}
\label{eqn: Gierer-Meinhardt-53}
\end{equation}
where $r$ and $\omega$ are constants.  
\end{shadebox}
\end{minipage}
\end{center}
We show the chaotic attractor, Poincar\'e map, and $i_{M}-v_{M}$ locus of Eq. (\ref{eqn: Gierer-Meinhardt-53}) in Figures \ref{fig:Gierer-attractor}, \ref{fig:Gierer-poincare}, and \ref{fig:Gierer-pinch}, respectively.   
The following parameters are used in our computer simulations:
\begin{equation}
  b = 0.65, \ c = 0.796, \ r = 0.2,  \ \omega = 0.5.  
\end{equation}
The $i_{M}-v_{M}$ locus in Figure \ref{fig:Gierer-pinch} lies in the first and the fourth quadrants. 
Thus, the extended memristor defined by Eq. (\ref{eqn: Gierer-Meinhardt-52}) is an active element.  
Let us next consider an instantaneous power defined by $p_{M}(t)=i_{M}(t)v_{M}(t)$. 
Then we obtain the $v_{M}-p_{M}$ locus in Figure \ref{fig:Gierer-power}.  
Observe that the locus is pinched at the origin, and the locus lies in the first and the third quadrants. 
Thus, when $v_{M}>0$, the instantaneous power delivered from the forced signal and the inductor is dissipated in the memristor.  
However, when $v_{M}<0$, the instantaneous power delivered from the forced signal and the inductor is not dissipated in the memristor. 
Note that the $v_{M}-p_{M}$ locus in Figure \ref{fig:Gierer-power} looks exactly like the $i_{M}-v_{M}$ locus of the ``\emph{passive}'' memristor, whose locus lies in the first and the third quadrants \cite{Chua(2012)}.  
Hence, the memristor switches between passive and active modes of operation, depending on its terminal voltage.  
We conclude as follow: 
\begin{center}
\begin{minipage}{.7\textwidth}
\begin{itembox}[l]{Switching behavior of the memristor}
Assume that Eq. (\ref{eqn: Gierer-Meinhardt-53}) exhibits chaotic oscillation.   
Then the extended memristor defined by Eq. (\ref{eqn: Gierer-Meinhardt-52}) can switch between ``passive'' and ``active'' modes of operation, depending on its terminal voltage.  
\end{itembox}
\end{minipage}
\end{center}
In order to obtain the results shown in Figures \ref{fig:Gierer-attractor}-\ref{fig:Gierer-power}, we have to choose the initial conditions carefully. 
It is due to the fact that a periodic orbit (drawn in magenta)\footnote{Without loss of generality, we can use the terminology \emph{``periodic orbit''} in order to describe a \emph{``periodic trajectory''} of the \emph{nonautonomous systems}, such as Eqs. (\ref{eqn: Brusselator-4}) and (\ref{eqn: Gierer-Meinhardt-53}) (see \emph{``Duffing's Equation''} in Sec. 2.2 of \cite{Holmes})). \label{orbit2}} coexists with a chaotic attractor (drawn in blue) as shown in Figure \ref{fig:Gierer-attractor-coexistence}.

As stated in Sec. \ref{sec: Brusselator}, we can reconstruct the chaotic attractor into two dimensional plane by using 
\begin{equation}
  (i(t), \ i'(t)).   
\end{equation}
Furthermore, the $i_{M}-v_{M}$ locus in Figure \ref{fig:Gierer-pinch} is considered to be the reconstruction of the chaotic attractor on the two-dimensional plane, since 
\begin{equation}
  \Bigl ( i_{M}(t), \,  v_{M}(t) \Bigr )  \equiv \Bigl ( i(t), \ - i'(t) + r \sin ( \omega t) \Bigr ), 
\end{equation}
where $i_{M}(t)=i(t)$.
We show their trajectories and Poincar\'e maps in Figures \ref{fig:Gierer-reconstruction} and \ref{fig:Gierer-reconstruction-poincare}, respectively.     
We can also reconstruct the chaotic attractor into the three-dimensional Euclidean space by using 
\begin{equation}
  (i(t), \ i'(t) \ i''(t)), 
\end{equation}
or 
\begin{equation}
\scalebox{0.9}{$\displaystyle  \Bigl ( i_{M}(t), \,  v_{M}(t), \, i_{M}''(t) \Bigr )  \equiv \Bigl ( i(t), \ - i'(t) + r \sin ( \omega t), \, i''(t) \Bigr ). $}
\end{equation}
We show the reconstructed three-dimensional attractors in Figure \ref{fig:Gierer-reconstruction-2}.

%---Fig. 9-------%
\begin{figure}[hpbt]
 \begin{center}
  \psfig{file=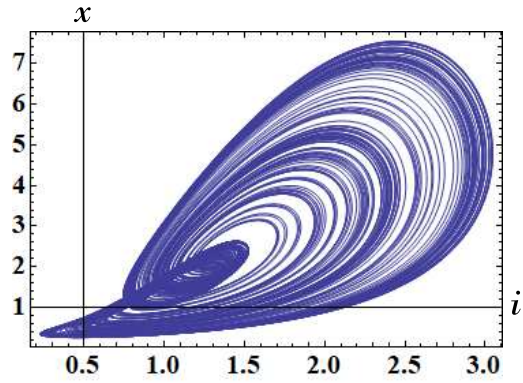, height=5.0cm} 
  \caption{Chaotic attractor of the forced memristor diffusion-less Gierer-Meinhardt equations (\ref{eqn: Gierer-Meinhardt-53}) 
  with time scaling.  
  Parameters: $b = 0.65, \ c = 0.796, \ r = 0.2,  \ \omega = 0.5$.    
  Initial conditions: $i(0)=0.5, \  x(0)=0.5$.}
  \label{fig:Gierer-attractor} 
 \end{center}
\end{figure}
%
%

%---Fig. 10-------%
\begin{figure}[hpbt]
 \begin{center}
  \psfig{file=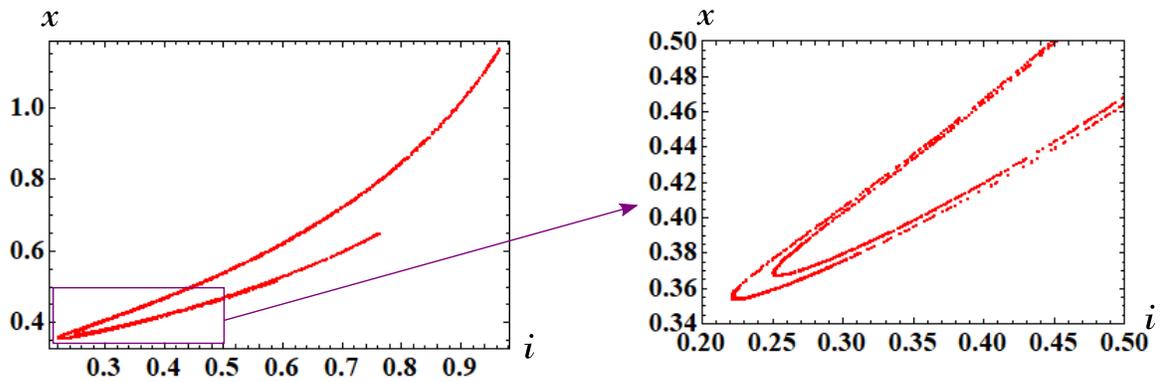, height=5.0cm}
  \caption{Poincar\'e map of the forced memristor diffusion-less Gierer-Meinhardt equations(\ref{eqn: Gierer-Meinhardt-53}) 
   with time scaling.   
   The partially enlarged view of the locus is shown on the right side of Figure \ref{fig:Gierer-poincare}.
   Observe the folding action of the chaotic attractor. 
   Parameters: $b = 0.65, \ c = 0.796, \ r = 0.2,  \ \omega = 0.5$.  
   Initial conditions: $i(0)=0.5, \  x(0)=0.5$.}
  \label{fig:Gierer-poincare} 
 \end{center}
\end{figure}
%
%

%---Fig. 11-------%
\begin{figure}[hpbt]
 \begin{center}
  \psfig{file=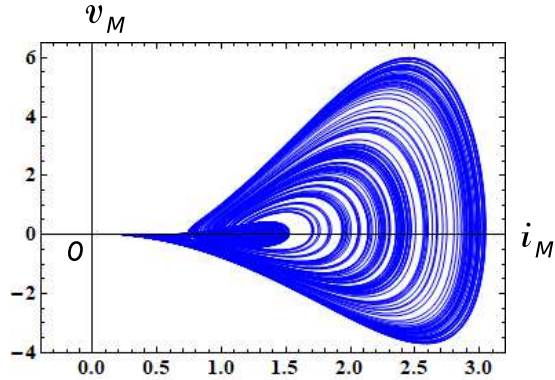, height=5.0cm}
  \caption{ The $i_{M}-v_{M}$ locus of the forced memristor diffusion-less  Gierer-Meinhardt equations (\ref{eqn: Gierer-Meinhardt-53}) 
   with time scaling.   
   Here, $v_{M}$ and  $i_{M}$ denote the terminal voltage and the terminal current of the current-controlled extended memristor.  
   Observe that the extended memristor defined by Eq. (\ref{eqn: Gierer-Meinhardt-52}) is an active element.  
   Parameters: $b = 0.65, \ c = 0.796, \ r = 0.2,  \ \omega = 0.5$.   
   Initial conditions: $i(0)=0.5, \  x(0)=0.5$.}
  \label{fig:Gierer-pinch} 
 \end{center}
\end{figure}
%
%

%---Fig. 12-------%
\begin{figure}[hpbt]
 \begin{center}
  \psfig{file=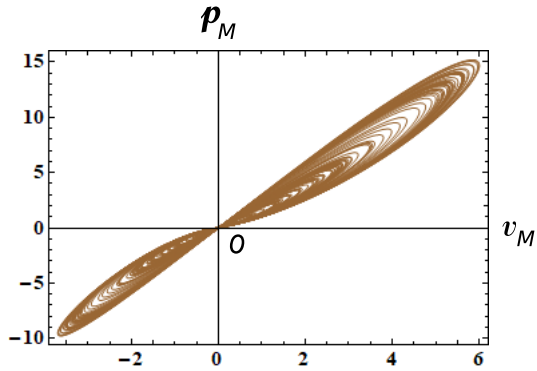, width=7cm}
  \caption{ The $v_{M}-p_{M}$ locus of the forced memristor diffusion-less Gierer-Meinhardt equations (\ref{eqn: Gierer-Meinhardt-53}) 
   with time scaling.   
   Here, $p_{M}(t)$ is an instantaneous power defined by $p_{M}(t)=i_{M}(t)v_{M}(t)$,
   and $v_{M}(t)$ and $i_{M}(t)$ denote the terminal voltage and the terminal current of the current-controlled extended memristor.   
   Observe that the $v_{M}-p_{M}$ locus is pinched at the origin, and the locus lies in the first and the third quadrants.  
   The memristor switches between passive and active modes of operation, depending on its terminal voltage $v_{M}(t)$. 
   Parameters: $b = 0.65, \ c = 0.796, \ r = 0.2,  \ \omega = 0.5$.   
   Initial conditions: $i(0)=0.5, \  x(0)=0.5$.}  
  \label{fig:Gierer-power} 
 \end{center}
\end{figure}
%
%

%---Fig. 13-------%
\begin{figure}[hpbt]
 \begin{center}
  \psfig{file=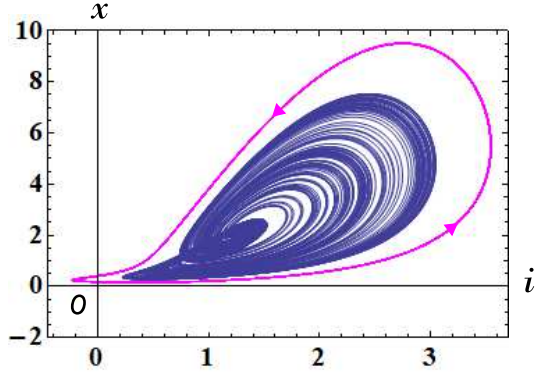, width=7cm} 
  \caption{A periodic orbit (magenta) coexists with a chaotic attractor (blue).
  Parameters: $b = 0.65, \ c = 0.796, \ r = 0.2,  \ \omega = 0.5$.    
  Initial conditions for a chaotic attractor: $i(0)=0.5, \  x(0)=0.5$.  \ \ 
  Initial conditions for a periodic orbit: $i(0)=0.5, \  x(0)=0.1$.}
  \label{fig:Gierer-attractor-coexistence} 
 \end{center}
\end{figure}
%
%

%---Fig. 14-------%
\begin{figure}[hpbt]
  \centering
   \begin{tabular}{cc}
   \psfig{file=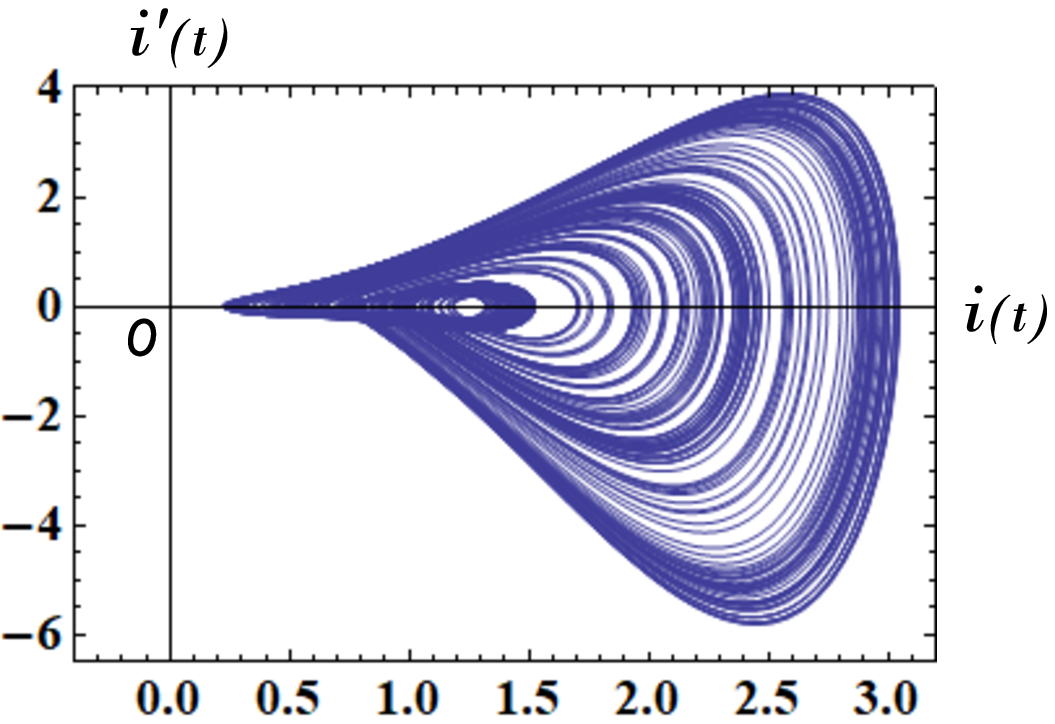, height=4.8cm} & 
   \psfig{file=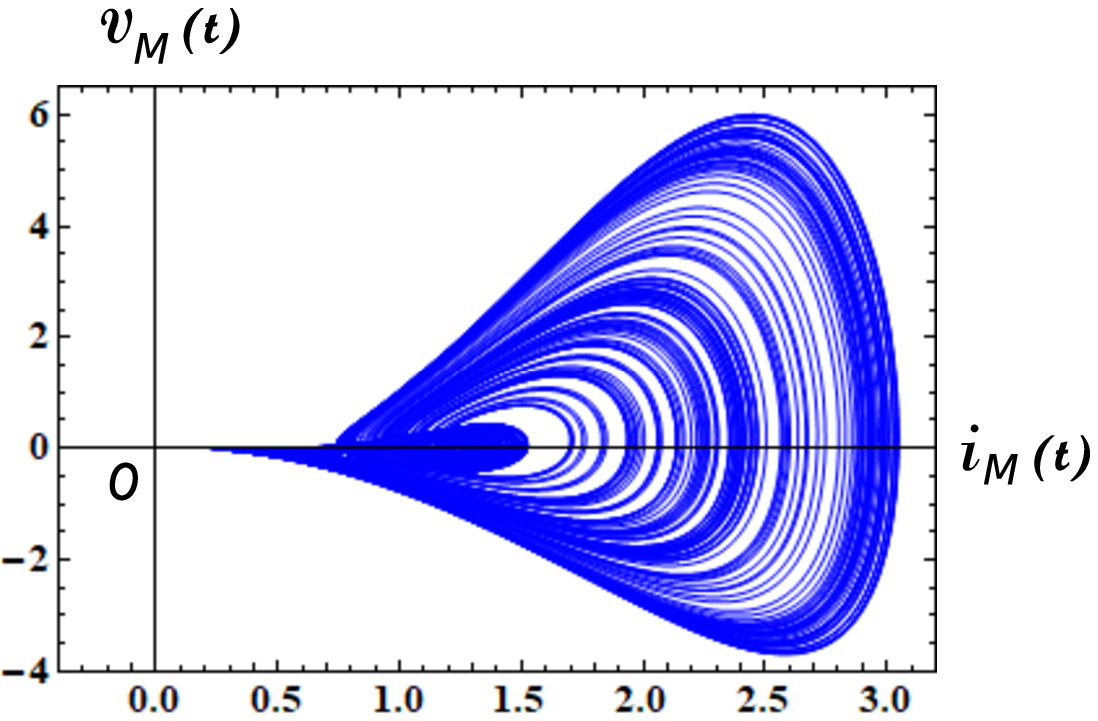,  height=4.8cm}  \\
   (a) $\bigl (i(t), \, i'(t) \bigr )$ reconstruction & (b) $\bigl ( i_{M}(t), \  v_{M}(t) \bigr )$ reconstruction \\
   \end{tabular}
  \caption{Reconstructed chaotic attractors using $\bigl ( i(t), \, i'(t) \bigr )$ and $\bigl ( i_{M}(t), \  v_{M}(t) \bigr )$,      
  where $v_{M}$ and  $i_{M}$ denote the terminal voltage and the terminal current of the current-controlled extended memristor. 
   Parameters: $b = 0.65, \ c = 0.796, \ r = 0.2,  \ \omega = 0.5$.  
   Initial conditions: $i(0)=0.5, \  x(0)=0.5$.}
  \label{fig:Gierer-reconstruction} 
\end{figure}
%
%

%---Fig. 15-------%
\begin{figure}[hpbt]
  \centering
   \begin{tabular}{cc}
   \psfig{file=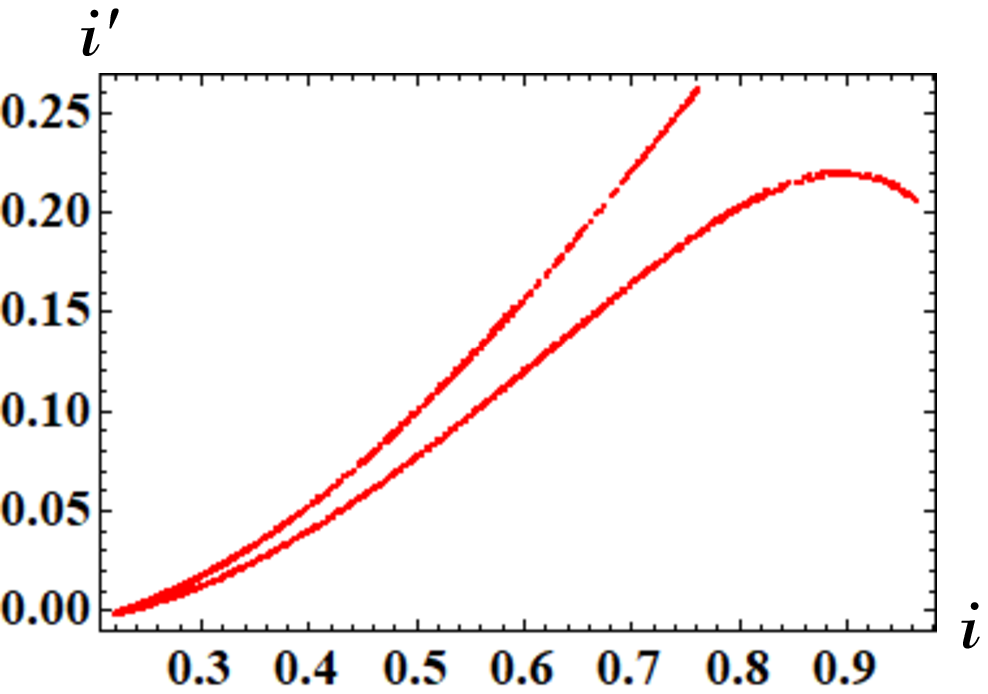, height=4.8cm} & 
   \psfig{file=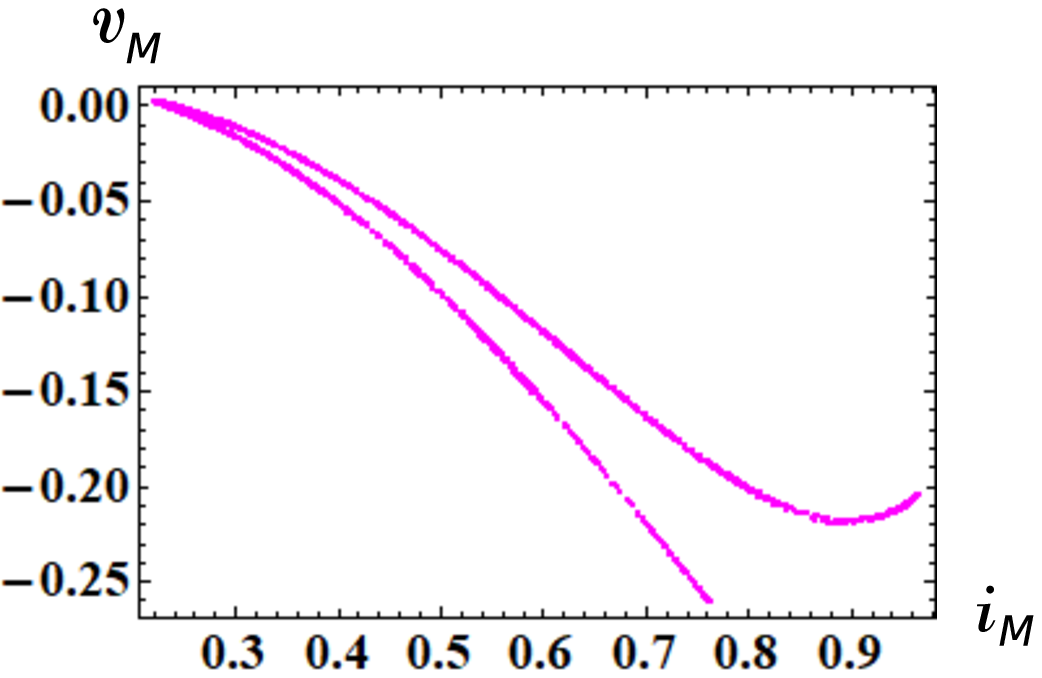, height=4.8cm} \\
   (a) Poincar\'e map for the attractor in Figure \ref{fig:Gierer-reconstruction}(a) & 
   (b) Poincar\'e map for the attractor in Figure \ref{fig:Gierer-reconstruction}(b)  \\
   \end{tabular}
  \caption{Poincar\'e maps for the reconstructed chaotic attractors in Figure \ref{fig:Gierer-reconstruction}.  
  Observe that these two Poincar\'e maps are quite similar.  
   Parameters: $b = 0.65, \ c = 0.796, \ r = 0.2,  \ \omega = 0.5$.  
   Initial conditions: $i(0)=0.5, \  x(0)=0.5$.}
  \label{fig:Gierer-reconstruction-poincare} 
\end{figure}
%
%

%---Fig. 16-------%
\begin{figure}[hpbt]
  \centering
   \begin{tabular}{cc}
   \psfig{file=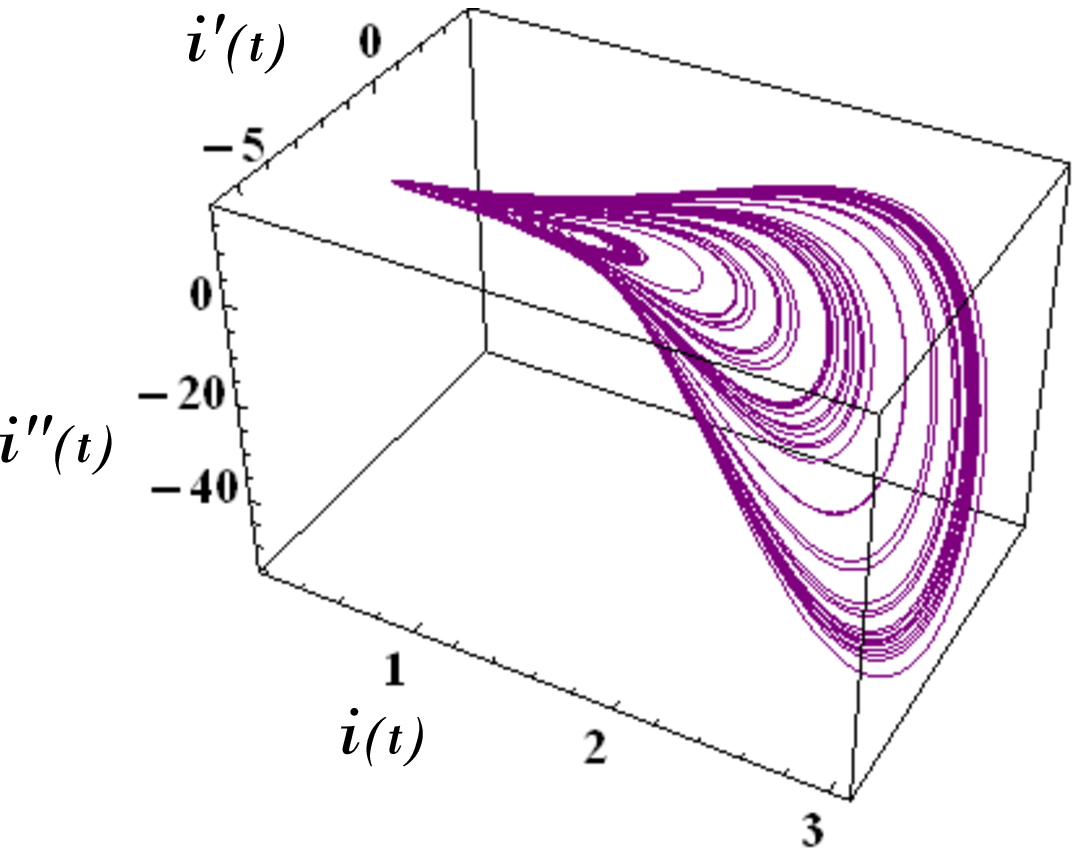,  height=5cm} & 
   \psfig{file=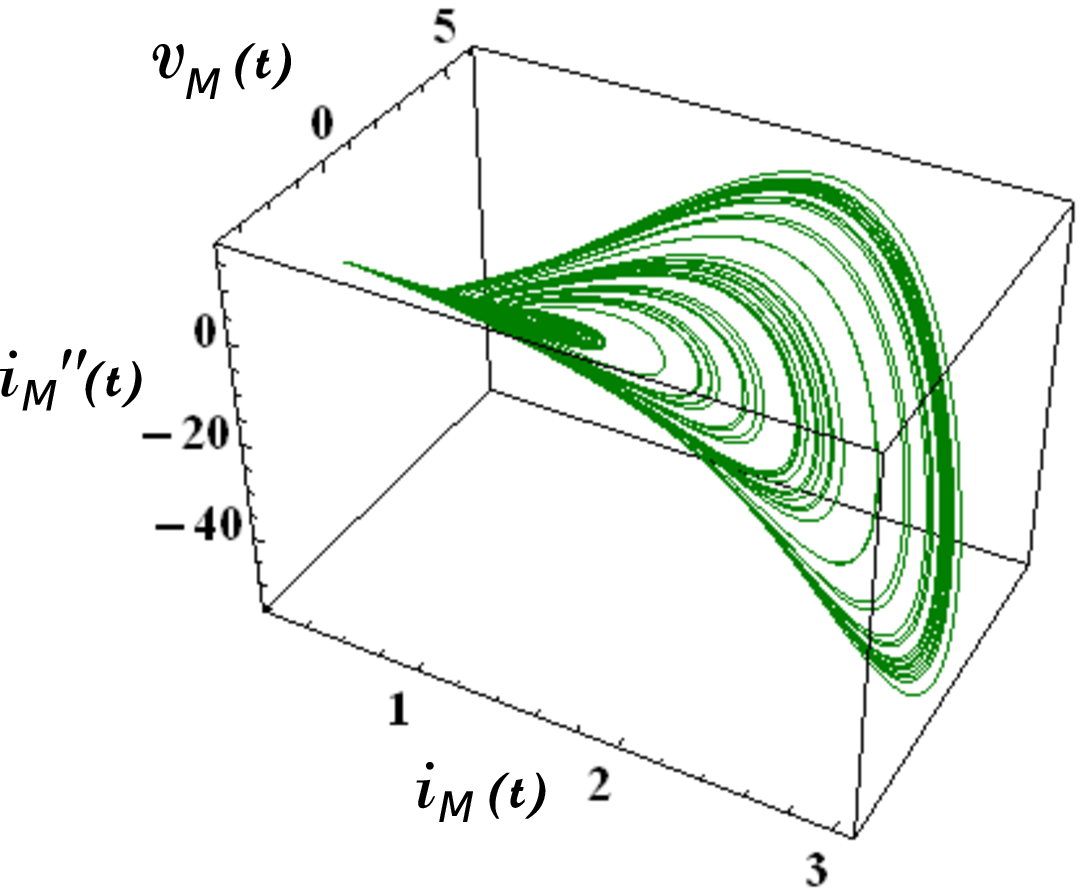,  height=5cm} \\
    (a) $\bigl ( i(t), \ i'(t), \ i''(t) \bigr )$ reconstruction &
    (b) $\bigl ( i_{M}(t), \  v_{M}(t), \ {i_{M}}''(t) \bigr )$ reconstruction \\
   \end{tabular}
  \caption{Reconstructed chaotic attractor using  
   $\bigl ( i(t), \ i'(t) \ i''(t) \bigr )$ and $\bigl ( i_{M}(t), \  v_{M}(t), \ {i_{M}}''(t) \bigr )$, 
   where $v_{M}$ and  $i_{M}$ denote the terminal voltage and the terminal current of the current-controlled extended memristor.   
   Parameters: $b = 0.65, \ c = 0.796, \ r = 0.2,  \ \omega = 0.5$.  
   Initial conditions: $i(0)=0.5, \  x(0)=0.5$.}
  \label{fig:Gierer-reconstruction-2} 
\end{figure}
%
%

%\clearpage
%==============================================================%
\subsection{Tyson-Kauffman equations}
%==============================================================%
%
The dynamics of the Tyson-Kauffman equations \cite{Tyson(1975)} can be described by 
\begin{center}
\begin{minipage}{8.7cm}
\begin{shadebox}
\underline{\emph{Tyson-Kauffman equations}}
\begin{equation}
\begin{array}{lll}
 \displaystyle \frac{d u}{dt} &=& \displaystyle A - B \, u - u \, v^{2} = A - ( B + v^{2} ) \, u , 
 \vspace{2mm} \\
 \displaystyle \frac{d v}{dt} &=& B \, u + u \, v^{2} - v,
\end{array}
\vspace{2mm}
\label{eqn: Tyson-Kauffman-1}
\end{equation}
where $A$ and $B$ are constants.  
\end{shadebox}
\end{minipage}
\end{center}
Consider the three-element memristor circuit in Figure \ref{fig:memristor-inductor-battery}.  
The dynamics of this circuit given by Eq. (\ref{eqn: dynamics-n-1}).  
Assume that Eq. (\ref{eqn: dynamics-n-1}) satisfies 
\begin{equation}
 \begin{array}{ccc}
  E &=& A, \vspace{2mm} \\
  \hat{R}(x, \, i) &=& ( B + x^{2} ),        \vspace{2mm} \\
  \tilde{f}_{1}(x, i) &=& B \, i + i \, x^{2} - x.
 \end{array}
\end{equation}
Then we obtain 
\begin{center}
\begin{minipage}{8.7cm}
\begin{shadebox}
\underline{\emph{Memristor Tyson-Kauffman equations}}
\begin{equation}
\begin{array}{lll}
 \displaystyle \frac{d i}{dt} &=& A - ( B + x^{2} ) \, i, 
 \vspace{2mm} \\
 \displaystyle \frac{d x}{dt} &=& B \, i + i \, x^{2} - x,
\end{array}
\vspace{2mm}
\label{eqn: Tyson-Kauffman-2}
\end{equation}
where $A$ and $B$ are constants
\end{shadebox}
\end{minipage}
\end{center}
Equations (\ref{eqn: Tyson-Kauffman-1}) and (\ref{eqn: Tyson-Kauffman-2}) are equivalent if we change the variables  
\begin{equation}
  i=u, \ x=v.  
\end{equation}
In this case, the extended memristor in Figure \ref{fig:memristor-inductor-battery} is replaced by the generic memristor (see Appendix A).
That is,
\begin{equation}
  \hat{R}(x, i) = \tilde{R}(x)=( B + x^{2} ).   
\end{equation}
The terminal voltage $v_{M}$ and the terminal current $i_{M}$ of the current-controlled generic memristor are described
by
\begin{center}
\begin{minipage}{8.7cm}
\begin{shadebox}
\underline{\emph{V-I characteristics of the generic memristor}}
\begin{equation}
\begin{array}{lll}
  v_{M} &=& \tilde{R}(x) \, i_{M} =  ( B + x^{2} ) \, i_{M},   
  \vspace{1mm} \\
 \displaystyle \frac{d x}{dt} &=& Bi_{M} - {i_{M}}^{2} - x,
\end{array}
\label{eqn: Tyson-Kauffman-3}
\end{equation}
where $\tilde{R}(x) =  ( B + x^{2} )$. \vspace{2mm}
\end{shadebox}
\end{minipage}
\end{center}
It follows that the Tyson-Kauffman equations (\ref{eqn: Tyson-Kauffman-1}) can be realized by 
the three-element memristor circuit in Figure \ref{fig:memristor-inductor-battery}.  
Equations (\ref{eqn: Tyson-Kauffman-1}) and (\ref{eqn: Tyson-Kauffman-2}) can exhibit periodic oscillation (limit cycle). 
When an external source is added as shown in Figure \ref{fig:memristive-inductor-battery-source}, 
the forced memristor Tyson-Kauffman equations can exhibit chaotic oscillation.   
The dynamics of the circuit is given by 
\begin{center}
\begin{minipage}{8.7cm}
\begin{shadebox}
\underline{\emph{Forced memristor Tyson-Kauffman equations}}
\begin{equation}
\begin{array}{lll}
 \displaystyle \frac{d i}{dt} &=& A - ( B + x^{2} ) \, i + r \sin ( \omega t), 
 \vspace{2mm} \\
 \displaystyle \frac{d x}{dt} &=& B \, i + i \, x^{2} - x,
\end{array}
\vspace{2mm}
\label{eqn: Tyson-Kauffman-4}
\end{equation}
where $r$ and $\omega$ are constants.  
\end{shadebox}
\end{minipage}
\end{center}
We show the chaotic attractors, Poincar\'e maps, and $i_{M}-v_{M}$ loci of Eq. (\ref{eqn: Tyson-Kauffman-4}) in Figures \ref{fig:Tyson-attractor}, \ref{fig:Tyson-poincare}, and \ref{fig:Tyson-pinch-2}, respectively.   
In our computer simulations, we used the following two kinds of the parameters:
\begin{equation}
\left.
\begin{array}{cccc}
 (a) &  A = 0.5, & B = 0.00803, &  C = 0.01, \vspace{1mm} \\ 
     & r = 0.5, & \omega = 0.5, &
\end{array}
\right \}
\end{equation}
and 
\begin{equation}
\left.
\begin{array}{cccc}
  (b) & A = 0.5, & B = 0.0079, &  C = 0.01, \vspace{1mm} \\ 
      & r = 0.55,  & \omega = 0.5. &
\end{array}
\right \}
\end{equation}
Note that the locus in Figure \ref{fig:Tyson-pinch-2}(a) moves in the first quadrant, 
and the locus in Figure \ref{fig:Tyson-pinch-2}(b) moves in the first and third quadrants.  
That is, they move in the \emph{passive} region, since the instantaneous power defined by 
\begin{equation}
  p_{M}(t) \stackrel{\triangle}{=} i_{M}(t) \, v_{M}(t), 
\end{equation}
is \emph{not} negative.  
In this case, the power is dissipated in the generic memristor, which is delivered from the forcing signal and the inductor. 

Let us define next the instantaneous power of the two circuit elements, as stated in Sec. \ref{sec: Brusselator}.   
That is, we define the instantaneous power of the extended memristor and the battery by 
\begin{equation}
  p_{ME}(t) \stackrel{\triangle}{=} i_{M}(t)\, v_{ME}(t), 
\end{equation}
where $v_{ME}(t) = v_{M}(t)-E$, and $E$ denotes the voltage of the battery. 
That is, $v_{ME}(t)$ denotes the voltage across the extended memristor and the battery.  
We show the $v_{ME}-p_{ME}$ locus in Figure \ref{fig:Tyson-power}.  
Observe that the locus is pinched at the origin, and it lies in the first and the third quadrants.  
Thus, the instantaneous power $p_{ME}(t)$ delivered from the forced signal and the inductor is dissipated when $v_{M}(t) - E > 0$.   
However, the instantaneous power $p_{ME}(t)$ is \emph{not} dissipated when $v_{M}(t) - E <0 $. 
Thus, we conclude as follow: \\
\begin{center}
\begin{minipage}{.8\textwidth}
\begin{itembox}[l]{Behavior of the generic memristor}
Assume that Eq. (\ref{eqn: Tyson-Kauffman-4}) exhibits chaotic oscillation.  
Then, we obtain the following results: 
\begin{enumerate}
\item
The generic memristor defined by Eq. (\ref{eqn: Tyson-Kauffman-3}) is operated as a \emph{passive} element.  
If we define the instantaneous power of the generic memristor by 
$p_{M}(t) \stackrel{\triangle}{=} i_{M}(t) v_{M}(t)$, 
then $p_{M}(t)$ is dissipated in this generic memristor, which is delivered from the forcing signal and the inductor.  \\
\item 
If we define the instantaneous power of the two elements, that is, the instantaneous power of the generic memristor and the battery, by $p_{ME}(t) \stackrel{\triangle}{=} i_{M}(t)(v_{M}(t)-E)$,  
then $p_{ME}(t)$ is \emph{not} dissipated when $v_{M}(t) - E <0$. 
However, $p_{ME}(t)$ is dissipated when $v_{M}(t) - E > 0$.       
\end{enumerate}
\end{itembox} 
\end{minipage}
\end{center}

As stated in Sec. \ref{sec: Brusselator}, we can reconstruct the chaotic attractor into two dimensional plane by using 
\begin{equation}
  (i(t), \ i'(t)).   
\end{equation}
Furthermore, the $i_{M}-v_{M}$ loci in Figure \ref{fig:Tyson-pinch-2} are considered to be the reconstruction of the chaotic attractor on the two-dimensional plane, since 
\begin{equation}
  \Bigl ( i_{M}(t), \,  v_{M}(t) \Bigr )  \equiv \Bigl ( i(t), \ - i'(t) + r \sin ( \omega t) \Bigr ), 
\end{equation}
where $i_{M}(t)=i(t)$.
We show their trajectories and Poincar\'e maps in Figures \ref{fig:Tyson-reconstruction} and \ref{fig:Tyson-reconstruction-poincare}, respectively.     
We can also reconstruct the chaotic attractor into the three-dimensional Euclidean space by using 
\begin{equation}
  (i(t), \ i'(t) \ i''(t)), 
\end{equation}
or 
\begin{equation}
\scalebox{0.9}{$\displaystyle  \Bigl ( i_{M}(t), \,  v_{M}(t), \, i_{M}''(t) \Bigr )  \equiv \Bigl ( i(t), \ - i'(t) + r \sin ( \omega t), \, i''(t) \Bigr ). $}
\end{equation}
We show the reconstructed three-dimensional attractors in Figure \ref{fig:Tyson-reconstruction-2}.

%---Fig. 17-------%
\begin{figure}[hpbt]
 \centering
   \begin{tabular}{cc}
    \psfig{file=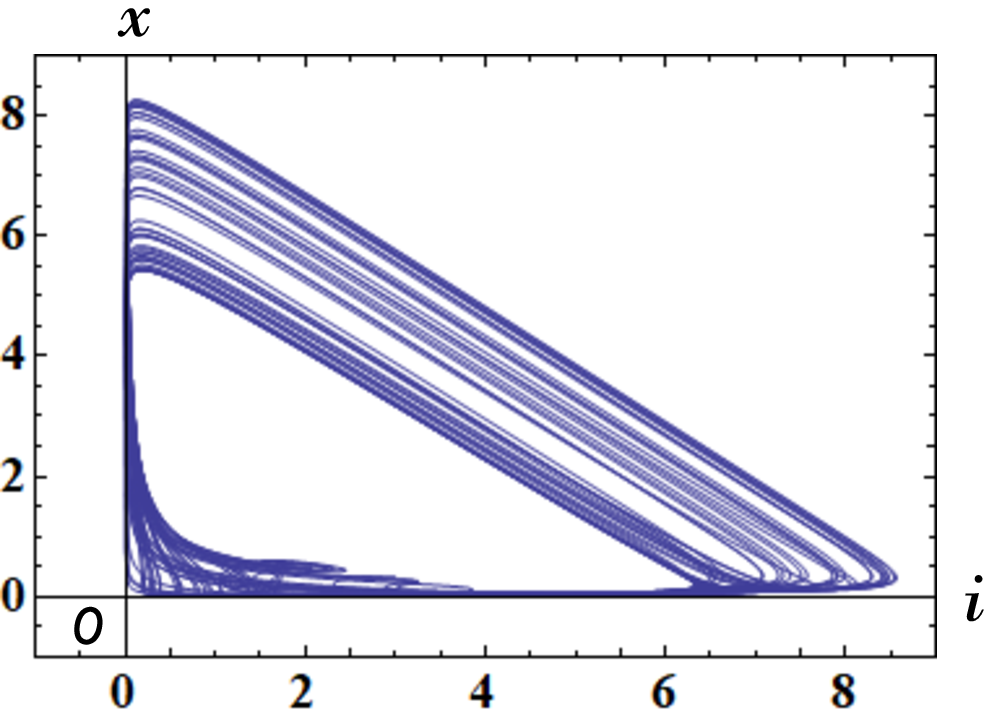, height=5cm}  & 
    \psfig{file=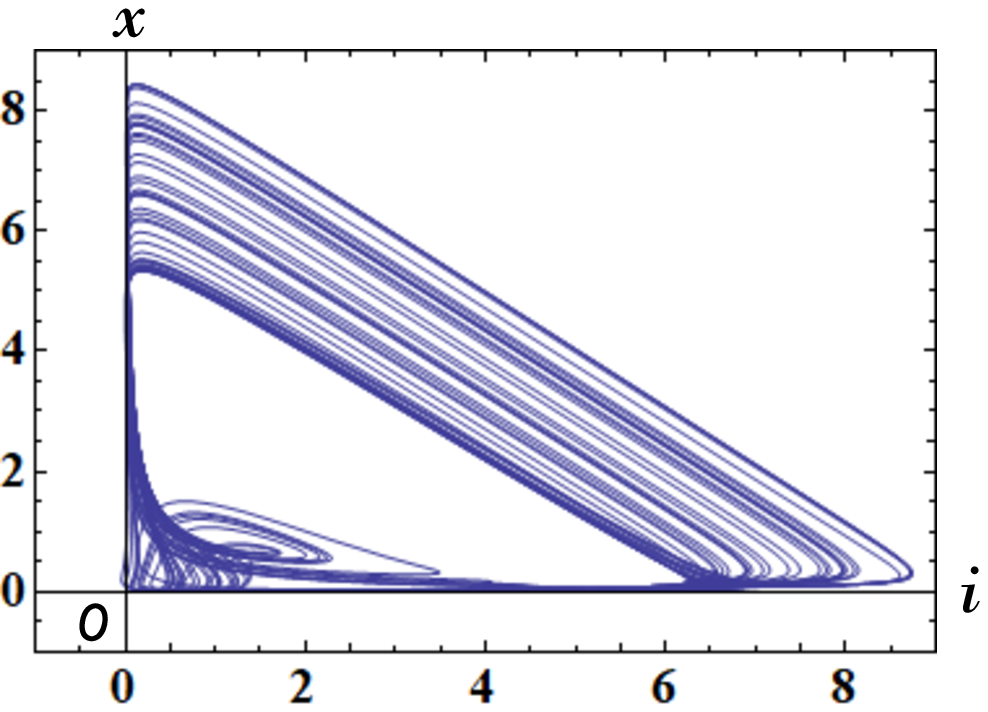, height=5cm}  \vspace{1mm} \\
   (a)  & (b)  \\ 
   \end{tabular}
  \caption{Chaotic attractors of the forced memristor Tyson-Kauffman equations (\ref{eqn: Tyson-Kauffman-4}).  
  Parameters: (a) $A = 0.5, \ B = 0.00803,  \ r = 0.5,  \ \omega = 0.5$. \ \
              (b) $A = 0.5, \ B = 0.0079, \ r = 0.55,  \ \omega = 0.5$.               
  Initial conditions: $i(0)=2.1, \  x(0)=2.1$.}
  \label{fig:Tyson-attractor} 
\end{figure}
%
%

%---Fig. 18-------%
\begin{figure}[hpbt]
 \centering
   \begin{tabular}{l}
    \psfig{file=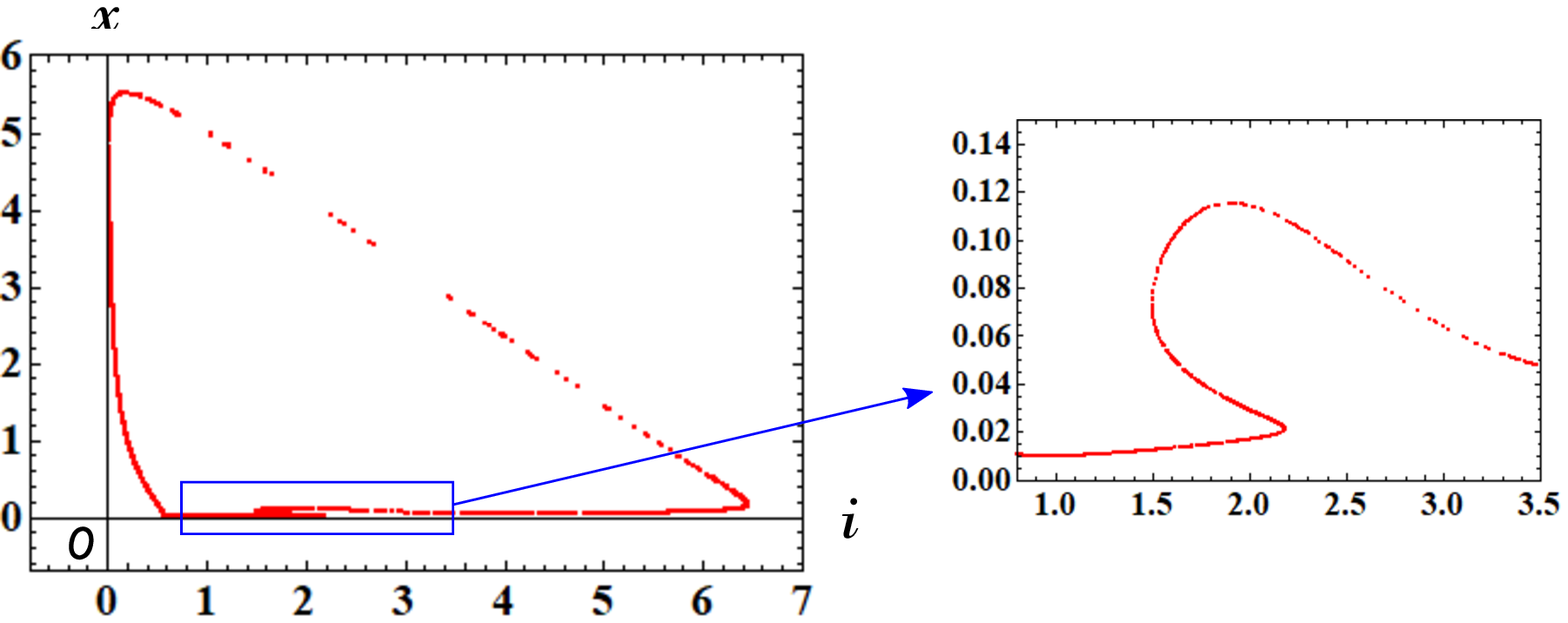, height=5cm}  \\
    (a) \vspace{1mm} \\
    \psfig{file=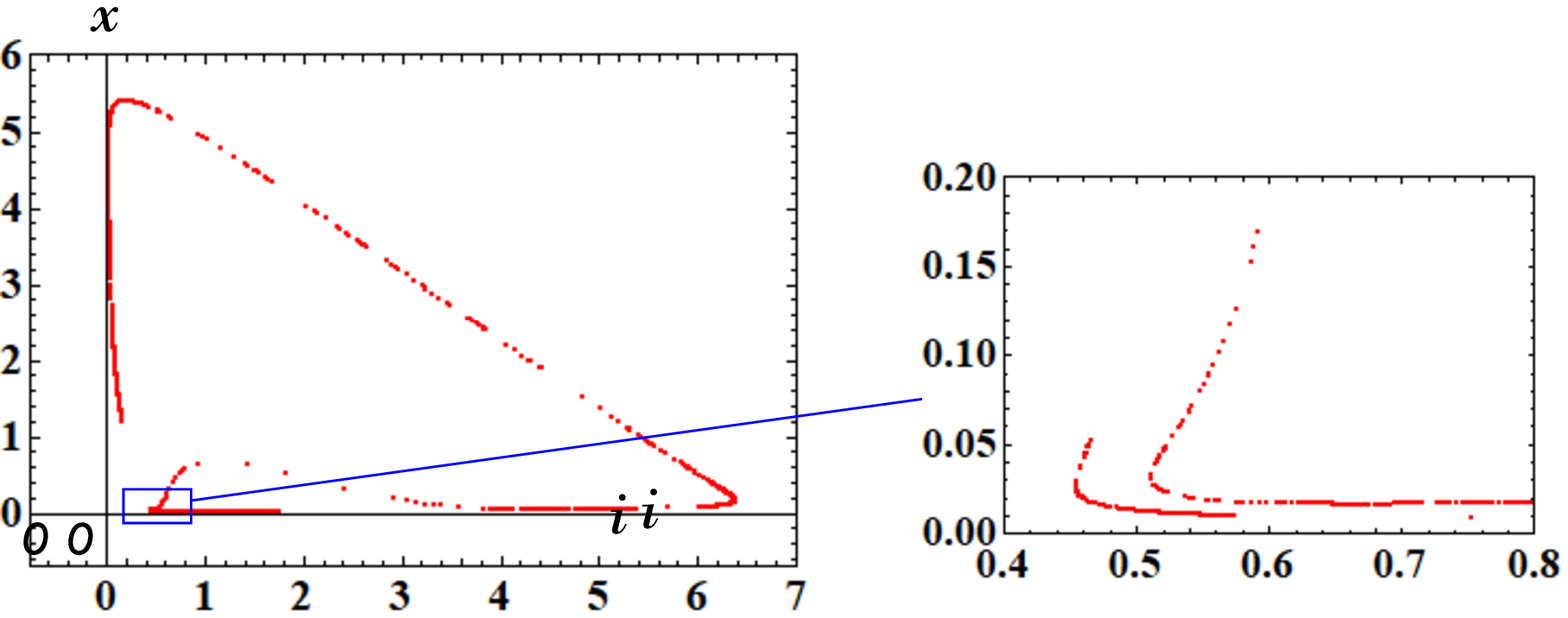, height=5cm}  \\
    (b)  
   \end{tabular}
  \caption{Poincar\'e maps of the forced memristor Tyson-Kauffman equations (\ref{eqn: Tyson-Kauffman-4}).  
  The partially enlarged view of the locus is shown on the right side of Figure\ref{fig:Tyson-poincare}.
  Parameters: (a) $A = 0.5, \ B = 0.00803,  \ r = 0.5,  \ \omega = 0.5$. \ \
              (b) $A = 0.5, \ B = 0.0079, \ r = 0.55,  \ \omega = 0.5$.               
  Initial conditions: $i(0)=2.1, \  x(0)=2.1$.}
  \label{fig:Tyson-poincare} 
\end{figure}
%
%

%---Fig. 19-------%
\begin{figure}[hpbt]
 \centering
   \begin{tabular}{c}
    \psfig{file=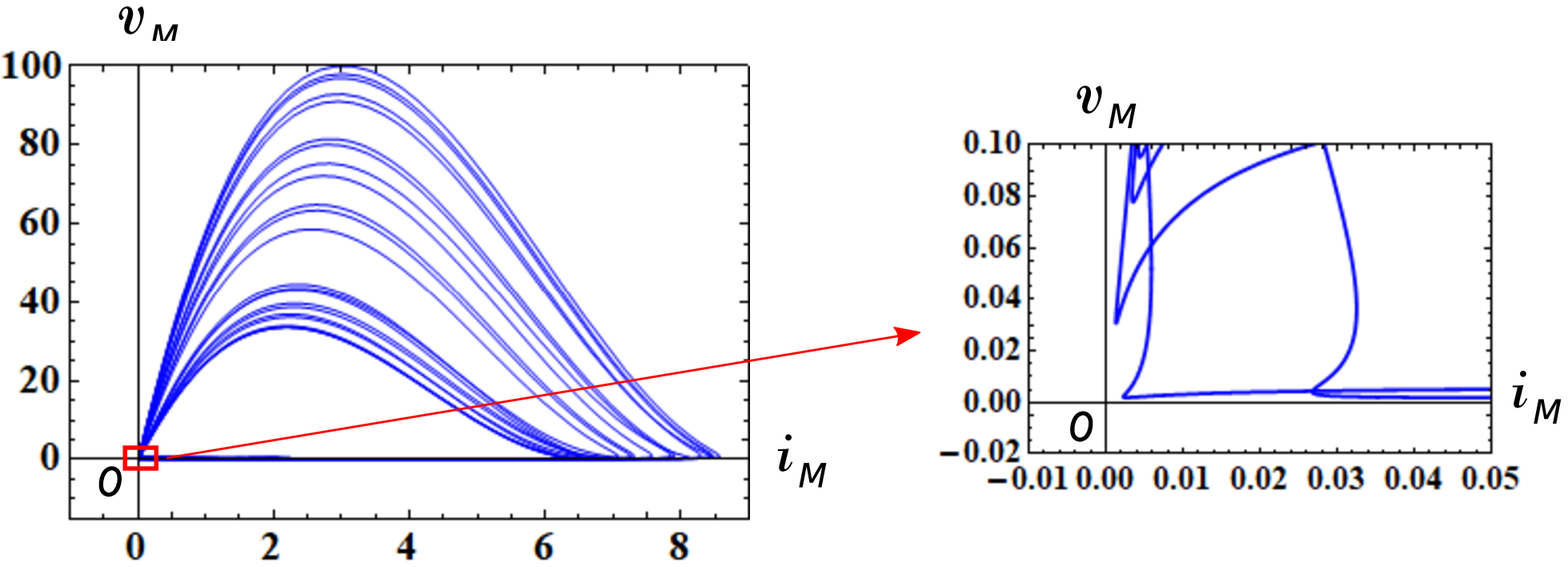, width=14cm}  \\
     (a)  \vspace{2mm} \\
    \psfig{file=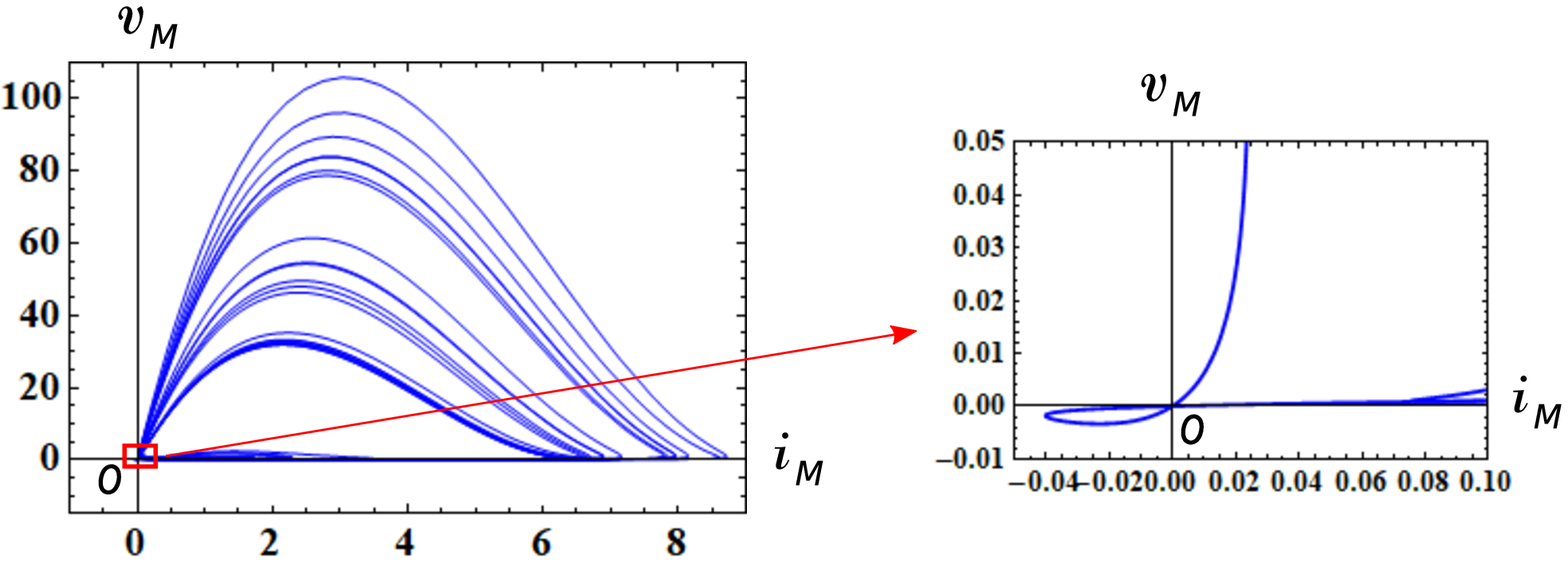, width=14cm}   \\
     (b)  
   \end{tabular}
  \caption{The $i_{M}-v_{M}$ loci of the forced memristor Tyson-Kauffman equations (\ref{eqn: Tyson-Kauffman-4}) (left) and   
   partially enlarged views of the $i_{M}-v_{M}$ loci (right).  
   Here, $v_{M}$ and  $i_{M}$ denote the terminal voltage and the terminal current of the current-controlled generic memristor. 
   Observe that the locus in Figure \ref{fig:Tyson-pinch-2}(a) moves in the first quadrant, 
   and the locus in Figure \ref{fig:Tyson-pinch-2}(b) moves in the first and third quadrants.  
   Thus, they move in the \emph{passive} region.  
   Parameters: (a) $A = 0.5, \ B = 0.00803,  \ r = 0.5,  \ \omega = 0.5$. \ \
               (b) $A = 0.5, \ B = 0.0079, \ r = 0.55,  \ \omega = 0.5$.               
   Initial conditions: $i(0)=2.1, \  x(0)=2.1$.}
  \label{fig:Tyson-pinch-2} 
\end{figure}
%
%

%---Fig. 20-------%
\begin{figure}[hpbt]
 \centering
   \begin{tabular}{c}
   \psfig{file=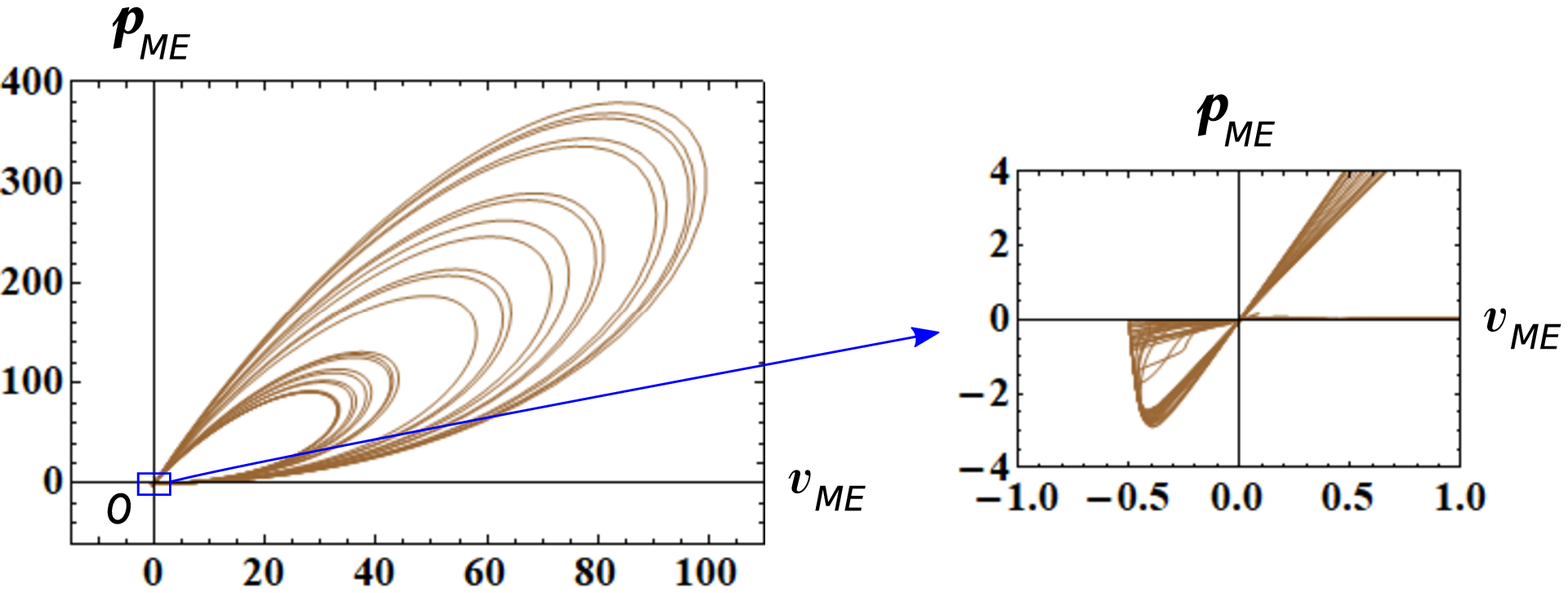, height=5cm} \\ 
   (a)  \\
   \psfig{file=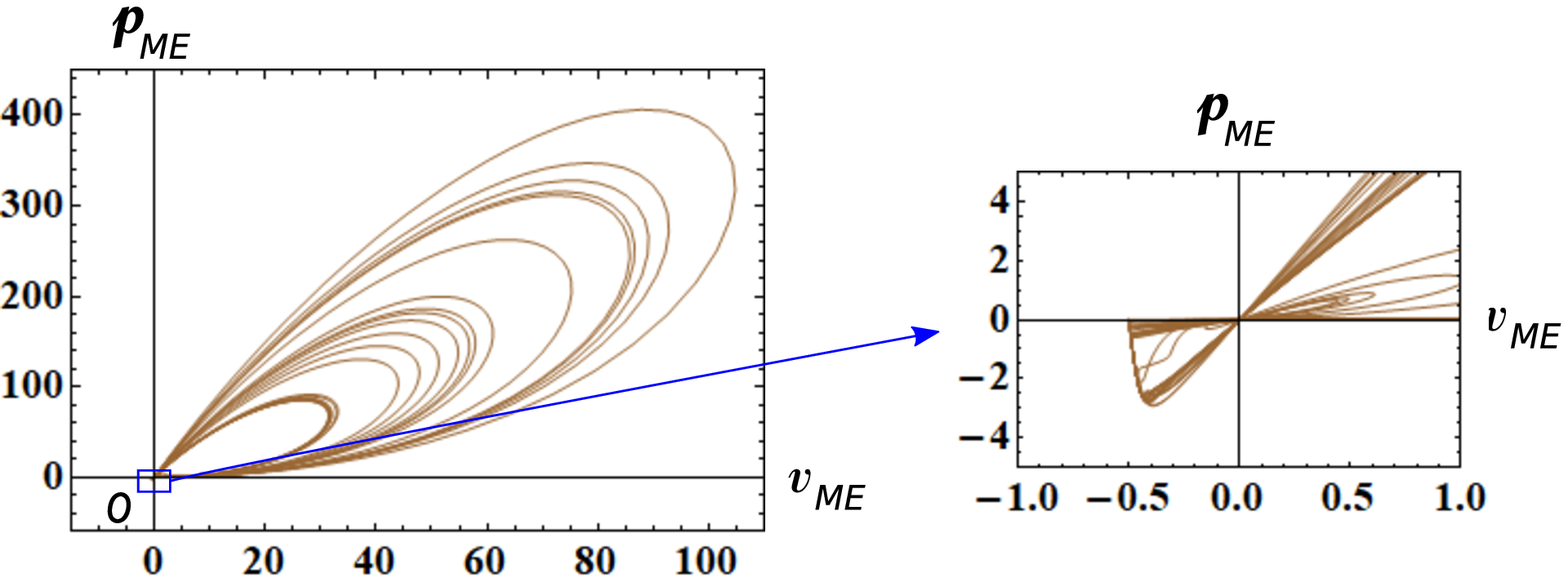, height=5cm} \\
   (b) 
   \end{tabular}  
  \caption{The $v_{ME}-p_{ME}$ loci of the forced memristor Tyson-Kauffman equations (\ref{eqn: Tyson-Kauffman-4}).  
   Their partially enlarged views are shown on the right side of Figure \ref{fig:Tyson-power}.  
   Here, 
   $v_{ME}(t) = v_{M}(t)- E$, and $p_{ME}(t)$ is an instantaneous powers defined by $p_{ME}(t)=i_{M}(t)\, v_{ME}(t)$, 
   $v_{M}$ and $i_{M}$ denote the terminal voltage and the terminal current of the current-controlled generic memristor, respectively,  
   and $E$ denotes the voltage of the battery. 
   Observe that the $v_{ME}-p_{ME}$ loci are pinched at the origin, and they lie in the first and the third quadrants.  
   Parameters: (a) $A = 0.5, \ B = 0.00803,  \ r = 0.5,  \ \omega = 0.5$. \ \
               (b) $A = 0.5, \ B = 0.0079, \ r = 0.55,  \ \omega = 0.5$.   
   Initial conditions: $i(0)=2.1, \  x(0)=2.1$.}
  \label{fig:Tyson-power} 
\end{figure}
%
%

%---Fig. 21-------%
\begin{figure}[hpbt]
  \centering
   \begin{tabular}{cc}
   \psfig{file=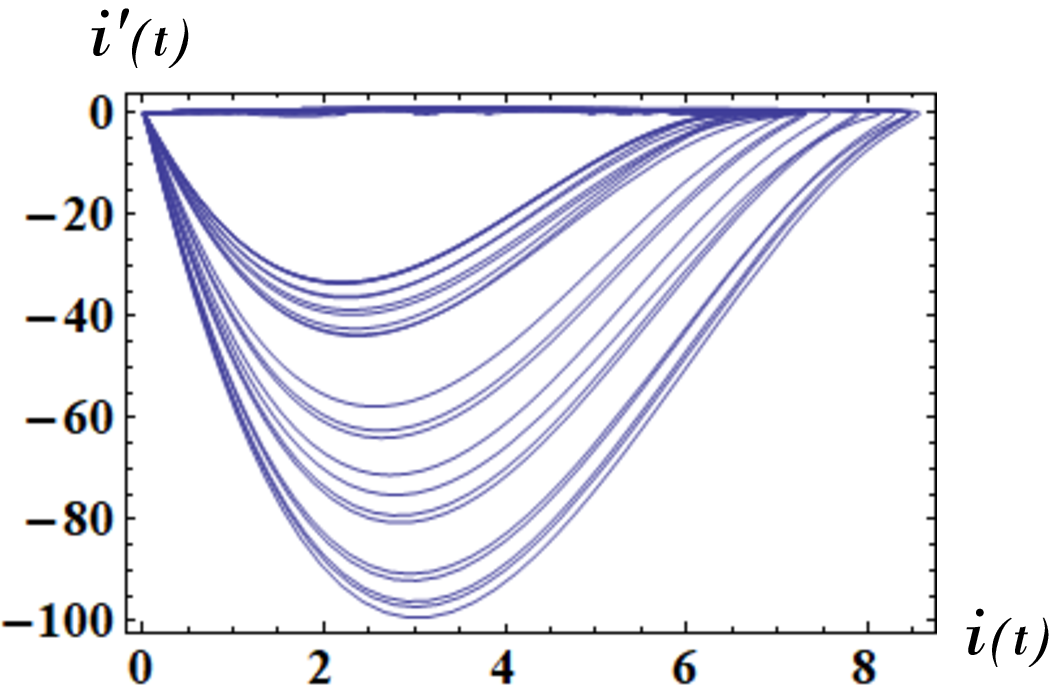, height=4.8cm} & 
   \psfig{file=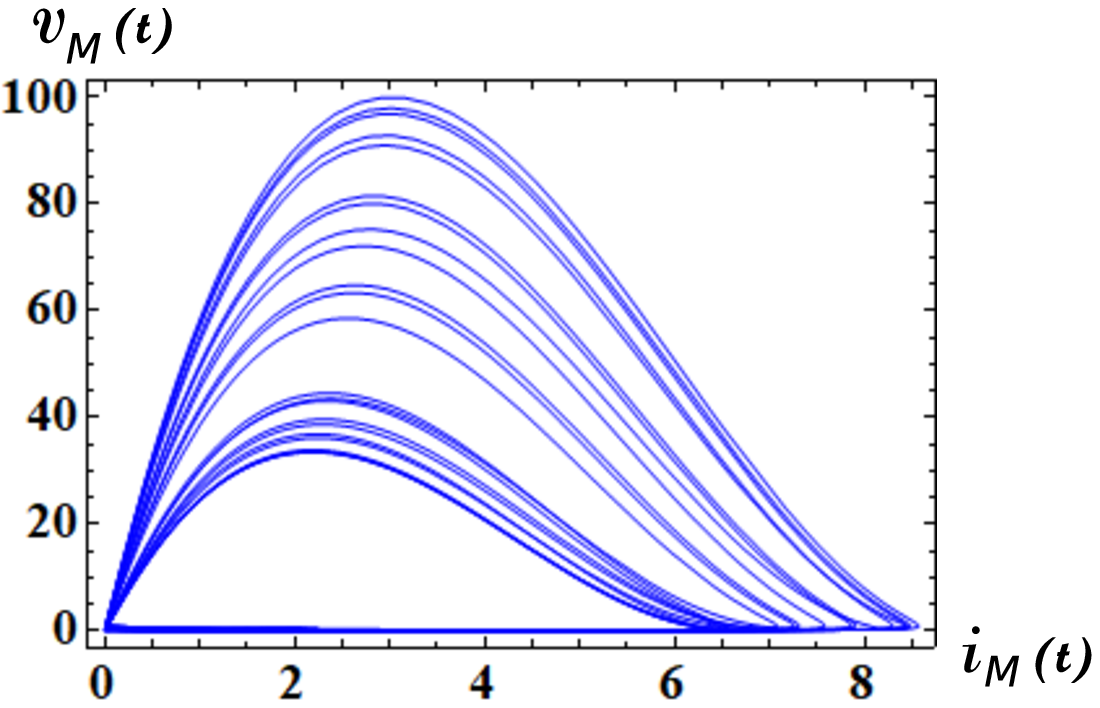,  height=4.8cm}  \\
   (a) $\bigl (i(t), \, i'(t) \bigr )$ reconstruction & (b) $\bigl ( i_{M}(t), \  v_{M}(t) \bigr )$ reconstruction \\
   \end{tabular}
  \caption{Reconstructed chaotic attractors using $\bigl ( i(t), \, i'(t) \bigr )$ and $\bigl ( i_{M}(t), \  v_{M}(t) \bigr )$,      
  where $v_{M}$ and  $i_{M}$ denote the terminal voltage and the terminal current of the current-controlled generic memristor. 
   Parameters:  $A = 0.5, \ B = 0.00803,  \ r = 0.5,  \ \omega = 0.5$. \
   Initial conditions: $i(0)=2.1, \  x(0)=2.1$.}
  \label{fig:Tyson-reconstruction} 
\end{figure}
%
%

%---Fig. 22-------%
\begin{figure}[hpbt]
  \centering
   \begin{tabular}{cc}
   \psfig{file=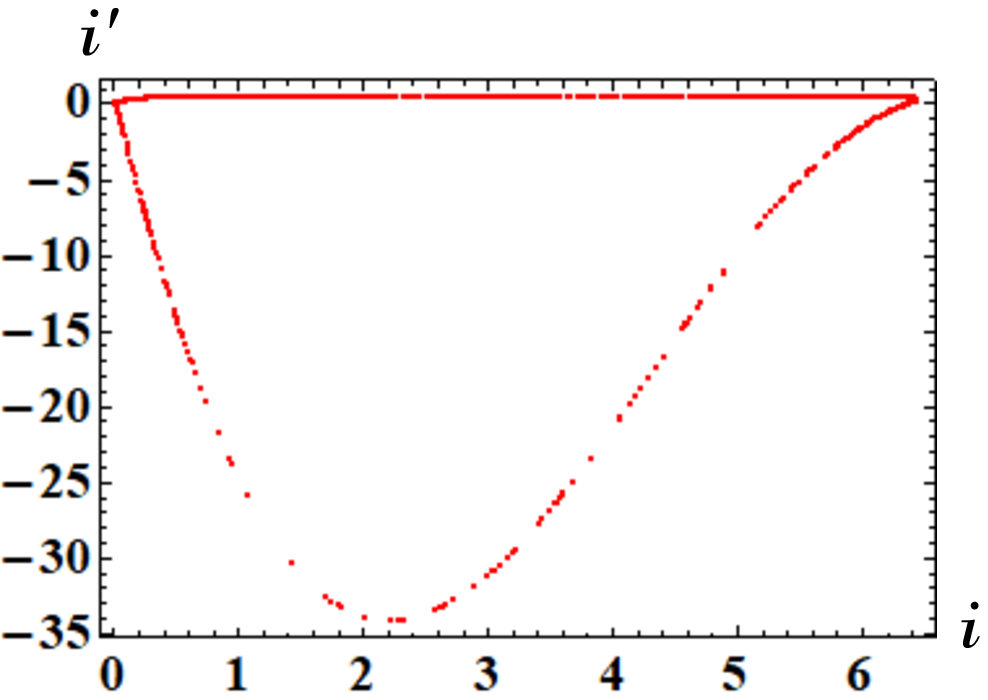, height=4.8cm} & 
   \psfig{file=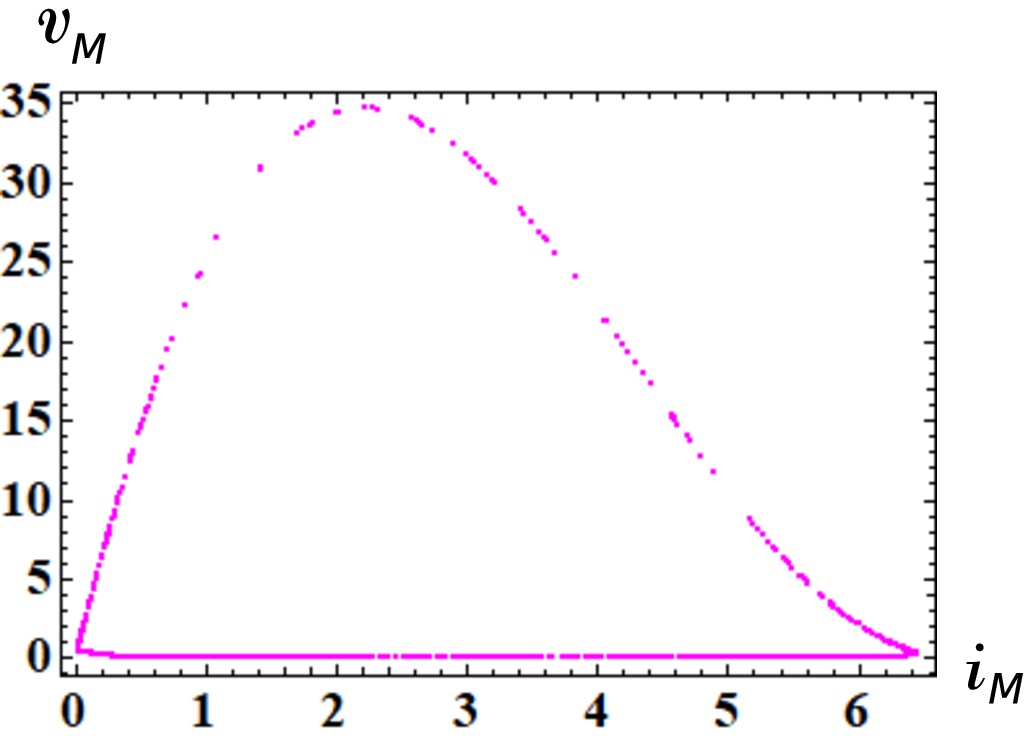, height=4.8cm} \\
   (a) Poincar\'e map for the attractor in Figure \ref{fig:Tyson-reconstruction}(a) & 
   (b) Poincar\'e map for the attractor in Figure \ref{fig:Tyson-reconstruction}(b)  \\
   \end{tabular}
  \caption{Poincar\'e maps for the reconstructed chaotic attractors in Figure \ref{fig:Tyson-reconstruction}.  
  Observe that these two Poincar\'e maps are quite similar.  
   Parameters:  $A = 0.5, \ B = 0.00803,  \ r = 0.5,  \ \omega = 0.5$. \
   Initial conditions: $i(0)=2.1, \  x(0)=2.1$.}
  \label{fig:Tyson-reconstruction-poincare} 
\end{figure}
%
%

%---Fig. 23-------%
\begin{figure}[hpbt]
  \centering
   \begin{tabular}{cc}
   \psfig{file=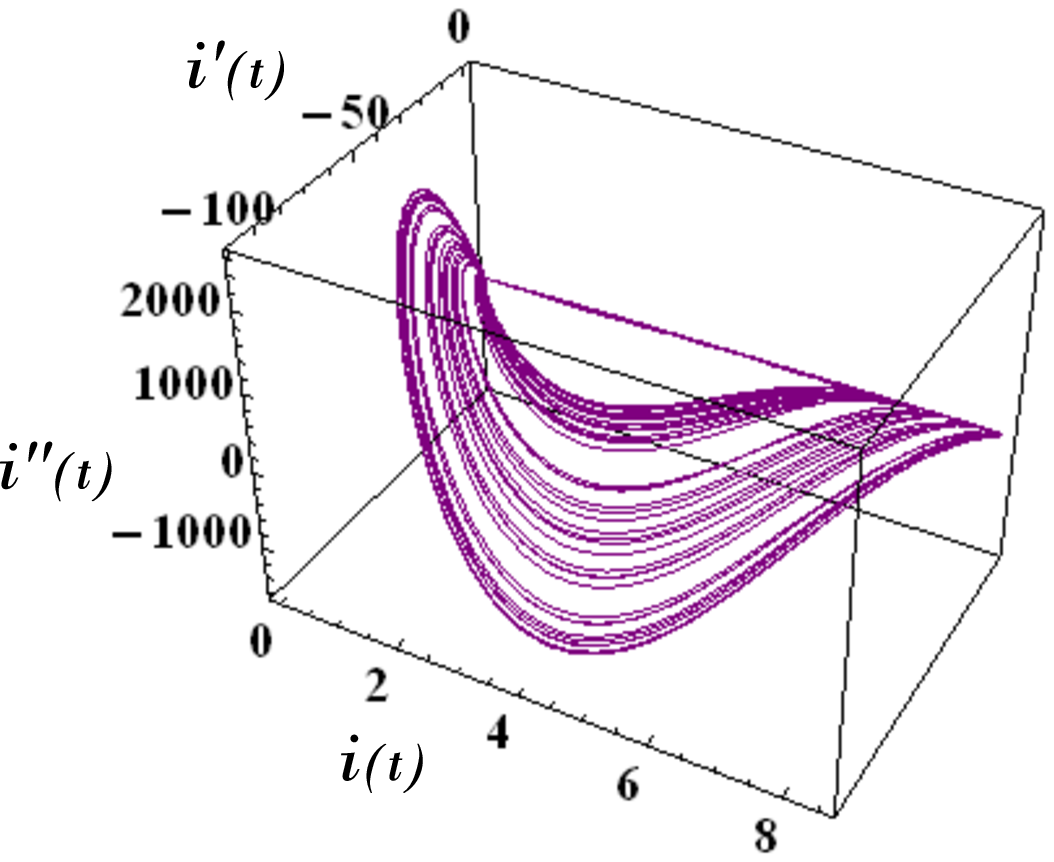,  height=4.8cm} & 
   \psfig{file=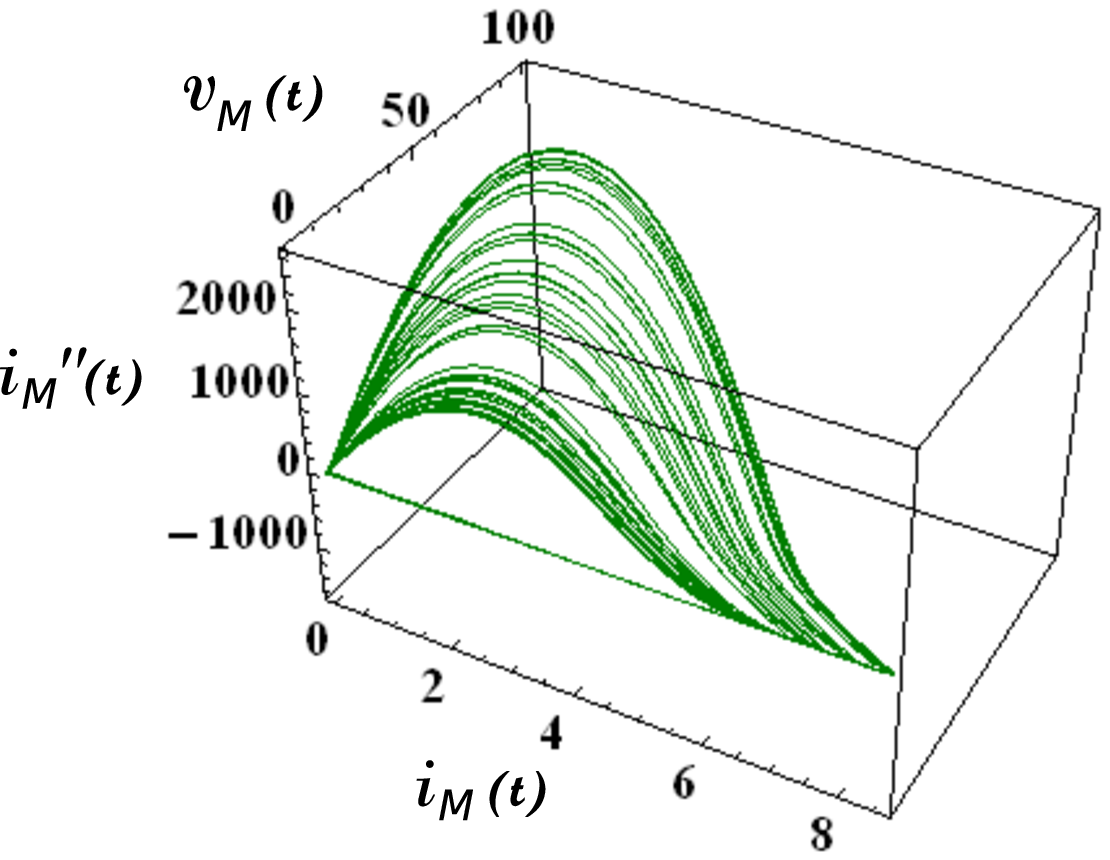,  height=4.8cm} \\
    (a) $\bigl ( i(t), \ i'(t), \ i''(t) \bigr )$ reconstruction &
    (b) $\bigl ( i_{M}(t), \  v_{M}(t), \ {i_{M}}''(t) \bigr )$ reconstruction \\
   \end{tabular}
  \caption{Reconstructed chaotic attractor using  $\bigl ( i(t), \ i'(t) \ i''(t) \bigr )$ and $\bigl ( i_{M}(t), \  v_{M}(t), \ {i_{M}}''(t) \bigr )$, 
  where $v_{M}$ and  $i_{M}$ denote the terminal voltage and the terminal current of the current-controlled generic memristor.   
   Parameters:  $A = 0.5, \ B = 0.00803,  \ r = 0.5,  \ \omega = 0.5$. \
   Initial conditions: $i(0)=2.1, \  x(0)=2.1$.}
  \label{fig:Tyson-reconstruction-2} 
\end{figure}
\newpage

%-------------------------------------%
\subsection{Lotka-Volterra equations}
\label{sec: Lotka-Volterra}
%-------------------------------------%

Consider Hamilton's Equations defined by 
\begin{center}
\begin{minipage}{8.7cm}
\begin{shadebox}
\underline{\emph{Hamilton's equations}}
\begin{equation}
\begin{array}{ccc}
 \displaystyle {\frac {d q}{d t}} 
  &=& \displaystyle {\frac {\partial {\mathcal {H}}(q, \, p)}{\partial p}},
 \vspace{2mm} \\
 \displaystyle {\frac {d p}{d t}} 
  &=& \displaystyle -{\frac {\partial {\mathcal {H}(q, \, p)}}{\partial q}},  
\end{array}
%\vspace{2mm}
\label{eqn: hamilton-10}
\end{equation}
where $q$ and $p$ denote the coordinate and the momentum and $\mathcal {H}(q, \, p)$ is the Hamiltonian.  
\end{shadebox}
\end{minipage}
\end{center}
Let us define the Hamiltonian:
\begin{center}
\begin{minipage}{8.7cm}
\begin{shadebox}
\underline{\emph{Hamiltonian}}
\begin{equation}
  \mathcal {H}(q, \, p) = - b \, p + a \ln p  - c \, q + d \ln q, 
\label{eqn: hamiltonian-10}
\end{equation}
where $a, \ b, \ c, \ d$ are constants.  
\end{shadebox}
\end{minipage}
\end{center}
From Eq. (\ref{eqn: hamilton-10}), we obtain 
\begin{equation}
\begin{array}{ccc}
 \displaystyle {\frac {d q}{d t}} 
  &=& \displaystyle {\frac {\partial {\mathcal {H}}}{\partial p}} =  - b + \frac{a}{p},
 \vspace{2mm} \\
 \displaystyle {\frac {d p}{d t}} 
  &=& \displaystyle -{\frac {\partial {\mathcal {H}}}{\partial q}} = c -  \frac{d}{q}.  
\end{array}
\vspace{2mm}
\label{eqn: hamilton-11}
\end{equation}
After time scaling by $ d\tau = p q \, dt$,
we obtain the associated Pfaff's equation \cite{Andronov}
\begin{equation}
\begin{array}{ccc}
 \displaystyle { \frac {d q}{d \tau}} 
  &=& \displaystyle  \left ( - b + \frac{a}{p} \right ) pq,
 \vspace{2mm} \\
 \displaystyle { \frac {d p}{d \tau}} 
  &=& \displaystyle \left ( c -  \frac{d}{q} \right ) pq.  
\end{array}
\vspace{2mm}
\label{eqn: hamilton-12}
\end{equation}
Equation (\ref{eqn: hamilton-12}) can be recast into the Lotka-Volterra equations \cite{Hirsch(2003)}
\begin{center}
\begin{minipage}{8.7cm}
\begin{shadebox}
\underline{\emph{Lotka-Volterra equations}}
\begin{equation}
\begin{array}{ccc}
 \displaystyle { \frac {d q}{d \tau}} 
  &=& \displaystyle  \left ( a - b \, p \right ) q,
 \vspace{2mm} \\
 \displaystyle { \frac {d p}{d \tau}} 
  &=& \displaystyle \left ( c \, q -  d \right ) p,  
\end{array}
\vspace{2mm}
\label{eqn: hamilton-13}
\end{equation}
where $a, \ b, \ c, \ d$ are constants. 
\end{shadebox}
\end{minipage}
\end{center}
Equation (\ref{eqn: hamilton-13}) has the Hamiltonian (\ref{eqn: hamiltonian-10}) as its integral invariant, that is,
\begin{equation}
  \frac{d \mathcal {H}(q, \, p) }{d \tau} = 0.
\label{eqn: hamiltonian-11}
\end{equation}

Consider next the three-element memristor circuit in Figure \ref{fig:memristor-inductor-battery}.  
The dynamics of this circuit given by Eq. (\ref{eqn: dynamics-n-1}).  
Assume that Eq. (\ref{eqn: dynamics-n-1}) satisfies 
\begin{equation}
 \begin{array}{ccc}
  E &=& 0, \vspace{2mm} \\
  \hat{R}(x, i) &=& - (c \, x - d), \vspace{2mm} \\
  \tilde{f}_{1}(x, \, i) &=& ( a - b \, i) x.  
 \end{array}
\end{equation}
Then we obtain 
\begin{center}
\begin{minipage}{8.7cm}
\begin{shadebox}
\underline{\emph{Memristor Lotka-Volterra equations}}
\begin{equation}
\begin{array}{ccc}
 \displaystyle { \frac {d i}{d \tau}} 
  &=& \displaystyle \left ( c \, x -  d \right ) i,  
 \vspace{2mm} \\
 \displaystyle { \frac {d x}{d \tau}} 
  &=& \displaystyle  \left ( a - b \, i \right ) x,
\end{array}
\vspace{2mm}
\label{eqn: hamilton-14}
\end{equation}
where $a, \ b, \ c, \ d$ are constants.  
\end{shadebox}
\end{minipage}
\end{center}
Equations (\ref{eqn: hamilton-13}) and (\ref{eqn: hamilton-14}) are equivalent if we change the variables  
\begin{equation}
  i=p, \ x=q.  
\end{equation}
In this case, the extended memristor in Figure \ref{fig:memristor-inductor-battery} is replaced by the \emph{generic} memristor (see Appendix A). 
That is,
\begin{equation}
  \hat{R}(x, i) = \tilde{R}(x)= - (cx -d).
\end{equation}
Thus, the terminal voltage $v_{M}$ and the terminal current $i_{M}$ of the current-controlled generic memristor are given by
\begin{center}
\begin{minipage}{8.7cm}
\begin{shadebox}
\underline{\emph{V-I characteristics of the generic memristors}}
\begin{equation}
\begin{array}{lll}
  v_{M} &=& \tilde{R}(x) \, i_{M} = - (c \, x -d) \, i_{M},   
  \vspace{3mm} \\
 \displaystyle \frac{d x}{dt} &=& ( a - b \, i_{M}) x,
\end{array}
\vspace{2mm}
\label{eqn: hamilton-15}
\end{equation}
where $\hat{R}(x) = - (cx -d)$.
\end{shadebox}
\end{minipage}
\end{center}
It follows that the Lotka-Volterra equations (\ref{eqn: hamilton-14}) can be realized by 
the three-element memristor circuit in Figure \ref{fig:memristor-inductor-battery}.

The Lotka-Volterra equations (\ref{eqn: hamilton-14}) can exhibit a periodic orbit (one-dimensional curve), 
since they have 
\begin{center}
\begin{minipage}{.5\textwidth}
\begin{itembox}[l]{Integral}
\begin{equation}
  \mathcal {H}(x, \, i) = - b \, i + a \ln i  - c \, x + d \ln x, 
\label{eqn: hamiltonian-12}
\end{equation}
\end{itembox}
\end{minipage}
\end{center}
as its integral invariant.    
When an external source is added as shown in Figure \ref{fig:memristive-inductor-battery-source}, the forced  Lotka-Volterra equations can exhibit a quasi-periodic or a non-periodic response,\footnote{In this paper, we use the terminology ``non-periodic response'' in order to describe ``chaotic-like but non-attracting response''.} which depends on initial conditions.  
The dynamics of the circuit is given by 
\begin{center}
\begin{minipage}{8.7cm}
\begin{shadebox}
\underline{\emph{Forced memristor Lotka-Volterra equations}}
\begin{equation}
\begin{array}{ccl}
 \displaystyle { \frac {d i}{d \tau}} 
  &=& \displaystyle \left ( c x -  d \right ) i + r \sin ( \omega \tau),  
 \vspace{2mm} \\
 \displaystyle { \frac {d x}{d \tau}} 
  &=& \displaystyle  \left ( a - b \, i \right ) x,
\end{array}
\vspace{2mm}
\label{eqn: hamilton-16}
\end{equation}
where $r$ and $\omega$ are constants.  
\end{shadebox}
\end{minipage}
\end{center}
We show their non-periodic and quasi-periodic responses, Poincar\'e maps, and $i_{M}-v_{M}$ loci in Figures \ref{fig:Lotka-attractor}, \ref{fig:Lotka-poincare}(a), and \ref{fig:Lotka-pinch}, respectively.  
The following parameters are used in our computer simulations:
\begin{equation}
\begin{array}{l}
  a = 2/3, \ b = 4/3, \ c = 1, \  d = 1, \vspace{2mm} \\
  r = 0.04, \  \omega = 1, \ \text{or} \ 1.01.  
\end{array}
\end{equation}
The $i_{M}-v_{M}$ loci in Figure \ref{fig:Lotka-pinch} lie in the first and the fourth quadrants. 
Thus, the generic memristor defined by Eq. (\ref{eqn: hamilton-15}) is an active element.  
Let us next show the $v_{M}-p_{M}$ locus in Figure \ref{fig:Lotka-power}, where $p_{M}(t)$ is an instantaneous power defined by $p_{M}(t)=i_{M}(t)v_{M}(t)$.  
Observe that the $v_{M}-p_{M}$ locus is pinched at the origin, and the locus lies in the first and the third quadrants. 
Thus, when $v_{M}>0$, the instantaneous power delivered from the forced signal and the inductor is dissipated in the generic memristor.  
However, when $v_{M}<0$, the instantaneous power delivered from the forced signal  and the inductor is not dissipated in the generic memristor. 
Hence, the memristor switches between passive and active modes of operation, depending on its terminal voltage. 
We conclude as follow: 
\begin{center}
\begin{minipage}{.7\textwidth}
\begin{itembox}[l]{Switching behavior of the memristor}
Assume that Eq. (\ref{eqn: hamilton-16}) exhibits non-periodic response.  
Then the generic memristor defined by Eq. (\ref{eqn: hamilton-15}) can switch between ``passive'' and ``active'' modes of operation, depending on its terminal voltage.  
\end{itembox}
\end{minipage}
\end{center}
%
%

%---Fig. 24-------%
\begin{figure}[hpbt]
 \centering
   \begin{tabular}{cc}
    \psfig{file=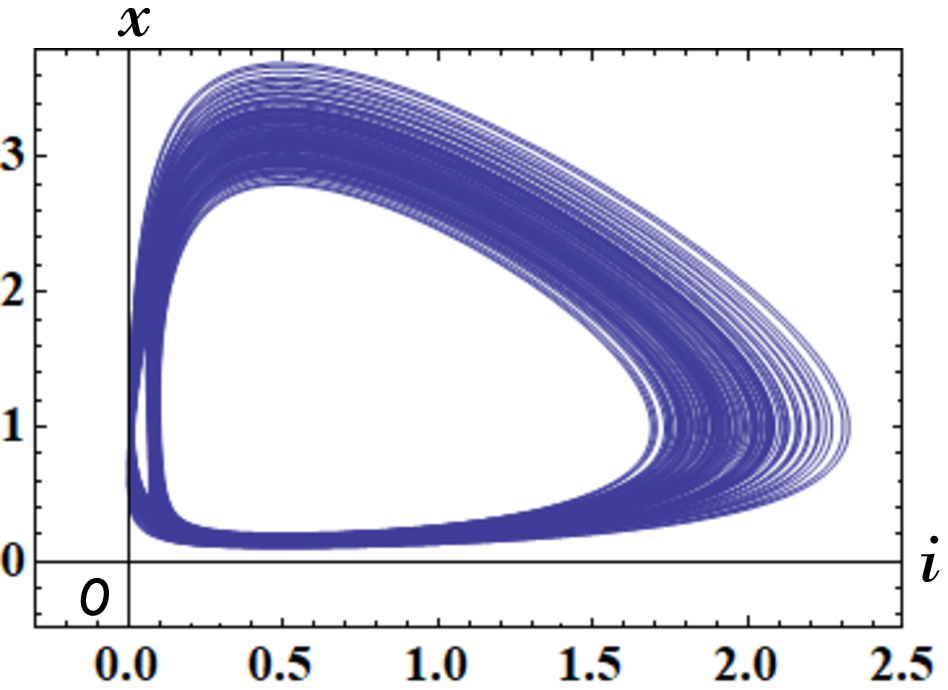, height=4.6cm}  & 
    \psfig{file=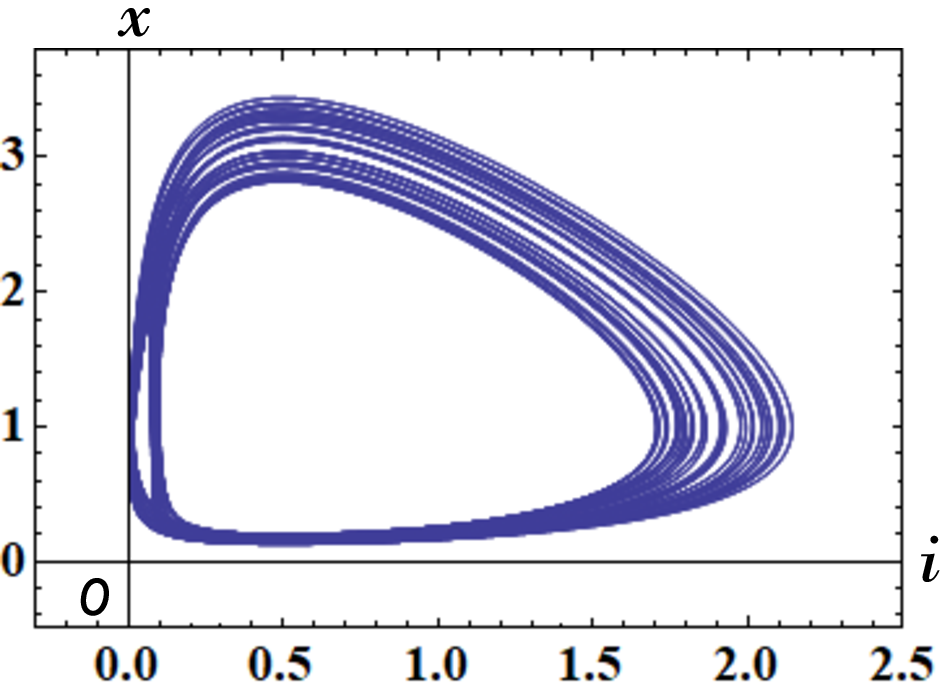, height=4.6cm}   \\
         (a) non-periodic ($\omega = 1$) & (b)  quasi-periodic ($\omega = 1.01$)
   \end{tabular}
  \caption{Non-periodic and quasi-periodic responses of the forced memristor Lotka-Volterra equations (\ref{eqn: hamilton-16}). 
  Parameters: (a) $a = 2/3, \ b = 4/3, \ c = 1, \  d = 1, \ r = 0.04, \  \omega = 1$. \  \
              (b) $a = 2/3, \ b = 4/3, \ c = 1, \  d = 1, \ r = 0.04, \  \omega = 1.01$.
  Initial conditions: $i(0)=0.19, \  x(0)=0.18$.}
  \label{fig:Lotka-attractor} 
\end{figure}
%
%
%---Fig. 25-------%
\begin{figure}[hpbt]
 \centering
   \begin{tabular}{c}
   \psfig{file=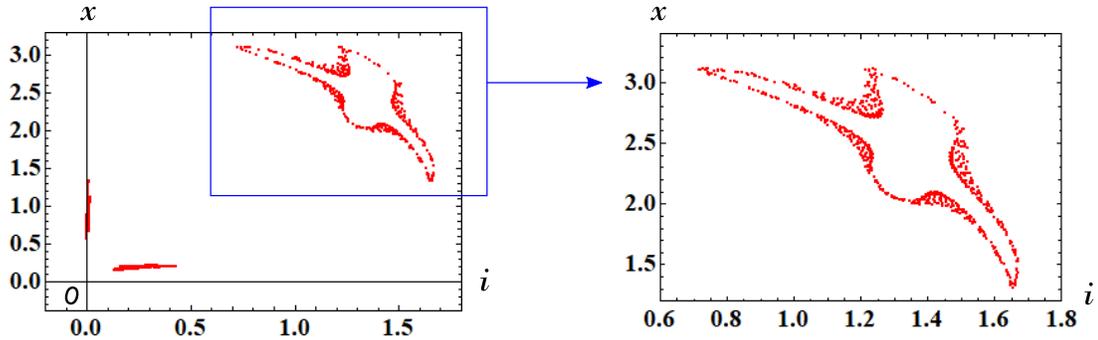, height=4.5cm} \\ 
   (a) Poincar\'e map for a non-periodic response. $i(0)=0.19$,  \  $x(0)=0.18$, \ $r = 0.04$, \ $\omega = 1$.  \vspace{1mm}\\   
   \psfig{file=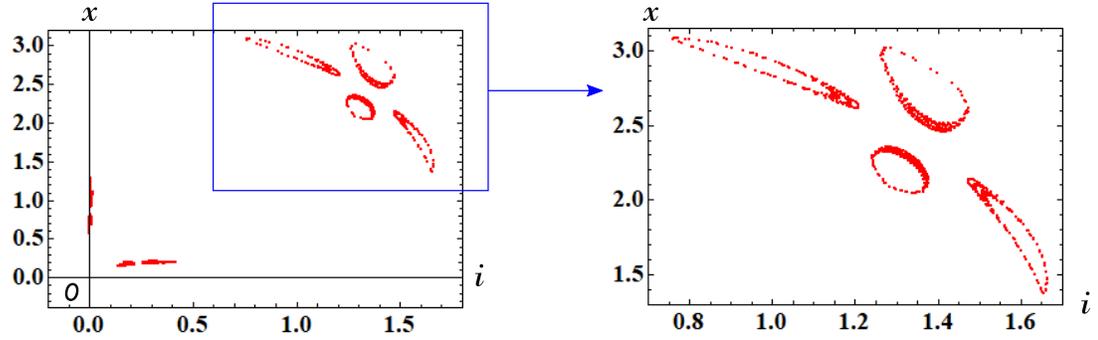, height=4.5cm} \\
   (b) Poincar\'e map for a quasi-periodic response. $i(0)=0.185$, \  $x(0)=0.185$, \ $r = 0.04$, \ $\omega = 1$.  \vspace{1mm} \\
   \psfig{file=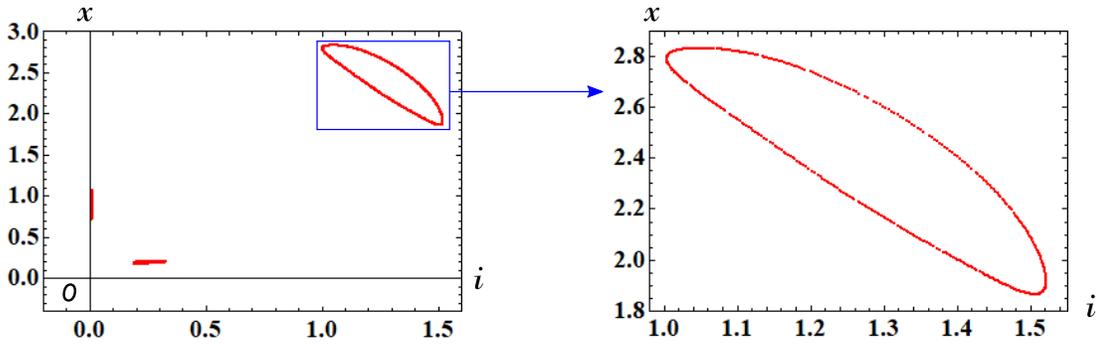, height=4.5cm}  \\
   (c) Poincar\'e map for a quasi-periodic response. $i(0)=0.19$,  \  $x(0)=0.18$, \ $r = 0.04$, \ $\omega = 1.01$.       
  \end{tabular}
  \caption{Poincar\'e maps of the forced memristor Lotka-Volterra equations (\ref{eqn: hamilton-16}).  
   Compare the three Poincar\'e maps in Figure \ref{fig:Lotka-poincare}.   
   In order to generate the non-periodic Poincar\'e map in Figure \ref{fig:Lotka-poincare}(a), 
   we have to choose the initial conditions and parameters carefully.  
   Parameters: $a = 2/3, \ b = 4/3, \ c = 1, \  d = 1, \ r = 0.04, \ \omega = 1 \ \text{or} \ 1.01$. }
  \label{fig:Lotka-poincare} 
\end{figure}
%
%

%---Fig. 26-------%
\begin{figure}[hpbt]
 \centering
   \begin{tabular}{cc}
    \psfig{file=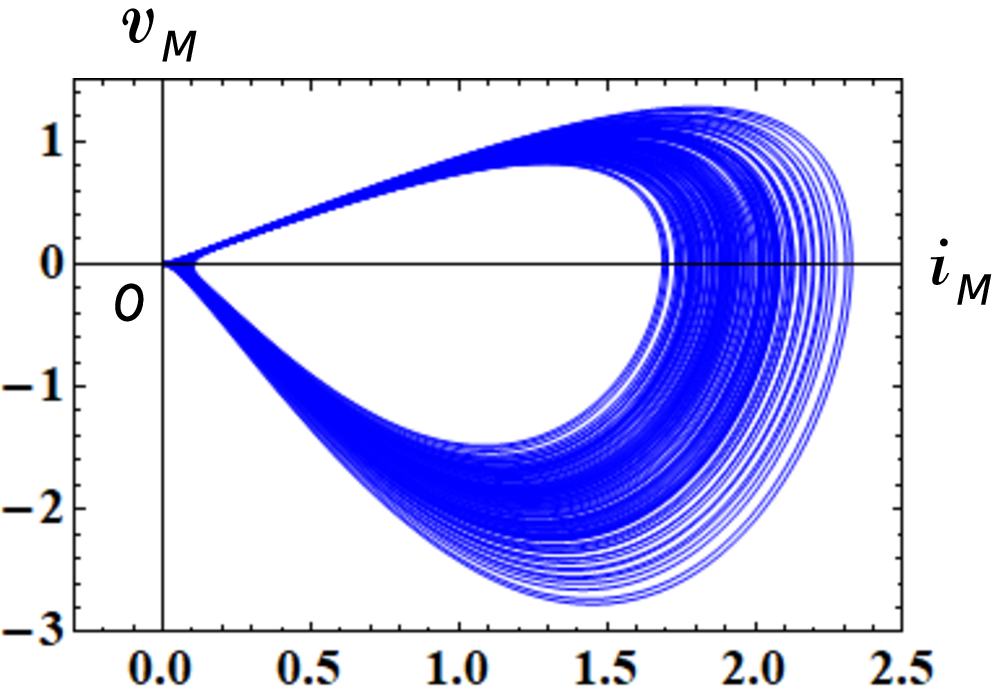, height=4.6cm}  & 
    \psfig{file=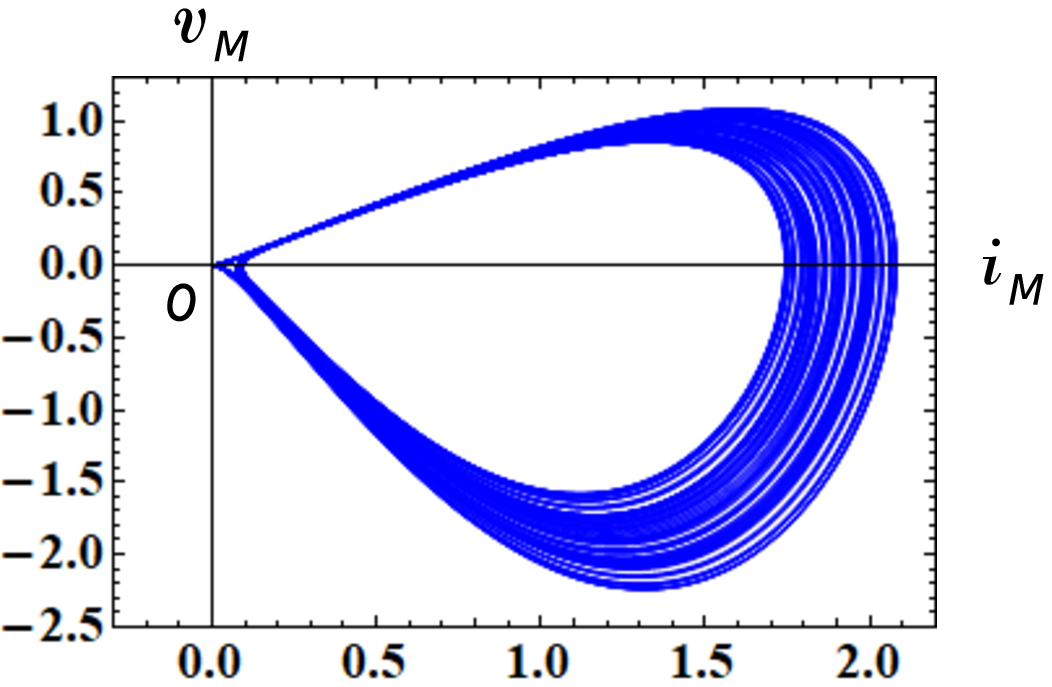, height=4.6cm}  \\
         (a) non-periodic ($\omega = 1$) & (b)  quasi-periodic ($\omega = 1.01$)
   \end{tabular}
  \caption{ The $i_{M}-v_{M}$ loci of the forced memristor Lotka-Volterra equations (\ref{eqn: hamilton-16}).  
  Here, $v_{M}$ and  $i_{M}$ denote the terminal voltage and the terminal current of the current-controlled generic memristor.  
  Parameters: (a) $a = 2/3, \ b = 4/3, \ c = 1, \  d = 1, \ r = 0.04, \  \omega = 1$. \  \
              (b) $a = 2/3, \ b = 4/3, \ c = 1, \  d = 1, \ r = 0.04, \  \omega = 1.01$.
   Initial conditions: $i(0)=0.19, \  x(0)=0.18$.}
  \label{fig:Lotka-pinch} 
\end{figure}

In order to generate the non-periodic Poincar\'e map in Figure \ref{fig:Lotka-poincare}(a), 
we have to choose the initial conditions and parameters carefully, 
and the maximum step size $h$ of the numerical integration must be sufficiently small ($h=0.002$).\footnote{We used ``NDSolve'' in Mathematica to solve differential equations numerically.   
Numerical integration tool, like NDSolve, can specify the maximum size of a single step used in generating a result.  For most differential equations, the results given by NDSolve are quite accurate.}  
Compare the three Poincar\'e maps in Figure \ref{fig:Lotka-poincare}.   
In order to view these Poincar\'e maps from a different perspective, 
let us project the trajectories into the $( \xi, \, \eta, \, \zeta )$-space via the transformation 
\begin{equation}
 \begin{array}{lll}
   \xi (\tau)   &=& (i(\tau) + 5) \cos \,( \omega \tau ), \vspace{2mm} \\ 
   \eta (\tau)  &=& (i(\tau) + 5) \sin \,( \omega \tau ), \vspace{2mm} \\ 
   \zeta (\tau) &=& x(\tau).
 \end{array}  
\label{eqn: LV-projection}
\end{equation}
Then the trajectory on the $(i, \, x)$-plane is transformed into the trajectory in the three-dimensional $( \xi, \, \eta, \, \zeta )$-space, 
as shown in Figure \ref{fig:Lotka-torus}.\footnote{For example, if $(i(t), \, x(t))$ moves on the unit circle, that is, 
$(i(t), \, x(t)) = ( \, \cos  ( 3 t ), \,  \sin ( 3 t ) \, )$,
then the projected trajectory $( \xi (t), \, \eta (t), \, \zeta (t) )$ moves on a torus (for more details on the transformation (\ref{eqn: LV-projection}), see Appendix in \cite{Itoh(2017b)}).}$^{, }$ \footnote{If we plot the intersection of the points with the plane defined by $\{ ( \xi, \, \eta, \, \zeta ) \in  \mathbb{R}^{3} \, | \, \eta = 0, \, \xi \ge 0 \}$, we obtain similar Poincar\'e maps (see Appendix in \cite{Itoh(2017b)}).}
Observe the difference among the three trajectories.

Note that the maximum step size limitation for $h$ is important for numerical stability, otherwise an overflow (outside the range of data) is likely to occur. 
We show its example in Figure \ref{fig:Lotka-trajectory}.   
Observe that if $h=0.002$, then the trajectory rapidly grows for $t \ge 1148$, and an overflow occurs as shown in Figure \ref{fig:Lotka-trajectory}(a). 
However, if $h=0.001$, then the trajectory stays in a finite region of the first-quadrant of the $(i, \ x)$-plane as shown in Figure \ref{fig:Lotka-trajectory}(b). 
The above numerical instability in long-time simulations is partially caused by the fact that if $i(\tau)$ takes sufficiently small \emph{negative} values at some time $\tau_{0}$ by the low-accuracy computation, then we obtain 
\begin{equation}
  \frac{dx(\tau)}{d \tau} = \bigl ( a - b \, i(\tau) \bigr ) x (\tau) \, >0,    
\end{equation}
for $\tau \approx \tau_{0}$ and $0 < - b \, i(\tau) \ll 1$, where $b>0$.  
Thus, $x(\tau)$ is approximated by 
\begin{equation}
  x(\tau) \approx x(\tau_{0})e^{a \tau},   
\end{equation}
and it grows rapidly, where $a>0$. 
Consequently, $x(\tau)$ would overflow.  
Thus, noise may considerably affect the behavior of the above memristor circuit.

As stated in Sec. \ref{sec: Brusselator}, we can reconstruct the non-periodic trajectory into two dimensional plane by using 
\begin{equation}
  (i(t), \ i'(t)),    
\end{equation}
or 
\begin{equation}
  \Bigl ( i_{M}(t), \,  v_{M}(t) \Bigr )  \equiv \Bigl ( i(t), \ - i'(t) + r \sin ( \omega t) \Bigr ), 
\end{equation}
where $i_{M}(t)=i(t)$.
Their trajectories and Poincar\'e maps are shown in Figures \ref{fig:Lotka-reconstruction} and \ref{fig:Lotka-reconstruction-poincare}, respectively.     
We can also reconstruct the non-periodic trajectory into the three-dimensional Euclidean space by using 
\begin{equation}
  (i(t), \ i'(t), \ i''(t)), 
\end{equation}
or 
\begin{equation}
\scalebox{0.9}{$\displaystyle  \Bigl ( i_{M}(t), \,  v_{M}(t), \, i_{M}''(t) \Bigr )  \equiv \Bigl ( i(t), \ - i'(t) + r \sin ( \omega t), \, i''(t) \Bigr ). $}
\end{equation}
These reconstructed trajectories are shown in three-dimensional space in Figure \ref{fig:Lotka-reconstruction-2}.

%---Fig. 27-------%
\begin{figure}[hpbt]
 \centering
   \begin{tabular}{cc}
    \psfig{file=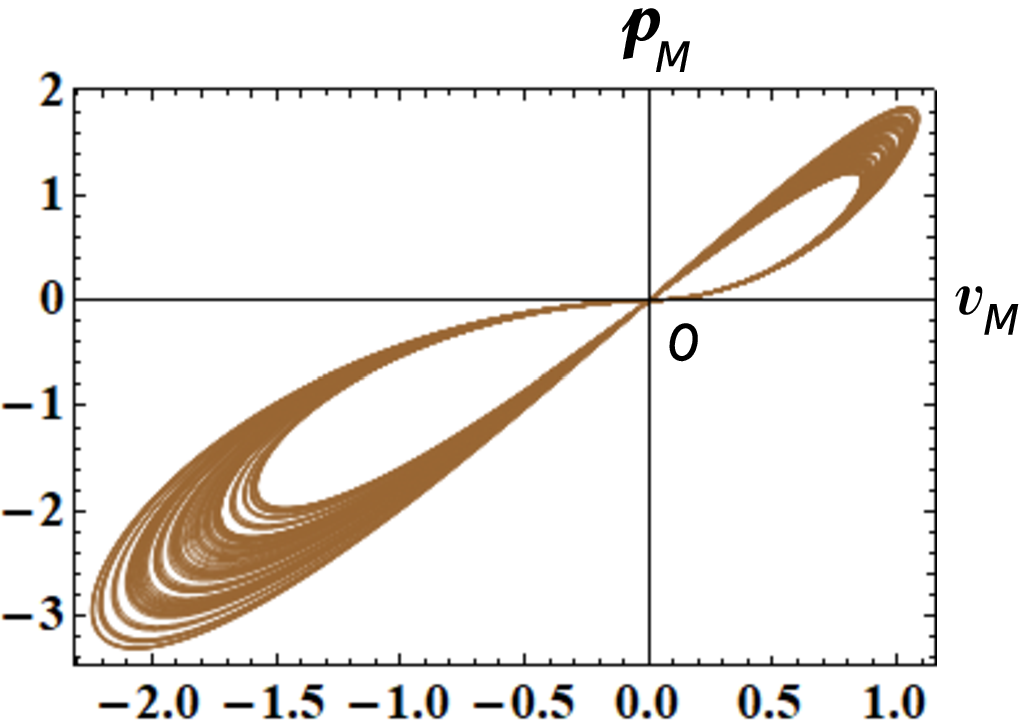, height=4.6cm}  & 
    \psfig{file=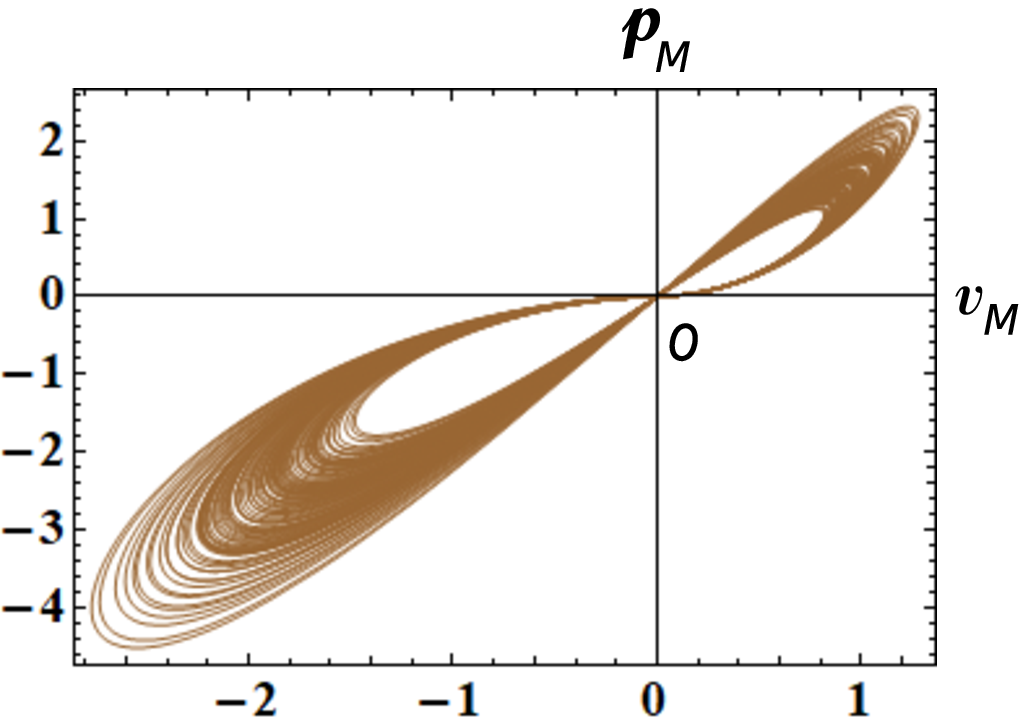, height=4.6cm}  \\
         (a) non-periodic ($\omega = 1$) & (b)  quasi-periodic ($\omega = 1.01$)
   \end{tabular}
  \caption{ The $v_{M}-p_{M}$ loci of the forced memristor Lotka-Volterra equations (\ref{eqn: hamilton-16}). 
   Here, $p_{M}(t)$ is an instantaneous power defined by $p_{M}(t)=i_{M}(t)v_{M}(t)$, 
   and $v_{M}(t)$ and $i_{M}(t)$ denote the terminal voltage and the terminal current of the current-controlled generic memristor.  
   Observe that the $v_{M}-p_{M}$ locus is pinched at the origin, and the locus lies in the first and the third quadrants.  
   The memristor switches between passive and active modes of operation, depending on its terminal voltage $v_{M}(t)$.
   Parameters: (a) $a = 2/3, \ b = 4/3, \ c = 1, \  d = 1, \ r = 0.04, \  \omega = 1$. \  \
               (b) $a = 2/3, \ b = 4/3, \ c = 1, \  d = 1, \ r = 0.04, \  \omega = 1.01$.
   Initial conditions: $i(0)=0.19, \  x(0)=0.18$.}
  \label{fig:Lotka-power} 
\end{figure}
%
%

%---Fig. 28-------%
\begin{figure}[hpbt]
 \centering
   \begin{tabular}{c}
   \psfig{file=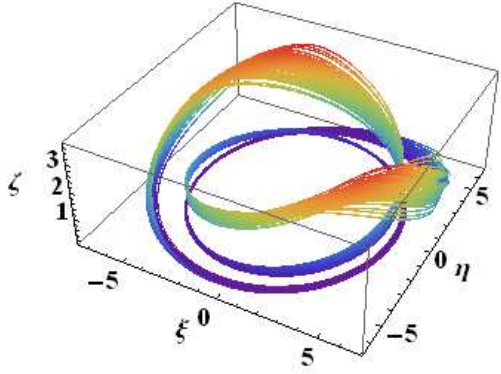, height=5.5cm} \\ 
   (a) A non-periodic response with $i(0)=0.19$,  \  $x(0)=0.18$, \ $r = 0.04$, and $\omega = 1$.  \vspace{1mm}\\   
   \psfig{file=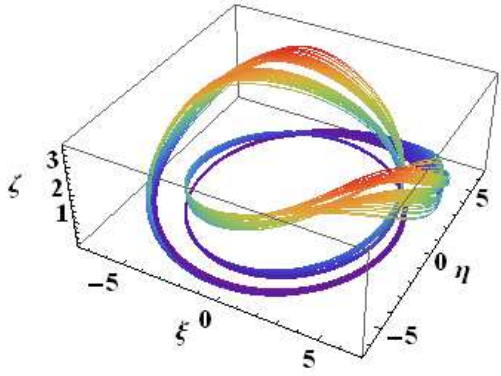, height=5.5cm} \\
   (b) A quasi-periodic response with $i(0)=0.185$, \  $x(0)=0.185$, \ $r = 0.04$, and $\omega = 1$.  \vspace{1mm} \\
   \psfig{file=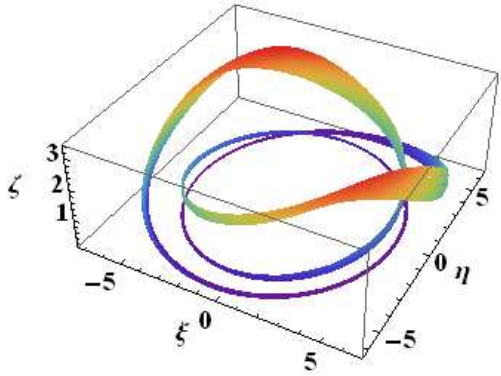, height=5.5cm}  \\
   (c) A quasi-periodic response with $i(0)=0.19$,  \  $x(0)=0.18$, \ $r = 0.04$, and $\omega = 1.01$.       
  \end{tabular}
  \caption{Three trajectories of the forced memristor Lotka-Volterra equations (\ref{eqn: hamilton-16}), 
   which are projected into the $( \xi, \, \eta, \, \zeta )$-space via the coordinate transformation (\ref{eqn: LV-projection}). 
   Compare these three trajectories in Figure \ref{fig:Lotka-torus} with the three Poincar\'e maps in Figure \ref{fig:Lotka-poincare}. 
   The trajectories are colored with the \emph{Rainbow} color code in Mathematica, 
   that is, the color evolves through violet, indigo, blue, green, yellow, orange and red, 
   as $\zeta$ varies from its minimum to its maximum value. 
   Parameters: $a = 2/3, \ b = 4/3, \ c = 1, \  d = 1, \ r = 0.04, \ \omega = 1 \ \text{or} \ 1.01$. }
  \label{fig:Lotka-torus} 
\end{figure}

\newpage
%---Fig. 29-------%
\begin{figure}[hpbt]
 \centering
   \begin{tabular}{cc}
    \psfig{file=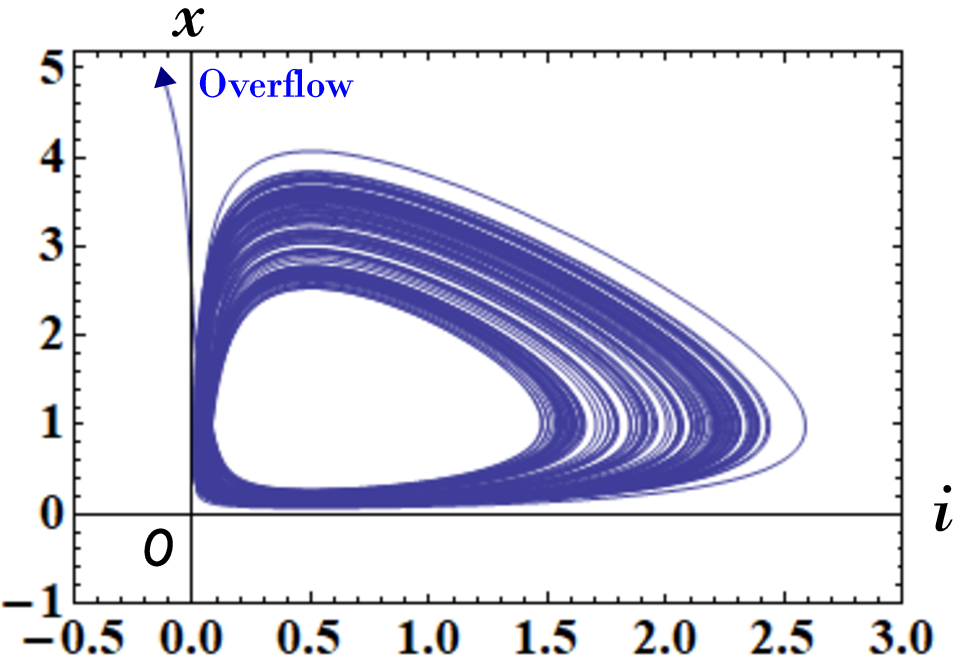, height=5cm}  & 
    \psfig{file=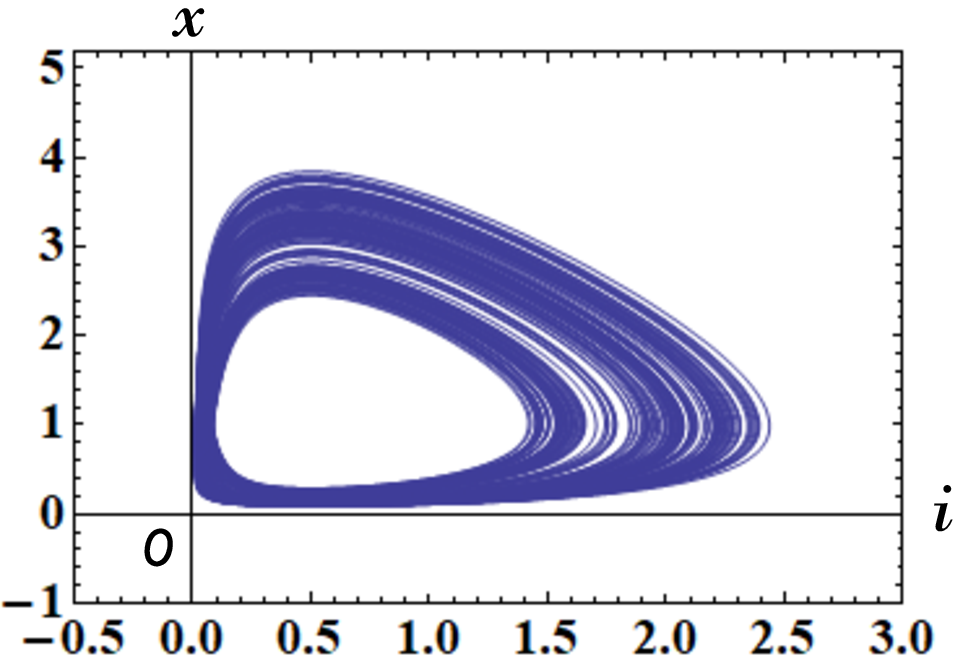, height=5cm}  \vspace{1mm} \\
   (a) $h=0.002$ & (b) $h=0.001$  \\ 
   \end{tabular}
  \caption{Behavior of the forced memristor Lotka-Volterra equations (\ref{eqn: hamilton-16}). 
  If $h=0.002$, then the trajectory rapidly grows for $t \ge 1148$, 
  and an overflow occurs as shown in Figure \ref{fig:Lotka-trajectory}(a). 
  However, if $h=0.001$, then the trajectory stays in a finite region of the first-quadrant of the $(i, \ x)$-plane 
  as shown in Figure \ref{fig:Lotka-trajectory}(b).   
  Here, $h$ denotes the maximum step size of the numerical integration.  
  Parameters: $a = 2/3, \ b = 4/3, \ c = 1, \  d = 1, \ r = 0.04, \  \omega = 1.056$.
  Initial conditions: $i(0)=0.18, \  x(0)=0.18$.}
  \label{fig:Lotka-trajectory} 
\end{figure}
%
%

%\clearpage
%---Fig. 30-------%
\begin{figure}[hpbt]
  \centering
   \begin{tabular}{cc}
   \psfig{file=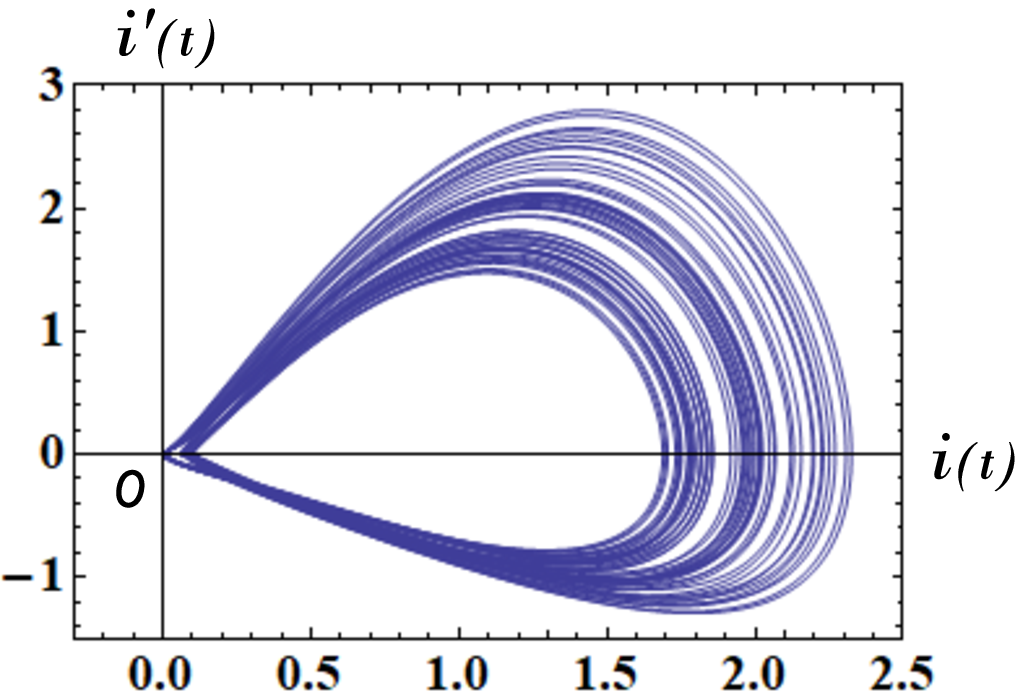, height=4.8cm} & 
   \psfig{file=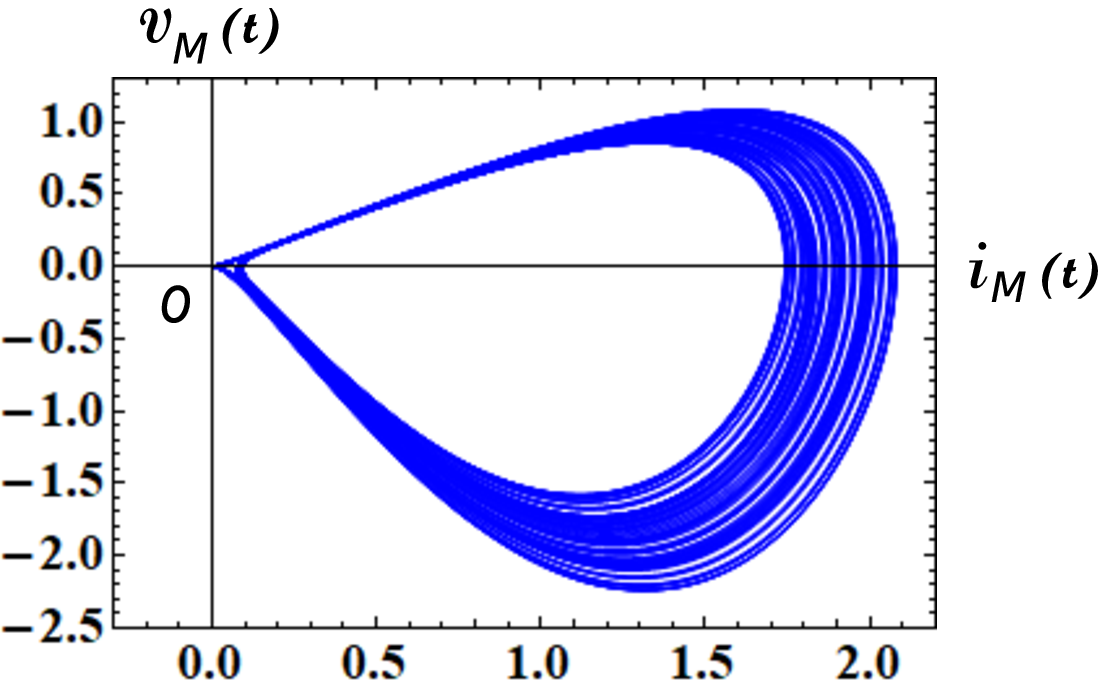,  height=5.0cm}  \\
   (a) $\bigl (i(t), \, i'(t) \bigr )$ reconstruction & (b) $\bigl ( i_{M}(t), \  v_{M}(t) \bigr )$ reconstruction \\
   \end{tabular}
  \caption{Reconstructed non-periodic trajectories using $\bigl ( i(t), \, i'(t) \bigr )$ and $\bigl ( i_{M}(t), \  v_{M}(t) \bigr )$,      
   where $v_{M}$ and  $i_{M}$ denote the terminal voltage and the terminal current of the current-controlled generic memristor. 
   Parameters:  $a = 2/3, \ b = 4/3, \ c = 1, \  d = 1, \ r = 0.04, \  \omega = 1$. \ \
   Initial conditions: $i(0)=0.19, \  x(0)=0.18$.}
  \label{fig:Lotka-reconstruction} 
\end{figure}
%
%

%---Fig. 31-------%
\begin{figure}[hpbt]
  \centering
   \begin{tabular}{cc}
   \psfig{file=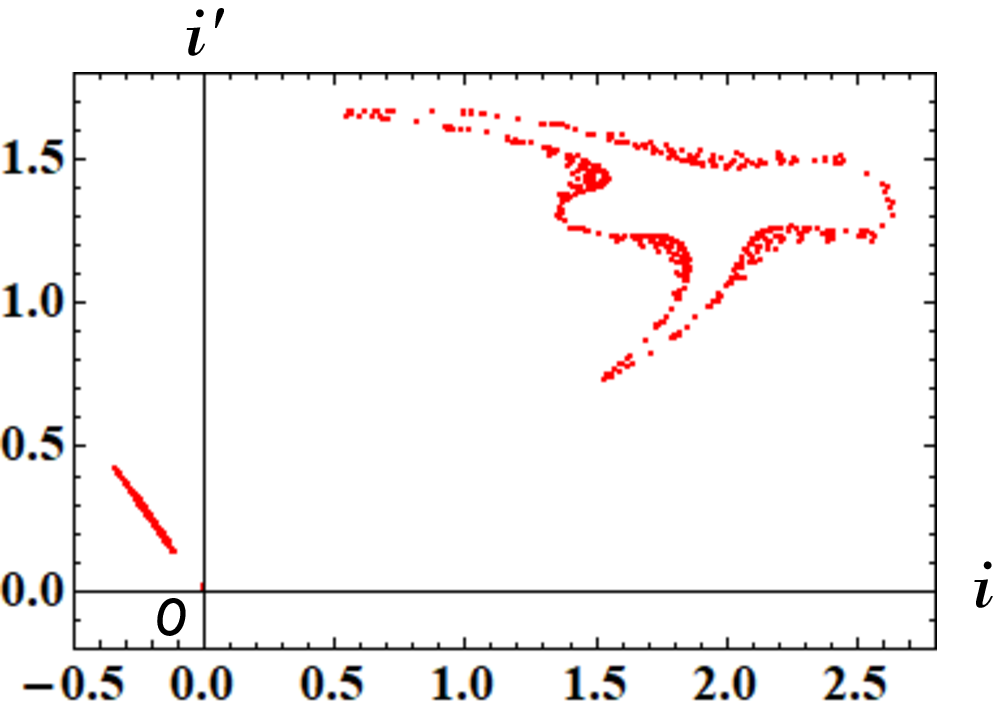, height=4.8cm} & 
   \psfig{file=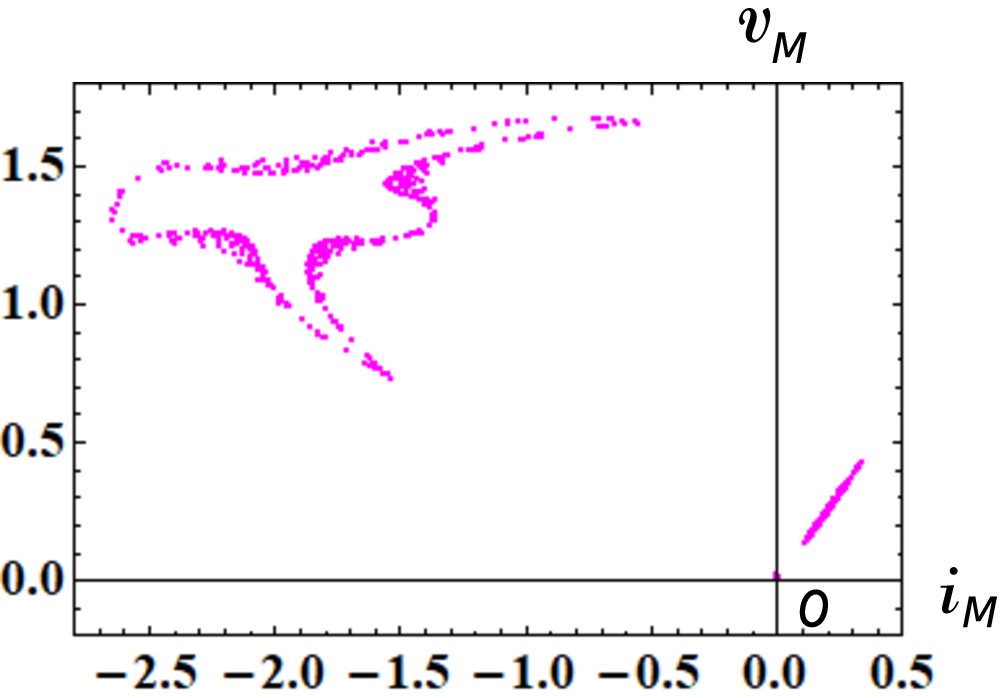, height=4.8cm} \\
   (a) Poincar\'e map for the attractor in Figure \ref{fig:Lotka-reconstruction}(a) & 
   (b) Poincar\'e map for the attractor in Figure \ref{fig:Lotka-reconstruction}(b)  \\
   \end{tabular}
  \caption{Poincar\'e maps for the reconstructed non-periodic trajectories in Figure \ref{fig:Lotka-reconstruction} .  
  Observe that these two Poincar\'e maps are quite similar.  
   Parameters:  $a = 2/3, \ b = 4/3, \ c = 1, \  d = 1, \ r = 0.04, \  \omega = 1$. \ \
   Initial conditions: $i(0)=0.19, \  x(0)=0.18$.}
  \label{fig:Lotka-reconstruction-poincare} 
\end{figure}
%
%

%---Fig. 32-------%
\begin{figure}[hpbt]
  \centering
   \begin{tabular}{cc}
   \psfig{file=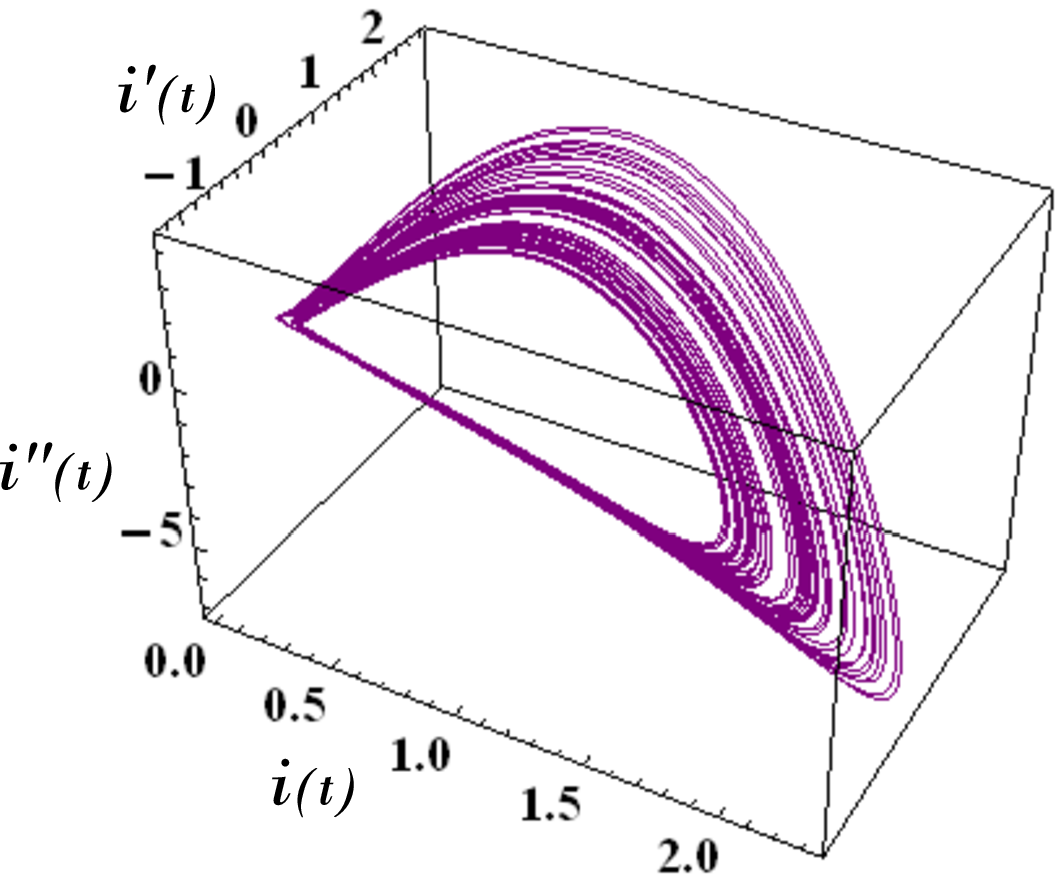,  height=4.8cm} & 
   \psfig{file=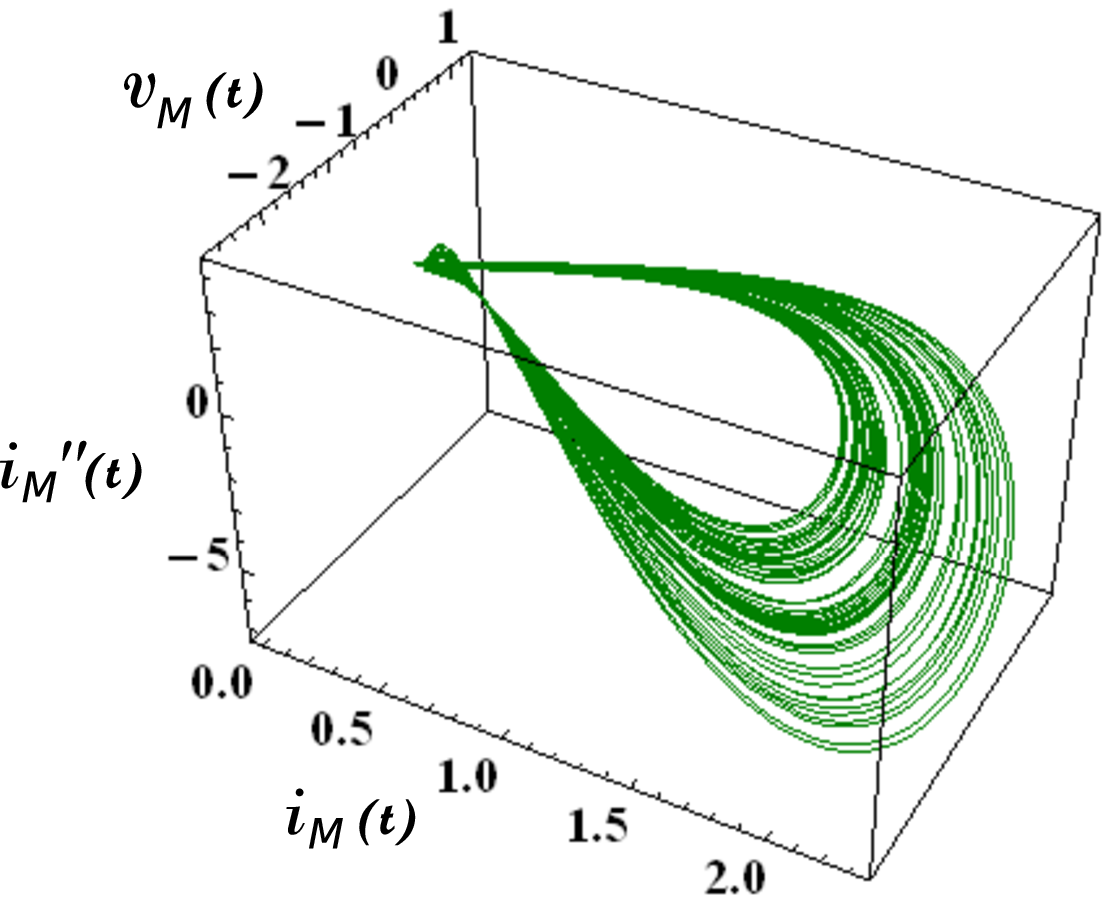,  height=4.8cm} \\
    (a) $\bigl ( i(t), \ i'(t), \ i''(t) \bigr )$ reconstruction &
    (b) $\bigl ( i_{M}(t), \  v_{M}(t), \ {i_{M}}''(t) \bigr )$ reconstruction \\
   \end{tabular}
  \caption{Reconstructed non-periodic trajectory using  $\bigl ( i(t), \ i'(t) \ i''(t) \bigr )$ 
   and $\bigl ( i_{M}(t), \  v_{M}(t), \ {i_{M}}''(t) \bigr )$, 
   where $v_{M}$ and  $i_{M}$ denote the terminal voltage and the terminal current of the current-controlled generic memristor.   
   Parameters:  $a = 2/3, \ b = 4/3, \ c = 1, \  d = 1, \ r = 0.04, \  \omega = 1$. \ \
   Initial conditions: $i(0)=0.19, \  x(0)=0.18$.}
  \label{fig:Lotka-reconstruction-2} 
\end{figure}
\newpage

%==============================================================%
\subsection{R\"ossler system}
%==============================================================%
%
The dynamics of the \emph{R\"ossler system} \cite{Peitgen, Letellie(2006)} is defined by 
\begin{center}
\begin{minipage}{8.7cm}
\begin{shadebox}
\underline{\emph{R\"ossler system}}
\begin{equation} 
 \left.
  \begin{array}{lll}
     \displaystyle \frac{d x_{1}}{dt} &=&  - x_{2} - x_{3} ,
      \vspace{2mm} \\
     \displaystyle \frac{d x_{2}}{dt} &=&  x_{1} + a \, x_{2},
      \vspace{2mm} \\    
     \displaystyle \frac{d x_{3}}{dt} &=&  b +  x_{3} \,(x_{1} - c ),    
  \end{array}
 \right \}
\vspace{2mm} 
\label{eqn: rossler-1}
\end{equation} 
where $a = 0.2$, $b = 0.2$, and $c = 5.7$. 
\end{shadebox}
\end{minipage}
\end{center}
The behavior of Eq. (\ref{eqn: rossler-1}) is chaotic for certain ranges of the three parameters, $a$, $b$ and $c$.

Consider the three-element memristor circuit in Figure \ref{fig:memristor-inductor-battery}.  
The dynamics of this circuit given by Eq. (\ref{eqn: dynamics-1}).  
Assume that Eq. (\ref{eqn: dynamics-1}) satisfies 
\begin{equation}
 \begin{array}{ccc}
  E =b, &&  L=1,                                         \vspace{2mm} \\
  \hat{R}(x_{1}, \, x_{2}, \, i) &=& - (x_{1} - c ),     \vspace{2mm} \\
  \tilde{f}_{1}(x_{1}, \, x_{2}, \, i) &=& - x_{2} - i,  \vspace{2mm} \\
  \tilde{f}_{2}(x_{1}, \, x_{2}, \, i) &=&  x_{1} + a \, x_{2}.   
 \end{array}
\end{equation}
Then we obtain  
\begin{center}
\begin{minipage}{8.7cm}
\begin{shadebox}
\underline{\emph{Memristor R\"ossler system}}
\begin{equation} 
 \left.
  \begin{array}{lll}
     \displaystyle \frac{d i}{dt} &=&  E +  (x_{1} - c ) \, i,     
      \vspace{2mm} \\     
     \displaystyle \frac{d x_{1}}{dt} &=&  - x_{2} - i,
      \vspace{2mm} \\
     \displaystyle \frac{d x_{2}}{dt} &=&  x_{1} + a \, x_{2},
  \end{array}
 \right \}
\vspace{2mm} 
\label{eqn: rossler-2}
\end{equation} 
where  $a = 0.2$, $E = b = 0.2$, $c = 5.7$.
\end{shadebox}
\end{minipage}
\end{center}
Equations (\ref{eqn: rossler-1}) and (\ref{eqn: rossler-2}) are equivalent if we change the variables  and the parameter 
\begin{equation}
  i = x_{3}, \ \ E = b.   
\end{equation}
In this case, the extended memristor in Figure \ref{fig:memristor-inductor-battery} is replaced by the \emph{generic} memristor (see Appendix A).
That is,
\begin{equation}
  \hat{R}(x_{1}, \, x_{2}, \, i) = \tilde{R}(x_{1}, \, x_{2})= - (x_{1} - c ).   
\end{equation}
The terminal voltage $v_{M}$ and the terminal current $i_{M}$ of the current-controlled generic memristor are described
by
\begin{center}
\begin{minipage}{8.7cm}
\begin{shadebox}
\underline{\emph{V-I characteristics of the generic memristor}}
\begin{equation}
\begin{array}{lll}
  v_{M} &=& \tilde{R}( x_{1}, \, x_{2}) \, i_{M} = - (x_{1} - c ) \, i_{M},   
  \vspace{3mm} \\
     \displaystyle \frac{d x_{1}}{dt} &=&  - x_{2} - i_{M},
      \vspace{2mm} \\
     \displaystyle \frac{d x_{2}}{dt} &=&  x_{1} + a \, x_{2},
\end{array}
\label{eqn: rossler-3}
\end{equation}
where $\tilde{R} ( x_{1}, \, x_{2}) = - (x_{1} - c ) $. \vspace{2mm}
\end{shadebox}
\end{minipage}
\end{center}
It follows that the  R\"ossler system (\ref{eqn: rossler-1}) can be realized by 
the three-element memristor circuit in Figure \ref{fig:memristor-inductor-battery}.  
The memristor  R\"ossler equations (\ref{eqn: rossler-2}) also exhibit chaotic oscillation.  
Thus, an external periodic forcing is unnecessary to generate chaotic or non-periodic oscillation. 
We show their chaotic attractor, Poincar\'e map, and $i_{M}-v_{M}$ locus Figures \ref{fig:Rossler-attractor}, \ref{fig:Rossler-poincare}, and \ref{fig:Rossler-pinch}, respectively.  
The following parameters are used in our computer simulations:
\begin{equation}
\begin{array}{l}
   a = 0.2, \ b = E = 0.2, \ c = 5.7.
\end{array}
\end{equation}
Observe that a chaotic attractor is a simple stretched and folded ribbon (see Figures \ref{fig:Rossler-attractor} and \ref{fig:Rossler-poincare}). 
The $i_{M}-v_{M}$ locus in Figure \ref{fig:Rossler-pinch} lies in the first and the fourth quadrants.
Thus, the extended memristor defined by Eq. (\ref{eqn: rossler-3}) is an active element.  
Let us show the $v_{M}-p_{M}$ locus in Figure \ref{fig:Rossler-power}, where $p_{M}(t)$ is an instantaneous power defined by $p_{M}(t)=i_{M}(t)v_{M}(t)$.  
Observe that the $v_{M}-p_{M}$ locus is pinched at the origin, and the locus lies in the first and the third quadrants. 
Thus, when $v_{M}>0$, the instantaneous power is dissipated in the memristor.  
However, when $v_{M}<0$, the instantaneous power is not dissipated in the memristor. 
Hence, the memristor switches between passive and active modes of operation, depending on its terminal voltage. 
We conclude as follow: 
\begin{center}
\begin{minipage}{.7\textwidth}
\begin{itembox}[l]{Switching behavior of the memristor}
Assume that Eq. (\ref{eqn: rossler-2}) exhibits chaotic oscillation.  
Then the generic memristor defined by Eq. (\ref{eqn: rossler-3}) can switch between ``passive'' and ``active'' modes of operation, depending on its terminal voltage.  
\end{itembox}
\end{minipage}
\end{center}

Finally, we reconstruct a chaotic attractor by using the current $i(t)$ (see \cite{Packard} ), that is, 
\begin{equation}
  \Bigl ( i(t), \ i'(t), \ i''(t) \Bigr ), 
\end{equation}
and by using the $i_{M}$ and $v_{M}$, that is, 
\begin{equation}
  \Bigl ( i_{M}(t), \  v_{M}(t), \ i_{M}''(t) \Bigr ) \equiv  \Bigl ( i(t), \ -i'(t) + E, \ i''(t) \Bigr ), 
\end{equation}
where $i_{M}(t)=i(t)$. 
We show the reconstructed attractors in Figure \ref{fig:Rossler-reconstruction}.  
%---Fig. 33-------%
\begin{figure}[hpbt]
 \begin{center}
  \psfig{file=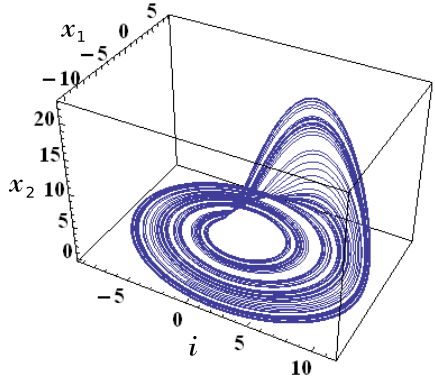, height=5.5cm}
  \caption{Chaotic attractor of the memristor R\"ossler equations (\ref{eqn: rossler-2}). 
  Parameters: $a = 0.2, \ b = E = 0.2, \ c = 5.7$.
  Initial conditions: $i(0)=0.1, \  x_{1}(0)=0.1, \  x_{2}(0)=0.1$.}
  \label{fig:Rossler-attractor} 
 \end{center}
\end{figure}
%
%

%---Fig. 34-------%
\begin{figure}[hpbt]
 \begin{center}
  \psfig{file=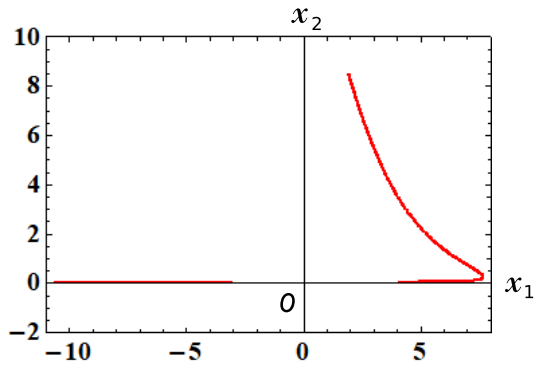, height=4.8cm}
  \caption{Poincar\'e map of the memristor R\"ossler equations (\ref{eqn: rossler-2}).    
  Its Poincar\'e cross-section (plane) is defined by $\{(i, \, x_{1}, \, x_{2}) \in R^{3} \ | \ i=0 \}$.  
  The trajectory of Eq. (\ref{eqn: rossler-2}) crosses the above Poincar\'e cross-section (plane) many times.   
  From Figures \ref{fig:Rossler-attractor} and \ref{fig:Rossler-poincare}, 
  we can observe that a chaotic attractor is a simple stretched and folded ribbon. 
  Parameters: $a = 0.2, \ b = E = 0.2, \ c = 5.7$.
  Initial conditions: $i(0)=0.1, \  x_{1}(0)=0.1, \  x_{2}(0)=0.1$.}
  \label{fig:Rossler-poincare} 
 \end{center}
\end{figure}
%
%

%---Fig. 35-------%
\begin{figure}[hpbt]
 \begin{center}
  \psfig{file=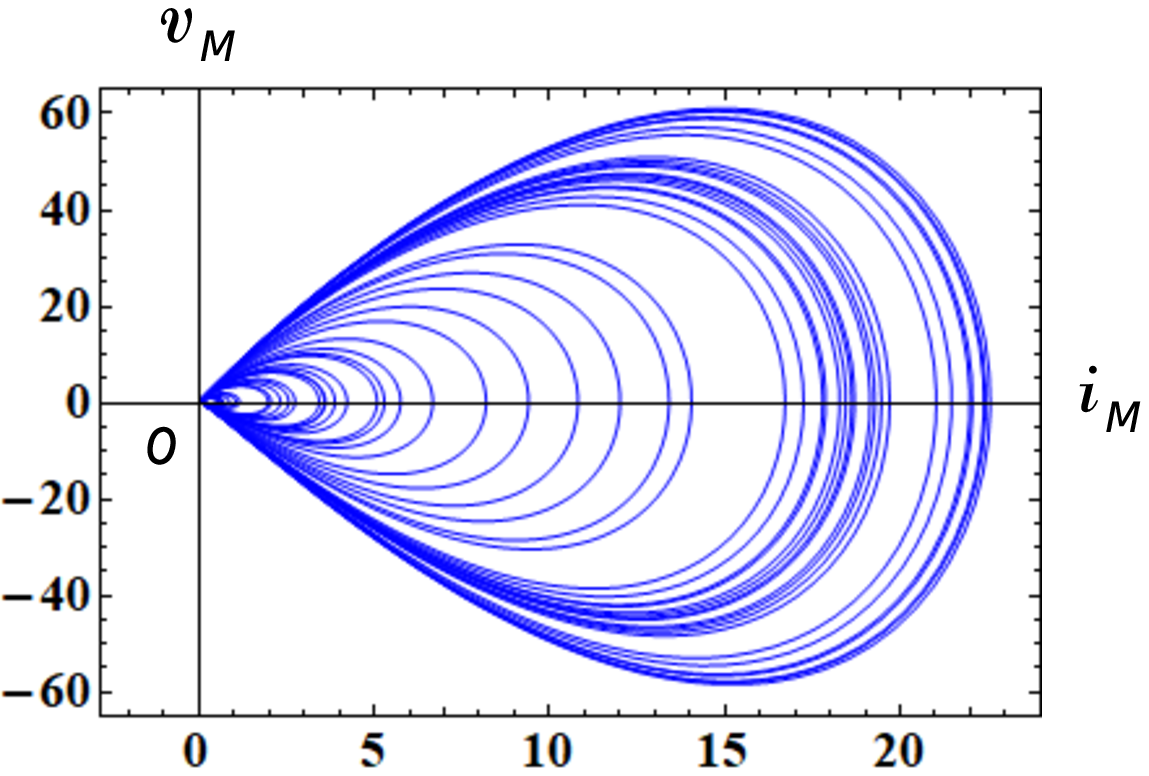, height=4.8cm}
  \caption{ The $i_{M}-v_{M}$ locus of the memristor R\"ossler equations (\ref{eqn: rossler-2}).  
   Here, $v_{M}$ and  $i_{M}$ denote the terminal voltage and the terminal current of the current-controlled generic memristor.  
   Parameters: $a = 0.2, \ b = E = 0.2, \ c = 5.7$.
   Initial conditions: $i(0)=0.1, \  x_{1}(0)=0.1, \  x_{2}(0)=0.1$.}
  \label{fig:Rossler-pinch} 
 \end{center}
\end{figure}
%
%

%---Fig. 36-------%
\begin{figure}[hpbt]
  \centering
  \psfig{file=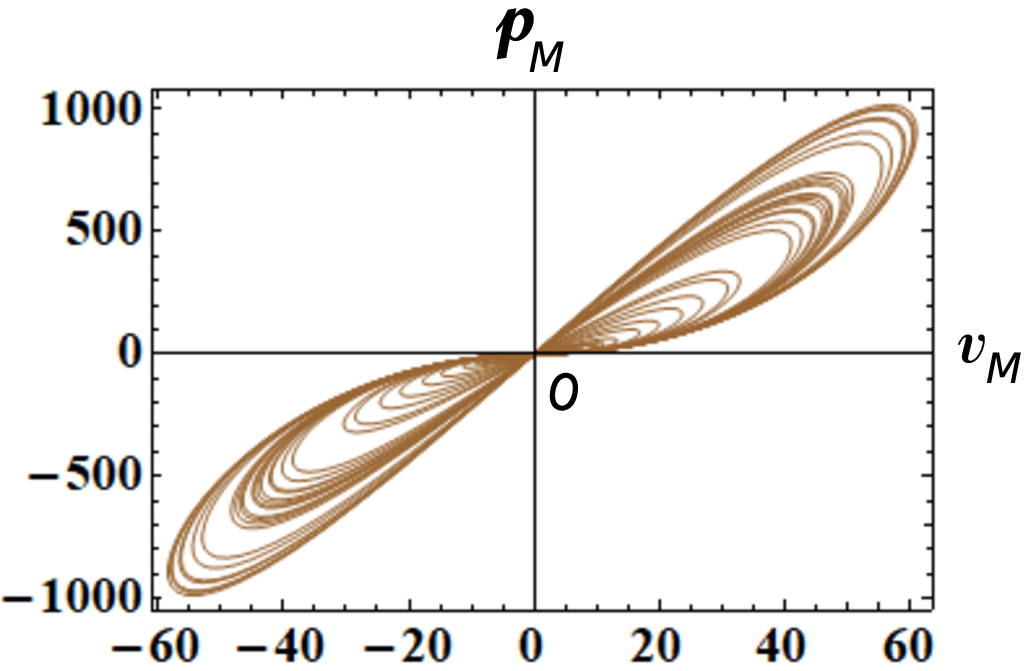, height=5cm}
  \caption{ The $v_{M}-p_{M}$ locus of the R\"ossler equations (\ref{eqn: rossler-2}).  
   Here, $p_{M}(t)$ is an instantaneous power defined by $p_{M}(t)=i_{M}(t)v_{M}(t)$, 
   and $v_{M}(t)$ and $i_{M}(t)$ denote the terminal voltage and the terminal current of the current-controlled generic memristor.   
   Observe that the $v_{M}-p_{M}$ locus is pinched at the origin, and the locus lies in the first and the third quadrants.  
   The memristor switches between passive and active modes of operation, depending on its terminal voltage $v_{M}(t)$.
   Parameters: $a = 0.2, \ b = E = 0.2, \ c = 5.7$.
   Initial conditions: $i(0)=0.1, \  x_{1}(0)=0.1, \  x_{2}(0)=0.1$.}
  \label{fig:Rossler-power} 
\end{figure}
%
%

%---Fig. 37-------%
\begin{figure}[hpbt]
 \centering
   \begin{tabular}{cc}
    \psfig{file=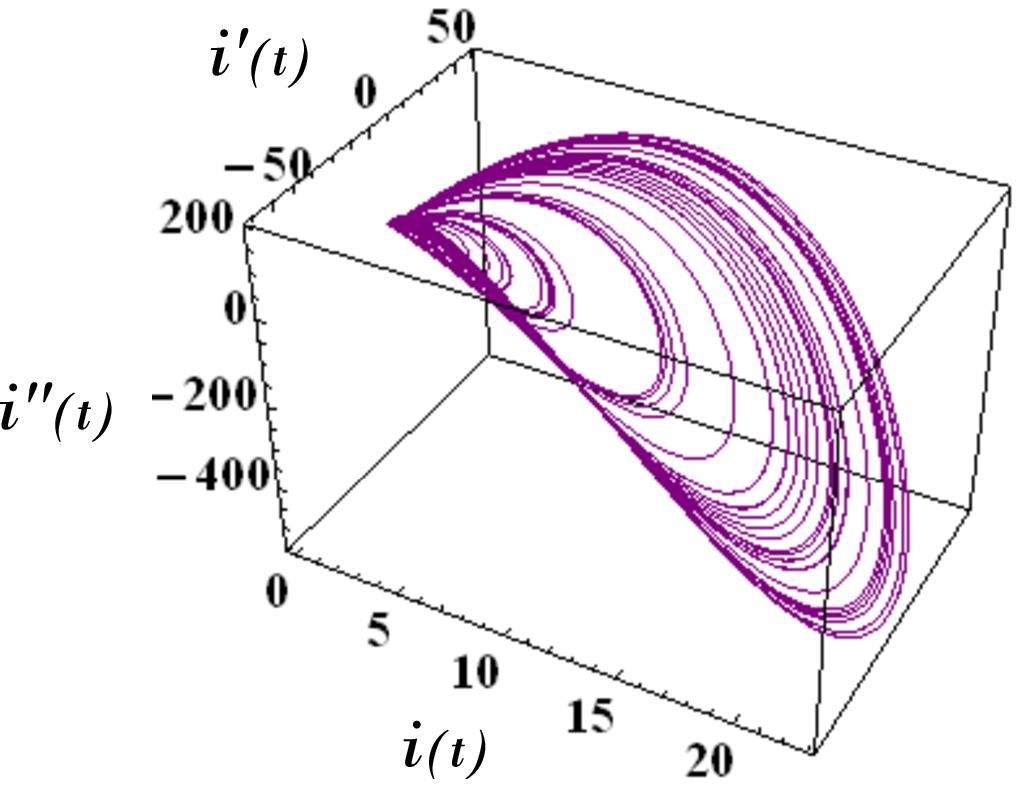, height=6cm}  & 
    \psfig{file=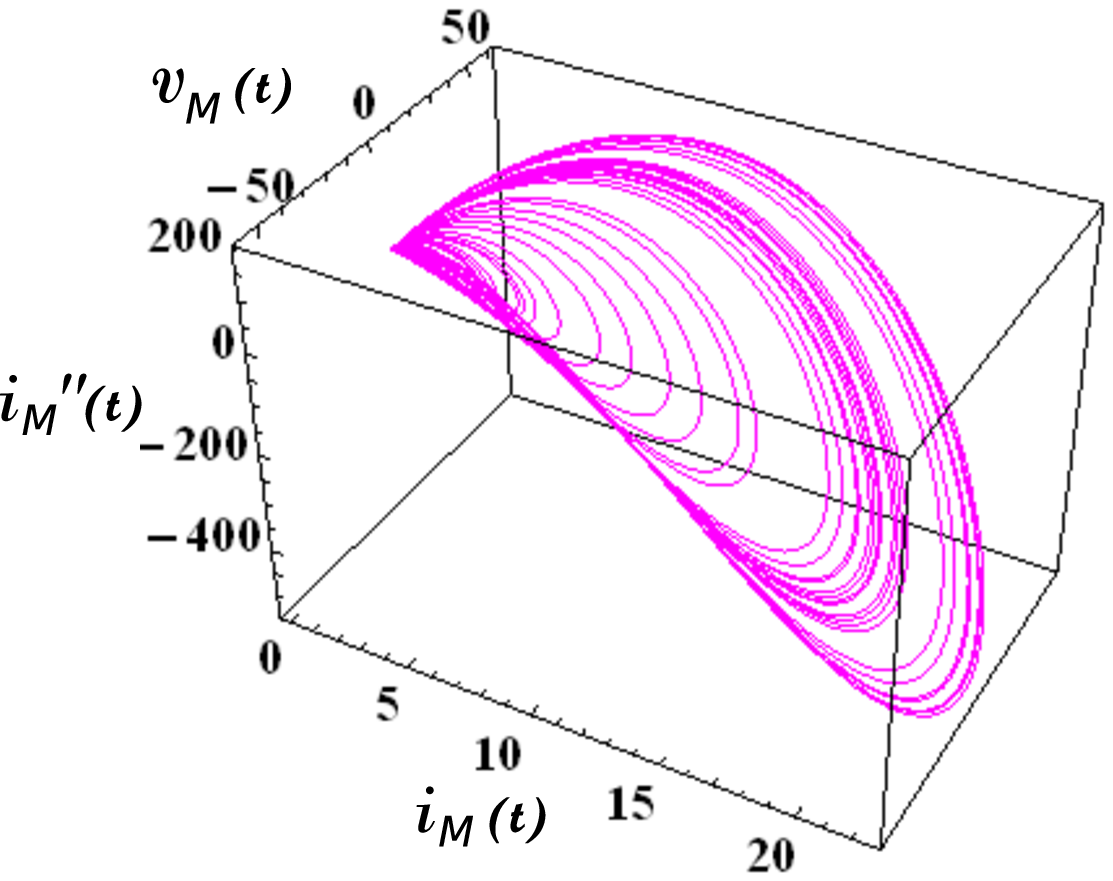, height=6cm}  \vspace{1mm} \\
   (a)  $\Bigl ( i(t), \ i'(t), \ i''(t) \Bigr )$ reconstruction & 
   (b)  $\Bigl ( i_{M}(t), \  v_{M}(t), \ {i_{M}}''(t) \Bigr )$ reconstruction \\ 
   \end{tabular}
  \caption{Two reconstructed chaotic attractors using $\Bigl ( i(t), \ i'(t), \ i''(t) \Bigr )$ 
   and  $\Bigl ( i_{M}(t), \  v_{M}(t), \ {i_{M}}''(t) \Bigr )$.   
   Parameters: $a = 0.2, \ b = E = 0.2, \ c = 5.7$.
   Initial conditions: $i(0)=0.1, \  x_{1}(0)=0.1, \  x_{2}(0)=0.1$.}
  \label{fig:Rossler-reconstruction} 
\end{figure}
\newpage

%==============================================================%
\subsection{$CO_{2}$ laser model}
%==============================================================%
%
The dynamics of the $CO_{2}$ laser model is given by \cite{Pisarchik(2001)}
\begin{center}
\begin{minipage}{10.5cm}
\begin{shadebox}
\underline{\emph{ $CO_{2}$ laser model equations}}
\begin{equation}
\left.
\begin{array}{lll}
\displaystyle  \frac{dX_{1}}{dt}  &=& k_{0} X_{1}  \left \{  X_{2}  - 1 - k_{1} \sin^{2} X_{6} \right \},	\vspace{1mm} \\	
 \displaystyle  \frac{dX_{2}}{dt}  &=& -\Gamma_{1} X_{2}  - 2 k_{0} X_{1}  X_{2}  + \gamma X_{3}  + X_{4}  + P_{0}, \vspace{1mm} \\
 \displaystyle  \frac{dX_{3}}{dt}  &=& -\Gamma_{1} X_{3}  + X_{5}  + \gamma X_{2}  + P_{0}, \vspace{1mm} \\
 \displaystyle  \frac{dX_{4}}{dt}  &=& -\Gamma_{2} X_{4}  + \gamma X_{5}  + z X_{2}  + z P_{0}, \vspace{1mm} \\
 \displaystyle  \frac{dX_{5}}{dt}  &=& -\Gamma_{2} X_{5}  + z X_{3}  + \gamma X_{4}  + z P_{0}, \vspace{1mm} \\
 \displaystyle  \frac{dX_{6}}{dt}  &=&  \displaystyle -b X_{6}  + b B_{0} - \frac{b R X_{1}}{ 1 + a X_{1} }.  
\end{array}
\right \}
\label{eqn: CO2-1}
\end{equation}
\end{shadebox}
\end{minipage}
\end{center}
Here, $X_{1}$ is the photon number proportional to the laser intensity, $X_{2}$ is proportional to the laser inversion, $X_{3}$ is proportional to the sum of the populations of the laser resonant levels, $X_{4}$ and $X_{5}$ are, respectively, proportional to the difference and sum of the populations of the rotational manifolds coupled to the resonant levels, and $X_{6}$ is a term proportional to the feedback voltage which acts on the cavity loss \cite{Pisarchik(2001)}. 
The parameters are chosen as follows: 
\begin{equation}
\left. 
\begin{array}{lll}
  k_{0} = 28.5714, &  k_{1} = 4.5556,  & \gamma = 0.05,   \vspace{2mm} \\
  P_{0} = 0.016,   & a = 32.8767,      & b = 0.4286,      \vspace{2mm} \\
  G_{1} = 10.0643, & G_{2} = 1.0643,   & z = 10.0,        \vspace{2mm} \\
  B_{0} = 0.1026,  &  R = 159, \, 160. & 
\end{array}
\right \}
\label{eqn: CO2-parameter}
\end{equation}

Let us consider the three-element memristor circuit in Figure \ref{fig:memristor-inductor-battery}.  
The dynamics of this circuit is given by Eq. (\ref{eqn: dynamics-1}).  
Assume that Eq. (\ref{eqn: dynamics-1}) satisfies 
\begin{equation}
 \left.
 \begin{array}{lll}
  E = 0, &&  L=1,    \vspace{2mm} \\
  \hat{R}(\bd{x}, \, i) &=& \displaystyle - k_{0} \left \{  x_{1}  - 1 - k_{1} \sin^{2} x_{5} \right \},                             \vspace{2mm} \\
  \tilde{f}_{1}(\bd{x}, \, i) &=&  -\Gamma_{1} x_{1}  - 2 k_{0} x_{1}  x_{1}  + \gamma x_{2}  + x_{3}  + P_{0},  \vspace{2mm} \\
  \tilde{f}_{2}(\bd{x}, \, i) &=&  -\Gamma_{1} x_{2}  + x_{4}  + \gamma x_{1}  + P_{0}, \vspace{2mm} \\
  \tilde{f}_{3}(\bd{x}, \, i) &=&  -\Gamma_{2} x_{3}  + \gamma x_{4}  + z x_{1}  + z P_{0}, \vspace{2mm} \\
  \tilde{f}_{4}(\bd{x}, \, i) &=&  -\Gamma_{2} x_{4}  + z x_{2}  + \gamma x_{3}  + z P_{0}, \vspace{2mm} \\
  \tilde{f}_{5}(\bd{x}, \, i) &=&  \displaystyle-b x_{5}  + b B_{0} - \frac{b R \, i}{ 1 + a \, i }.  
 \end{array}
 \right \} 
\end{equation}
Then we obtain 
\begin{center}
\begin{minipage}{10.5cm}
\begin{shadebox}
\underline{\emph{Memristor $CO_{2}$ laser model equations}}
\begin{equation}
\left. 
\begin{array}{lll}
 \displaystyle  \frac{di}{dt}  &=& k_{0} \left \{  x_{1}  - 1 - k_{1} \sin^{2} x_{5}  \right \} i ,	\vspace{2mm} \\	
 \displaystyle  \frac{dx_{1}}{dt}  &=& -\Gamma_{1} x_{1}  - 2 k_{0} x_{1}  x_{1}  + \gamma x_{2}  + x_{3}  + P_{0},  \vspace{2mm} \\
 \displaystyle  \frac{dx_{2}}{dt}  &=& -\Gamma_{1} x_{2}  + x_{4}  + \gamma x_{1}  + P_{0}, \vspace{2mm} \\
 \displaystyle  \frac{dx_{3}}{dt}  &=& -\Gamma_{2} x_{3}  + \gamma x_{4}  + z x_{1}  + z P_{0}, \vspace{2mm} \\
 \displaystyle  \frac{dx_{4}}{dt}  &=& -\Gamma_{2} x_{4}  + z x_{2}  + \gamma x_{3}  + z P_{0}, \vspace{2mm} \\
 \displaystyle  \frac{dx_{5}}{dt}  &=& \displaystyle-b x_{5}  + b B_{0} - \frac{b R \, i}{ 1 + a \, i },
\end{array}
\right \}
\label{eqn: CO2-2}
\end{equation}
\end{shadebox}
\end{minipage}
\end{center}
where
\begin{equation}
\left. 
\begin{array}{lll}
  X_{1}=i,     & X_{2}=x_{1}, &  X_{3}=x_{2},  \vspace{2mm} \\	
  X_{4}=x_{3}, & X_{5}=x_{4}, &  X_{6}=x_{5}.  
\end{array}
\right \}
\end{equation}
In this case, the extended memristor in Figure \ref{fig:memristor-inductor-battery} is replaced by the \emph{generic} memristor.   
That is,
\begin{equation}
  \hat{R}(\bd{x}, \, i)  = \tilde{R}(\bd{x})= - k_{0} \left \{  x_{1}  - 1 - k_{1} \sin^{2} x_{5} \right \}.
\end{equation}
The terminal voltage $v_{M}$ and the terminal current $i_{M}$ of the generic memristor are described
by
\begin{center}
\begin{minipage}{10.5cm}
\begin{shadebox}
\underline{\emph{V-I characteristics of the generic memristor}}
\begin{equation}
\left. 
\begin{array}{l}
  \scalebox{0.93}{$\displaystyle 
    v_{M} = \tilde{R}(\bd{x}) \, i_{M} = - k_{0} \left \{  x_{1}  - 1 - k_{1} \sin^{2} x_{5}  \right \} \, i_{M}, 
   $}   
  \vspace{3mm} \\
 \displaystyle  \frac{dx_{1}}{dt}  = -\Gamma_{1} x_{1}  - 2 k_{0} x_{1}  x_{1}  + \gamma x_{2}  + x_{3}  + P_{0},  \vspace{2mm} \\
 \displaystyle  \frac{dx_{2}}{dt}  = -\Gamma_{1} x_{2}  + x_{4}  + \gamma x_{1}  + P_{0}, \vspace{2mm} \\
 \displaystyle  \frac{dx_{3}}{dt}  = -\Gamma_{2} x_{3}  + \gamma x_{4}  + z x_{1}  + z P_{0}, \vspace{2mm} \\
 \displaystyle  \frac{dx_{4}}{dt}  = -\Gamma_{2} x_{4}  + z x_{2}  + \gamma x_{3}  + z P_{0}, \vspace{2mm} \\
 \displaystyle  \frac{dx_{5}}{dt}  = \displaystyle-b x_{5}  + b B_{0} - \frac{b R \, i}{ 1 + a \, i },
\end{array}
\right \}
\label{eqn: CO2-3}
\vspace{2mm}
\end{equation}
where $\tilde{R}(\bd{x})= - k_{0} \left \{  x_{1}  - 1 - k_{1} \sin^{2}  x_{5} \right \}$. \vspace{2mm}
\end{shadebox}
\end{minipage}
\vspace{10mm}
\end{center}
Thus, the $CO_{2}$ laser model (\ref{eqn: CO2-1}) can be realized by 
the three-element memristor circuit in Figure \ref{fig:memristor-inductor-battery}.  
Equations (\ref{eqn: CO2-1}) and (\ref{eqn: CO2-2}) can exhibit chaotic oscillation \cite{Pisarchik(2001)}.  
Thus, an external periodic forcing is unnecessary to generate chaotic or non-periodic oscillation. 

We next show the chaotic attractors and $i_{M}-v_{M}$ loci in Figures \ref{fig:CO2-attractor} and \ref{fig:CO2-pinch}, respectively.  
The $i_{M}-v_{M}$ locus in Figure \ref{fig:CO2-pinch} lies in the first and the fourth quadrants.  
Thus, the generic memristor defined by Eq. (\ref{eqn: CO2-3}) is an active element. 
Let us show the $v_{M}-p_{M}$ locus in Figure \ref{fig:CO2-power}, where $p_{M}(t)$ is an instantaneous power defined by $p_{M}(t)=i_{M}(t)v_{M}(t)$.  
Observe that the $v_{M}-p_{M}$ locus is pinched at the origin, and the locus lies in the first and the third quadrants. 
Thus, when $v_{M}>0$, the instantaneous power is dissipated in the memristor.  
However, when $v_{M}<0$, the instantaneous power is not dissipated in the memristor. 
Hence, the memristor switches between passive and active modes of operation, depending on its terminal voltage. 
In order to obtain these figures, we have to choose the parameters and the initial conditions carefully and the maximum step size $h$ of the numerical integration must be sufficiently small ($h=0.005$). 
It is due to the fact that a stable limit cycle (drawn in red) also coexists with a chaotic attractor (drawn in blue) as shown in Figure \ref{fig:CO2-cycle}.   
We conclude as follow: 
\begin{center}
\begin{minipage}{.7\textwidth}
\begin{itembox}[l]{Switching behavior of the memristor}
Assume that Eq. (\ref{eqn: CO2-2}) exhibits chaotic oscillation.  
Then the generic memristor defined by Eq. (\ref{eqn: CO2-3}) can switch between ``passive'' and ``active'' modes of operation, depending on its terminal voltage.  
\end{itembox}
\end{minipage}
\end{center}
%
%

%\newpage
We next show the Lorenz maps\footnote{The Lorenz map is defined as follow: Consider a chaotic solution $i(t)$ of the memristor $CO_{2}$ laser model equations (\ref{eqn: CO2-2}).  Extract the local maxima in $i(t)$ and label the $n$th local maximum $i_{n}$.  Plot of the maximum $i_{n+1}$ versus the maximum $i_{n}$.  The above constructed map is called Lorenz map, and it gives a well-defined relation between successive peaks, that is, we can estimate the peak knowing the peak.  
Furthermore, we need only one state variable to construct the Lorenz map.}  in Figure \ref{fig:CO2-map}, instead of Poincar\'e maps, since Eq.(\ref{eqn: CO2-2}) is the high-dimensional differential equations.
The Lorenz map in Figure \ref{fig:CO2-map}(b) represents a repeated folding and stretching of the space on which it is defined.   
In order to view the above folding action from a different perspective, let us plot the point $(x_{2}, \, x_{3})$ when $x_{6} = K$ ($K$ is a constant).   
This is the simplified version of Poincar\'e map.
We show the above plot in Figure \ref{fig:CO2-poincare}.   
Observe that Figure \ref{fig:CO2-poincare}(b) exhibits the repeated folding and stretching action.   

Finally, we reconstruct the chaotic attractors by using the time-delayed current signal (For more details, see \cite{Packard} ), that is, 
\begin{equation}
  \Bigl ( i(t), \ i(t-1) \Bigr ).   
\end{equation}
The reconstructed chaotic attractors are shown in Figure \ref{fig:CO2-reconstruction}.  
Observe that the chaotic attractors in Figure \ref{fig:CO2-reconstruction} are similar to those in Figure \ref{fig:CO2-attractor}.  
As stated in Sec. \ref{sec: Brusselator}, the $i_{M}-v_{M}$ locus in Figure \ref{fig:CO2-pinch} is considered to be the reconstruction of the chaotic attractor on the two-dimensional plane.

%---Fig. 38-------%
\begin{figure}[hpbt]
 \centering
   \begin{tabular}{ccc}
    \psfig{file=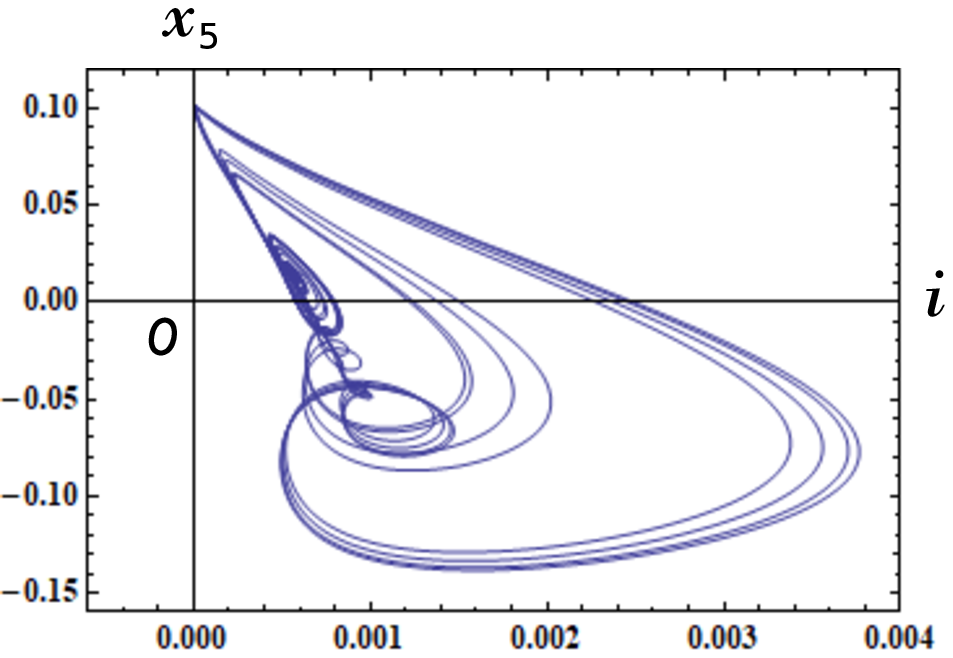, height=4.7cm}  & \hspace{5mm} & 
    \psfig{file=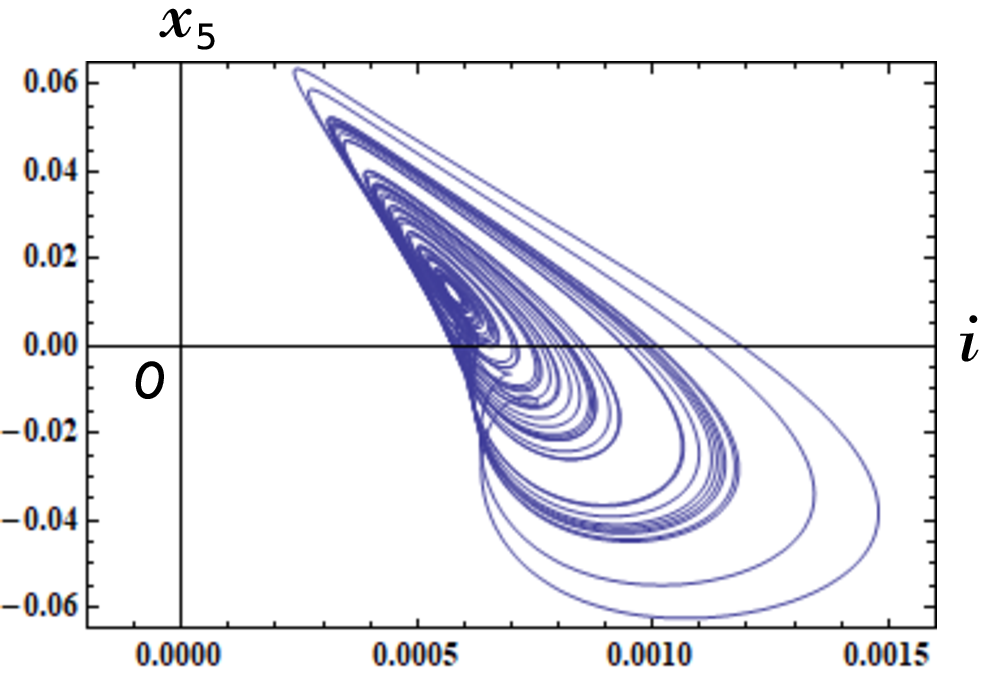, height=4.7cm}  \vspace{1mm} \\
    (a) $(i, \, x_{6})$-plane  && (d)   $(i, \, x_{6})$-plane   \vspace{2mm} \\
    \psfig{file=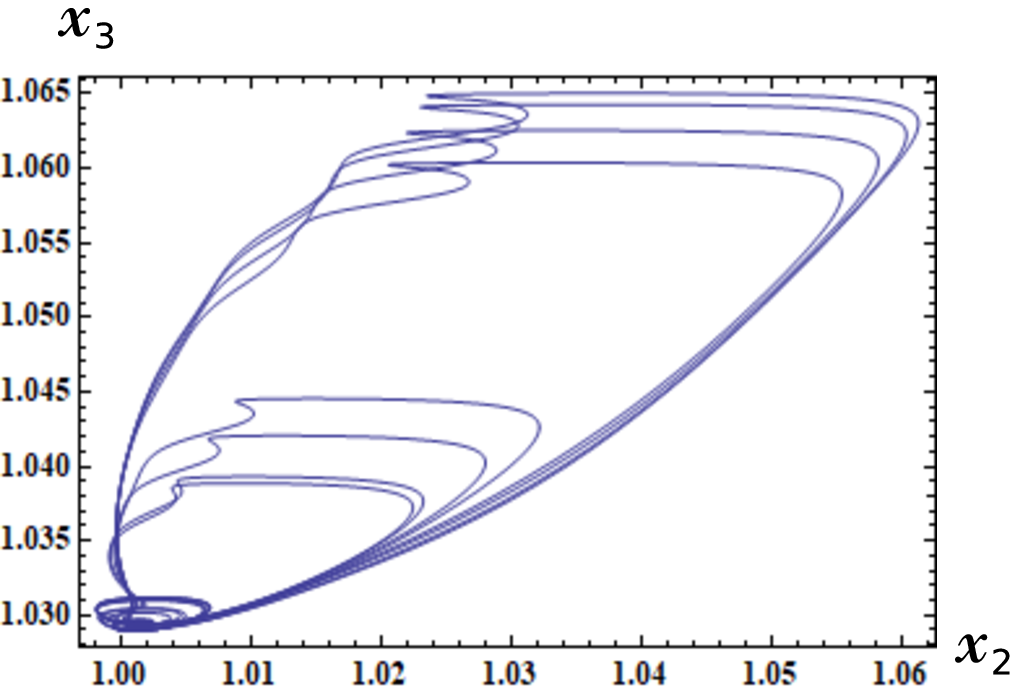, height=4.7cm} & \hspace{5mm} & 
    \psfig{file=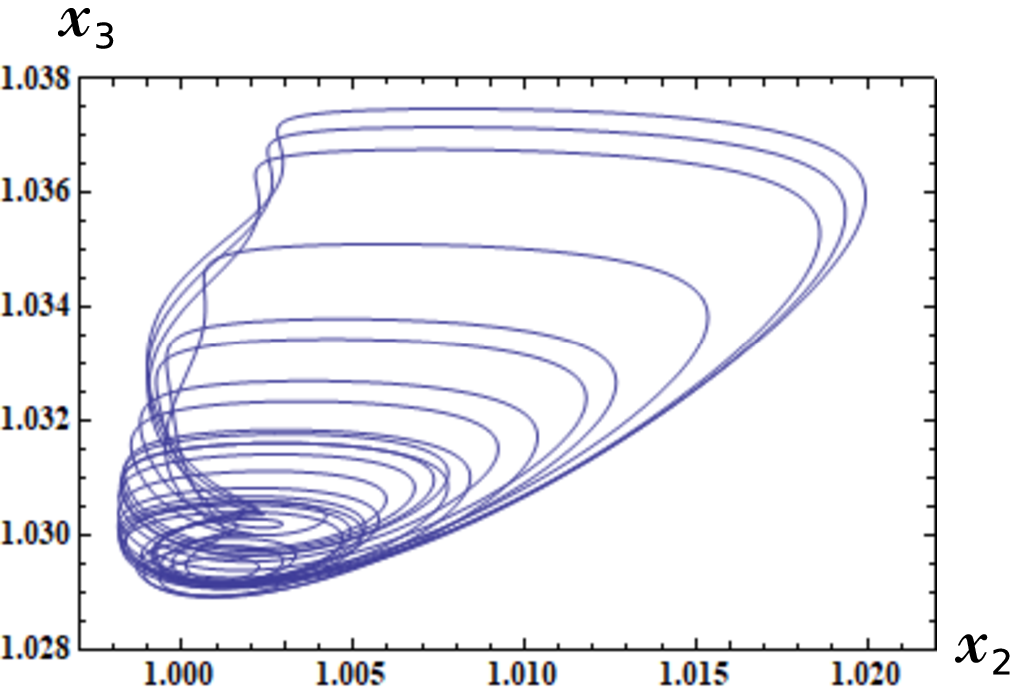, height=4.7cm}  \vspace{1mm} \\
    (b) $(x_{2}, \, x_{3})$-plane && (e)  $(x_{2}, \, x_{3})$-plane  \vspace{2mm} \\
    \psfig{file=./figure/CO2-attractor-10-2.eps, height=4.7cm}  & \hspace{5mm} & 
    \psfig{file=./figure/CO2-attractor-11-2.eps, height=4.7cm}  \vspace{1mm} \\
    (c) $(x_{4}, \, x_{5})$-plane && (f) $(x_{4}, \, x_{4})$-plane  \\
   \end{tabular}
  \caption{Chaotic attractors of the memristor CO2 laser model equations (\ref{eqn: CO2-2}).  
  All parameters except for $R$ are given in Eq. (\ref{eqn: CO2-parameter}).  
  Parameters: (a)-(c) $R=159$, \ \ (d)-(f) $ R=160$.
  Initial conditions: $i(0) = x_{1}(0) = x_{2}(0) = x_{3}(0) =x_{4}(0)= x_{5}(0)= 0.1$.}
  \label{fig:CO2-attractor} 
\end{figure}
%
%

%---Fig. 39-------%
\begin{figure}[hpbt]
 \centering
   \begin{tabular}{cc}
    \psfig{file=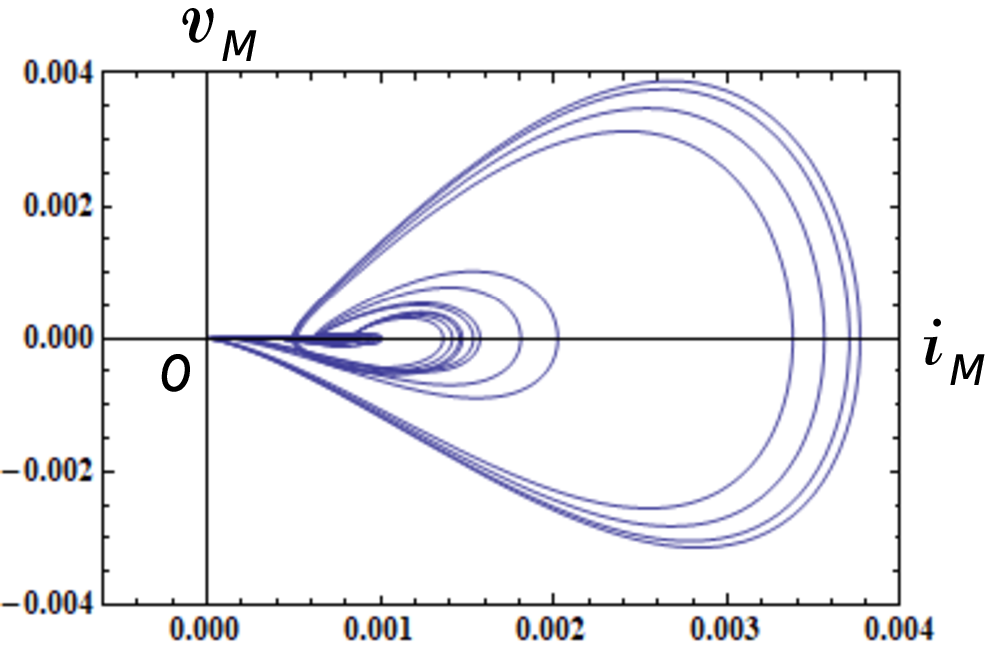, height=4.7cm}  & 
    \psfig{file=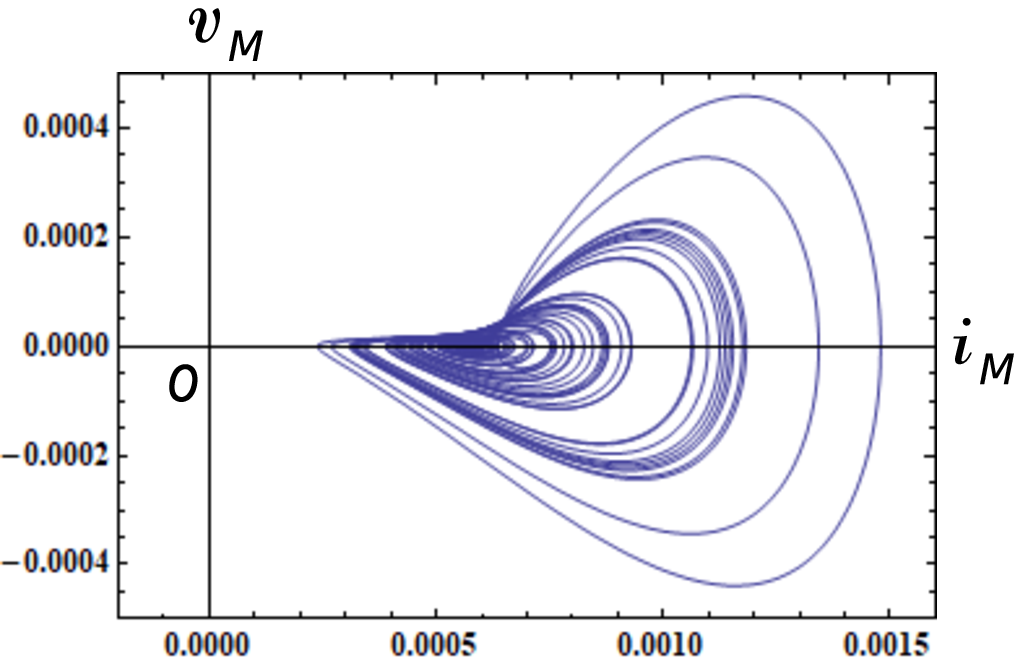, height=4.7cm}  \vspace{1mm} \\
   (a) $R=159$ & (b)  $R=160$ 
   \end{tabular}
  \caption{ The $i_{M}-v_{M}$ locus of the memristor CO2 laser model equations (\ref{eqn: CO2-2}).  
   Here, $v_{M}$ and  $i_{M}$ denote the terminal voltage and the terminal current of the current-controlled generic memristor.  
   All parameters except for $R$ are given in Eq. (\ref{eqn: CO2-parameter}). 
   Initial conditions: $i(0) = x_{1}(0) = x_{2}(0) = x_{3}(0) =x_{4}(0)= x_{5}(0)= 0.1$.}
  \label{fig:CO2-pinch} 
\end{figure}

%---Fig. 40-------%
\begin{figure}[hpbt]
 \centering
   \begin{tabular}{cc}
    \psfig{file=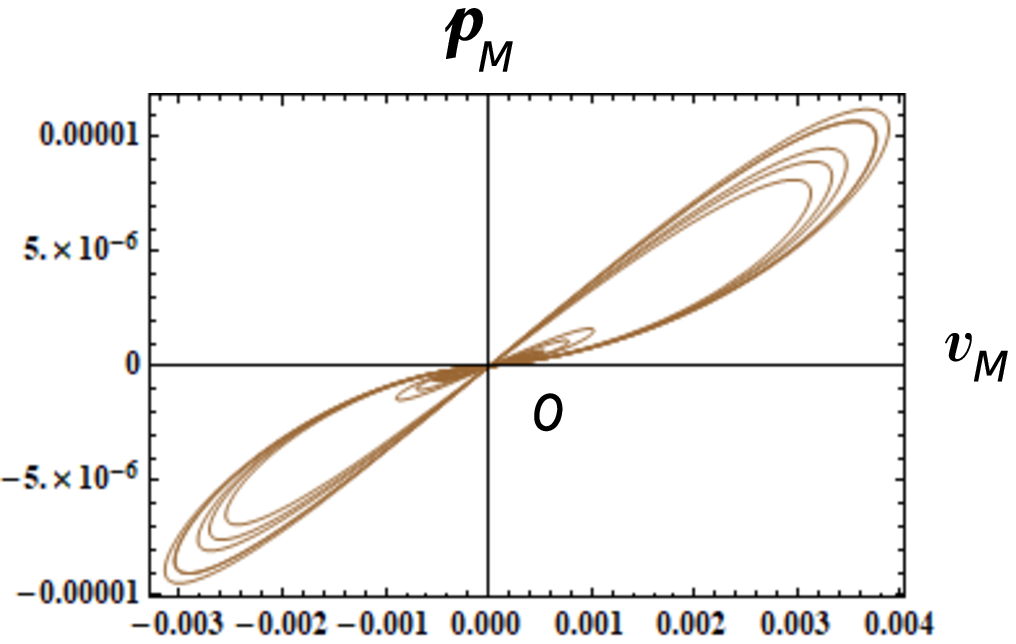, height=4.7cm}  & 
    \psfig{file=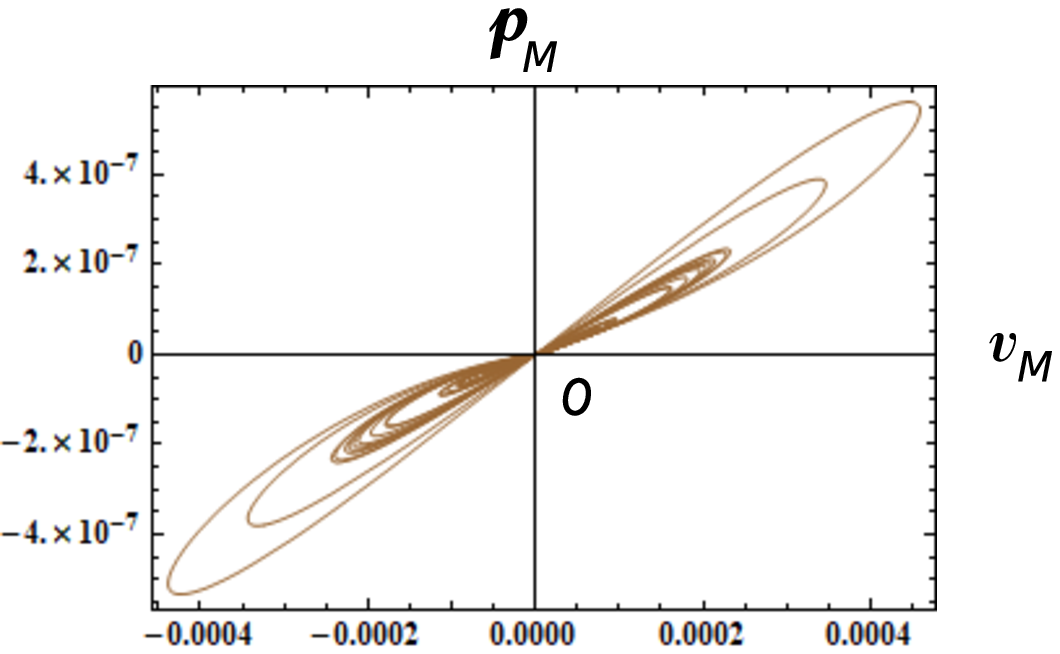, height=4.7cm}  \vspace{1mm} \\
    (a) $R=159$ & (b)  $R=160$  
   \end{tabular}
  \caption{ The $v_{M}-p_{M}$ locus of the memristor CO2 laser model equations (\ref{eqn: CO2-2}).  
   Here, $p_{M}(t)$ is an instantaneous power defined by $p_{M}(t)=i_{M}(t)v_{M}(t)$,   
   and $v_{M}(t)$ and $i_{M}(t)$ denote the terminal voltage and the terminal current of the current-controlled generic memristor.  
   Observe that the $v_{M}-p_{M}$ locus is pinched at the origin, and the locus lies in the first and the third quadrants. 
   The memristor switches between passive and active modes of operation, depending on its terminal voltage $v_{M}(t)$.
   All parameters except for $R$ are given in Eq. (\ref{eqn: CO2-parameter}).  
   Initial conditions: $i(0) = x_{1}(0) = x_{2}(0) = x_{3}(0) =x_{4}(0)= x_{5}(0)= 0.1$.}
  \label{fig:CO2-power} 
\end{figure}
%
%

%---Fig. 41-------%
\begin{figure}[hpbt]
 \centering
   \begin{tabular}{cc}
    \psfig{file=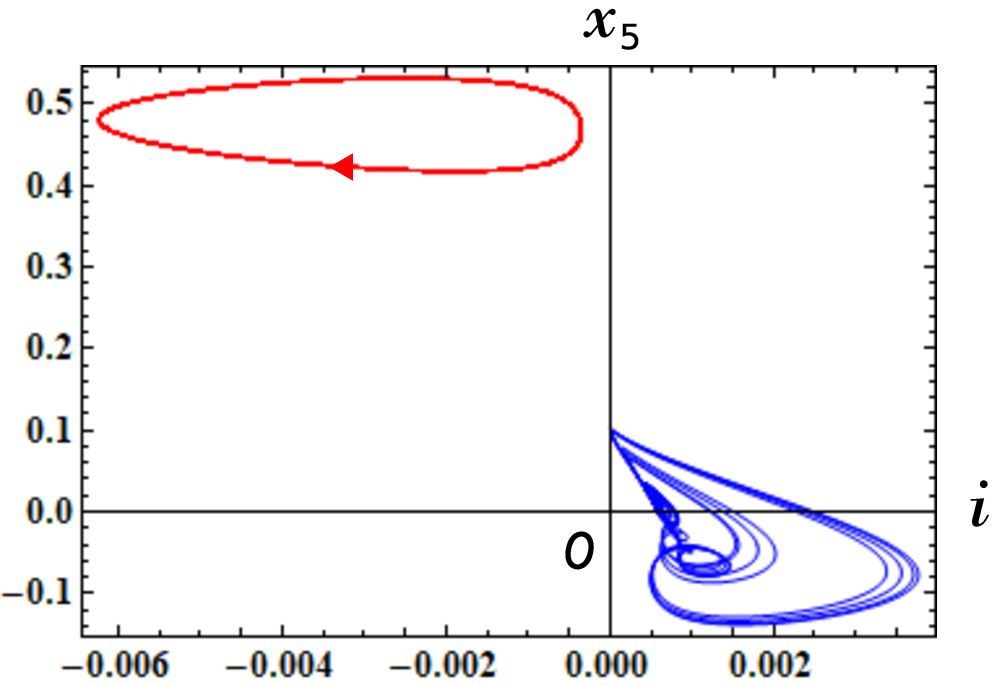, height=5cm}  & 
    \psfig{file=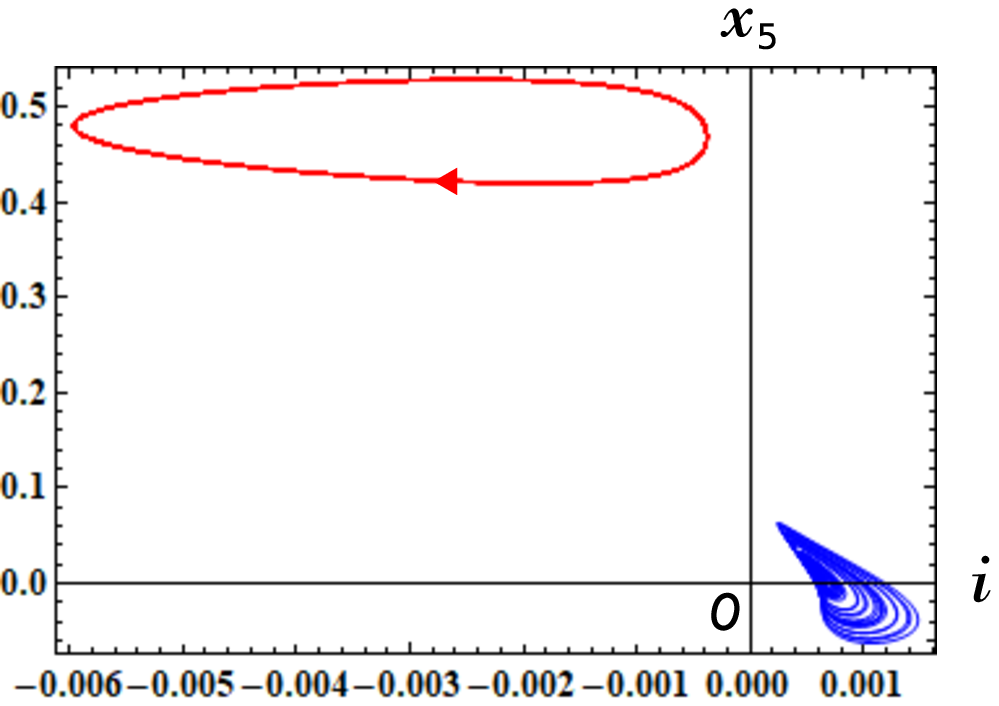, height=5cm}  \vspace{1mm} \\
   (a) $R=159$ & (b)  $R=160$ \\ 
   \end{tabular}
  \caption{A stable limit cycle (red) coexists with a chaotic attractor (blue).  
   All parameters except for $R$ are given in Eq. (\ref{eqn: CO2-parameter}). 
   Initial conditions for chaotic attractors: $i(0) = x_{1}(0) = x_{2}(0) = x_{3}(0) =x_{4}(0)= x_{5}(0)= 0.1$. \ \  
   Initial conditions for a limit cycles: $i(0) = x_{1}(0) = x_{2}(0) = x_{3}(0) =x_{4}(0)= 0.1, \, x_{5}(0)= 5.1$.  }
  \label{fig:CO2-cycle} 
\end{figure}
%
%

%---Fig. 42-------%
\begin{figure}[hpbt]
 \centering
   \begin{tabular}{cc}
    \psfig{file=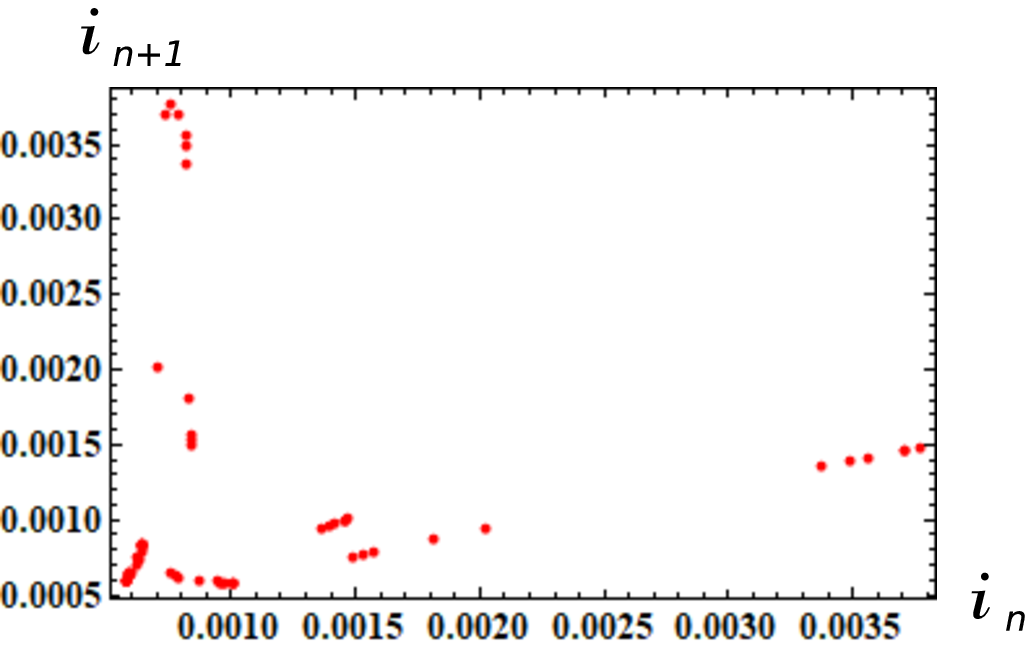, height=4.8cm}  & 
    \psfig{file=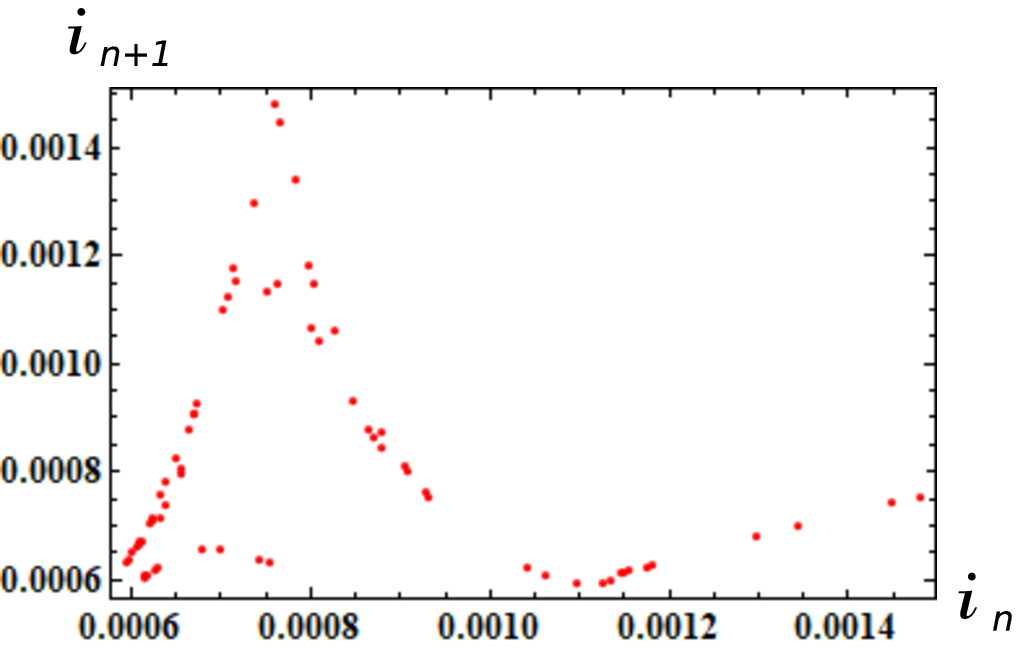, height=4.8cm}  \vspace{1mm} \\
     (a) $R=159$ & (b)  $R=160$ \\ 
   \end{tabular}
  \caption{Lorenz maps of the memristor $CO_{2}$ laser model equations (\ref{eqn: CO2-2}).  
   All parameters except for $R$ are given in Eq. (\ref{eqn: CO2-parameter}). 
   Initial conditions: $i(0) = x_{1}(0) = x_{2}(0) = x_{3}(0) =x_{4}(0)= x_{5}(0)= 0.1$.}
  \label{fig:CO2-map} 
\end{figure}
%
%

%---Fig. 43-------%
\begin{figure}[hpbt]
 \centering
   \begin{tabular}{c}
   \psfig{file=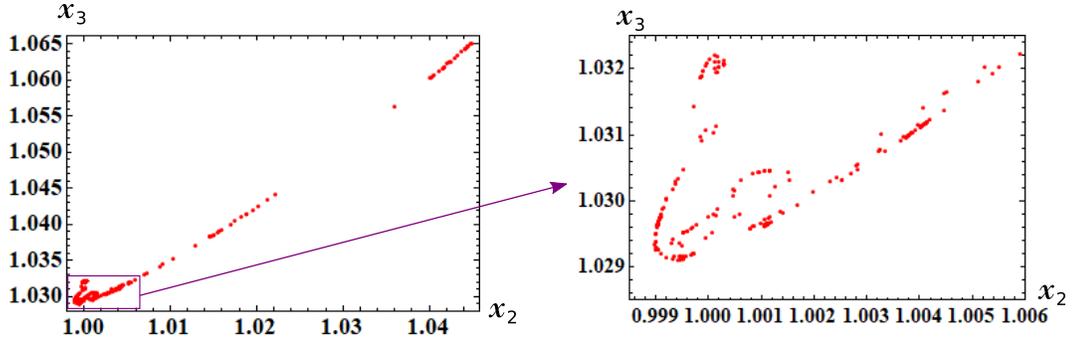, height=4.5cm} \\ 
   (a) Poincar\'e map for $x_{6}=0.01$ and $R=159$.  \vspace{5mm}\\   
   \psfig{file=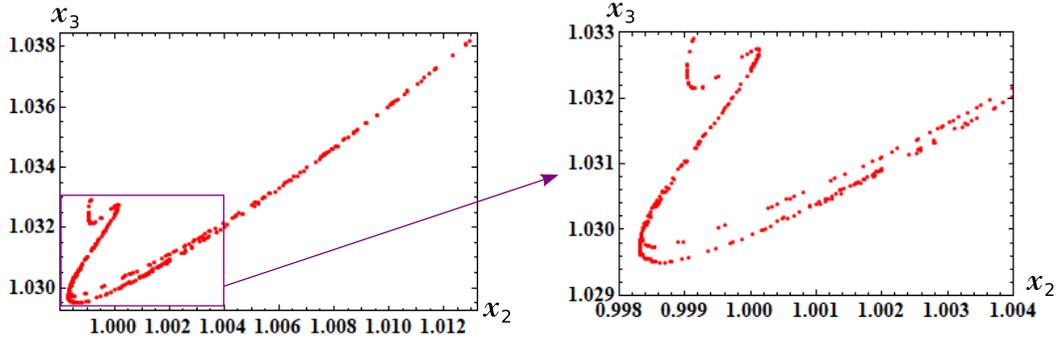, height=4.5cm} \\
   (b) Poincar\'e map for $x_{6}=0$ and $R=160$. 
  \end{tabular}
  \caption{Two-dimensional plot of the memristor CO2 laser model equations (\ref{eqn: CO2-2}).  
  (a)  The point $(x_{2}, \, x_{3})$ is plotted when $x_{6}=0.01$.  
  (b)  The point $(x_{2}, \, x_{3})$ is plotted when $x_{6}=0$.  
  Observe that Figure \ref{fig:CO2-poincare}(b) exhibits the repeated folding and stretching action, 
  and Figure \ref{fig:CO2-poincare}(a) exhibits the distorted folding action.  
  All parameters except for $R$ are given in Eq. (\ref{eqn: CO2-parameter}).  
  Initial conditions: $i(0) = x_{1}(0) = x_{2}(0) = x_{3}(0) =x_{4}(0)= x_{5}(0)= 0.1$.}
  \label{fig:CO2-poincare} 
\end{figure}
%
%

%---Fig. 44-------%
\begin{figure}[hpbt]
 \centering
   \begin{tabular}{ccc}
    \psfig{file=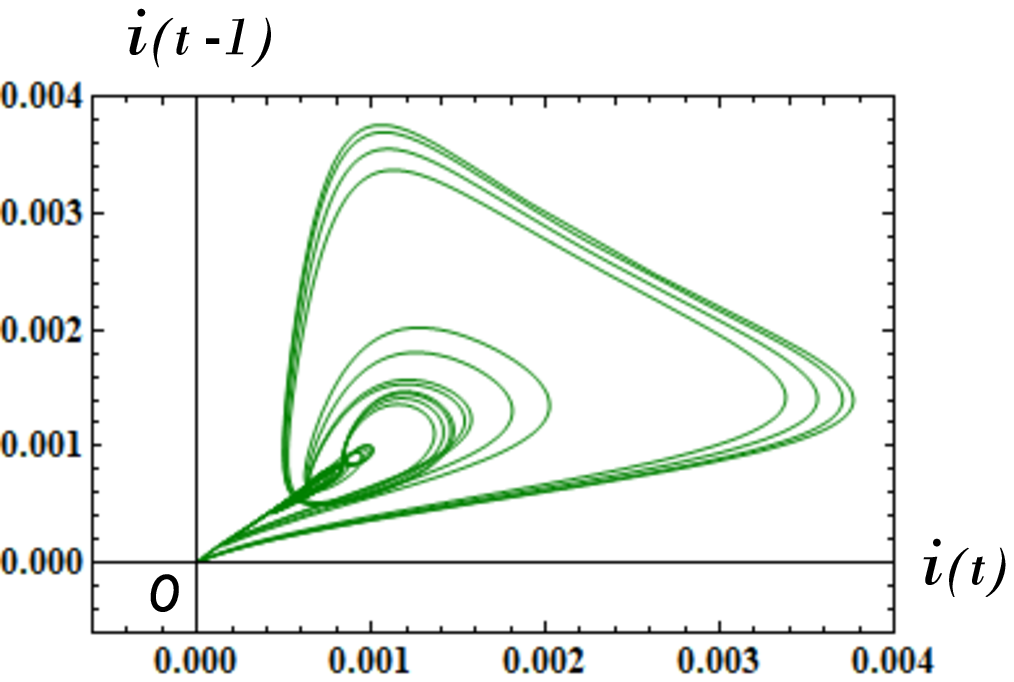, height=4.5cm}  & \hspace{2mm} &
    \psfig{file=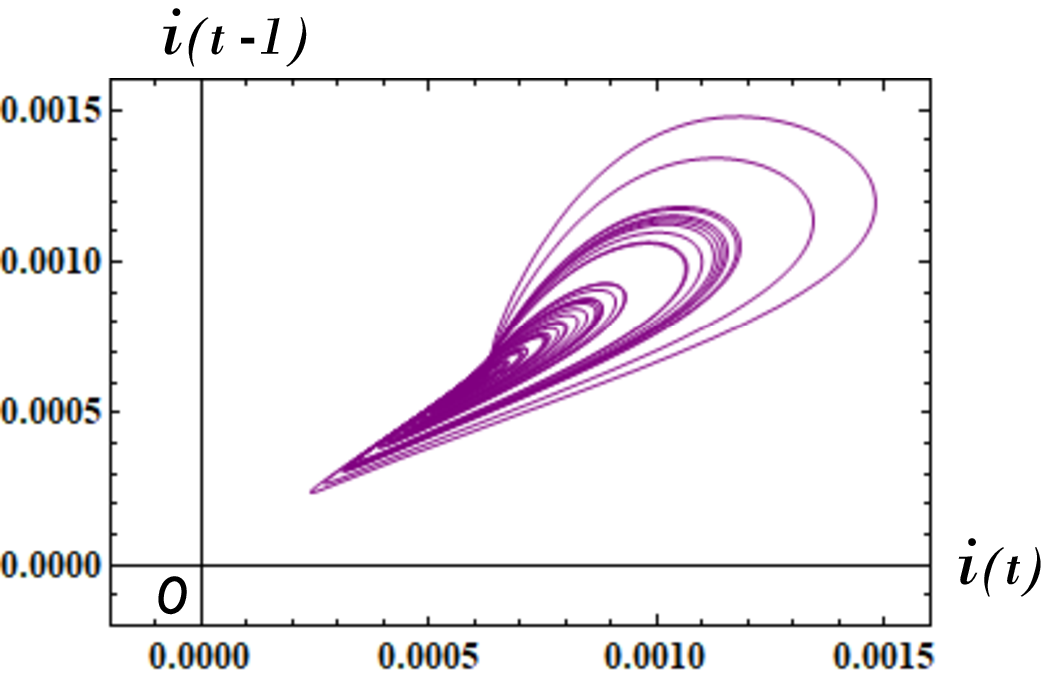, height=4.5cm}  \vspace{1mm} \\
    (a) $R=159$ & &(b)  $ R=160$ \\ 
   \end{tabular}
  \caption{Reconstructed chaotic attractors using 
   $\Bigl ( i(t), \ i(t-1) \Bigr )$.   
   Observe that the chaotic attractors in Figure \ref{fig:CO2-reconstruction} are similar to those in Figure \ref{fig:CO2-attractor}.  
   All parameters except for $R$ are given in Eq. (\ref{eqn: CO2-parameter}). 
   Initial conditions: $i(0) = x_{1}(0) = x_{2}(0) = x_{3}(0) =x_{4}(0)= x_{5}(0)= 0.1$.}
  \label{fig:CO2-reconstruction} 
\end{figure}
\newpage

%\clearpage
%
%
%%%%%%%%%%%%%%%%%%%%%%%%%%%%%%%%%%%%%%%%%%%%%%%%%%%%%%%%%%%%%%%%%%%%%%%
\section{$2N$-element memristor circuit}
\label{sec: N-memristors}
%%%%%%%%%%%%%%%%%%%%%%%%%%%%%%%%%%%%%%%%%%%%%%%%%%%%%%%%%%%%%%%%%%%%%%%
%
%
%
Let us consider the $2N$-element memristor circuit shown in Figure \ref{fig:memristor-inductor-N}.   
It consists of $N$ inductors with the inductance $L_{n} \ (n=1, \ 2, \cdots \ N)$ and $N$ extended memristors described by 
\begin{center}
\begin{minipage}{8.7cm}
\begin{shadebox}
\underline{\emph{V-I characteristics of the extended memristors}} 
\vspace{1mm} 
\begin{equation}
\begin{array}{cll}
 v_{n} &=& \hat{R}( \bd{x}, \ i_{n} ) \, i_{n}, \vspace{2mm} \\
  \displaystyle \frac{d\bd{x}}{dt} &=& \tilde{\bd{f}}(\bd{x}, \ \bd{i}).
\end{array}
\vspace{2mm}
\label{eqn: extended-N}
\end{equation}
\end{shadebox}
\end{minipage}
\end{center}
Here, 
$\bd{x} = (x_{1}, \, x_{2}, \, \cdots, \, x_{N}) \in \mathbb{R}^{N}$,  $\bd{i} = (i_{1}, \, i_{2}, \, \cdots, \, i_{N}) \in \mathbb{R}^{N}$,
$\hat{R}( \bd{x}, \ i_{n} )$ is a continuous scalar-valued function, 
and $\tilde{\bd{f}} = (\tilde{f}_{1}, \, \tilde{f}_{2}, \, \cdots, \, \tilde{f}_{N}): \mathbb{R}^{N} \rightarrow \mathbb{R}^{N}$.  
Even though the extended memristors in Figure \ref{fig:memristor-inductor-N} appear to be disconnected, their dynamics are coupled via the memristance equation involving the same state variables $\bd{x} = (x_{1}, \ x_{2}, \ \cdots, \ x_{N})$ and $\bd{i} = (i_{1}, \, i_{2}, \, \cdots, \, i_{N})$.   
Note the memristor defined by Eq. (\ref{eqn: extended-N}) is considered to be a special case of the \emph{extended memristor} defined in Appendix A.  
That is, we modified the current $i$ of Eq. (\ref{eqn: extended-2}) into the vector form $\bd{i} = (i_{1}, \, i_{2}, \, \cdots, \, i_{N})$.    

The dynamics of the above $2N$-element memristor circuit is given by 
\begin{center}
\begin{minipage}{8.7cm}
\begin{shadebox}
\underline{\emph{$2N$-element memristor circuit equations}}
\begin{equation}
\begin{array}{rll}
  \displaystyle L_{n} \frac{di_{n}}{dt} &=& - v_{n} = - \hat{R}( \bd{x}, \ i_{n} ) \, i_{n},
  \vspace{2mm} \\
   \displaystyle \frac{d\bd{x}}{dt} &=& \tilde{\bd{f}}(\bd{x}, \ \bd{i}), 
\end{array}
\vspace{2mm}
\label{eqn: dynamics-N}
\end{equation}
where $n=1, \ 2, \cdots, \ N$ and we usually assume $L_{n}=1$.  
\end{shadebox}
\end{minipage}
\end{center}
%
%

%---Fig. 45-------%
\begin{figure}[hpbt]
 \centering
  \psfig{file=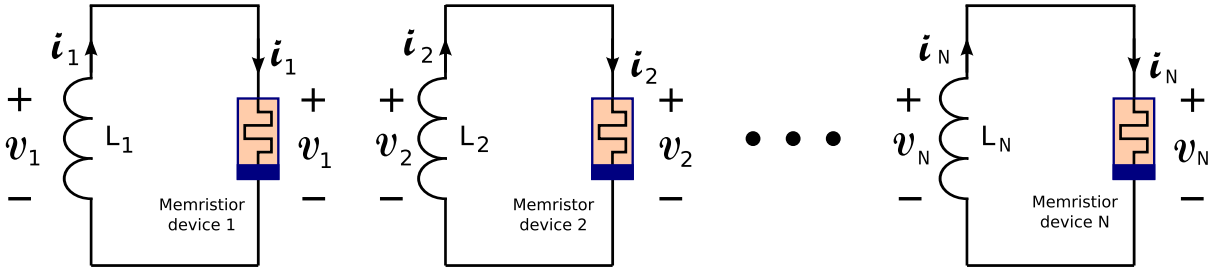, width=16cm} 
  \caption{A $2N$-element extended memristor circuit, 
  which consists of $N$ inductors with the inductance $L_{k} \ (k=1, \ 2, \cdots N)$ 
  and $N$ memristor devices described by 
  $ \displaystyle v_{k} = \hat{R}_{k}(\bd{x}, \, i_{k}) \ i_{k}$ $(k=1, \ 2, \cdots N) $.  \ \ 
  $\frac{d\bd{x}}{dt} = \tilde{\bd{f}}(\bd{x}, \ \bd{i})$, 
  Here, $\hat{R}_{k}(\bd{x}, \, i_{k})$ denotes the memristance of the $k$th extended memristor device,  
  $\bd{x} = (x_{1}, \, x_{2}, \, \cdots, \, x_{N}) \in \mathbb{R}^{n}$,  
  $\bd{i} = (i_{1}, \, i_{2}, \, \cdots, \, i_{N}) \in \mathbb{R}^{N}$, 
  and $\tilde{\bd{f}} = (\tilde{f}_{1}, \, \tilde{f}_{2}, \, \cdots, \, \tilde{f}_{N}): \mathbb{R}^{N} \rightarrow \mathbb{R}^{N}$.  
  Even though the extended memristor devices appear to be disconnected, 
  their dynamics are coupled via the memristance equation involving the same state variables 
  $\bd{x} = (x_{1}, \ x_{2}, \ \cdots, \ x_{N})$ and $\bd{i} = (i_{1}, \ i_{2}, \ \cdots, \ i_{N})$. }
  \label{fig:memristor-inductor-N} 
\end{figure}
%
%

%---Fig. 46-------%
\begin{figure}[hpbt]
 \centering
  \psfig{file=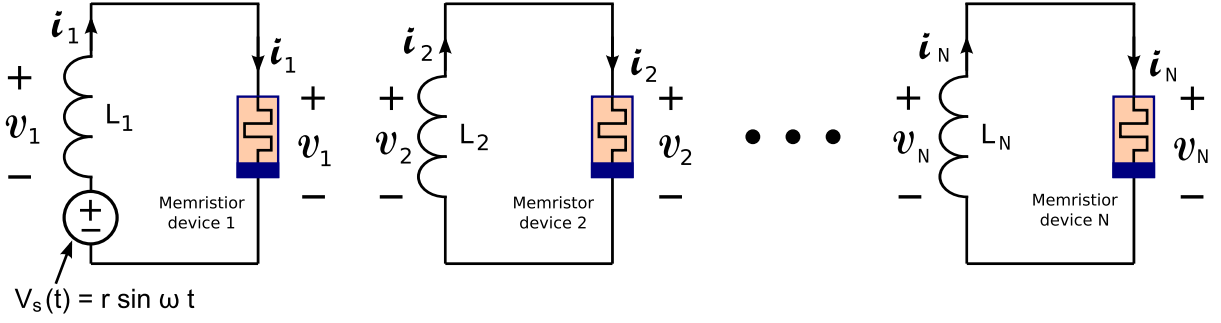, width=16cm} 
  \caption{A $2N$-element extended memristor circuit driven by a periodic voltage source $v_{s}(t) = r \sin ( \omega t)$, 
  where $r$ and $\omega$ are constants.}
  \label{fig:memristor-inductor-source-N} 
\end{figure}
%
%

%-------------------------------------%
\subsection{Toda lattice equations}
\label{sec: Toda-lattice}
%-------------------------------------%

Consider the Hamiltonian for a chain of particles with nearest neighbor exponential interaction \cite{Toda(1986), Toda(1989)} 
\begin{equation}
  \mathcal {H} = \sum_{n} \, \frac{1}{2} \, {p_{n}}^{2} + \sum_{n}  e^{- ( \, q_{n+1} - q_{n} \, )},   
\label{eqn: hamiltonian-toda}
\end{equation}
where $q_{n}$ is the displacement of the  $n$-th particle from its equilibrium position, 
and $p_{n}$  is its momentum  (mass $m = 1$).
Then, the Hamilton's Equations are given by 
\begin{center}
\begin{minipage}{8.7cm}
\begin{shadebox}
\underline{\emph{Toda lattice equations A}}
\begin{equation}
 \begin{array}{lll}
 \displaystyle \frac{d q_{n}}{dt}&=& \displaystyle p_{n}, \vspace{2mm} \\
 \displaystyle \frac{d p_{n}}{dt}&=& \displaystyle e^{-( \, q_{n}-q_{n-1} \, )}-e^{-( \, q_{n+1}-q_{n} \, )}, 
 \end{array}
\label{eqn: toda-a1}
\end{equation}
where $n=1, \ 2, \cdots, \ N$ and we consider the case of a periodic lattice of the length $N$: $q_{n} = q_{n + N}$. 
\end{shadebox}
\end{minipage}
\end{center}
%
%

%-------------------------------------%
\subsubsection{\underline{Toda lattice equations $B$}}
\label{sec: Toda-lattice-B}
%-------------------------------------%

Let us define a new variable 
\begin{equation}
  X_{n} = e^{-  ( \, q_{n+1} - q_{n} \, ) }.
\end{equation}
Then, Eq. (\ref{eqn: toda-a1}) can be recast into the form \cite{Henon(1974)}
\begin{center}
\begin{minipage}{8.7cm}
\begin{shadebox}
\underline{\emph{Toda lattice equations $B$}}
\begin{equation}
 \begin{array}{lll}
 \displaystyle \frac{d X_{n}}{dt}&=& \displaystyle (p_{n} - p_{n+1})X_{n}, \vspace{2mm} \\
 \displaystyle \frac{d p_{n}}{dt}&=& \displaystyle X_{n-1} - X_{n}, 
 \end{array}
\label{eqn: toda-B-1}
\end{equation}
where $n=1, \ 2, \cdots, \ N$ and we consider the case of a periodic lattice of the length $N$: $p_{n} = p_{n + N}$, $X_{n} = X_{n + N}$.   
\end{shadebox}
\end{minipage}
\end{center}

Consider the $2N$-element memristor circuit in Figure \ref{fig:memristor-inductor-N}.  
The dynamics of this circuit given by Eq. (\ref{eqn: dynamics-N}).  
Assume that Eq. (\ref{eqn: dynamics-N}) satisfies 
\begin{equation}
\begin{array}{cll}
  L_{n} &=& 1, \vspace{2mm} \\
  \hat{R}( \bd{x}, \ i_{n} ) &=& - (x_{n} - x_{n+1})  \vspace{2mm} \\
  \tilde{f}_{n}(\bd{x}, \bd{i}) &=& i_{n-1} - i_{n}.  
\end{array}
\end{equation}
Then we obtain 
\begin{center}
\begin{minipage}{9.5cm}
\begin{shadebox}
\underline{\emph{Memristor Toda lattice equations $B$}}
\begin{equation}
 \begin{array}{lll}
 \displaystyle \frac{d i_{n}}{dt}&=& \displaystyle (x_{n} - x_{n+1}) \, i_{n}, \vspace{2mm} \\
 \displaystyle \frac{d x_{n}}{dt}&=& \displaystyle i_{n-1} - i_{n},  
 \end{array}
\vspace{1mm}
\label{eqn: toda-B-3}
\end{equation}
where $n=1, \ 2, \cdots, \ N$ and we consider the case of a periodic lattice of the length $N$: $x_{n} = x_{n + N}$, $i_{n} = i_{n + N}$. 
\end{shadebox}
\end{minipage}
\end{center}
Equations (\ref{eqn: toda-B-1})  and (\ref{eqn: toda-B-3}) are equivalent if we change the variables  
\begin{equation}
  i_{n} = X_{n},  \  x_{n} = p_{n}.   
\end{equation}
In this case, the extended memristors in Figure \ref{fig:memristor-inductor-N} are replaced by the \emph{generic} memristors, that is,
\begin{equation}
  \hat{R}_{n}( \bd{x}, \ i_{n} ) = \tilde{R}_{n}(\bd{x}) = - (x_{n} - x_{n+1}), 
\end{equation}
though the current $i$ of Eq. (\ref{eqn: generic2-2}) is modified into the vector form $\bd{i} = (i_{1}, \, i_{2}, \, \cdots, \, i_{n})$.  
Thus, their terminal voltage $v_{n}$ and the terminal current $i_{n}$ are described by
\begin{center}
\begin{minipage}{9.5cm}
\begin{shadebox}
\underline{\emph{V-I characteristics of the generic memristors}} \vspace{1mm} 
\begin{equation}
\begin{array}{cll}
\displaystyle v_{n} &=& \tilde{R}_{n}(x_{n}, \, x_{n+1}) \ i_{n}  
                     = - (x_{n} - x_{n+1})i_{n}, \vspace{2mm} \\
\displaystyle \frac{dx_{n}}{dt} 
                    &=& \tilde{f}_{n}(i_{n-1}, \, i_{n}) 
                     =  i_{n-1} - i_{n},
\end{array}
\vspace{1mm}
\label{eqn: toda-B-4}
\end{equation}
where $x_{n} = x_{n + N}$, $i_{n} = i_{n + N}$, and $n=1, \ 2, \cdots, \ N$.
\end{shadebox}
\end{minipage}
\end{center}
It follows that Eq. (\ref{eqn: toda-B-1}) can be realized by the $2N$-element memristor circuit in Figure \ref{fig:memristor-inductor-N}.  

For $N=3$, Eq. (\ref{eqn: toda-B-1}) is given by 
\begin{center}
\begin{minipage}{8.7cm}
\begin{shadebox}
\underline{\emph{Toda lattice equations $B$ with $N=3$}}
\begin{equation}
\left.
 \begin{array}{lll}
 \displaystyle \frac{d X_{1}}{dt}&=& \displaystyle (p_{1} - p_{2})X_{1}, \vspace{1mm} \\
 \displaystyle \frac{d p_{1}}{dt}&=& \displaystyle X_{3} - X_{1},        \vspace{1mm} \\
 \displaystyle \frac{d X_{2}}{dt}&=& \displaystyle (p_{2} - p_{3})X_{2}, \vspace{1mm} \\
 \displaystyle \frac{d p_{2}}{dt}&=& \displaystyle X_{1} - X_{2},        \vspace{1mm} \\
 \displaystyle \frac{d X_{3}}{dt}&=& \displaystyle (p_{3} - p_{1})X_{3}, \vspace{1mm} \\
 \displaystyle \frac{d p_{3}}{dt}&=& \displaystyle X_{2} - X_{3}.          
 \end{array}
\right \}
\label{eqn: toda-B-1-1}
\end{equation}
\end{shadebox}
\end{minipage}
\end{center}
Equation (\ref{eqn: toda-B-1-1}) has the three integrals \cite{Henon(1974), Toda(1989)}, since the solution satisfies 
\begin{center}
\begin{minipage}{.5\textwidth}
\begin{itembox}[l]{Integrals}
\begin{equation}
 \begin{array}{l}
  \displaystyle \frac{d}{dt} \bigl ( p_{1} + p_{2} + p_{3} \bigr )  = 0, \vspace{2mm} \\
  \displaystyle 
    \frac{d}{dt} \biggl \{  p_{1} p_{2} + p_{1} p_{2} + p_{3} p_{1} - X_{1} - X_{2} - X_{3} \biggr \} = 0, 
  \vspace{2mm} \\
  \displaystyle 
    \frac{d}{dt} \biggl ( p_{1} p_{2} p_{3} - p_{1} X_{2} - p_{2} X_{3} - p_{3} X_{1} \biggr ) = 0. 
 \end{array}
\label{eqn: toda-a2}
\end{equation}
\end{itembox}
\end{minipage}
\end{center}

The corresponding memristor circuit equations for Eq. (\ref{eqn: toda-B-1-1}) are given by  
\begin{center}
\begin{minipage}{8.7cm}
\begin{shadebox}
\underline{\emph{Memristor Toda lattice equations $B$ with $N=3$}}
\begin{equation}
\left.
\begin{array}{ccl}
 \displaystyle \frac{d i_{1}}{dt}&=& \displaystyle (x_{1} - x_{2})i_{1},  \vspace{1mm} \\
 \displaystyle \frac{d x_{1}}{dt}&=& \displaystyle i_{3} - i_{1},         \vspace{1mm} \\
 \displaystyle \frac{d i_{2}}{dt}&=& \displaystyle (x_{2} - x_{3})i_{2},  \vspace{1mm} \\
 \displaystyle \frac{d x_{2}}{dt}&=& \displaystyle i_{1} - i_{2},         \vspace{1mm} \\
 \displaystyle \frac{d i_{3}}{dt}&=& \displaystyle (x_{3} - x_{1})i_{3},  \vspace{1mm} \\
 \displaystyle \frac{d x_{3}}{dt}&=& \displaystyle i_{2} - i_{3},                              
\end{array}
\right \}
\vspace{2mm}
\label{eqn: toda-B-1-3}
\end{equation}
where $i_{1}$, $i_{2}$, and $i_{3}$ denote the current through the generic memristor. 
\end{shadebox}
 \end{minipage}
\end{center}
The terminal voltage $v_{n}$ and the terminal current $i_{n}$ of the generic memristors are described by
\begin{center}
\begin{minipage}{8.7cm}
\begin{shadebox}
\underline{\emph{V-I characteristics of the $3$ generic memristors}} \vspace{1mm} 
\begin{equation}
\begin{array}{c}
\left.
\begin{array}{cll}
\displaystyle v_{1} &=& \tilde{R}_{1}(x_{1}, \, x_{2}) \ i_{1}
                     = - (x_{1} - x_{2})i_{1}, \vspace{1mm} \\
\displaystyle \frac{dx_{1}}{dt} 
                    &=& \tilde{f}_{1}(i_{3}, \, i_{1}) 
                     =  i_{3} - i_{1},
\end{array} 
\right \} 
\vspace{2mm} \\
\left.
\begin{array}{cll}
\displaystyle v_{2} &=& \tilde{R}_{2}(x_{2}, \, x_{3}) \ i_{2}
                     = - (x_{2} - x_{3})i_{2}, \vspace{1mm} \\
\displaystyle \frac{dx_{1}}{dt} 
                    &=& \tilde{f}_{2}(i_{1}, \, i_{2}) 
                     =  i_{1} - i_{2},
\end{array} 
\right \}
\vspace{2mm} \\
\left.
\begin{array}{cll}
\displaystyle v_{3} &=& \tilde{R}_{3}(x_{3}, \, x_{1}) \ i_{3} 
                     = - (x_{3} - x_{1})i_{3}, \vspace{1mm} \\
\displaystyle \frac{dx_{1}}{dt} 
                    &=& \tilde{f}_{3}(i_{2}, \, i_{3}) 
                     =  i_{2} - i_{3},
\end{array} 
\right \}
\end{array}
\label{eqn: toda-B-v-i}
\end{equation}
where $x_{4} = x_{1}$, $i_{0} = i_{3}$.
\end{shadebox}
\end{minipage}
\vspace{10mm}
\end{center}
Equation (\ref{eqn: toda-B-1-3}) can exhibit periodic behavior.  
If an external source is added to the memristor circuit as shown in Figure \ref{fig:memristor-inductor-source-N},
then the forced memristor circuit can exhibit a non-periodic response.  
The dynamics of this circuit is given by 
\begin{center}
\begin{minipage}{9.5cm}
\begin{shadebox}
\underline{\emph{Forced memristor Toda lattice equations $B$ {with $N=3$}}}
\begin{equation}
\left.
\begin{array}{ccl}
 \displaystyle \frac{d i_{1}}{dt}&=& \displaystyle (x_{1} - x_{2})i_{1} + r \sin ( \omega t),  \vspace{2mm} \\
 \displaystyle \frac{d x_{1}}{dt}&=& \displaystyle i_{3} - i_{1},                              \vspace{2mm} \\
 \displaystyle \frac{d i_{2}}{dt}&=& \displaystyle (x_{2} - x_{3})i_{2},                       \vspace{2mm} \\
 \displaystyle \frac{d x_{2}}{dt}&=& \displaystyle i_{1} - i_{2},                              \vspace{2mm} \\
 \displaystyle \frac{d i_{3}}{dt}&=& \displaystyle (x_{3} - x_{1})i_{3},                       \vspace{2mm} \\
 \displaystyle \frac{d x_{3}}{dt}&=& \displaystyle i_{2} - i_{3},                             
\end{array}
\right \}
%\vspace{1mm}
\label{eqn: toda-B-1-2}
\end{equation}
where $r$ and $\omega$ are constants.  
\end{shadebox}
\end{minipage}
\vspace{10mm}
\end{center}
We show their non-periodic and quasi-periodic responses, Poincar\'e maps, and $i_{j}-v_{j}$ loci in Figures \ref{fig:Toda-B-attractor}, \ref{fig:Toda-B-poincare}, and \ref{fig:Toda-B-pinch}, respectively ($j=1, \, 2, \, 3$).  
In order to obtain these figures, we have to choose the parameters and the initial conditions carefully and the maximum step size $h$ of the numerical integration must be sufficiently small ($h=0.006$).  
The following parameters are used in our computer simulations:
\begin{equation}
 \ r = 0.25,  \ \omega = 0.925.
\end{equation}

\clearpage
The $i_{j}-v_{j}$ loci in Figure \ref{fig:Toda-B-pinch} lie in the first and the fourth quadrants.  
Thus, the three generic memristors are active element.  
Let us show the $v_{j}-p_{j}$ locus in Figure \ref{fig:Toda-B-power}, where $p_{j}(t)$ is an instantaneous power defined by $p_{j}(t)=i_{j}(t)v_{j}(t)$ ($j=1, \, 2, \, 3$).
Observe that the $v_{j}-p_{j}$ loci are pinched at the origin, and the loci lie in the first and the third quadrants. 
Thus, when $v_{j}>0$, the instantaneous power $p_{j}(t)$ delivered from the forced signal and the inductor is dissipated in the memristor.  
However, when $v_{j}<0$, the instantaneous power $p_{j}(t)$ is \emph{not} dissipated in the memristor. 
Hence, the memristors switch between passive and active modes of operation, depending on its terminal voltage. 
Thus, we conclude as follow: \\
\begin{center}
\begin{minipage}{.7\textwidth}
\begin{itembox}[l]{Switching behavior of the memristor}
Assume that Eq. (\ref{eqn: toda-B-1-2}) exhibits non-periodic or quasi-periodic oscillation.  
Then the generic memristors defined by Eq. (\ref{eqn: toda-B-v-i}) can switch between ``passive'' and ``active'' modes of operation, depending on its terminal voltage.  
\end{itembox}
\end{minipage}
\end{center}
%
%

%---Fig. 47-------%
\begin{figure}[hpbt]
 \centering
   \begin{tabular}{cc}
    \psfig{file=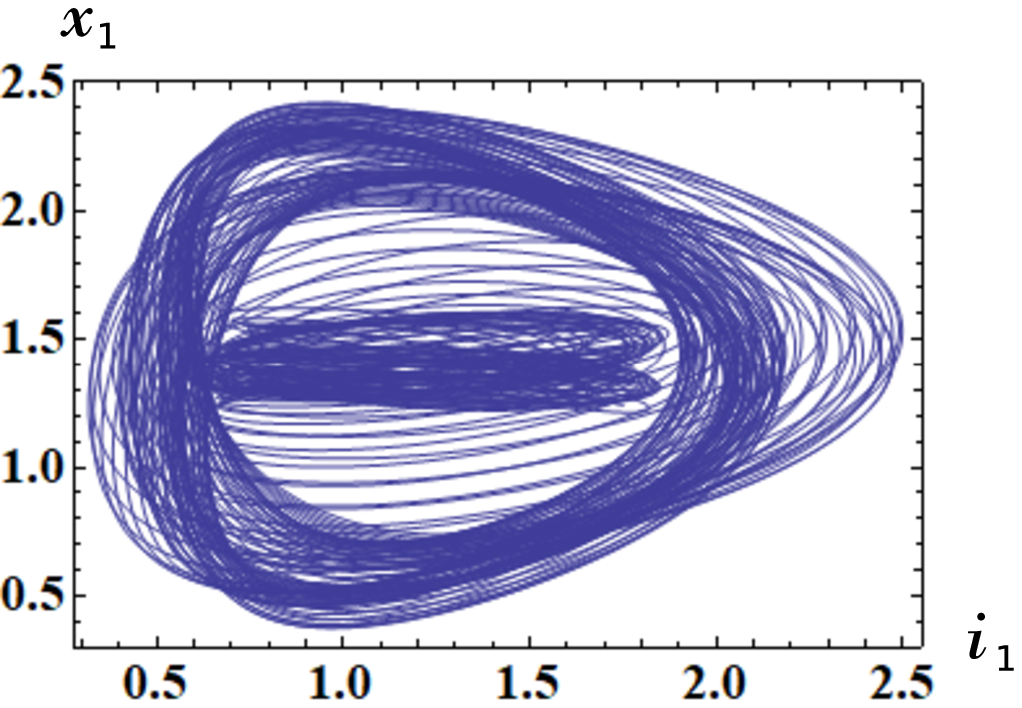, height=4.8cm}  & 
    \psfig{file=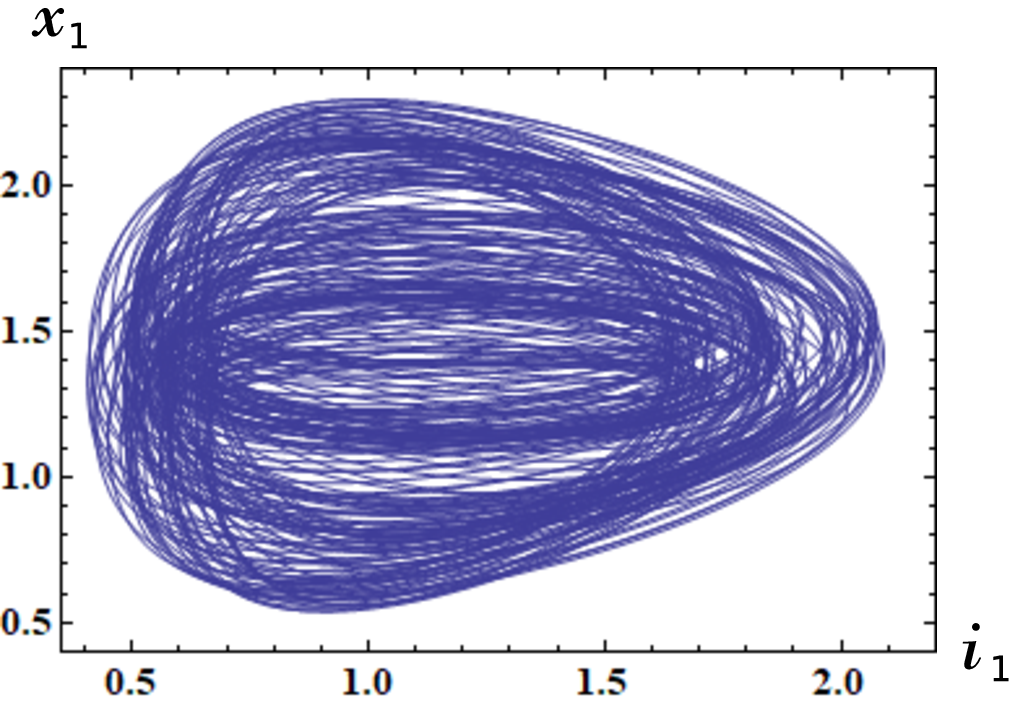, height=4.8cm}  \\
    (a) non-periodic  ($i_{1}(0) = 0.7$) & (d)  quasi-periodic ($i_{1}(0) = 0.4$) \vspace{5mm} \\
    \psfig{file=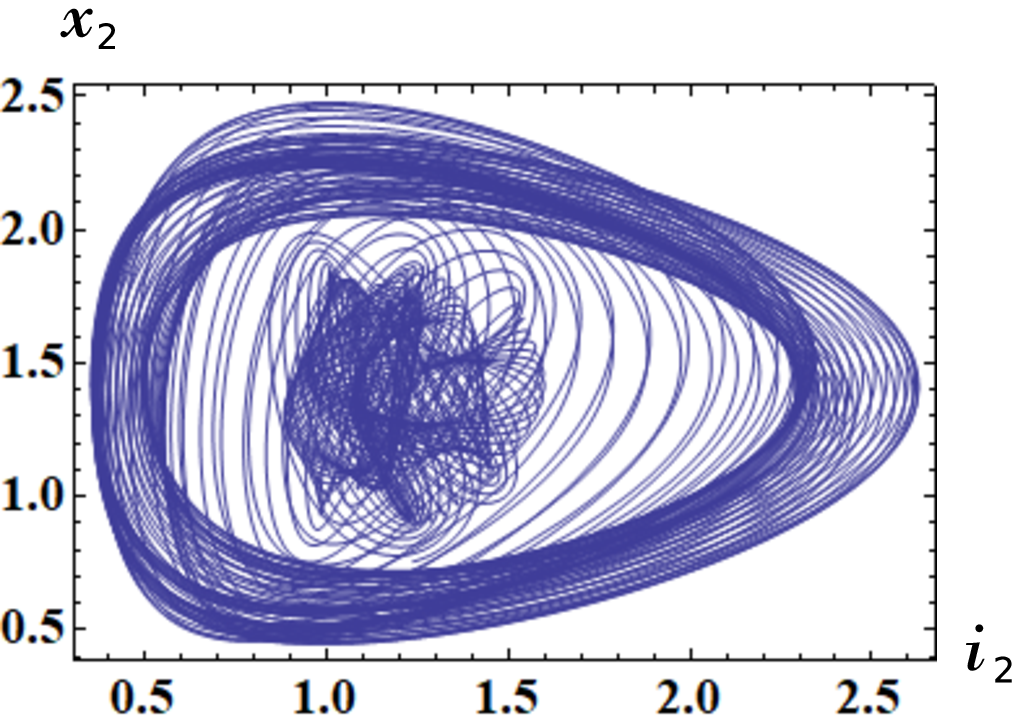, height=4.8cm}  & 
    \psfig{file=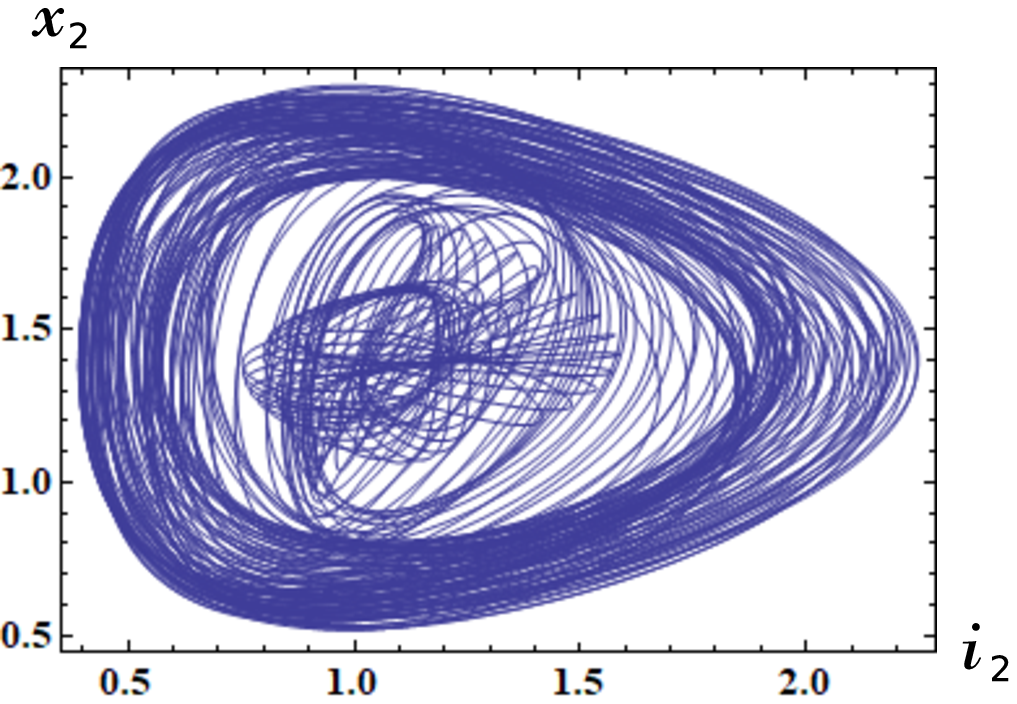, height=4.8cm}  \\
    (b) non-periodic ($i_{1}(0) = 0.7$)& (e)  quasi-periodic  ($i_{1}(0) = 0.4$)\vspace{5mm} \\    
    \psfig{file=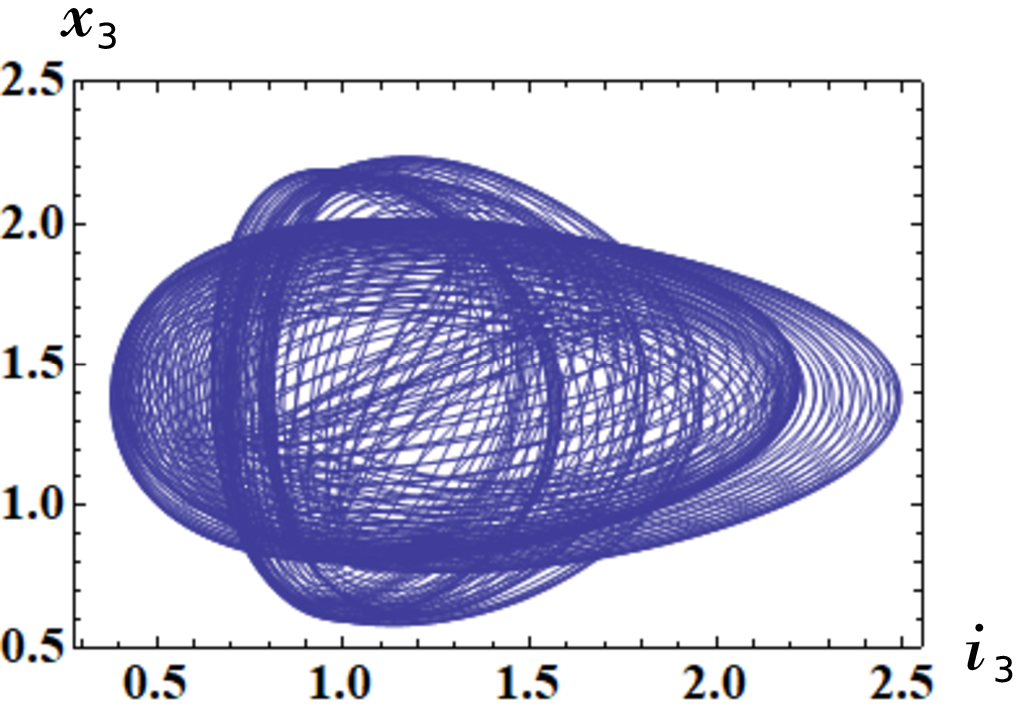, height=4.8cm}  & 
    \psfig{file=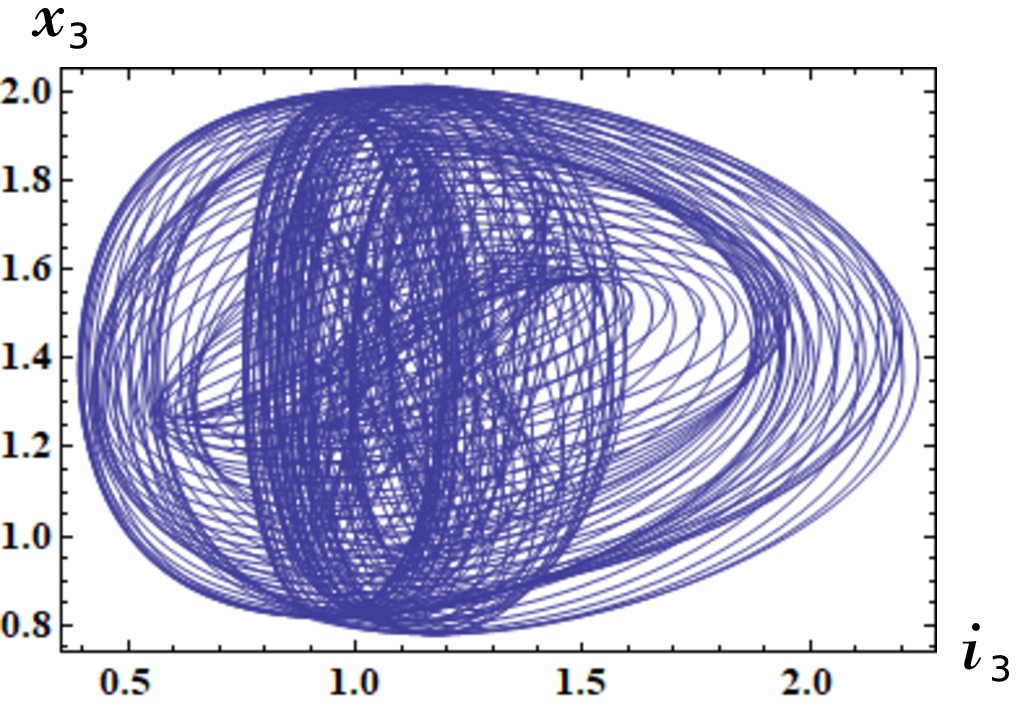, height=4.8cm}  \\
    (c) non-periodic ($i_{1}(0) = 0.7$)& (f)  quasi-periodic  ($i_{1}(0) = 0.4$)    
   \end{tabular}
  \caption{Non-periodic and quasi-periodic responses of the forced memristor Toda lattice equations B, 
  which are defined by Eq. (\ref{eqn: toda-B-1-2}). 
  Here, $i_{j}$ and $x_{j}$ denote the terminal current and the internal state of the $j$-th generic memristor, respectively  
  $(j=1, \, 2, \, 3)$. 
  Parameters:  $\ r = 0.25,  \ \omega = 0.925$.
  Initial conditions: 
  (a) $i_{1}(0) = 0.7, \, x_{1}(0)= 1.2, \, i_{2}(0) = 1.3, \, x_{2}(0)= 1.4, \, i_{3}(0) = 1.5, \, x_{3}(0) = 1.6$. \ \ 
  (b) $i_{1}(0) = 0.4, \, x_{1}(0)= 1.2, \, i_{2}(0) = 1.3, \, x_{2}(0)= 1.4, \, i_{3}(0) = 1.5, \, x_{3}(0) = 1.6$.}
  \label{fig:Toda-B-attractor} 
\end{figure}
%
%

%---Fig. 48-------%
\begin{figure}[hpbt]
 \centering
   \begin{tabular}{cc}
    \psfig{file=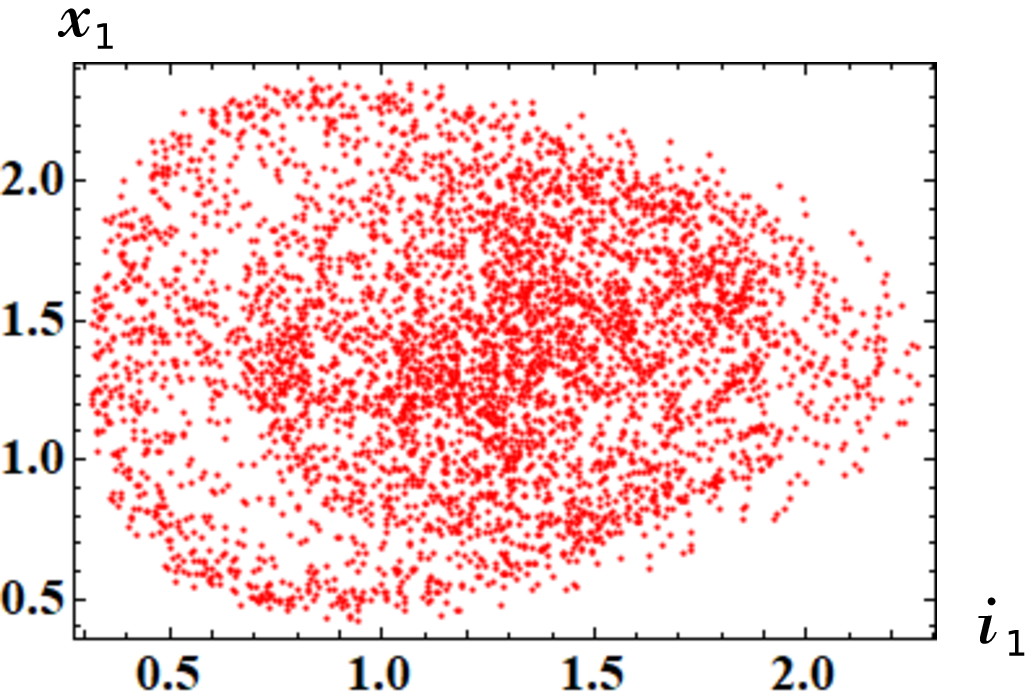, height=4.8cm}  & 
    \psfig{file=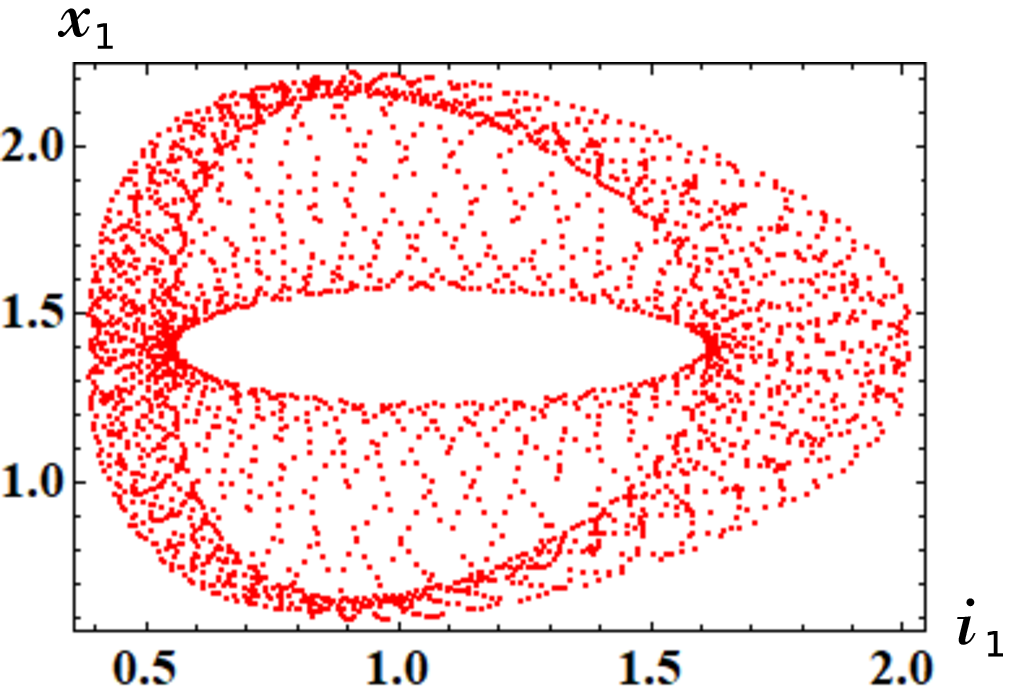, height=4.8cm}  \\
    (a) non-periodic  ($i_{1}(0) = 0.7$) & (d)  quasi-periodic ($i_{1}(0) = 0.4$) \vspace{5mm} \\
    \psfig{file=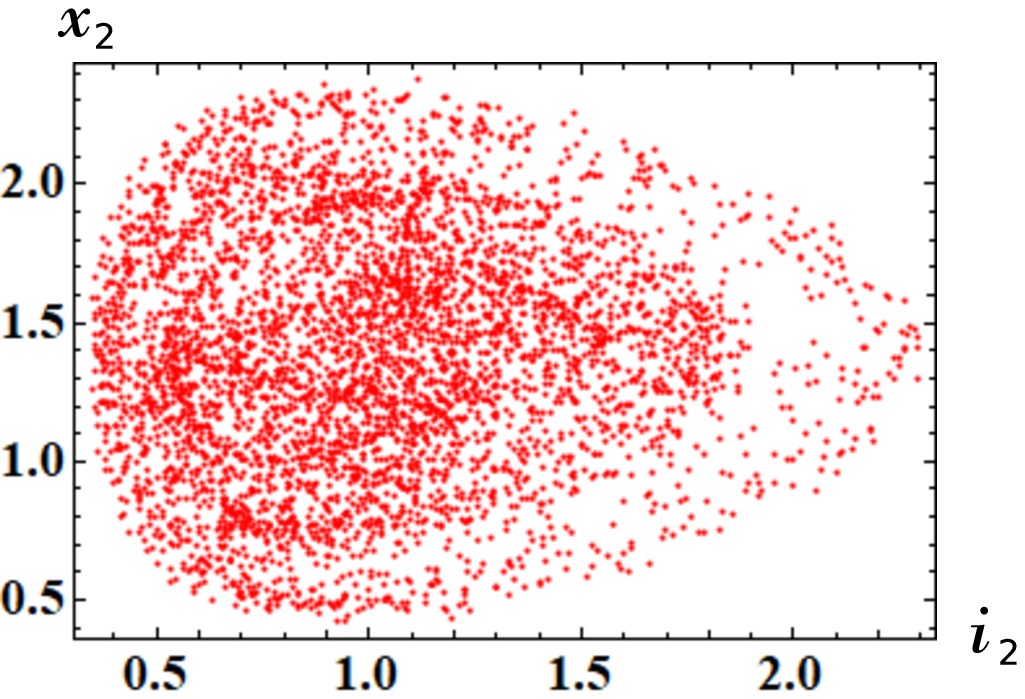, height=4.8cm}  & 
    \psfig{file=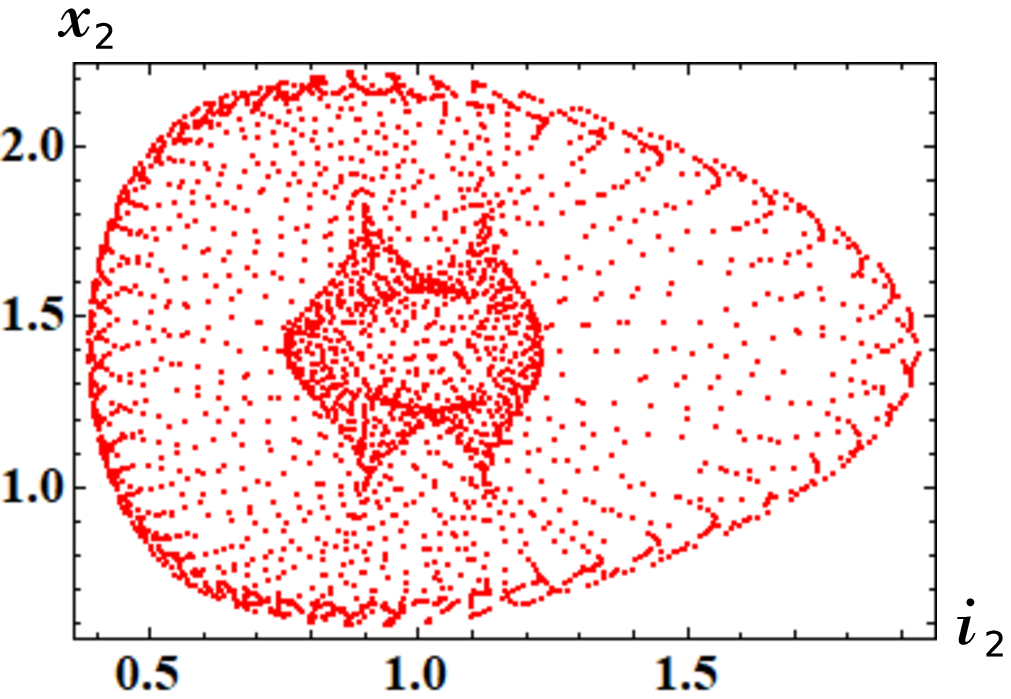, height=4.8cm}  \\
    (b) non-periodic ($i_{1}(0) = 0.7$)& (e)  quasi-periodic  ($i_{1}(0) = 0.4$)\vspace{5mm} \\    
    \psfig{file=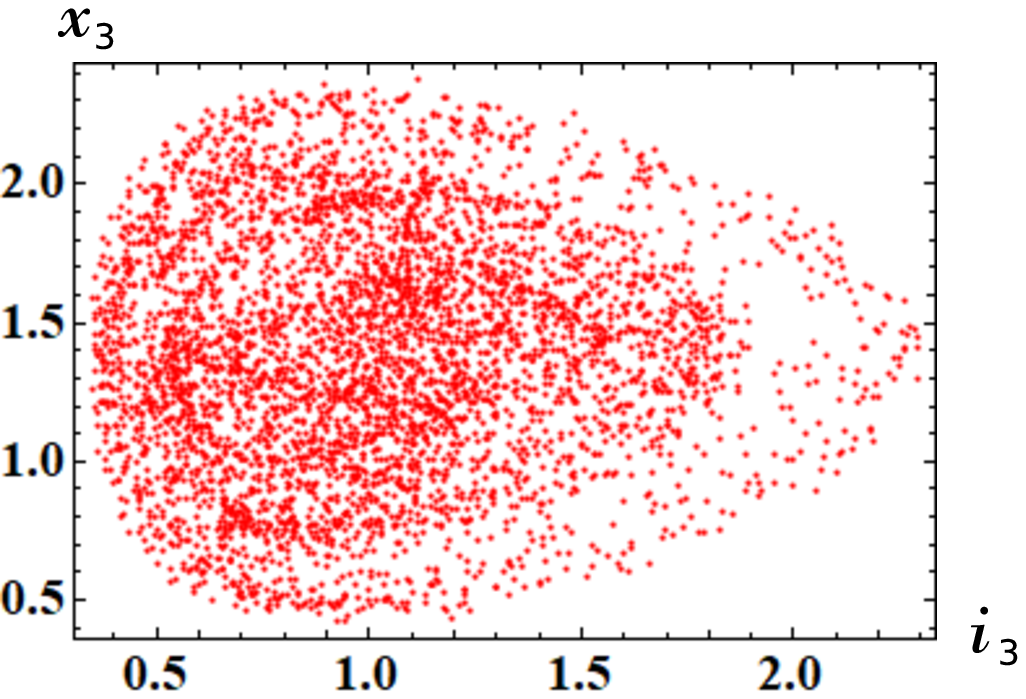, height=4.8cm}  & 
    \psfig{file=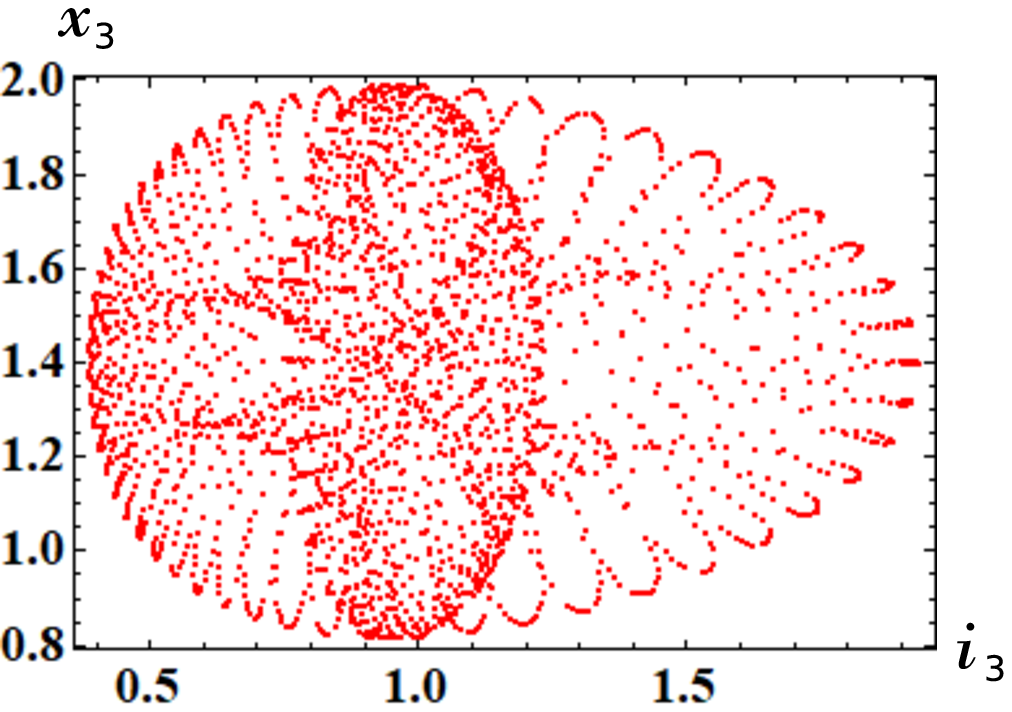, height=4.8cm}  \\
    (c) non-periodic ($i_{1}(0) = 0.7$)& (f)  quasi-periodic  ($i_{1}(0) = 0.4$)    
   \end{tabular}
  \caption{Poincar\'e maps of the forced memristor Toda lattice equations B, which are defined by Eq. (\ref{eqn: toda-B-1-2}). 
  Here, $i_{j}$ and $x_{j}$ denote the terminal current and the internal state of the $j$-th generic memristor, respectively 
  $(j=1, \, 2, \, 3)$. 
  Parameters:  $\ r = 0.25,  \ \omega = 0.925$.
  Initial conditions: 
  (a) $i_{1}(0) = 0.7, \, x_{1}(0)= 1.2, \, i_{2}(0) = 1.3, \, x_{2}(0)= 1.4, \, i_{3}(0) = 1.5, \, x_{3}(0) = 1.6$. \ \ 
  (b) $i_{1}(0) = 0.4, \, x_{1}(0)= 1.2, \, i_{2}(0) = 1.3, \, x_{2}(0)= 1.4, \, i_{3}(0) = 1.5, \, x_{3}(0) = 1.6$.}
  \label{fig:Toda-B-poincare} 
\end{figure}
%
%

%---Fig. 49-------%
\begin{figure}[hpbt]
 \centering
   \begin{tabular}{cc}
    \psfig{file=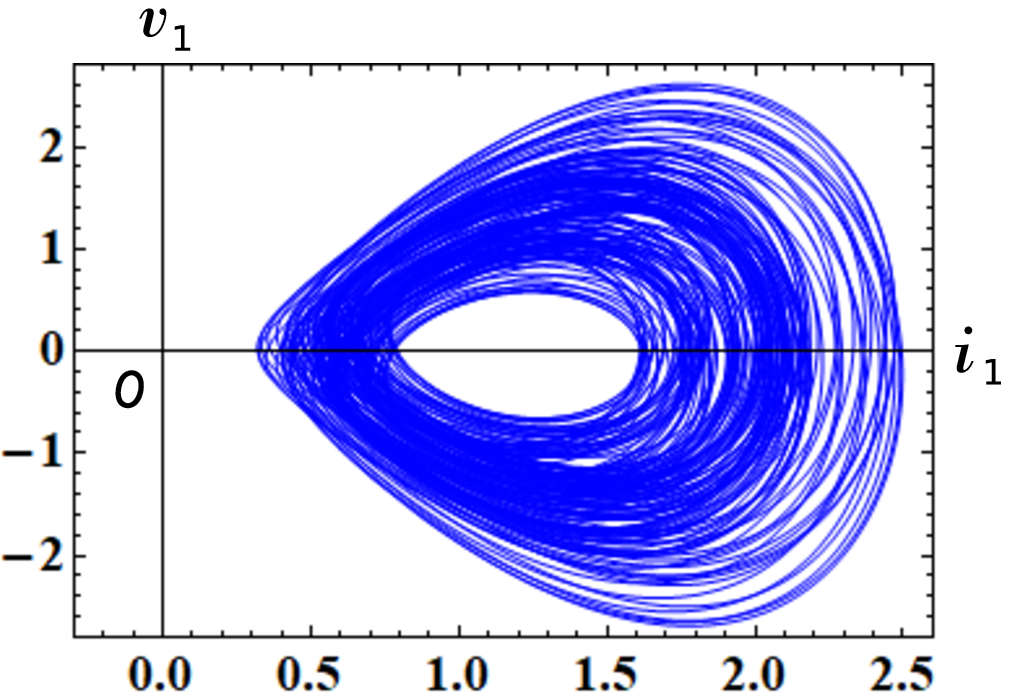, height=4.8cm}  & 
    \psfig{file=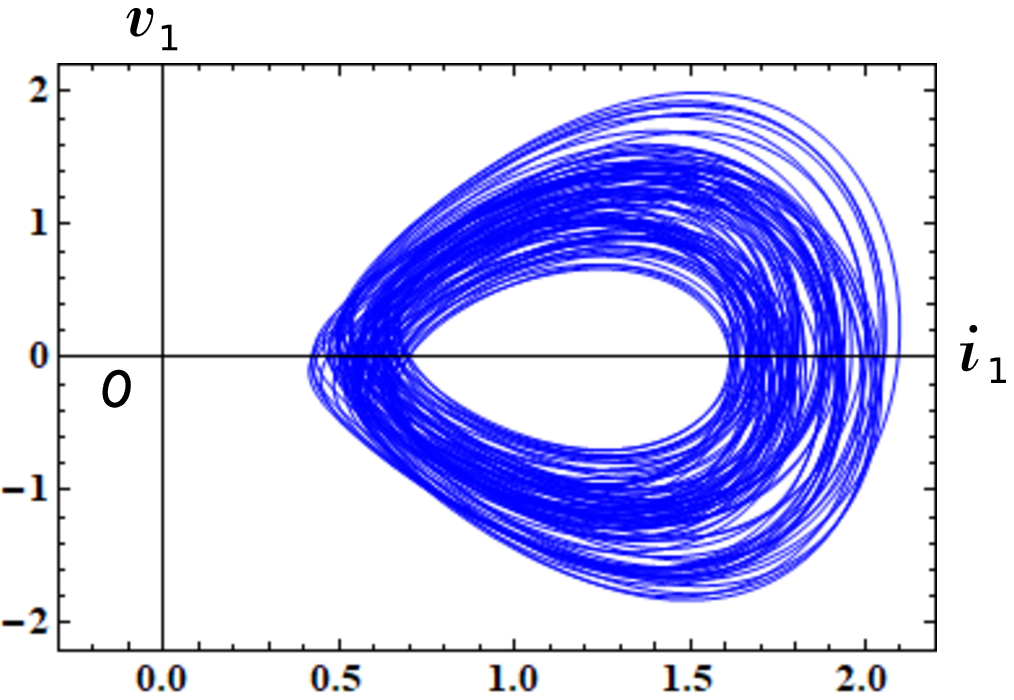, height=4.8cm}  \\
     (a) non-periodic $(i_{1}(0) = 0.7)$ & (d)  quasi-periodic $(i_{1}(0) = 0.4)$ \vspace{5mm} \\
    \psfig{file=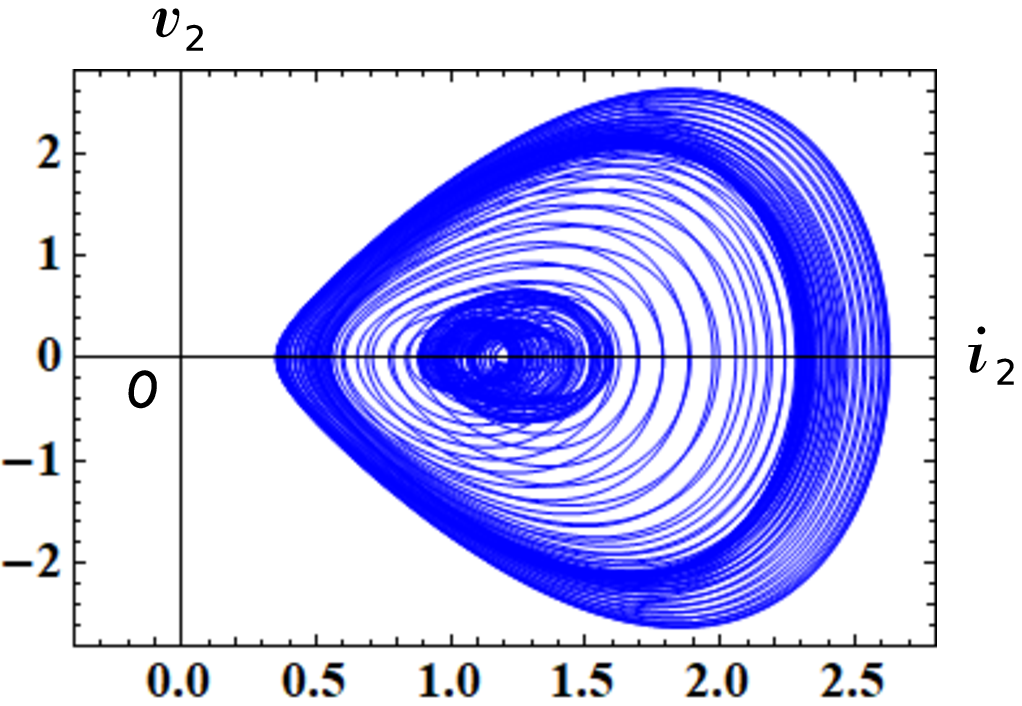, height=4.8cm}  & 
    \psfig{file=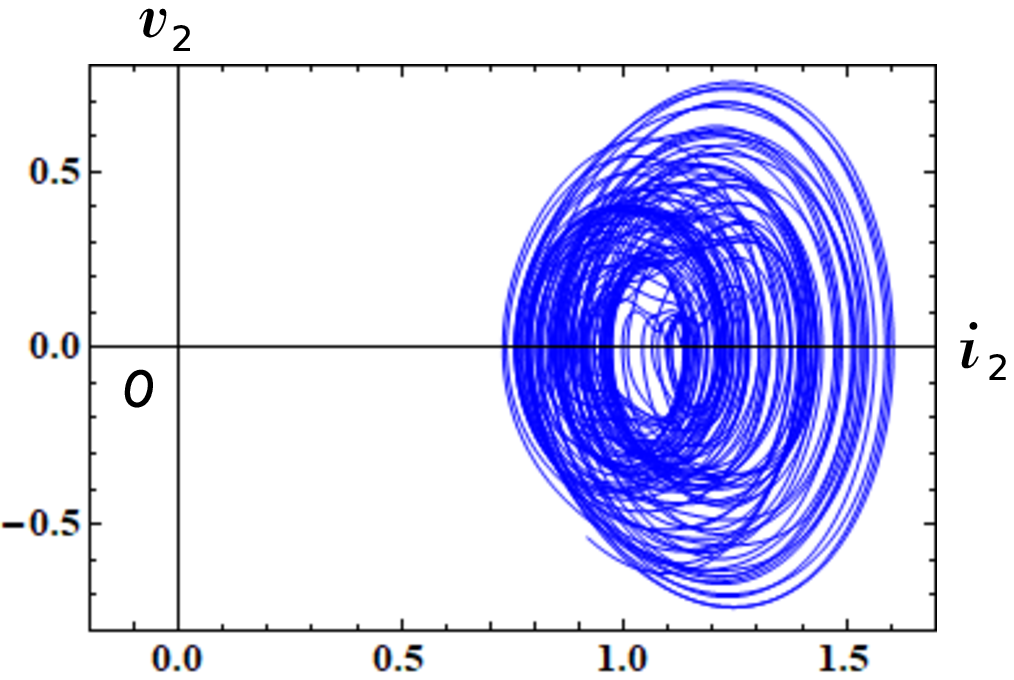, height=4.8cm}  \\
     (b) non-periodic $(i_{1}(0) = 0.7)$ & (e)  quasi-periodic $(i_{1}(0) = 0.4)$ \vspace{5mm} \\
    \psfig{file=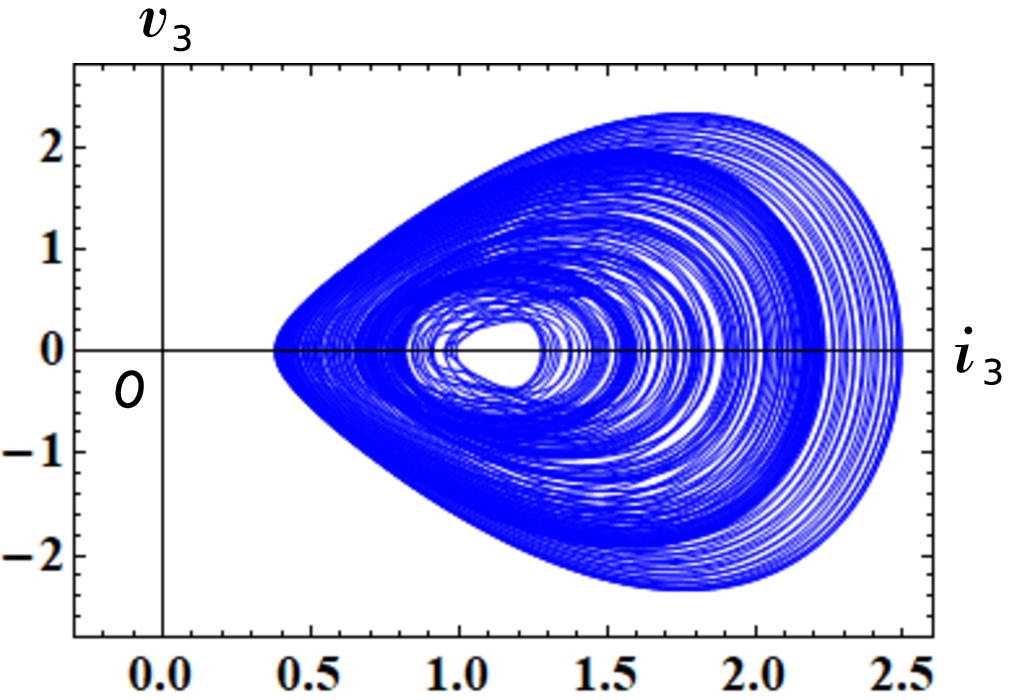, height=4.8cm}  & 
    \psfig{file=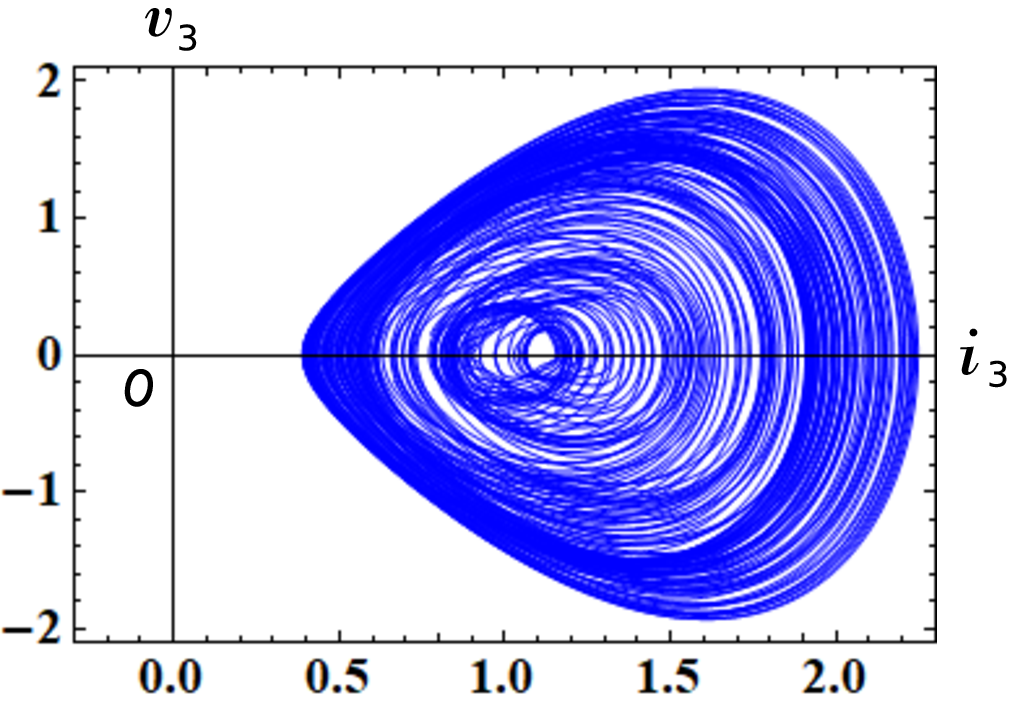, height=4.8cm}  \\
     (c) non-periodic $(i_{1}(0) = 0.7)$ & (f)  quasi-periodic $(i_{1}(0) = 0.4)$
    \end{tabular}
  \caption{ The $i_{j}-v_{j}$ loci of the forced memristor Toda lattice equations B, which are defined by Eq. (\ref{eqn: toda-B-1-2}).  
  Here, $i_{j}$ and $v_{j}$ denote the terminal current and the voltage of the $j$-th generic memristor, respectively 
  $(j=1, \, 2, \, 3)$. 
  Parameters:  $\ r = 0.25,  \ \omega = 0.925$.
  Initial conditions: 
  (a)-(c) $i_{1}(0) = 0.7, \, x_{1}(0)= 1.2, \, i_{2}(0) = 1.3, \, x_{2}(0)= 1.4, \, i_{3}(0) = 1.5, \, x_{3}(0) = 1.6$. \ \ 
  (d)-(f) $i_{1}(0) = 0.4, \, x_{1}(0)= 1.2, \, i_{2}(0) = 1.3, \, x_{2}(0)= 1.4, \, i_{3}(0) = 1.5, \, x_{3}(0) = 1.6$.}
  \label{fig:Toda-B-pinch} 
\end{figure}
%
%

%---Fig. 50-------%
\begin{figure}[hpbt]
 \centering
   \begin{tabular}{cc}
    \psfig{file=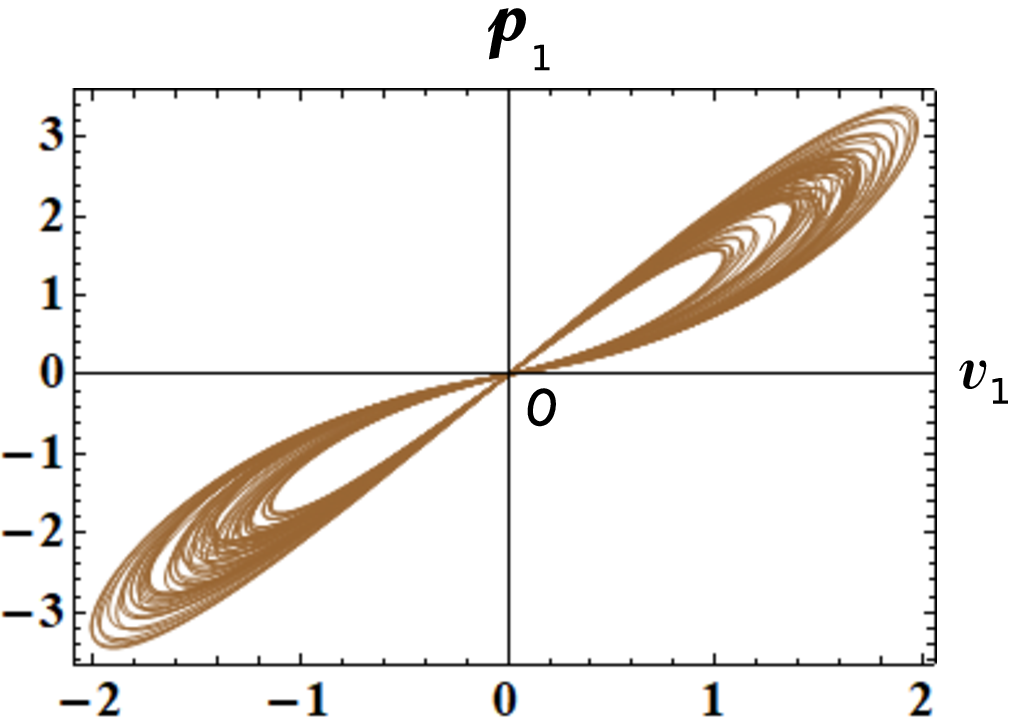, height=4.8cm}  & 
    \psfig{file=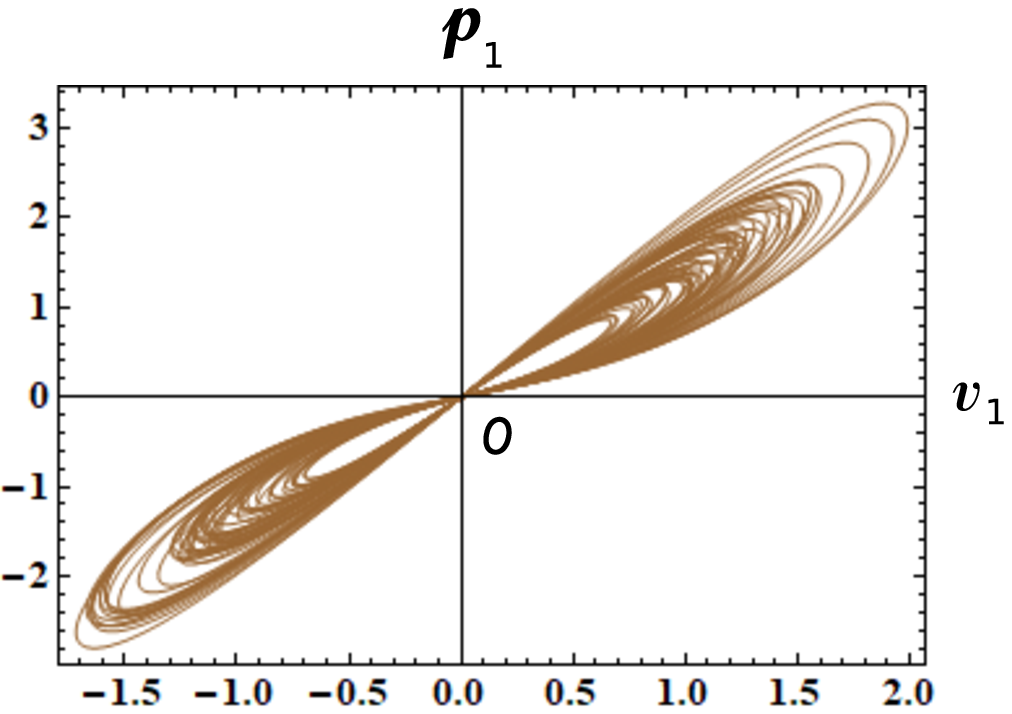, height=4.8cm}  \\
     (a) non-periodic $(i_{1}(0) = 0.7)$ & (d)  quasi-periodic $(i_{1}(0) = 0.4)$ \vspace{5mm} \\
    \psfig{file=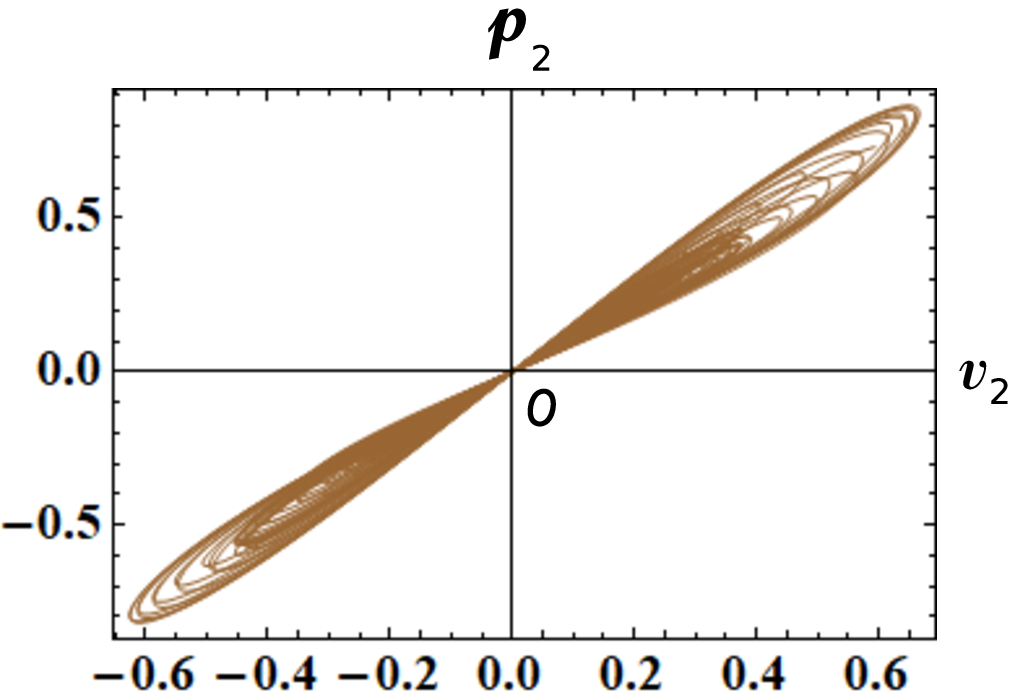, height=4.8cm}  & 
    \psfig{file=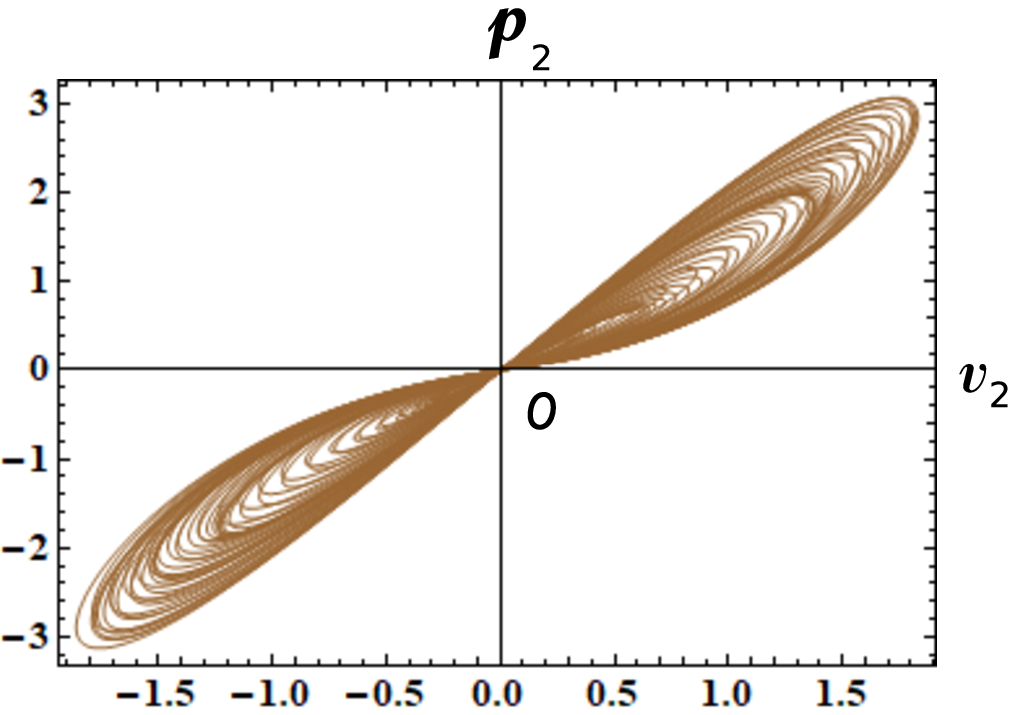, height=4.8cm}  \\
     (b) non-periodic $(i_{1}(0) = 0.7)$ & (e)  quasi-periodic $(i_{1}(0) = 0.4)$ \vspace{5mm} \\
    \psfig{file=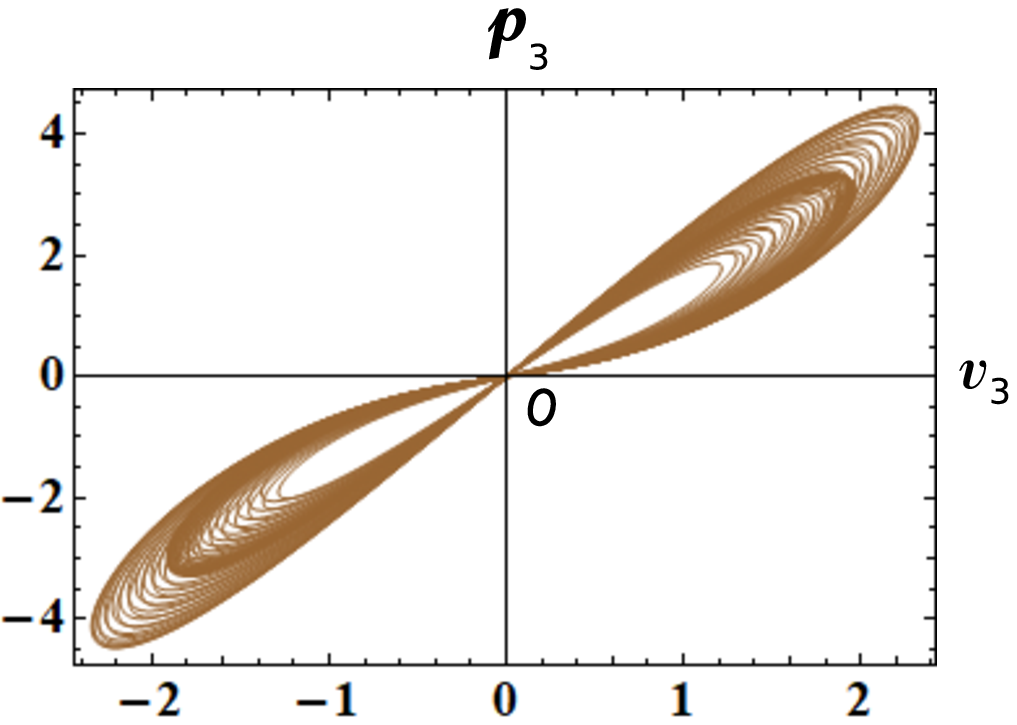, height=4.8cm}  & 
    \psfig{file=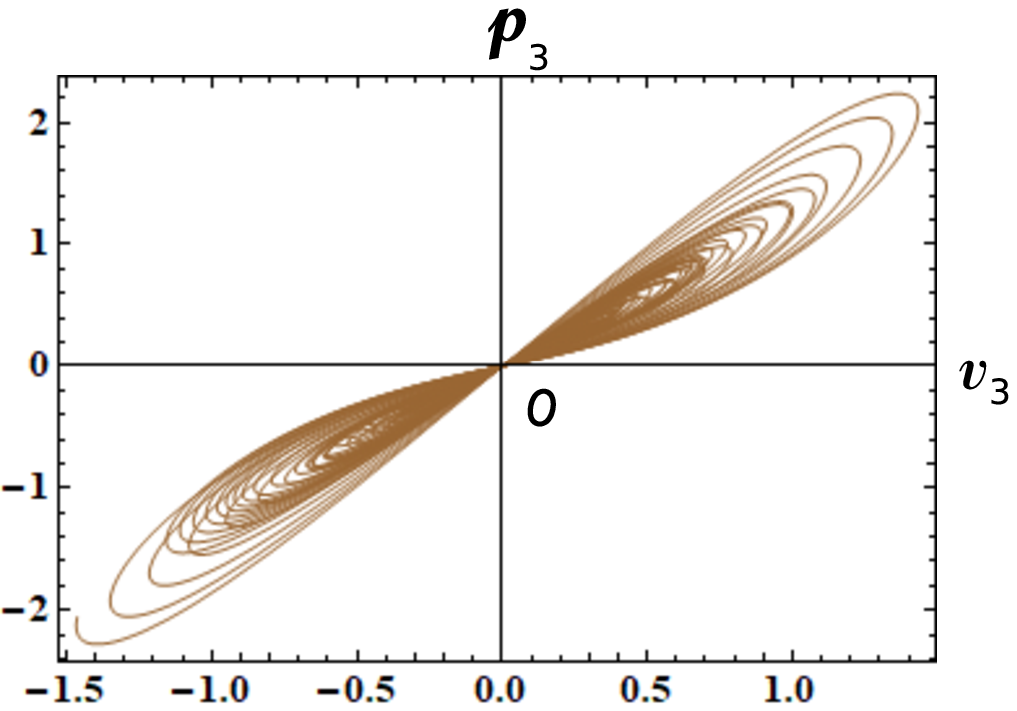, height=4.8cm}  \\
     (c) non-periodic $(i_{1}(0) = 0.7)$ & (f)  quasi-periodic $(i_{1}(0) = 0.4)$
   \end{tabular}
  \caption{ The $v_{j}-p_{j}$ loci of the forced memristor Toda lattice equations B, which are defined by Eq. (\ref{eqn: toda-B-1-2}). 
   Here, $p_{j}(t)$ is an instantaneous power defined by $p_{j}(t)=i_{j}(t)v_{j}(t)$,   
   and $v_{j}(t)$ and $i_{j}(t)$ denote the terminal voltage and the terminal current of the $j$-th generic memristor, respectively 
   $(j=1, \, 2, \, 3)$.  
   Observe that the $v_{j}-p_{j}$ locus is pinched at the origin, and the locus lies in the first and the third quadrants. 
   The memristor switches between passive and active modes of operation, depending on its terminal voltage $v_{j}(t)$.
   Parameters:  $\ r = 0.25,  \ \omega = 0.925$.
   Initial conditions: 
   (a)-(c) $i_{1}(0) = 0.7, \, x_{1}(0)= 1.2, \, i_{2}(0) = 1.3, \, x_{2}(0)= 1.4, \, i_{3}(0) = 1.5, \, x_{3}(0) = 1.6$. \ \ 
   (d)-(f) $i_{1}(0) = 0.4, \, x_{1}(0)= 1.2, \, i_{2}(0) = 1.3, \, x_{2}(0)= 1.4, \, i_{3}(0) = 1.5, \, x_{3}(0) = 1.6$.}
  \label{fig:Toda-B-power} 
\end{figure}
\clearpage

%-------------------------------------%
\subsubsection{\underline{Toda lattice equations $C$}}
\label{sec: Toda-lattice-C}
%-------------------------------------%
Consider Eq.(\ref{eqn: toda-a1}) and define new variables 
\begin{equation}
 \begin{array}{lll}
 \displaystyle a_{n} &=& \displaystyle \frac{1}{2} e^{- ( \, q_{n+1} - q_{n} \, )/2 }, \vspace{2mm} \\
 \displaystyle b_{n} &=& \displaystyle \frac{1}{2} p_{n}, 
 \end{array}
\label{eqn: toda-c1}
\end{equation}
where $n=1, \ 2, \cdots, \ N$.  
Then Eq.(\ref{eqn: toda-a1}) can be recast into the form \cite{{Toda(1989), Teschl(2001)}} 
\begin{center}
\begin{minipage}{8.7cm}
\begin{shadebox}
\underline{\emph{Toda lattice equations $C$}}
\begin{equation}
 \begin{array}{lll}
   \displaystyle \frac{d a_{n}}{dt}&=& \displaystyle (b_{n} - b_{n+1})a_{n}, \vspace{2mm} \\
   \displaystyle \frac{d b_{n}}{dt}&=& \displaystyle  2( {a_{n-1}}^{2} - {a_{n}}^{2} ), 
 \end{array}
\label{eqn: toda-c2}
\end{equation}
where $n=1, \ 2, \cdots, \ N$ and we consider the case of a periodic lattice of the length $N$: $a_{n} = a_{n + N}$, $b_{n} = b_{n + N}$.
\end{shadebox}
\end{minipage}
\end{center}

Consider the $2N$-element memristor circuit in Figure \ref{fig:memristor-inductor-N}.  
The dynamics of this circuit given by Eq. (\ref{eqn: dynamics-N}).  
Assume that Eq. (\ref{eqn: dynamics-N}) satisfies 
\begin{equation}
\begin{array}{cll}
  L_{n} &=& 1, \vspace{2mm} \\
  \hat{R}( \bd{x}, \ i_{n} )  &=&  - (x_{n} - x_{n+1})  \vspace{2mm} \\
  \tilde{f}_{n}(\bd{x}, \bd{i}) &=& 2 ( {i_{n-1}}^{2} - {i_{n}}^{2} ).  
\end{array}
\end{equation}
Then we obtain 
\begin{center}
\begin{minipage}{8.7cm}
\begin{shadebox}
\underline{\emph{Memristor Toda lattice equations $C$}}
\begin{equation}
 \begin{array}{lll}
 \displaystyle \frac{d i_{n}}{dt}&=& \displaystyle (x_{n} - x_{n+1})i_{n}, \vspace{2mm} \\
 \displaystyle \frac{d x_{n}}{dt}&=& \displaystyle 2 ( {i_{n-1}}^{2} - {i_{n}}^{2} ), 
 \end{array}
\label{eqn: toda-c4}
\end{equation}
where $n=1, \ 2, \cdots, \ N$ and we consider the case of a periodic lattice of the length $N$: $i_{n} = i_{n + N}$, $x_{n} = x_{n + N}$. 
\end{shadebox}
\end{minipage}
\end{center}
Equations (\ref{eqn: toda-c2}) and (\ref{eqn: toda-c4}) are equivalent if we change the variables  
\begin{equation}
  i_{n} = a_{n}, \ x_{n}=b_{n}.
\end{equation}
In this case, the extended memristors in Figure \ref{fig:memristor-inductor-N} are replaced by the \emph{generic} memristors, that is, 
\begin{equation}
  \hat{R}( \bd{x}, \ i_{n} ) = \tilde{R}_{n}(\bd{x}) = - (x_{n} - x_{n+1}), 
\end{equation}
though the current $i$ of Eq. (\ref{eqn: generic2-2}) is modified into the vector form $\bd{i} = (i_{1}, \, i_{2}, \, \cdots, \, i_{n})$.  
Thus, their terminal voltage $v_{n}$ and the terminal current $i_{n}$ of the current-controlled memristor are described
by
\begin{center}
\begin{minipage}{9.5cm}
\begin{shadebox}
\underline{\emph{V-I characteristics of the generic memristors}} \vspace{1mm} 
\begin{equation}
\begin{array}{cll}
\displaystyle v_{n} &=& \tilde{R}_{n}(\bd{x}) = \tilde{R}_{n}(x_{n}, \, x_{n+1}) \ i_{n}  \vspace{2mm} \\
                    &=& - (x_{n} - x_{n+1})i_{n}, \vspace{2mm} \\
\displaystyle \frac{dx_{n}}{dt} 
                    &=& \tilde{f}_{n}(\bd{x}, \bd{i}) = \tilde{f}(i_{n-1}, \, i_{n})  \vspace{2mm} \\
                    &=& 2 ( {i_{n-1}}^{2} - {i_{n}}^{2} ),
\end{array}
\vspace{1mm}
\label{eqn: toda-c5}
\end{equation}
where $i_{n} = i_{n + N}$, $x_{n} = x_{n + N}$, and $n=1, \ 2, \cdots, \ N$.
\end{shadebox}
\end{minipage}
\end{center}
It follows that Eq. (\ref{eqn: toda-c2}) can be realized by the $2N$-element memristor circuit in Figure \ref{fig:memristor-inductor-N}.  

For $n=3$, Eq. (\ref{eqn: toda-c2}) is given by 
\begin{center}
\begin{minipage}{8.7cm}
\begin{shadebox}
\underline{\emph{Toda lattice equations $C$ with $N=3$}}
\begin{equation}
\left.
 \begin{array}{lll}
   \displaystyle \frac{d a_{1}}{dt}&=& \displaystyle (b_{1} - b_{2})a_{1}, \vspace{2mm} \\    
   \displaystyle \frac{d b_{1}}{dt}&=& \displaystyle  2( {a_{3}}^{2} - {a_{1}}^{2} ), \vspace{2mm} \\  
   \displaystyle \frac{d a_{2}}{dt}&=& \displaystyle (b_{2} - b_{3})a_{2}, \vspace{2mm} \\    
   \displaystyle \frac{d b_{2}}{dt}&=& \displaystyle  2( {a_{1}}^{2} - {a_{2}}^{2} ), \vspace{2mm} \\  
   \displaystyle \frac{d a_{3}}{dt}&=& \displaystyle (b_{3} - b_{1})a_{3}, \vspace{2mm} \\    
   \displaystyle \frac{d b_{3}}{dt}&=& \displaystyle  2( {a_{2}}^{2} - {a_{3}}^{2} ), 
 \end{array}
\right \}
\label{eqn: toda-c10}
\end{equation}
\end{shadebox}
\end{minipage}
\end{center}
Equation (\ref{eqn: toda-c10}) has the three integrals \cite{Henon(1974), Toda(1989)}, since the solution satisfies 
\begin{center}
\begin{minipage}{.6\textwidth}
\begin{itembox}[l]{Integrals}
\begin{equation}
 \begin{array}{l}
  \displaystyle \frac{d}{dt} \bigl ( b_{1} + b_{2} + b_{3} \bigr )  = 0, \vspace{4mm} \\
  \displaystyle 
    \frac{d}{dt} \biggl \{{b_{1}}^{2} + {b_{2}}^{2} + {b_{3}}^{2}+ 2 \, ({a_{1}}^{2} + {a_{2}}^{2} + {a_{3}}^{2}) \biggr \} = 0, 
  \vspace{4mm} \\
  \displaystyle 
    \frac{d}{dt} \biggl ( b_{1} b_{2} b_{3} - b_{1}{a_{2}}^{2} - b_{2}{a_{3}}^{2} - b_{3}{a_{1}}^{2} + 2 a_{1} a_{2} a_{3}  \biggr ) = 0, 
  \vspace{1mm} \\
 \end{array}
\label{eqn: toda-c-11}
\end{equation}
where $\displaystyle 2 a_{1} a_{2} a_{3} = \frac{1}{4}$.  
\end{itembox}
\end{minipage}
\end{center}

The corresponding memristor circuit equations for Eq. (\ref{eqn: toda-c10}) are given by 
\begin{center}
\begin{minipage}{8.7cm}
\begin{shadebox}
\underline{\emph{Memristor Toda lattice equations $C$ with $N=3$}}
\begin{equation}
 \left.
 \begin{array}{lll}
 \displaystyle \frac{d i_{1}}{dt}&=& \displaystyle (x_{1} - x_{2})i_{1},   \vspace{2mm} \\
 \displaystyle \frac{d x_{1}}{dt}&=& \displaystyle 2 ( {i_{3}}^{2} - {i_{1}}^{2} ),             \vspace{2mm} \\
 \displaystyle \frac{d i_{2}}{dt}&=& \displaystyle (x_{2} - x_{3})i_{2},                        \vspace{2mm} \\
 \displaystyle \frac{d x_{2}}{dt}&=& \displaystyle 2 ( {i_{1}}^{2} - {i_{2}}^{2} ),             \vspace{2mm} \\
 \displaystyle \frac{d i_{3}}{dt}&=& \displaystyle (x_{3} - x_{1})i_{3},                        \vspace{2mm} \\
 \displaystyle \frac{d x_{3}}{dt}&=& \displaystyle 2 ( {i_{2}}^{2} - {i_{3}}^{2} ). 
 \end{array}
\right \}
\label{eqn: toda-c-13}
\end{equation}
\end{shadebox}
\end{minipage}
\end{center}
The terminal voltage $v_{n}$ and the terminal current $i_{n}$ of the three generic memristors are described by 
\begin{center}
\begin{minipage}{8.7cm}
\begin{shadebox}
\underline{\emph{V-I characteristics of the $3$ generic memristors}} \vspace{1mm} 
\begin{equation}
\begin{array}{c}
\left.
\begin{array}{cll}
\displaystyle v_{1} &=& \tilde{R}_{1}(x_{1}, \, x_{2}) \ i_{1}
                     = - (x_{1} - x_{2})i_{1}, \vspace{2mm} \\
\displaystyle \frac{dx_{1}}{dt} 
                    &=& \tilde{f}_{1}(i_{3}, \, i_{1}) 
                     =  2 ( {i_{3}}^{2} - {i_{1}}^{2} ),  
\end{array} 
\right \} 
\vspace{2mm} \\
\left.
\begin{array}{cll}
\displaystyle v_{2} &=& \tilde{R}_{2}(x_{2}, \, x_{3}) \ i_{2}
                     = - (x_{2} - x_{3})i_{2}, \vspace{2mm} \\
\displaystyle \frac{dx_{1}}{dt} 
                    &=& \tilde{f}_{2}(i_{1}, \, i_{2}) 
                     =  2 ( {i_{1}}^{2} - {i_{2}}^{2} ),
\end{array} 
\right \}
\vspace{2mm} \\
\left.
\begin{array}{cll}
\displaystyle v_{3} &=& \tilde{R}_{3}(x_{3}, \, x_{1}) \ i_{3} 
                     = - (x_{3} - x_{1})i_{3}, \vspace{2mm} \\
\displaystyle \frac{dx_{1}}{dt} 
                    &=& \tilde{f}_{3}(i_{2}, \, i_{3}) 
                     =  2 ( {i_{2}}^{2} - {i_{3}}^{2} ). 
\end{array} 
\right \}
\end{array}
\label{eqn: toda-C-v-i}
\end{equation}
where $x_{4} = x_{1}$, $i_{0} = i_{3}$.
\end{shadebox}
\end{minipage}
\end{center}
Equations (\ref{eqn: toda-c-13}) can exhibit periodic behavior. 
If an external source is added as shown in Figure \ref{fig:memristor-inductor-source-N},
then the forced memristor circuit can exhibit non-periodic response.  
The dynamics of this circuit is given by 
\begin{center}
\begin{minipage}{9.5cm}
\begin{shadebox}
\underline{\emph{Forced memristor Toda lattice equations $C$ with $N=3$}}
\begin{equation}
 \left.
 \begin{array}{lll}
 \displaystyle \frac{d i_{1}}{dt}&=& \displaystyle (x_{1} - x_{2})i_{1} + r \sin ( \omega t),   \vspace{2mm} \\
 \displaystyle \frac{d x_{1}}{dt}&=& \displaystyle 2 ( {i_{3}}^{2} - {i_{1}}^{2} ),             \vspace{3mm} \\
 \displaystyle \frac{d i_{2}}{dt}&=& \displaystyle (x_{2} - x_{3})i_{2},                        \vspace{2mm} \\
 \displaystyle \frac{d x_{2}}{dt}&=& \displaystyle 2 ( {i_{1}}^{2} - {i_{2}}^{2} ),             \vspace{3mm} \\
 \displaystyle \frac{d i_{3}}{dt}&=& \displaystyle (x_{3} - x_{1})i_{3},                        \vspace{2mm} \\
 \displaystyle \frac{d x_{3}}{dt}&=& \displaystyle 2 ( {i_{2}}^{2} - {i_{3}}^{2} ), 
 \end{array}
\right \}
\label{eqn: toda-c-12}
\vspace{1mm}
\end{equation}
where $r$ and $\omega$ are constants.  
\end{shadebox}
\end{minipage}
%\vspace*{10mm}
\end{center}
We show their non-periodic and quasi-periodic responses, Poincar\'e maps, and $i_{j}-v_{j}$ loci in Figures \ref{fig:Toda-C-attractor}, \ref{fig:Toda-C-poincare}, and \ref{fig:Toda-C-pinch}, respectively ($j=1, \, 2, \, 3$).  
In order to obtain these figures, we have to choose the parameters and the initial conditions carefully, and the maximum step size $h$ of the numerical integration must be sufficiently small ($h=0.006$). 
The following parameters are used in our computer simulations:
\begin{equation}
 \ r = 0.12,  \ \omega = 1.
\end{equation}
%
%
%

%
%-------------------------------------%
\subsubsection{Complexity order}
%-------------------------------------%
%
Consider the case where  Eq. (\ref{eqn: toda-c-12}) exhibits the quasi-periodic response.
Then, the $i_{j}-v_{j}$ loci  lie in the first and the fourth quadrants as shown in Figure \ref{fig:Toda-C-pinch}(d)-(f). 
We show next the $v_{j}-p_{j}$ locus in Figure \ref{fig:Toda-C-power}(d)-(f), where $p_{j}(t)$ is an instantaneous power defined by $p_{j}(t)=i_{j}(t)v_{j}(t)$ ($j=1, \, 2, \, 3$).
Observe that the $v_{j}-p_{j}$ loci are pinched at the origin, and the loci lie in the first and the third quadrants. 
Thus, when $v_{j}>0$, the instantaneous power $p_{j}(t)$ delivered from the forced signal and the inductor is dissipated in the memristor.  
However, when $v_{j}<0$, the instantaneous power $p_{j}(t)$ is \emph{not} dissipated in the memristor. 
Hence, the memristors switch between passive and active modes of operation, depending on its terminal voltage. 
They switches between two modes of operation:  
\begin{equation}
  (v_{j}, \, p_{j}) = ( +, \, + ), \ ( -, \, -), \ \  (j=1, \, 2, \, 3).     
\label{eqn: mode-quasi}
\end{equation}
Here, $(v_{j}, \, p_{j}) = (+, \, +)$ is read as $v_{1}>0$ and $p_{1}>0$,
$(v_{j}, \, p_{j}) = (-, \, -)$  is read as $v_{1}<0$ and $p_{1}<0$, and 
we excluded the special case where $(v_{1}, \, p_{1}) = (0, \, 0)$.  
Thus, we conclude as follow: \\
\begin{center}
\begin{minipage}{.7\textwidth}
\begin{itembox}[l]{Switching behavior of the memristor}
Assume that Eq. (\ref{eqn: toda-c-12}) exhibits the \emph{quasi-periodic} response.  
Then the generic memristors defined by Eq. (\ref{eqn: toda-C-v-i}) can switch between ``passive'' and ``active'' modes of operation, depending on its terminal voltage.  
\end{itembox} 
\end{minipage}
\end{center}

In the case of \emph{non-periodic} response, the above property does not hold as shown in Figure \ref{fig:Toda-C-pinch}(a) and Figure \ref{fig:Toda-C-power}(a).  
The generic memristor connected across the periodic source exhibits more complicated behavior. 
It switches between four modes of operation:  
\begin{equation}
  (v_{1}, \, p_{1}) = ( +, \, + ), \ ( +, \, -), \  ( -, \, +), \  ( -, \, -).     
\label{eqn: mode-chaotic}
\end{equation}
Here, $v_{1}$ and $p_{1}$ denote the terminal voltage and the instantaneous power of the generic memristor connected across the periodic source.   
That is,  $p_{1}(t)$ is defined by $p_{1}(t)=i_{1}(t)v_{1}(t)$, 
and $(v_{1}, \, p_{1}) = (+, \, -)$ is read as $v_{1}>0$ and $p_{1}<0$.   
Here, we exclude the special case where $(v_{1}, \, p_{1}) = (0, \, 0)$.  
Note that the other two memristors switch between passive and active modes of operation, depending on its terminal voltage, as shown in Figures Figure \ref{fig:Toda-C-pinch}(b)-(c) and Figure \ref{fig:Toda-C-power}(b)-(c).   
That is, they switch between two modes of operation:  
\begin{equation}
  (v_{1}, \, p_{1}) = ( +, \, + ), \ ( -, \, -),   
\label{eqn: mode-chaotic-2}
\end{equation}
and
\begin{equation}
  (v_{2}, \, p_{2}) = ( +, \, + ), \ ( -, \, -).     
\label{eqn: mode-chaotic-3}
\end{equation}
Thus, the memristor's operation modes (\ref{eqn: mode-chaotic}) can be coded by two bits: 
\begin{equation}
  ( 0, \, 0 ), \ ( 0, \,1), \  ( 1, \, 0), \  (1, \, 1),  
\end{equation}
where $+$ is coded to a binary number $0$ and $-$ to $1$.  
The operation modes (\ref{eqn: mode-quasi}), (\ref{eqn: mode-chaotic-2}), and (\ref{eqn: mode-chaotic-3}) are coded by one bit: 
\begin{equation}
   ( 0 ), \  ( 1 ),  
\end{equation}
which are equivalent to $( 0, \, 0 ), \ ( 1, \,1)$, respectively.  
Hence, we can measure the \emph{complexity order} of the memristor's operation modes, by using the above binary coding.  
Thus, the memristor's operation have the higher complexity when Eq. (\ref{eqn: toda-c-12}) exhibits \emph{non-periodic} response.   
We conclude as follow: 
\begin{center}
\begin{minipage}{.7\textwidth}
\begin{itembox}[l]{Switching behavior of the memristor}
\begin{enumerate}
\item Assume that Eq. (\ref{eqn: toda-c-12}) exhibits \emph{non-periodic} response.  
Then the generic memristor defined by Eq. (\ref{eqn: toda-C-v-i}) switches randomly between four modes of operation, that is,
\[
  (v_{1}, \, p_{1}) = ( +, \, + ), \ ( +, \, -), \  ( -, \, +), \  ( -, \, -).   
\]
They can be coded by two bits: 
\[
  ( 0, \, 0 ), \ ( 0, \,1), \  ( 1, \, 0), \  (1, \, 1),  
\]
where $+$ is coded to a binary number $0$ and $-$ to $1$.  

\item Assume that Eq. (\ref{eqn: toda-c-12}) exhibits \emph{quasi-periodic} response.  
Then the generic memristor defined by Eq. (\ref{eqn: toda-C-v-i})  switches randomly between two modes of operation, that is,
\[
  (v_{1}, \, p_{1}) = ( +, \, + ), ( -, \, -).   
\]
They can be coded by
\[
  ( 0, \, 0 ), \ ( 1, \,1), 
\]
respectively, which are equivalent to the one-bit coding $( 0 ), \  ( 1 )$. 
\end{enumerate}
\end{itembox} 
\end{minipage}
\end{center}
Note that if the forced memristor circuits have different kinds of elements, for example, capacitors, then more complicated modes may appear.

%---Fig. 51-------%
\begin{figure}[hpbt]
 \centering
   \begin{tabular}{cc}
    \psfig{file=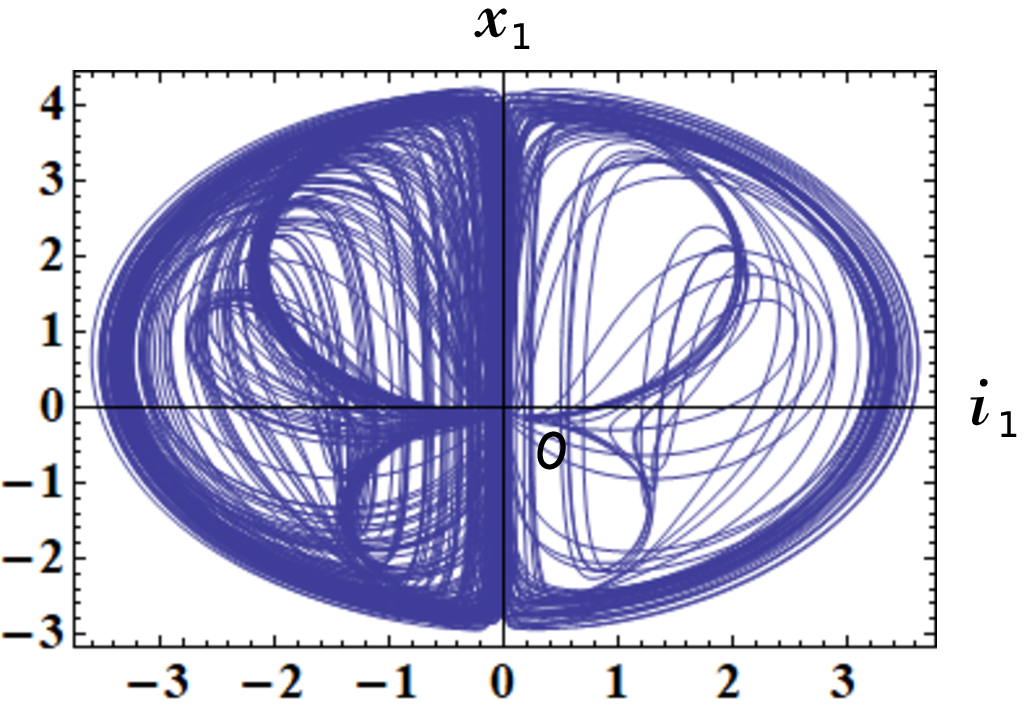, height=4.8cm}  & 
    \psfig{file=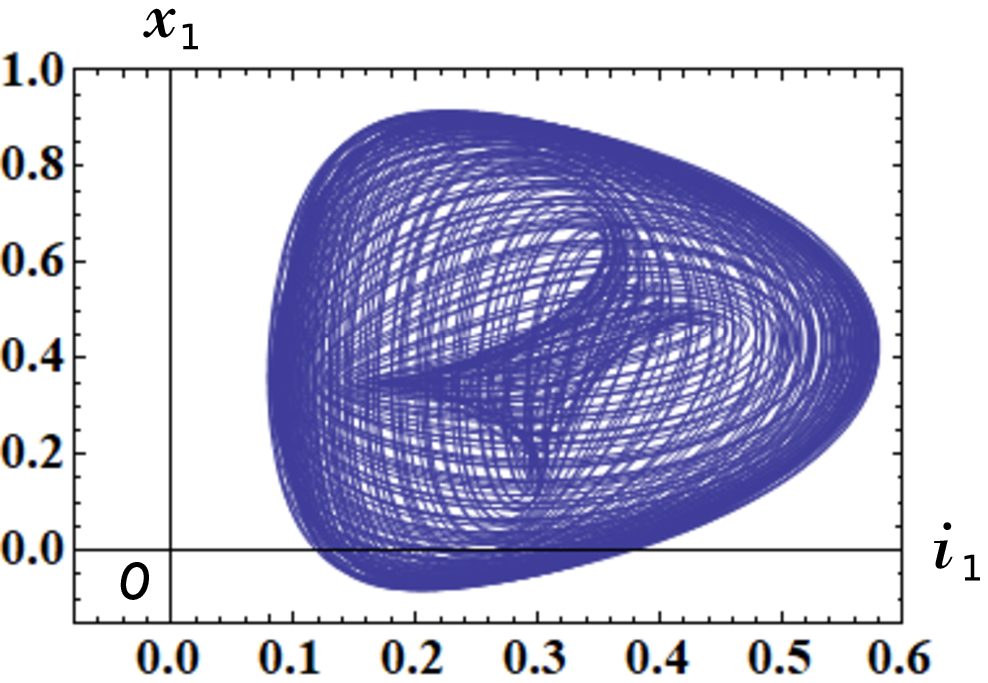, height=4.8cm}  \\
     (a) non-periodic ($r = 0.12,  \ \omega = 1$) & (d)  quasi-periodic ($r = 0.001,  \ \omega = 1$) \vspace{5mm} \\
    \psfig{file=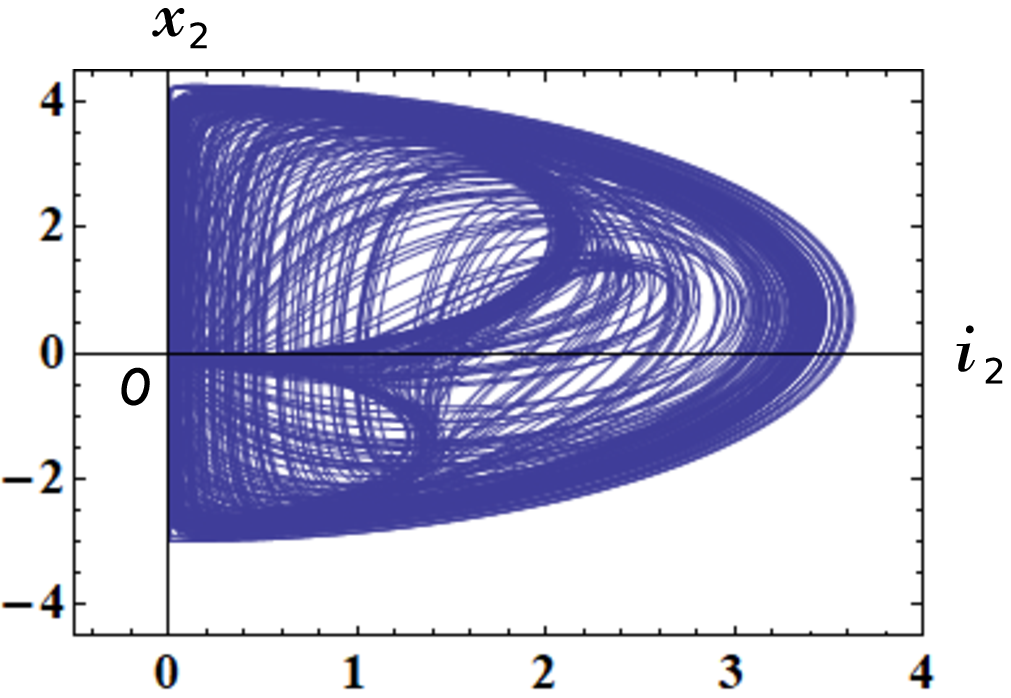, height=4.8cm}  & 
    \psfig{file=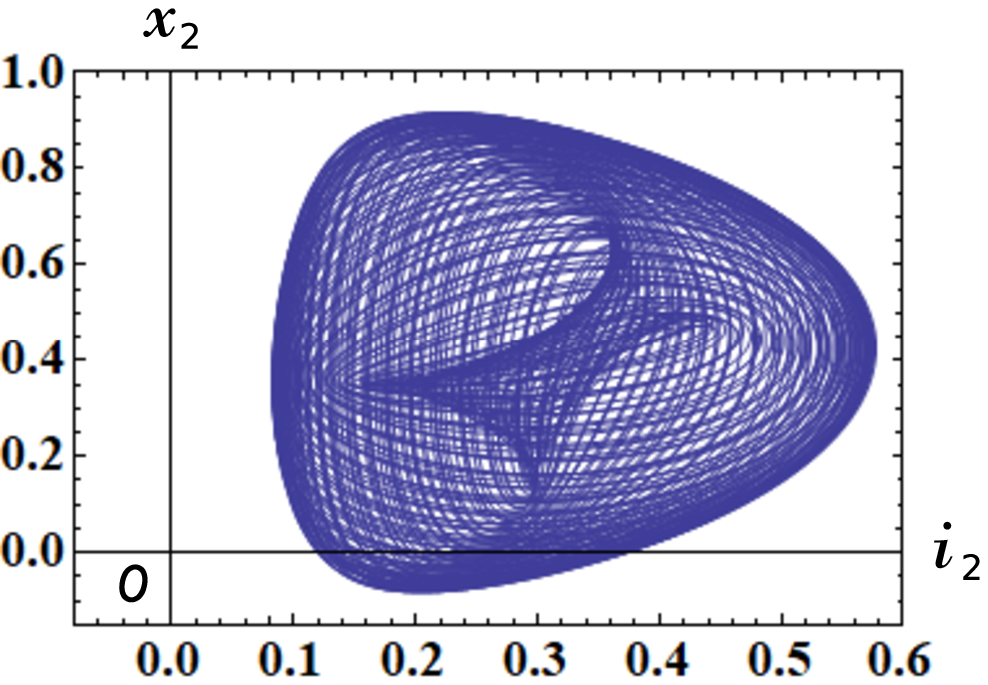, height=4.8cm}  \\
     (b) non-periodic ($r = 0.12,  \ \omega = 1$) & (e)  quasi-periodic ($r = 0.001,  \ \omega = 1$) \vspace{5mm} \\
    \psfig{file=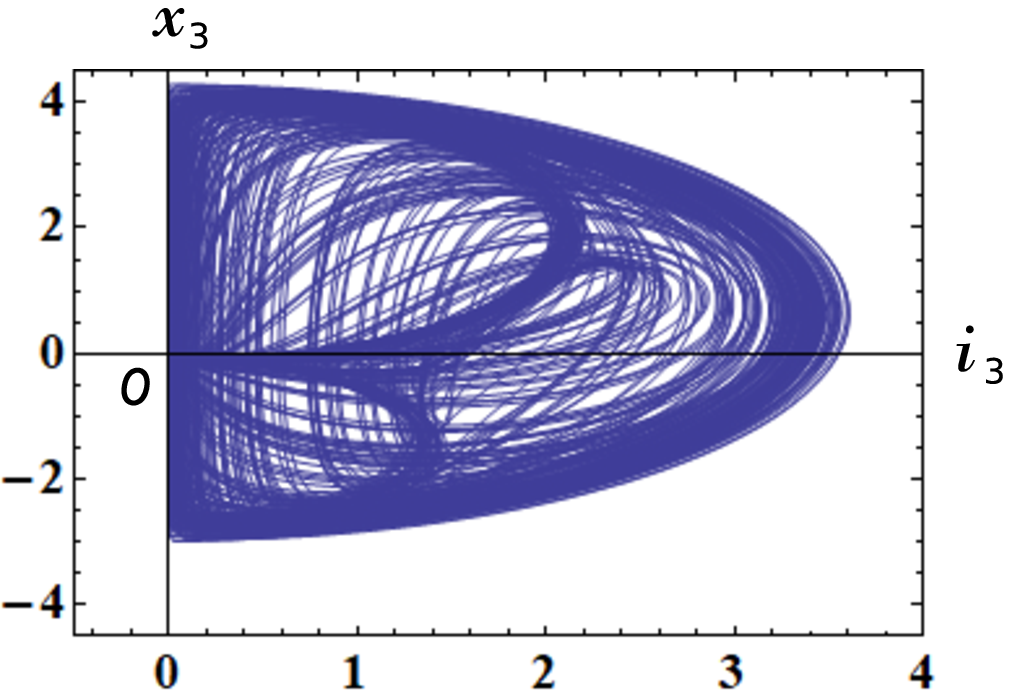, height=4.8cm}  & 
    \psfig{file=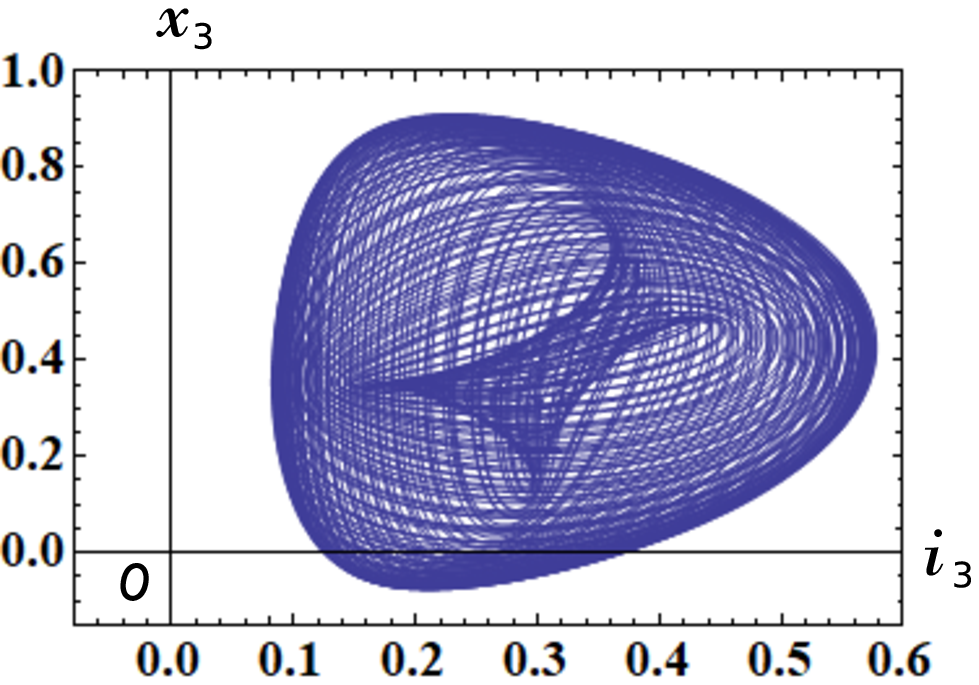, height=4.8cm}  \\
     (c) non-periodic ($r = 0.12,  \ \omega = 1$) & (f)  quasi-periodic ($r = 0.001,  \ \omega = 1$) \vspace{5mm} \\
    \end{tabular}
  \caption{Non-periodic and quasi-periodic responses of the forced memristor Toda lattice equations C, 
   which are defined by Eq. (\ref{eqn: toda-c-12}). 
   Here, $i_{j}$ and $x_{j}$ denote the terminal current and the internal state of the $j$-th generic memristor, respectively 
   $(j=1, \, 2, \, 3)$. 
   Parameters:  (a)-(c) $r = 0.12,  \ \omega = 1$. \ \ (b)-(c) $r = 0.001,  \ \omega = 1$.
   Initial conditions: $i_{1}(0) = 0.1, \, x_{1}(0)= 0.2, \, i_{2}(0) = 0.3, \, x_{2}(0)= 0.4, \, i_{3}(0) = 0.5, \, x_{3}(0) = 0.6$.}
  \label{fig:Toda-C-attractor} 
\end{figure}
%
%

%---Fig. 52-------%
\begin{figure}[hpbt]
 \centering
   \begin{tabular}{cc}
    \psfig{file=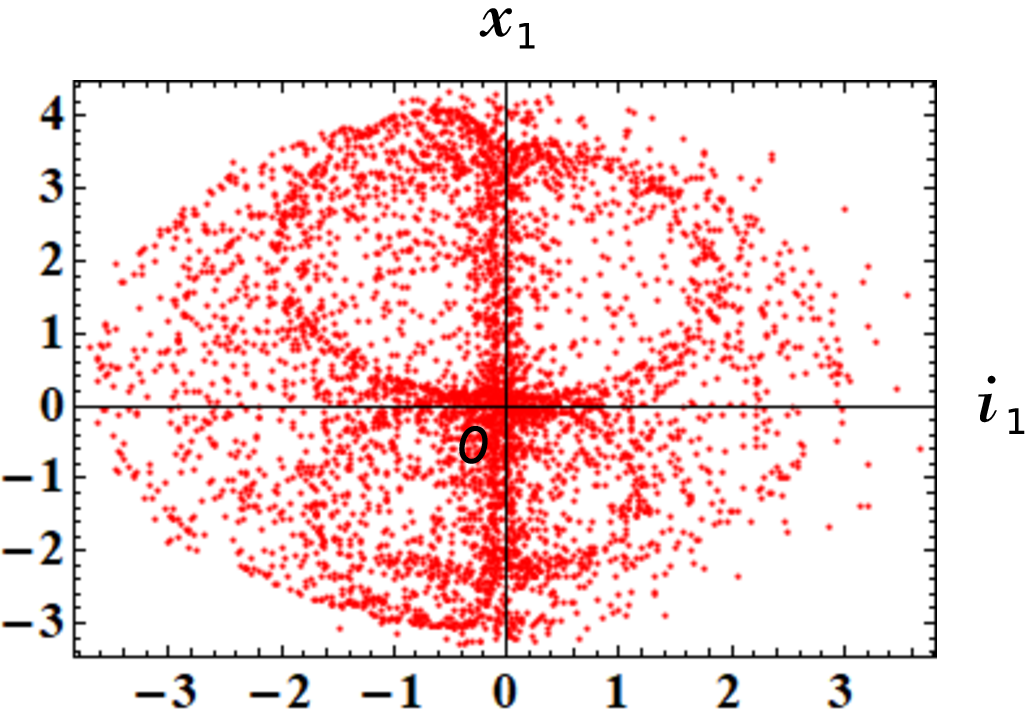, height=4.8cm}  & 
    \psfig{file=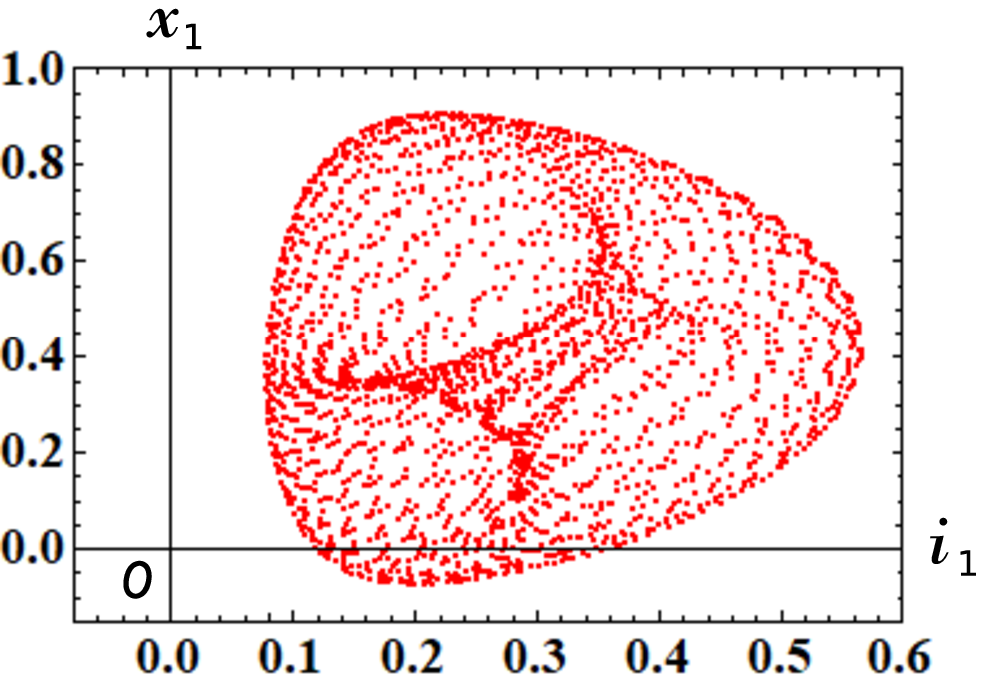, height=4.8cm}  \\
     (a) non-periodic ($r = 0.12,  \ \omega = 1$) & (d)  quasi-periodic ($r = 0.001,  \ \omega = 1$) \vspace{5mm} \\
    \psfig{file=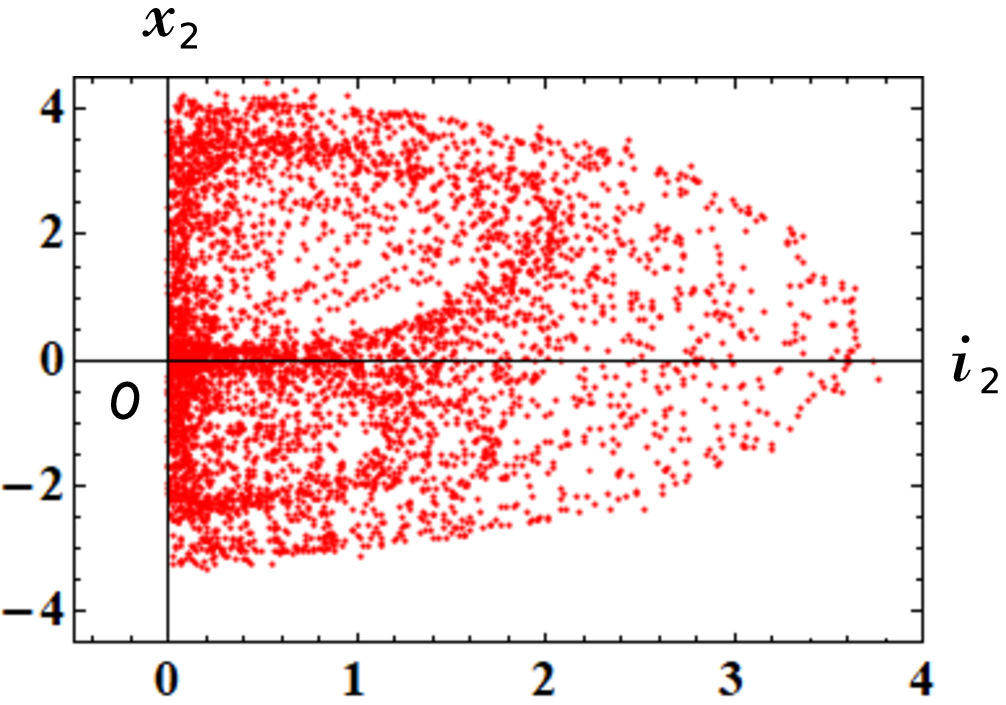, height=4.8cm}  & 
    \psfig{file=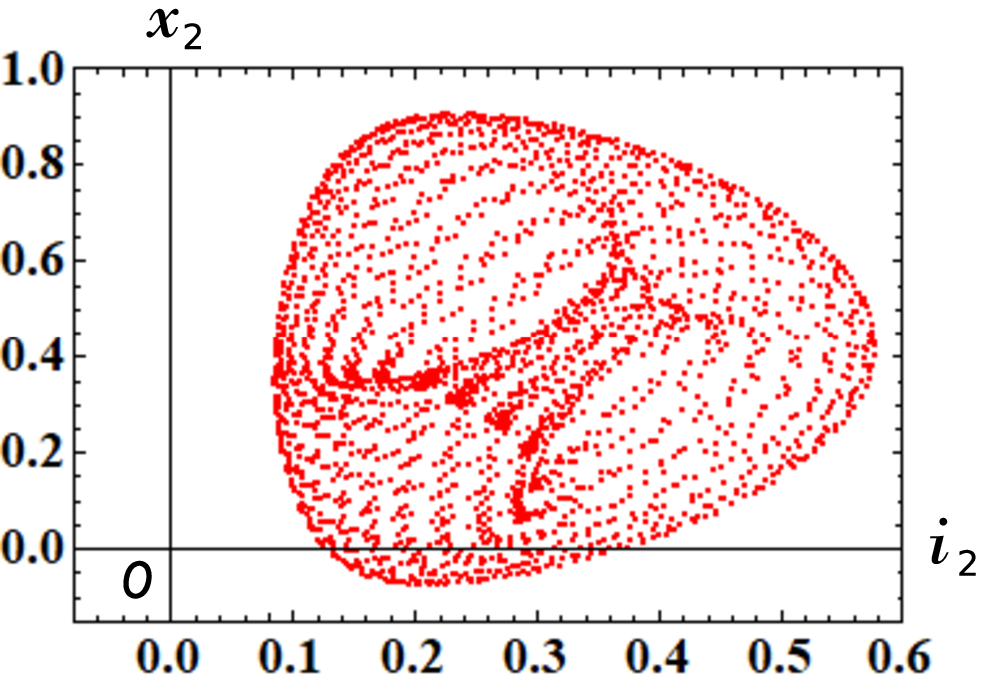, height=4.8cm}  \\
     (b) non-periodic ($r = 0.12,  \ \omega = 1$) & (e)  quasi-periodic ($r = 0.001,  \ \omega = 1$) \vspace{5mm} \\
    \psfig{file=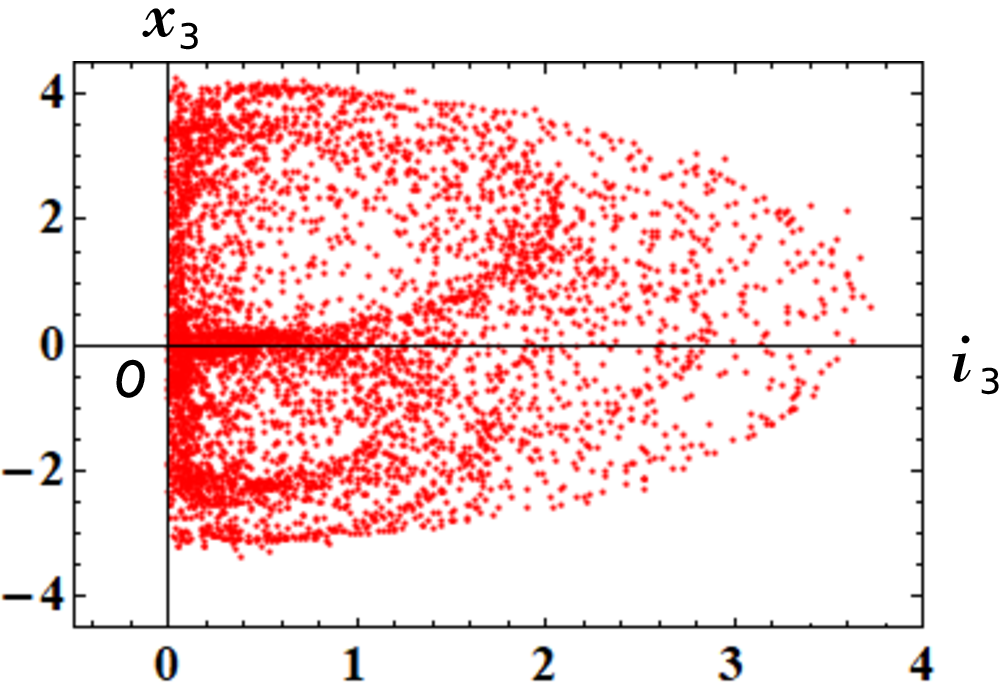, height=4.8cm}  & 
    \psfig{file=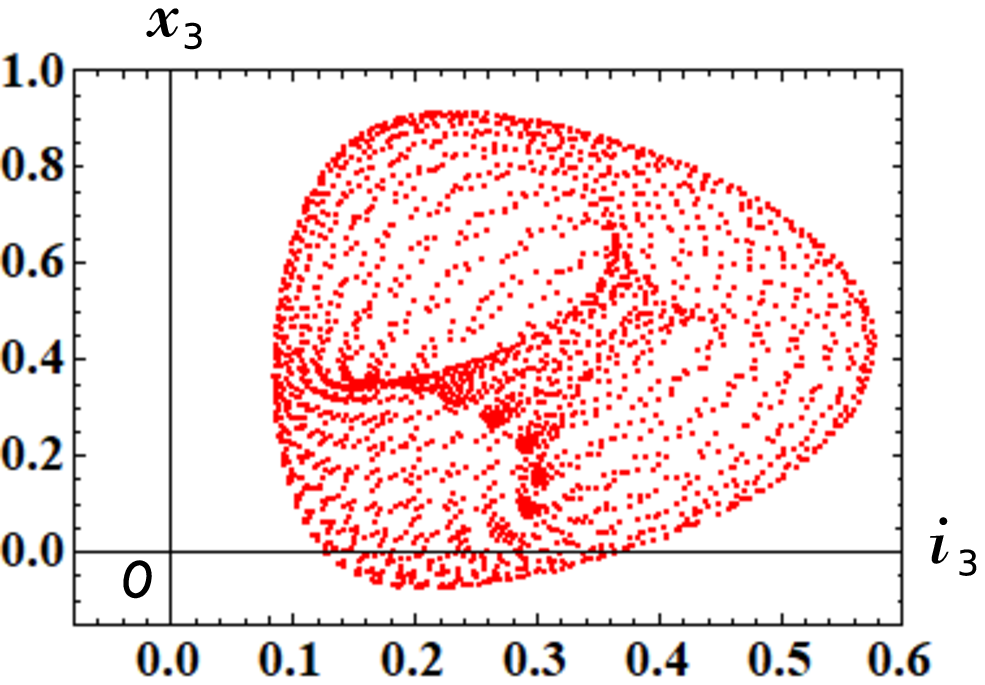, height=4.8cm}  \\
     (c) non-periodic ($r = 0.12,  \ \omega = 1$) & (f)  quasi-periodic ($r = 0.001,  \ \omega = 1$) \vspace{5mm} \\
    \end{tabular}
  \caption{Poincar\'e maps of the forced memristor Toda lattice equations C, which are defined by Eq. (\ref{eqn: toda-c-12}). 
   Here, $i_{j}$ and $x_{j}$ denote the terminal current and the internal state of the $j$-th generic memristor, respectively 
   $(j=1, \, 2, \, 3)$. 
   Parameters:  (a)-(c) $r = 0.12,  \ \omega = 1$. \ \ (b)-(c) $r = 0.001,  \ \omega = 1$.
   Initial conditions: $i_{1}(0) = 0.1, \, x_{1}(0)= 0.2, \, i_{2}(0) = 0.3, \, x_{2}(0)= 0.4, \, i_{3}(0) = 0.5, \, x_{3}(0) = 0.6$.}
  \label{fig:Toda-C-poincare} 
\end{figure}
%
%

%---Fig. 53-------%
\begin{figure}[hpbt]
 \centering
   \begin{tabular}{cc}
    \psfig{file=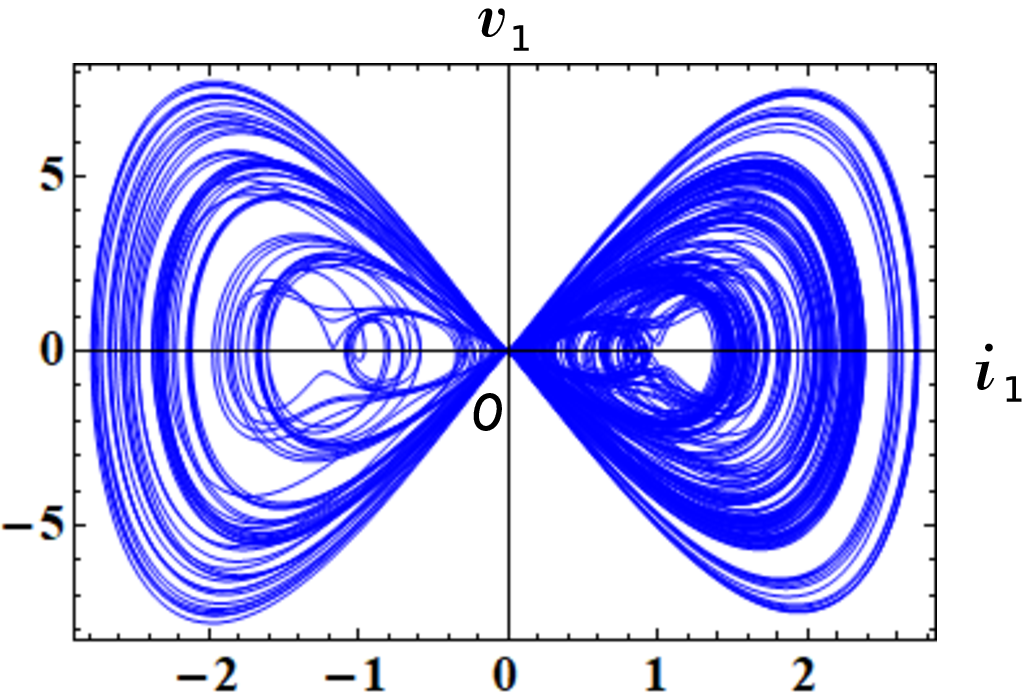, height=5cm}  & 
    \psfig{file=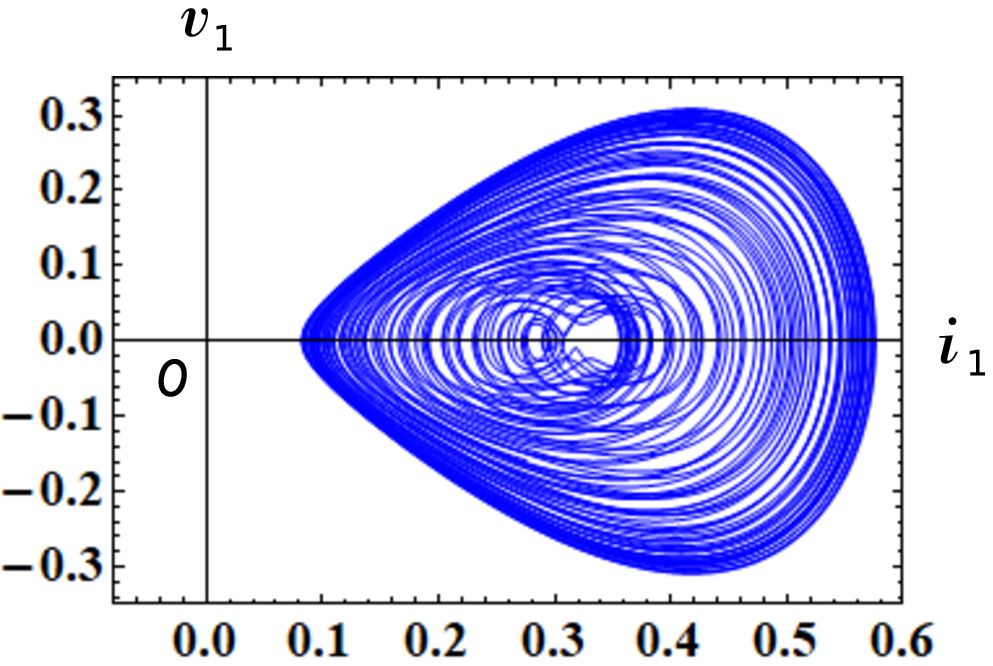, height=5cm}  \\
    (a) non-periodic $(r = 0.12,  \ \omega = 1)$ & (d) quasi-periodic $(r = 0.001,  \ \omega = 1)$ \vspace{5mm} \\   
    \psfig{file=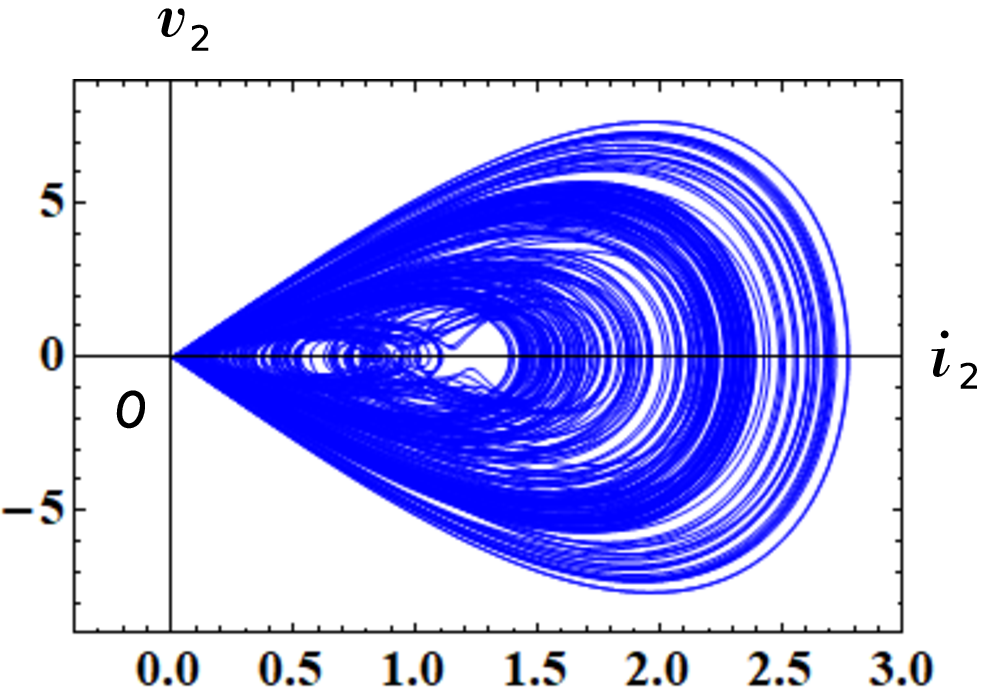, height=5cm}  & 
    \psfig{file=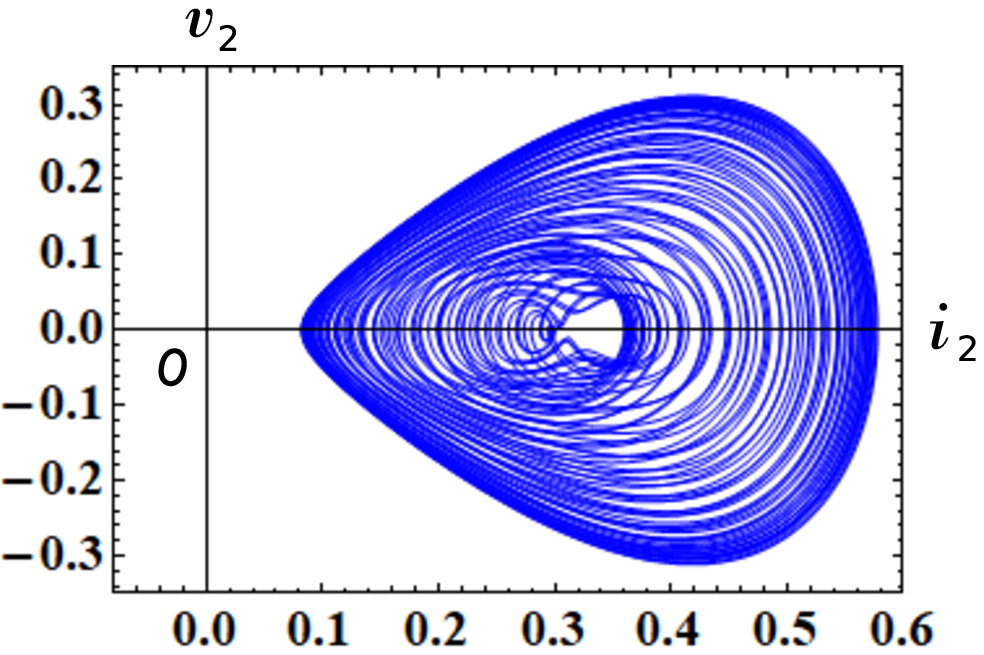, height=5cm}  \\    
    (b) non-periodic $(r = 0.12,  \ \omega = 1)$& (e) quasi-periodic  $(r = 0.001,  \ \omega = 1)$ \vspace{5mm}\\
    \psfig{file=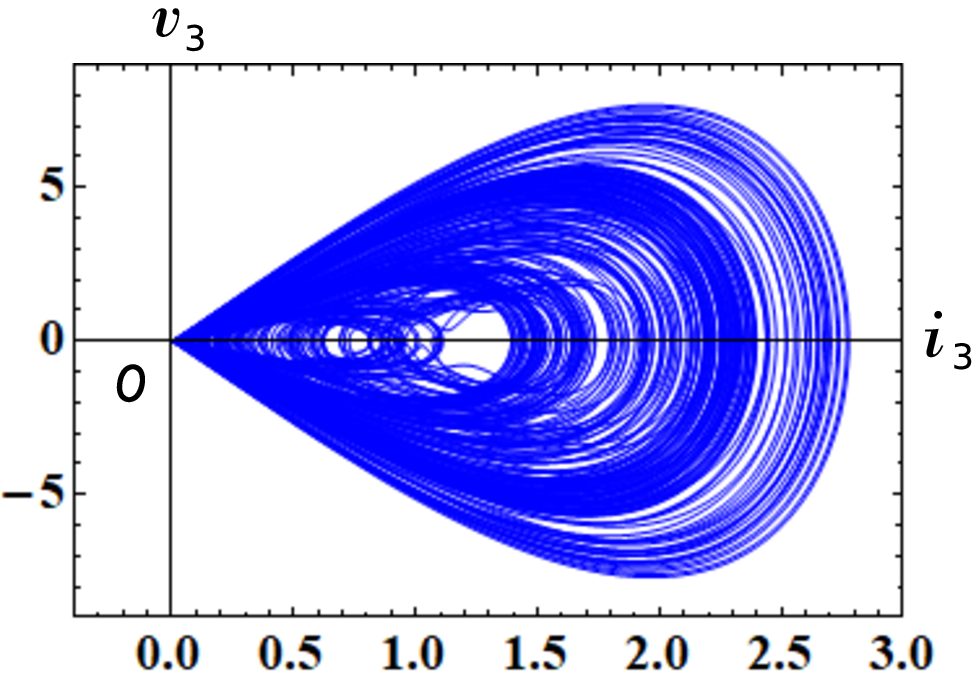, height=5cm}  & 
    \psfig{file=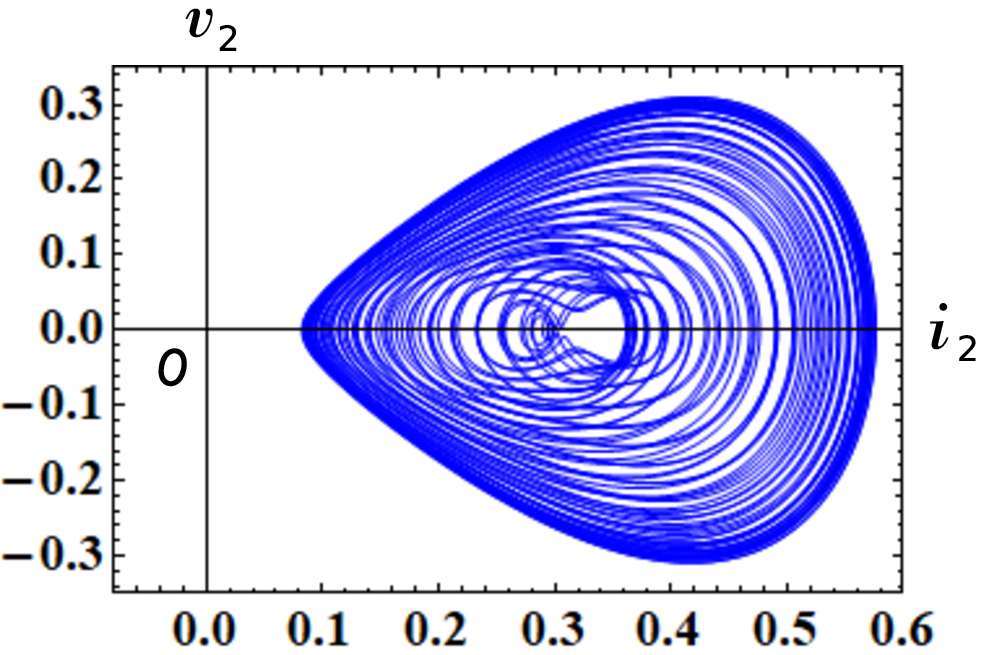, height=5cm}  \\    
    (c) non-periodic $(r = 0.12,  \ \omega = 1)$& (f) quasi-periodic  $(r = 0.001,  \ \omega = 1)$
   \end{tabular}
  \caption{ The $i_{j}-v_{j}$ loci of the forced memristor Toda lattice equations C, which are defined by Eq. (\ref{eqn: toda-c-12}).  
   Here, $i_{j}$ and $v_{j}$ denote the terminal current and the voltage of the $j$-th generic memristor, respectively 
   $(j=1, \, 2, \, 3)$.  
   Parameters:  (a)-(c) $\ r = 0.12,  \ \omega = 1$. \ \ (d)-(f) $\ r = 0.001,  \ \omega = 1$.
   Initial conditions: $i_{1}(0) = 0.1, \, x_{1}(0)= 0.2, \, i_{2}(0) = 0.3, \, x_{2}(0)= 0.4, \, i_{3}(0) = 0.5, \, x_{3}(0) = 0.6$.}
  \label{fig:Toda-C-pinch} 
\end{figure}
%
%

%---Fig. 54-------%
\begin{figure}[hpbt]
 \centering
   \begin{tabular}{cc}
    \psfig{file=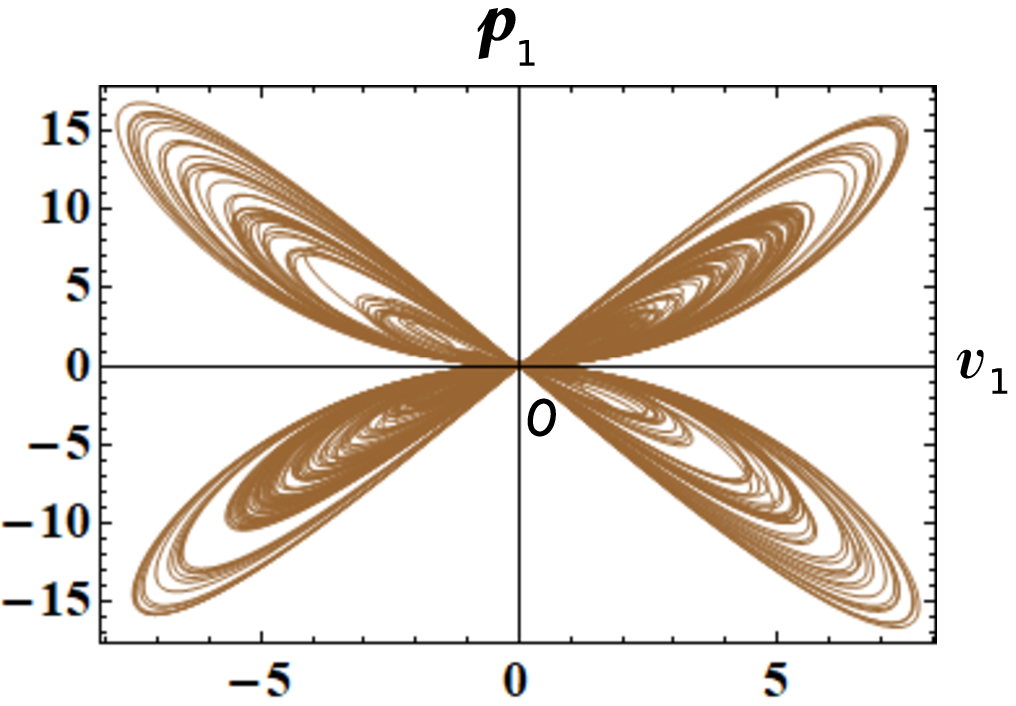, height=5cm}  & 
    \psfig{file=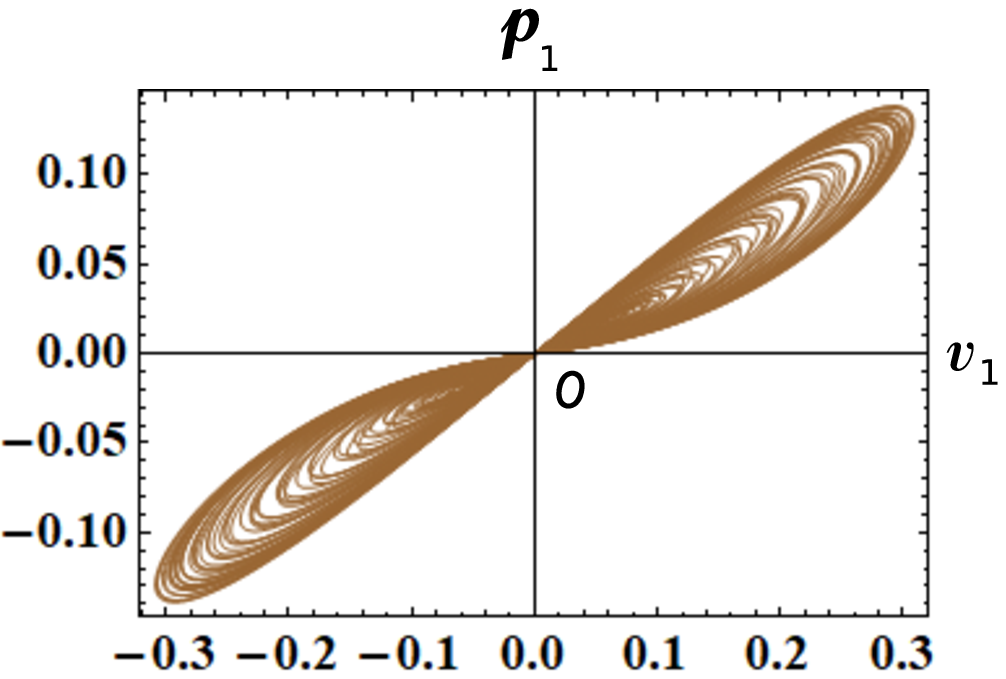, height=5cm}  \\
    (a) non-periodic $(r = 0.12,  \ \omega = 1)$ & (d) quasi-periodic $(r = 0.001,  \ \omega = 1)$ \vspace{5mm} \\   
    \psfig{file=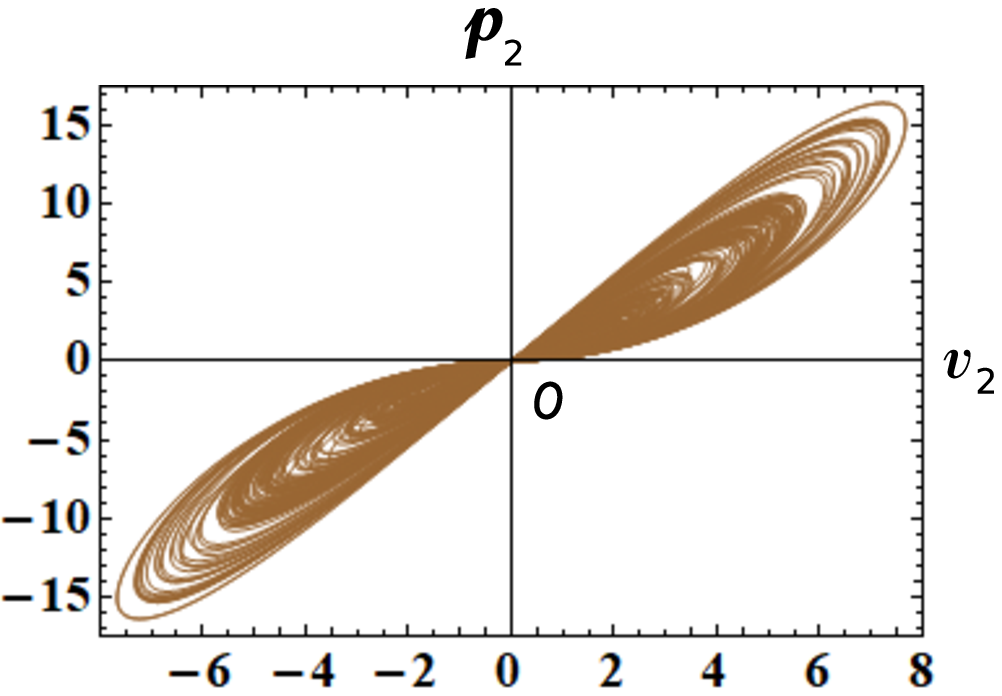, height=5cm}  & 
    \psfig{file=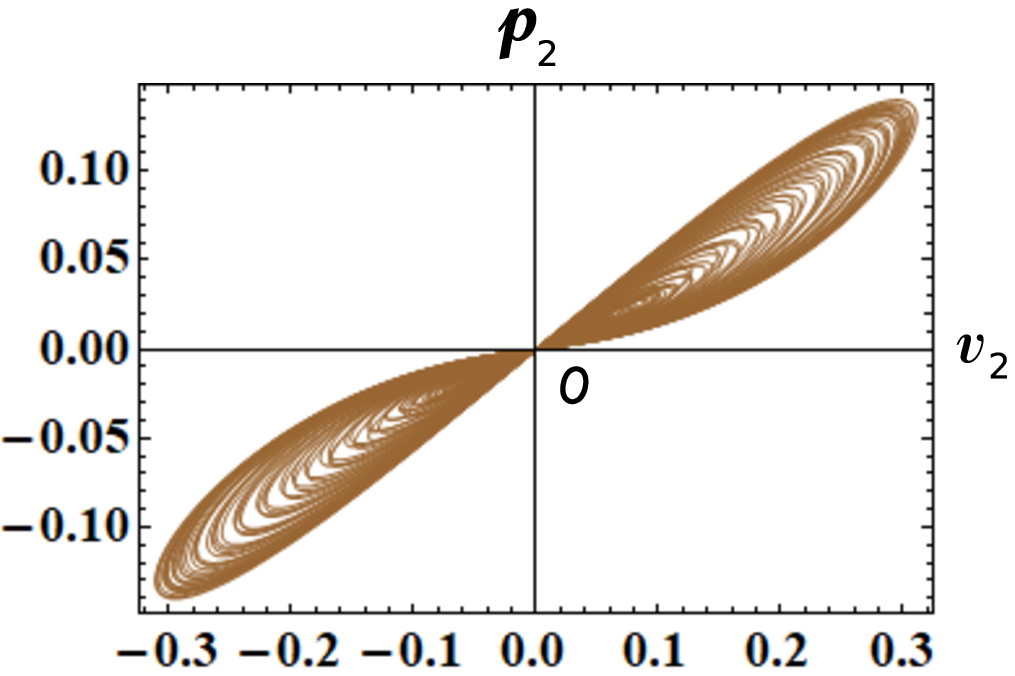, height=5cm}  \\    
    (b) non-periodic $(r = 0.12,  \ \omega = 1)$& (e) quasi-periodic  $(r = 0.001,  \ \omega = 1)$ \vspace{5mm}\\
    \psfig{file=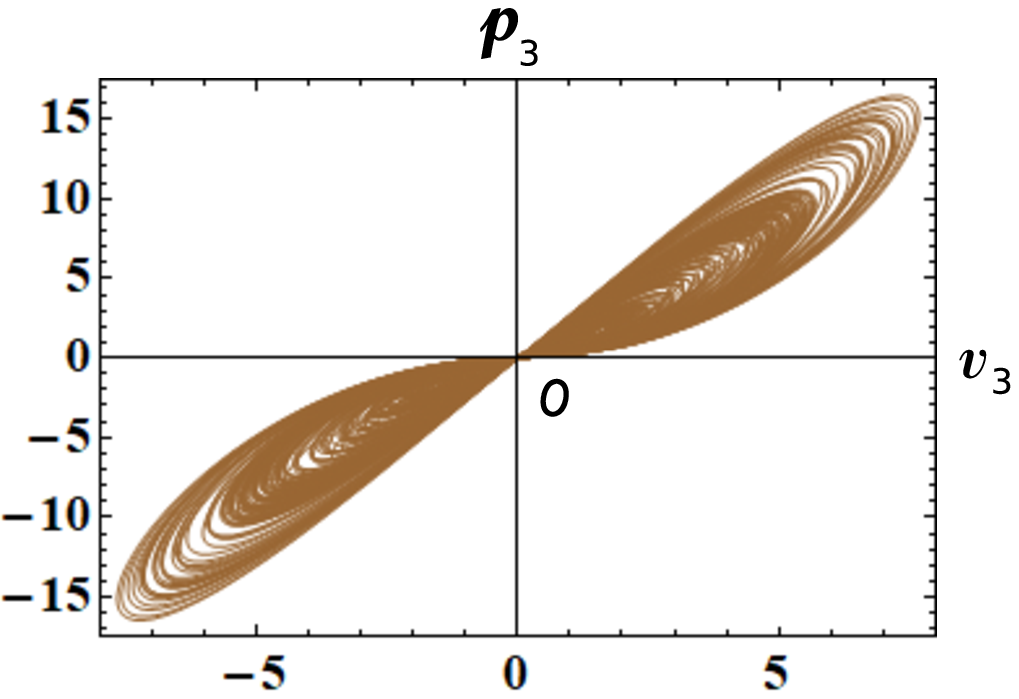, height=5cm}  & 
    \psfig{file=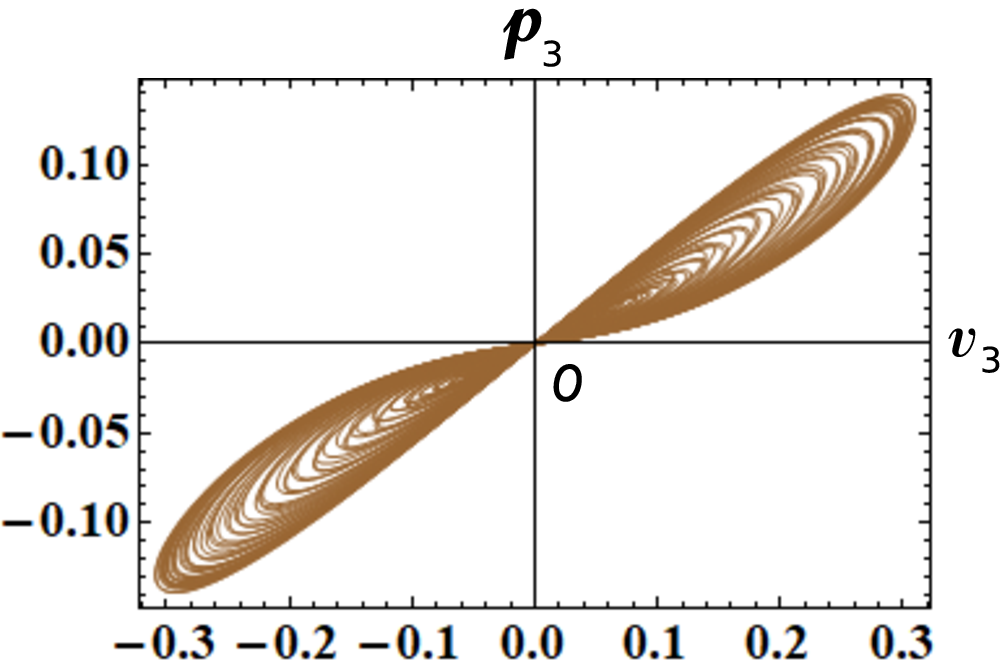, height=5cm}  \\    
    (c) non-periodic $(r = 0.12,  \ \omega = 1)$& (f) quasi-periodic  $(r = 0.001,  \ \omega = 1)$
   \end{tabular}
  \caption{ The $v_{j}-p_{j}$ loci of the forced memristor Toda lattice equations C, which are defined by Eq. (\ref{eqn: toda-c-12}).  
   Here, $p_{j}(t)$ is an instantaneous power defined by $p_{j}(t)=i_{j}(t)v_{j}(t)$ , 
   and $v_{j}(t)$ and $i_{j}(t)$ denote the terminal voltage and the terminal current of the $j$-th generic memristor, respectively 
   $(j=1, \, 2, \, 3)$.  
   Observe that the loci except for Figure \ref{fig:Toda-C-power}(a) are pinched at the origin, 
   and their loci lie in the first and the third quadrants. 
   These memristors switch between passive and active modes of operation, depending on its terminal voltage $v_{j}(t)$.  
   However, the $v_{1}-p_{1}$ locus in Figure \ref{fig:Toda-C-power}(a) does not satisfy this behavior.  
   Parameters:  (a)-(c) $\ r = 0.12,  \ \omega = 1$. \ \ (d)-(f) $\ r = 0.001,  \ \omega = 1$.
   Initial conditions: $i_{1}(0) = 0.1, \, x_{1}(0)= 0.2, \, i_{2}(0) = 0.3, \, x_{2}(0)= 0.4, \, i_{3}(0) = 0.5, \, x_{3}(0) = 0.6$.}
  \label{fig:Toda-C-power} 
\end{figure}
\newpage

%\clearpage
%-------------------------------------%
\subsection{$N$-dimensional Lotka-Volterra equations}
\label{sec: N-Lotka-Volterra}
%-------------------------------------%

Consider the Hamiltonian defined by \cite{Fernandes(1998)}
\begin{equation}
  \mathcal {H} = 
    - \sum_{j=1}^{N} \ e^{ \ \left ( P_{j} + \displaystyle \frac{1}{2} \sum_{k=1}^{N} a_{jk} \, Q_{k} \right )}, 
\label{eqn: hamiltonian-20}
\end{equation}
where $a_{jk}$ is skew-symmetric ($a_{jk}=-a_{kj}$).  
Let us define the Hamiltonian form 
\begin{equation}
\begin{array}{ccr}
 \displaystyle {\frac {d Q_{i}}{d t}} 
  &=& \displaystyle -{\frac {\partial {\mathcal {H}}}{\partial P_{i}}},
 \vspace{2mm} \\
 \displaystyle {\frac {d P_{i}}{d t}} 
  &=& \displaystyle  {\frac {\partial {\mathcal {H}}}{\partial Q_{i}}}.  
\end{array}
\vspace{2mm}
\label{eqn: hamilton-21}
\end{equation}
Remark the reversed sign of the Hamiltonian form of Eq. (\ref{eqn: hamilton-21}) \cite{Fernandes(1998)}.

From Eq. (\ref{eqn: hamilton-10}), we obtain 
\begin{equation}
\begin{array}{lll}
 \displaystyle {\frac {d Q_{i}}{d t}} 
  &=& \displaystyle - \frac {\partial {\mathcal {H}}}{\partial P_{i}} = e^{ \ \left ( P_{i} + \displaystyle \frac{1}{2} \sum_{k=1}^{N}a_{ik} \, Q_{k} \right )}, 
 \vspace{2mm} \\
 \displaystyle {\frac {d P_{i}}{d t}} 
  &=& \displaystyle \frac {\partial {\mathcal {H}}}{\partial Q_{i}} = - \sum_{j=1}^{N} \frac{a_{ji}}{2} \ e^{ \ \left ( P_{j} + \displaystyle \frac{1}{2} \sum_{k=1}^{N} a_{jk} \, Q_{k} \right )}.  
\end{array}
\vspace{2mm}
\label{eqn: hamilton-22}
\end{equation}
It can be recast into the form \cite{Fernandes(1998)}
\begin{equation}
  \ddot{Q_{i}} = \sum_{k=1}^{N} a_{ik} \, \dot{Q_{i}}\dot{Q_{k}}, 
\end{equation}
where $\dot{Q_{i}} = \frac{dQ_{i}}{dt}$ and $\ddot{Q_{i}} = \frac{d^{2}Q_{i}}{dt^{2}}$.  
If we set $X_{i} = \dot{Q_{i}}$, we obtain 
\begin{equation}
  \frac{d X_{i}}{dt} = \sum_{k=1}^{N} a_{ik} X_{i} X_{k}.   
\label{eqn: hamilton-23}
\end{equation}
As a special case of Eq. (\ref{eqn: hamilton-23}), we can define the following $N$-dimensional Lotka-Volterra equations \cite{Toda(1989)}.  

The dynamics of the $N$-dimensional Lotka-Volterra equations is given by  
\begin{center}
\begin{minipage}{8.7cm}
\begin{shadebox}
\underline{\emph{$N$-dimensional Lotka-Volterra equations}}
\begin{equation}
 \displaystyle \frac{d X_{n}}{dt} = (X_{n-1} - X_{n+1}) X_{n}, 
\label{eqn: N-Lotka-Volterra-1}
\vspace{1mm}
\end{equation}
where $n=1, \ 2, \cdots, \ N$ and we consider the case of a periodic lattice of the length $N$: $X_{n} = X_{n + N}$. 
\end{shadebox}
\end{minipage}
\end{center}
For example, if $n=3$, Eq. (\ref{eqn: N-Lotka-Volterra-1}) is written as 
\begin{center}
\begin{minipage}{8.7cm}
\begin{shadebox}
\underline{\emph{$3$-dimensional Lotka-Volterra equations}}
\begin{equation}
\left. 
\begin{array}{ccl}
 \displaystyle \frac{d X_{1}}{dt} &=& (X_{3} - X_{2}) X_{1},  \vspace{2mm} \\
 \displaystyle \frac{d X_{2}}{dt} &=& (X_{1} - X_{3}) X_{2},  \vspace{2mm} \\
 \displaystyle \frac{d X_{3}}{dt} &=& (X_{2} - X_{1}) X_{3}.   
\end{array}
\right \}
\label{eqn: 3-Lotka-Volterra}
\end{equation}
\end{shadebox}
\end{minipage}
\end{center}
Equation (\ref{eqn: 3-Lotka-Volterra}) has the two integrals \cite{He(2012)}, since the solution satisfies  
\begin{center}
\begin{minipage}{.5\textwidth}
\begin{itembox}[l]{Integrals}
\begin{equation}
 \begin{array}{l}
  \displaystyle \frac{d}{dt} \bigl ( X_{1} + X_{2} + X_{3} \bigr )  = 0, \vspace{4mm} \\
  \displaystyle \frac{d}{dt} \bigl ( X_{1} X_{2} X_{3} \bigr ) = 0.   
 \end{array}
\label{eqn: 3-Lotka-Volterra-integrals}
\end{equation}
\end{itembox}
\end{minipage}
\end{center}
%
%

%-------------------------------------%
\subsubsection{Three-element memristor circuit realization}
%-------------------------------------%
%
Consider first the three-element memristor circuit in Figure \ref{fig:memristor-inductor-battery}.  
The dynamics of this circuit given by Eq. (\ref{eqn: dynamics-1}). 
Assume that Eq. (\ref{eqn: dynamics-1}) satisfies 
\begin{equation}
\left.
 \begin{array}{cc}
  E = 0, &  L = 1,  \vspace{2mm} \\
  \hat{R}( x_{1}, \, x_{2}, \ i ) &=  - (x_{2} - x_{1}),        \vspace{2mm} \\
  \tilde{f}_{1}(x_{1}, \, x_{2}, \, i) &= (i - x_{2}) x_{1},  \vspace{2mm} \\
  \tilde{f}_{2}(x_{1}, \, x_{2}, \, i) &= (x_{1} - i) x_{2}. 
 \end{array}
\right \}
\end{equation}
Then, we obtain the following $3$-dimensional memristor Lotka-Volterra equations 
\begin{center}
\begin{minipage}{8.7cm}
\begin{shadebox}
\underline{\emph{$3$-dimensional memristor Lotka-Volterra}}
\underline{\emph{equations}}
\begin{equation}
\left. 
\begin{array}{ccl}
 \displaystyle \frac{d i}{dt} &=& (x_{2} - x_{1}) i,  \vspace{2mm} \\
 \displaystyle \frac{d x_{1}}{dt} &=& (i - x_{2}) x_{1},  \vspace{2mm} \\
 \displaystyle \frac{d x_{2}}{dt} &=& (x_{1} - i) x_{2}.   
\end{array}
\right \}
\label{eqn: 3-Lotka-Volterra-2}
\end{equation}
\end{shadebox}
\end{minipage}
\end{center}
Equations (\ref{eqn: 3-Lotka-Volterra}) and (\ref{eqn: 3-Lotka-Volterra-2}) are equivalent if we change the variables  
\begin{equation}
  X_{1}=i, \ X_{2}=x_{1}, \ X_{3}=x_{2}.  
\end{equation}
In this case, the extended memristor  in Figure \ref{fig:memristor-inductor-battery} is replaced by the \emph{generic} memristor (see Appendix A). 
Thus, 
\begin{equation}
  \hat{R}(x_{1}, \, x_{2}, \, i)= \tilde{R}(x_{1}, \, x_{2})= - (x_{2} - x_{1}).  
\end{equation}
The terminal voltage $v_{M}$ and the terminal current $i_{M}$ of the current-controlled generic memristor are given by
\begin{center}
\begin{minipage}{8.7cm}
\begin{shadebox}
\underline{\emph{V-I characteristics of the generic memristor}}
\begin{equation}
\begin{array}{lll}
  v_{M} &=& \tilde{R}( x_{1}, \, x_{2} ) \, i_{M} = - (x_{2} - x_{1}) \, i_{M},   
  \vspace{3mm} \\
     \displaystyle \frac{d x_{1}}{dt} &=& (i_{M} - x_{2}) x_{1},
      \vspace{2mm} \\
     \displaystyle \frac{d x_{2}}{dt} &=& (x_{1} - i_{M}) x_{2}, 
\end{array}
\label{eqn: 3-Lotka-Volterra-4}
\end{equation}
where $\tilde{R}( x_{1}, \, x_{2} ) = - (x_{2} - x_{1})$. \vspace{2mm}
\end{shadebox}
\end{minipage}
\end{center}
It follows that the $3$-dimensional memristor Lotka-Volterra equations (\ref{eqn: 3-Lotka-Volterra-2}) can be realized by 
the three-element memristor circuit in Figure \ref{fig:memristor-inductor-battery}.  
This circuit can exhibit periodic behavior.  
If an external source is added as shown in Figure \ref{fig:memristive-inductor-battery-source},
then the forced memristor circuit can exhibit a non-periodic response.  
The dynamics of this circuit is given by 
\begin{center}
\begin{minipage}{10cm}
\begin{shadebox}
\underline{\emph{Forced $3$-dimensional memristor Lotka-Volterra equations}}
\begin{equation}
\left. 
\begin{array}{ccl}
 \displaystyle \frac{d i}{dt} &=& (x_{2} - x_{1}) i + r \sin ( \omega t),  \vspace{2mm} \\
 \displaystyle \frac{d x_{1}}{dt} &=& (i - x_{2}) x_{1},  \vspace{2mm} \\
 \displaystyle \frac{d x_{2}}{dt} &=& (x_{1} - i) x_{2},   
\end{array}
\right \}
\vspace{1mm}
\label{eqn: 3-Lotka-Volterra-5}
\end{equation}
where $r$ and $\omega$ are constants.  
\end{shadebox}
\end{minipage}
\end{center}
The solution of Eq. (\ref{eqn: 3-Lotka-Volterra-5}) satisfies 
\begin{equation}
 i(t) + x_{1}(t) + x_{2}(t) + \frac{r}{\omega } \cos ( \omega t) = K,    
\label{eqn: 3-Lotka-Volterra-integral-2}
\end{equation}
where $K$ is a constant.  
Thus, by eliminating $x_{2}$ from Eq. (\ref{eqn: 3-Lotka-Volterra-5}), we obtain the second-order non-autonomous differential equations: 
\begin{equation}
\left. 
\begin{array}{ccl}
 \displaystyle \frac{d i}{dt} 
  &=& \displaystyle \biggl \{ -\frac{r}{\omega } \cos ( \omega t) + K - i - 2 x_{1} \biggr \} i + r \sin ( \omega t),  \vspace{2mm} \\
 \displaystyle \frac{d x_{1}}{dt} 
  &=& \displaystyle \biggl \{ 2 i + \frac{r}{\omega } \cos ( \omega t) - K + x_{1}) \biggr \} x_{1}.     
\end{array}
\right \}
\vspace{1mm}
\label{eqn: 3-Lotka-Volterra-6}
\end{equation}
Equations (\ref{eqn: 3-Lotka-Volterra-5}) and (\ref{eqn: 3-Lotka-Volterra-6}) can exhibit non-periodic behavior.  
We show the non-periodic and quasi-periodic responses, Poincar\'e maps, and $i_{M}-v_{M}$ loci of Eq. (\ref{eqn: 3-Lotka-Volterra-5}) in Figures \ref{fig:3-Lotka-attractor}, \ref{fig:3-Lotka-poincare}, and \ref{fig:3-Lotka-pinch}, respectively.  
The $i_{M}-v_{M}$ loci in Figure \ref{fig:3-Lotka-pinch} lie in the first and the fourth quadrants.
Thus, the generic memristor defined by Eq. (\ref{eqn: 3-Lotka-Volterra-4}) is an active element. 
We show the $v_{M}-p_{M}$ locus in Figure \ref{fig:3-Lotka-power}, 
where $p_{M}(t)$ is an instantaneous power defined by $p_{M}(t)=i_{M}(t)v_{M}(t)$.  
Observe that the $v_{M}-p_{M}$ locus is pinched at the origin, and the locus lies in the first and the third quadrants. 
Thus, the memristor switches between passive and active modes of operation, depending on its terminal voltage. 
We conclude as follow: \\
\begin{center}
\begin{minipage}{.7\textwidth}

\begin{itembox}[l]{Switching behavior of the memristor}
Assume that Eq. (\ref{eqn: 3-Lotka-Volterra-5}) exhibits non-periodic or quasi-periodic oscillation.  
Then the generic memristor defined by Eq. (\ref{eqn: 3-Lotka-Volterra-4}) can switch between ``passive'' and ``active'' modes of operation, depending on its terminal voltage.  
\end{itembox}
\end{minipage}
\end{center}
The following parameters are used in our computer simulations:
\begin{equation}
 \ r = 0.5,  \ \omega = 1.1.
\end{equation}
We also show Poincar\'e maps of Eq. (\ref{eqn: 3-Lotka-Volterra-6}) in Figure \ref{fig:3-Lotka-poincare-10}.  
Compare the Poincar\'e maps in Figure \ref{fig:3-Lotka-poincare}(a) and Figure \ref{fig:3-Lotka-poincare-10}(a).  
The rightmost part in these figures is not identical, since the small differences due to rounding errors in numerical computation result in differences in a later state. 

In order to view the Poincar\'e maps in Figure \ref{fig:3-Lotka-poincare} from a different perspective, 
let us project the trajectory into the $( \xi, \, \eta, \, \zeta )$-space via the transformation 
\begin{equation}
 \begin{array}{lll}
   \xi (\tau)   &=& (i(\tau) + 5) \cos \,( \omega \tau ), \vspace{2mm} \\ 
   \eta (\tau)  &=& (i(\tau) + 5) \sin \,( \omega \tau ), \vspace{2mm} \\ 
   \zeta (\tau) &=& x_{1}(\tau).
 \end{array}  
\label{eqn: 3-LV-projection}
\end{equation}
Then the trajectory on the $(i, \, x)$-plane is transformed into the trajectory in the three-dimensional $( \xi, \, \eta, \, \zeta )$-space, 
as shown in Figure \ref{fig:3-Lotka-torus}.  
Observe that the trajectory in Figure \ref{fig:3-Lotka-torus}(b) is less dense than Figure \ref{fig:3-Lotka-torus}(a).  
  
Note that in order to generate a non-periodic response, we have to choose the initial conditions and the maximum step size $h$ carefully.  
In our computer simulations, we choose $h=0.002$. 
It is important for numerical stability, otherwise an overflow (outside the range of data) is likely to occur.  
That is, the numerical instability in long-time simulations is likely to occur.  
We show its example in Figure \ref{fig:3-Lotka-trajectory}.  
Suppose that Eq. (\ref{eqn: 3-Lotka-Volterra-6}) has the following parameters and initial conditions:  
\begin{equation}
 \begin{array}{l}
  \text{Parameters: }  r = 0.5, \ \  \omega =1.1, \vspace{2mm} \\ 
  \text{Initial conditions: } i(0) = 1.12, \, x_{1}(0)= 1.21. 
 \end{array}  
\end{equation}
If we choose $h=0.005$, then $i(t)$ rapidly decreases for $t \ge 6848$, 
and an overflow (outside the range of data) occurs as shown in Figure \ref{fig:3-Lotka-trajectory}(a). 
However, if we choose $h=0.002$, then the trajectory stays in the first-quadrant of the $(i, \ x_{1})$-plane as shown in Figure \ref{fig:3-Lotka-trajectory}(b).
The maximum step size of the numerical integration greatly affects the behavior of Eq. (\ref{eqn: 3-Lotka-Volterra-6}).   
Thus, noise may considerably affect the behavior in the physical memristor circuits.  

Similarly, the $4$-dimensional Lotka-Volterra equations are given by 
\begin{center}
\begin{minipage}{8.7cm}
\begin{shadebox}
\underline{\emph{$4$-dimensional Lotka-Volterra equations}}
\begin{equation}
\left. 
\begin{array}{ccl}
 \displaystyle \frac{d X_{1}}{dt} &=& (X_{4} - X_{2}) X_{1},  \vspace{2mm} \\
 \displaystyle \frac{d X_{2}}{dt} &=& (X_{1} - X_{3}) X_{2},  \vspace{2mm} \\
 \displaystyle \frac{d X_{3}}{dt} &=& (X_{2} - X_{4}) X_{3},  \vspace{2mm} \\
 \displaystyle \frac{d X_{4}}{dt} &=& (X_{3} - X_{1}) X_{4}.  
\end{array}
\right \}
\label{eqn: 4-Lotka-Volterra}
\end{equation}
\end{shadebox}
\end{minipage}
\end{center}
Equation (\ref{eqn: 4-Lotka-Volterra}) can be realized by the circuit in Figure \ref{fig:memristor-inductor-battery}. 
The dynamics of this circuit is given by 
\begin{center}
\begin{minipage}{9.5cm}
\begin{shadebox}
\underline{\emph{$4$-dimensional memristor Lotka-Volterra equations}}
\begin{equation}
\left. 
\begin{array}{ccl}
 \displaystyle \frac{d i}{dt} &=& (x_{3} - x_{1}) i,  \vspace{2mm} \\
 \displaystyle \frac{d x_{1}}{dt} &=& (i - x_{2}) x_{1},  \vspace{2mm} \\
 \displaystyle \frac{d x_{2}}{dt} &=& (x_{1} - x_{3}) x_{2},  \vspace{2mm} \\
 \displaystyle \frac{d x_{3}}{dt} &=& (x_{2} - i) x_{3}. 
\end{array}
\right \}
\label{eqn: 4-Lotka-Volterra-2}
\end{equation}
\end{shadebox}
\end{minipage}
\end{center}
The terminal voltage $v_{M}$ and the terminal current $i_{M}$ of the generic memristor are given by
\begin{center}
\begin{minipage}{8.7cm}
\begin{shadebox}
\underline{\emph{V-I characteristics of the generic memristor}}
\begin{equation}
\begin{array}{lll}
  v_{M} &=& \tilde{R}( x_{1}, \, x_{3} ) \, i_{M} = - (x_{3} - x_{1}) \, i_{M},   
  \vspace{3mm} \\
     \displaystyle \frac{d x_{1}}{dt} &=& (i_{M} - x_{2}) x_{1},
      \vspace{2mm} \\
    \displaystyle \frac{d x_{2}}{dt} &=& (x_{1} - x_{3}) x_{2}, 
      \vspace{2mm} \\
     \displaystyle \frac{d x_{3}}{dt} &=& (x_{3} - i_{M}) x_{3}, 
\end{array}
\label{eqn: 3-Lotka-Volterra-40}
\end{equation}
where $\tilde{R}( x_{1}, \, x_{3} ) = - (x_{3} - x_{1})$. \vspace{2mm}
\end{shadebox}
\end{minipage}
\end{center}
The memristor circuit equations (\ref{eqn: 4-Lotka-Volterra-2}) exhibit periodic behavior.
If an external source is added as shown in Figure \ref{fig:memristive-inductor-battery-source},
then the forced memristor Lotka-Volterra equations can exhibit a non-periodic response.   
The dynamics of this circuit is given by 
\begin{center}
\begin{minipage}{10.5cm}
\begin{shadebox}
\underline{\emph{Forced $4$-dimensional memristor Lotka-Volterra equations}}
\begin{equation}
\left. 
\begin{array}{ccl}
 \displaystyle \frac{d i}{dt} &=& (x_{3} - x_{1}) i+ r \sin ( \omega t),  \vspace{2mm} \\
 \displaystyle \frac{d x_{1}}{dt} &=& (i - x_{2}) x_{1},  \vspace{2mm} \\
 \displaystyle \frac{d x_{2}}{dt} &=& (x_{1} - x_{3}) x_{2},  \vspace{2mm} \\
 \displaystyle \frac{d x_{3}}{dt} &=& (x_{2} - i) x_{3}, 
\end{array}
\right \}
\label{eqn: 4-Lotka-Volterra-3}
\end{equation}
where $r$ and $\omega$ are constants.  
\end{shadebox}
\end{minipage}
\end{center}
The solution of Eq. (\ref{eqn: 4-Lotka-Volterra-3}) satisfies 
\begin{equation}
 i(t) + x_{1}(t) + x_{2}(t) + x_{3}(t) + \frac{r}{\omega } \cos ( \omega t) = K, 
\label{eqn: 3-Lotka-Volterra-integral-3}
\end{equation}
where $K$ is a constant. 
Thus, by eliminating $x_{3}$ from Eq. (\ref{eqn: 4-Lotka-Volterra-3}), Eq. (\ref{eqn: 4-Lotka-Volterra-3}) can be recast into the third-order non-autonomous differential equations  
\begin{equation}
\left. 
\begin{array}{ccl}
 \displaystyle \frac{d i}{dt} &=& \displaystyle \left \{ K - i - 2 x_{1} - x_{2} -  \frac{r}{\omega } \cos ( \omega t) \right \} i \vspace{1mm} \\
   && + \ r \sin ( \omega t),  \vspace{2mm} \\
 \displaystyle \frac{d x_{1}}{dt} &=& (i - x_{2}) x_{1},  \vspace{2mm} \\
 \displaystyle \frac{d x_{2}}{dt} &=& \displaystyle \left \{ 2 x_{1} -  K + i + x_{2} +  \frac{r}{\omega } \cos ( \omega t)) \right \} x_{2}.   
\end{array}
\right \}
\label{eqn: 4-Lotka-Volterra-10}
\end{equation}
We show the non-periodic response, quasi-periodic response, Poincar\'e maps, and $i_{M}-v_{M}$ loci of Eq. (\ref{eqn: 4-Lotka-Volterra-3}) in Figures \ref{fig:4-Lotka-attractor}, \ref{fig:4-Lotka-attractor-2}, \ref{fig:4-Lotka-poincare}, and \ref{fig:4-Lotka-pinch}, respectively.  
The $i_{M}-v_{M}$ loci in Figure \ref{fig:4-Lotka-pinch} lie in the first and the fourth quadrants. 
Thus, the extended memristor defined by Eq. (\ref{eqn: 3-Lotka-Volterra-40}) is an active element.  
We show next the $v_{M}-p_{M}$ locus in Figure \ref{fig:4-Lotka-power}, 
where $p_{M}(t)$ is an instantaneous power defined by $p_{M}(t)=i_{M}(t)v_{M}(t)$.  
Observe that the $v_{M}-p_{M}$ locus is pinched at the origin, and the locus lies in the first and the third quadrants. 
Thus, the memristor switches between passive and active modes of operation, depending on its terminal voltage. 
We conclude as follow: \\
\begin{center}
\begin{minipage}{.7\textwidth}
\begin{itembox}[l]{Switching behavior of the memristor}
Assume that Eq.  (\ref{eqn: 4-Lotka-Volterra-3}) exhibits non-periodic or quasi-periodic oscillation.  
Then the generic memristor defined by Eq. (\ref{eqn: 3-Lotka-Volterra-40}) can switch between ``passive'' and ``active'' modes of operation, depending on its terminal voltage.  
\end{itembox}
\end{minipage}
\end{center}
We also show Poincar\'e maps of Eq. (\ref{eqn: 4-Lotka-Volterra-10}) in Figure \ref{fig:4-Lotka-poincare-10}.  
In order to view the Poincar\'e maps from a different perspective, 
let us project the trajectory of Eq. (\ref{eqn: 4-Lotka-Volterra-3}) into the $( \xi, \, \eta, \, \zeta )$-space via the transformation 
\begin{equation}
 \begin{array}{lll}
   \xi (\tau)   &=& (i(\tau) + 5) \cos \,( \omega \tau ), \vspace{2mm} \\ 
   \eta (\tau)  &=& (i(\tau) + 5) \sin \,( \omega \tau ), \vspace{2mm} \\ 
   \zeta (\tau) &=& x_{1}(\tau).
 \end{array}  
\label{eqn: 4-LV-projection}
\end{equation}
Observe that the trajectory in Figure \ref{fig:4-Lotka-torus}(a) is less dense than Figure \ref{fig:4-Lotka-torus}(b). 
The following parameters are used in our computer simulations:
\begin{equation}
  r = 0.1,  \ \omega = 2.
\end{equation}

Note that in order to generate a non-periodic response in Figure \ref{fig:4-Lotka-torus}(a), we have to choose the initial conditions and the maximum step size $h$, carefully. 
In our computer simulations, we choose $h=0.0015$.  
Furthermore, an overflow (outside the range of data) is likely to occur due to the numerical instability in long-time simulations. 
We show its example in Figure \ref{fig:4-Lotka-trajectory}.  
Suppose that Eq. (\ref{eqn: 4-Lotka-Volterra-3}) has the following parameters and initial conditions:  
\begin{equation}
 \begin{array}{ll}
  \text{Parameters: } & r = 0.107, \ \  \omega =2, \vspace{2mm} \\ 
  \text{Initial conditions: }& i(0) = 0.608, \, x_{1}(0)= 1.2, \vspace{2mm} \\ 
  &  x_{2}(0) =1.3, \, x_{3}(0) =1.3. 
 \end{array}  
\end{equation}
If we choose $h=0.0015$, then the trajectory rapidly grows for $t \ge 1443$, 
and an overflow  (outside the range of data) occurs as shown in Figure \ref{fig:4-Lotka-trajectory}(a). 
However, if we choose $h=0.001$, then the trajectory stays in a finite region of the $(r_{1}, \, r_{2})$-plane as shown in Figure \ref{fig:4-Lotka-trajectory}(b).
The maximum step size of the numerical integration greatly affects the behavior of Eq. (\ref{eqn: 4-Lotka-Volterra-3}).   
Therefore, noise may considerably affect the behavior of the physical memristor circuits.

%-------------------------------------%
\subsubsection{$2N$-element memristor circuit realization}
%-------------------------------------%
%
In order to realize Eq. (\ref{eqn: N-Lotka-Volterra-1})  
by the $2N$-element memristor circuit in Figure \ref{fig:memristor-inductor-N},   
let us group with odd or even indexes in Eq. (\ref{eqn: N-Lotka-Volterra-1}) separately, namely, 
\begin{center}
\begin{minipage}{8.7cm}
\begin{shadebox}
\underline{\emph{$2N$-dimensional Lotka-Volterra equations}}
\begin{equation}
\begin{array}{ccl}
  \displaystyle \frac{d X_{2n-1}}{dt} &=& (X_{2n-2} - X_{2n}) X_{2n-1},  \vspace{2mm} \\
  \displaystyle \frac{d X_{2n}}{dt} &=& (X_{2n-1} - X_{2n+1}) X_{2n}, 
\end{array} 
\label{eqn: N-Lotka-Volterra-101}
\end{equation}
where $n=1, \ 2, \cdots, \ N$.  
\end{shadebox}
\end{minipage}
\end{center}
We assume a periodic lattice of the length $2N$: $X_{n} = X_{n+2N}$, that is,  
\begin{equation}
   X_{2N+1} =  X_{1}, \  X_{2N+2} = X_{2}. 
\end{equation}
Consider next the $2N$-element memristor circuit in Figure \ref{fig:memristor-inductor-N}.  
The dynamics of this circuit, which is given by Eq. (\ref{eqn: dynamics-N}).  
Assume that Eq. (\ref{eqn: dynamics-N}) satisfies  
\begin{equation}
\begin{array}{cll}
  L_{n} &=& 1, \vspace{2mm} \\
  \hat{R}( \bd{x}, \ i_{n} ) &=& \tilde{R}_{n}(\bd{x}) = \tilde{R}_{n}(x_{n-1}, \, x_{n}) = - (x_{n} - x_{n+1})  \vspace{2mm} \\
  \tilde{f}_{n}(\bd{x}, \bd{i}) &=& \tilde{f}_{n}(x_{n}, \, i_{n}, \, i_{n+1}) = (i_{n} - i_{n+1}) x_{n}.
\end{array}
\end{equation}
Then we obtain 
\begin{center}
\begin{minipage}{9.5cm}
\begin{shadebox}
\underline{\emph{$2N$-dimensional memristor Lotka-Volterra equations}}
\begin{equation}
 \begin{array}{lll}
 \displaystyle \frac{d i_{n}}{dt}&=& \displaystyle  (x_{n-1} - x_{n}) i_{n},  \vspace{2mm} \\
 \displaystyle \frac{d x_{n}}{dt}&=& \displaystyle  (i_{n} - i_{n+1}) x_{n}, 
 \end{array}
\label{eqn: N-Lotka-Volterra-102}
\end{equation}
\end{shadebox}
\end{minipage}
\end{center}
Equations (\ref{eqn: N-Lotka-Volterra-101}) and (\ref{eqn: N-Lotka-Volterra-102}) are equivalent if we change the variables  
\begin{equation}
  i_{n} = X_{2n-1}, \ x_{n} = X_{2n}.  
\end{equation}
In this case, the extended memristors in Figure \ref{fig:memristor-inductor-N} are replaced by the \emph{generic} memristors, 
though the current $i$ of Eq. (\ref{eqn: generic2-2}) is modified into the vector form $\bd{i} = (i_{1}, \, i_{2}, \, \cdots, \, i_{n})$.  
Their terminal voltage $v_{n}$ and the terminal current $i_{n}$ of the current-controlled generic memristor are described
by
\begin{center}
\begin{minipage}{8.7cm}
\begin{shadebox}
\underline{\emph{V-I characteristics of the generic memristors}}
\begin{equation}
\begin{array}{lll}
  v_{n} &=& \tilde{R}_{n}(x_{n-1}, \, x_{n}) \, i_{n} = - (x_{n-1} - x_{n}) i_{n}, 
  \vspace{1mm} \\
  \displaystyle \frac{d x_{n}}{dt} 
   &=& \tilde{f}_{n}(x_{n}, \, i_{n}, \, i_{n+1}) = (i_{n} - i_{n+1}) x_{n}. \vspace{2mm} \\
\end{array}
\label{eqn: extended-103}
\end{equation}
\end{shadebox}
\end{minipage}
\end{center}

For $N=2$, Eq. (\ref{eqn: N-Lotka-Volterra-102}) can be written as 
\begin{center}
\begin{minipage}{9.5cm}
\begin{shadebox}
\underline{\emph{$4$-dimensional memristor Lotka-Volterra equations}}
\begin{equation}
\left. 
\begin{array}{ccl}
 \displaystyle \frac{d i_{1}}{dt} &=& (x_{2} - x_{1}) \, i_{1},  \vspace{2mm} \\
 \displaystyle \frac{d x_{1}}{dt} &=& (i_{1} - i_{2}) \, x_{1},  \vspace{2mm} \\
 \displaystyle \frac{d i_{2}}{dt} &=& (x_{1} - x_{2}) \, i_{2},  \vspace{2mm} \\
 \displaystyle \frac{d x_{2}}{dt} &=& (i_{2} - i_{1}) \, x_{2}. 
\end{array}
\right \}
\label{eqn: 4-Lotka-Volterra-2b}
\end{equation}
\end{shadebox}
\end{minipage}
\end{center}
Here $i_{1}$ and $i_{2}$ denote the currents of two generic memristors.   
The terminal voltage $v_{n}$ and the terminal current $i_{n}$ of these memristors are given by
\begin{center}
\begin{minipage}{8.7cm}
\begin{shadebox}
\underline{\emph{V-I characteristics of the $2$ generic memristors}}
\begin{equation}
\begin{array}{c}
\left.
\begin{array}{cll}
  v_{1} &=& \tilde{R}_{1}(x_{1}, \, x_{2}) \, i_{1} = - (x_{2} - x_{1}) \, i_{1}, 
  \vspace{1mm} \\
  \displaystyle \frac{d x_{1}}{dt} 
   &=& \tilde{f}_{1}(x_{1}, \, i_{1}, \, i_{2}) = (i_{1} - i_{2}) x_{1} 
\end{array}
\right \} 
\vspace{4mm} \\
\left. 
\begin{array}{cll}
  v_{2} &=& \tilde{R}_{2}(x_{1}, \, x_{2}) \, i_{2} = - (x_{1} - x_{2}) \, i_{2}, 
  \vspace{1mm} \\
  \displaystyle \frac{d x_{2}}{dt} 
   &=& \tilde{f}_{2}(x_{2}, \, i_{1}, \, i_{2}) = (i_{2} - i_{1}) x_{2}.  
\end{array}
\right \}
\end{array}
\label{eqn: extended-201}
\end{equation}
\end{shadebox}
\end{minipage}
\end{center}
Since (\ref{eqn: 4-Lotka-Volterra}) is equivalent to Eq. (\ref{eqn: 4-Lotka-Volterra-2b}),   
Eq. (\ref{eqn: 4-Lotka-Volterra}) can be realized by the $4$-element memristor circuit in Figure \ref{fig:memristor-inductor-N}.

%-------------------------------------%
\subsection{Ecological predator-prey model}
\label{sec: ecological-predator-prey}
%-------------------------------------%

Consider the ecological predator-prey model \cite{Hirota(1985)} defined by 
\begin{center}
\begin{minipage}{8.7cm}
\begin{shadebox}
\underline{\emph{Ecological predator-prey model equations}}
\begin{equation}
 \displaystyle \frac{d M_{n}}{dt} = (M_{n-1} - M_{n+1}) {M_{n}}^{2}, 
\label{eqn: ecological-1}
\end{equation}
\noindent
where $n=1, \ 2, \cdots, \ N$ and we consider the case of a periodic lattice of the length $N$: $M_{n} = M_{n + N}$. 
\end{shadebox}
\end{minipage}
\end{center}
Assume that $n=3$.  Then Eq. (\ref{eqn: ecological-1}) is written as
\begin{center}
\begin{minipage}{9.5cm}
\begin{shadebox}
\underline{\emph{$3$-dimensional ecological predator-prey model equations}}
\begin{equation}
\left. 
\begin{array}{ccl}
 \displaystyle \frac{d M_{1}}{dt} &=& (M_{3} - M_{2}) {M_{1}}^{2},  \vspace{2mm} \\
 \displaystyle \frac{d M_{2}}{dt} &=& (M_{1} - M_{3}) {M_{2}}^{2},  \vspace{2mm} \\
 \displaystyle \frac{d M_{3}}{dt} &=& (M_{2} - M_{1}) {M_{3}}^{2}.   
\end{array}
\right \}
\label{eqn: ecological-2}
\end{equation}
\end{shadebox}
\end{minipage}
\end{center}
Equation (\ref{eqn: ecological-2}) has the two integrals, since the solution satisfies 
\begin{center}
\begin{minipage}{.5\textwidth}
\begin{itembox}[l]{Integrals}
\begin{equation}
\left. 
 \begin{array}{l}
  \displaystyle \frac{d}{dt} \bigl ( M_{1} M_{2} M_{3} \bigr )  = 0, \vspace{4mm} \\
  \displaystyle \frac{d}{dt} \bigl ( M_{1} M_{2} + M_{2} M_{3} + M_{3} M_{1} \bigr ) = 0.   
 \end{array}
\right \} 
\label{eqn: ecological-integrals}
\end{equation}
\end{itembox}
\end{minipage}
\end{center}
Thus, Eq. (\ref{eqn: ecological-2}) can not exhibit chaotic oscillation nor a quasi-periodic oscillation.  
Furthermore, it can be recast into Eq. (\ref{eqn: N-Lotka-Volterra-1}) \cite{Hirota(1985)} 
if we set
\begin{equation}
  L_{n} = M_{n + \frac{1}{2}} M_{n -\frac{1}{2}}.  
\end{equation}

Consider first the three-element memristor circuit in Figure \ref{fig:memristor-inductor-battery}.  
The dynamics of this circuit given by Eq. (\ref{eqn: dynamics-1}).  
Assume that Eq. (\ref{eqn: dynamics-1}) satisfies 
\begin{equation}
\left.
 \begin{array}{ccc}
  E = 0, && L = 1, \vspace{2mm} \\
  \hat{R}(x_{1}, \, x_{2}, \, i) &=& - (x_{2} - x_{1}) \, i,           \vspace{2mm} \\
  \tilde{f}_{1}(x_{1}, \, x_{2}, \, i) &=& (i - x_{2}) {x_{1}}^{2},  \vspace{2mm} \\
  \tilde{f}_{2}(x_{1}, \, x_{2}, \, i) &=& (x_{1} - i) {x_{2}}^{2}.  
 \end{array}
\right \}
\end{equation}
Then we obtain 
\begin{center}
\begin{minipage}{11cm}
\begin{shadebox}
\underline{\emph{$3$-dimensional memristor ecological predator-prey model equations}}
\begin{equation}
\left. 
\begin{array}{ccc}
 \displaystyle \frac{d i}{dt} &=& (x_{2} - x_{1}) i^{2},      \vspace{2mm} \\
 \displaystyle \frac{d x_{1}}{dt} &=& (i - x_{2}) {x_{1}}^{2},  \vspace{2mm} \\
 \displaystyle \frac{d x_{2}}{dt} &=& (x_{1} - i) {x_{2}}^{2}.    
\end{array}
\right \}
\label{eqn: ecological-3}
\end{equation}
\end{shadebox}
\end{minipage}
\end{center}
Equations (\ref{eqn: ecological-2}) and (\ref{eqn: ecological-3}) are equivalent if we change the variables  
\begin{equation}
  M_{1}=i, \ M_{2}=x_{1}, \ M_{3}=x_{2}. 
\end{equation}
In this case, the small-signal \emph{memristance} of the extended memristor in Figure \ref{fig:memristor-inductor-battery} is defined by   
\begin{equation}
  \hat{R}(\bd{x}, \, i) = \hat{R}( x_{1}, \, x_{2}, \,  i)  = - (x_{2} - x_{1}) \, i.  
\end{equation}
where $\bd{x} = ( x_{1}, \, x_{2} )$.  
The terminal voltage $v_{M}$ and the terminal current $i_{M}$ of the extended memristor are described
by
\begin{center}
\begin{minipage}{8.7cm}
\begin{shadebox}
\underline{\emph{V-I characteristics of the extended memristor}}
\begin{equation}
\begin{array}{lll}
  v_{M} &=& \hat{R}( x_{1}, \, x_{2}, \, i_{M}) \, i_{M} = - (x_{2} - x_{1}) \, {i_{M}}^{2},   
  \vspace{3mm} \\
     \displaystyle \frac{d x_{1}}{dt} &=& (i_{M} - x_{2}) {x_{1}}^{2},
      \vspace{2mm} \\
     \displaystyle \frac{d x_{2}}{dt} &=& (x_{1} - i_{M}) {x_{2}}^{2}, 
\end{array}
\label{eqn: ecological-4}
\end{equation}
where $\hat{R}( x_{1}, \, x_{2} \, i_{M}) = - (x_{2} - x_{1}) \, i_{M} $ and 
$i_{M}=i$. 
\end{shadebox}
\end{minipage}
\end{center}

\clearpage
%---Fig. 55-------%
\begin{figure}[hpbt]
 \centering
   \begin{tabular}{ccc}
    \psfig{file=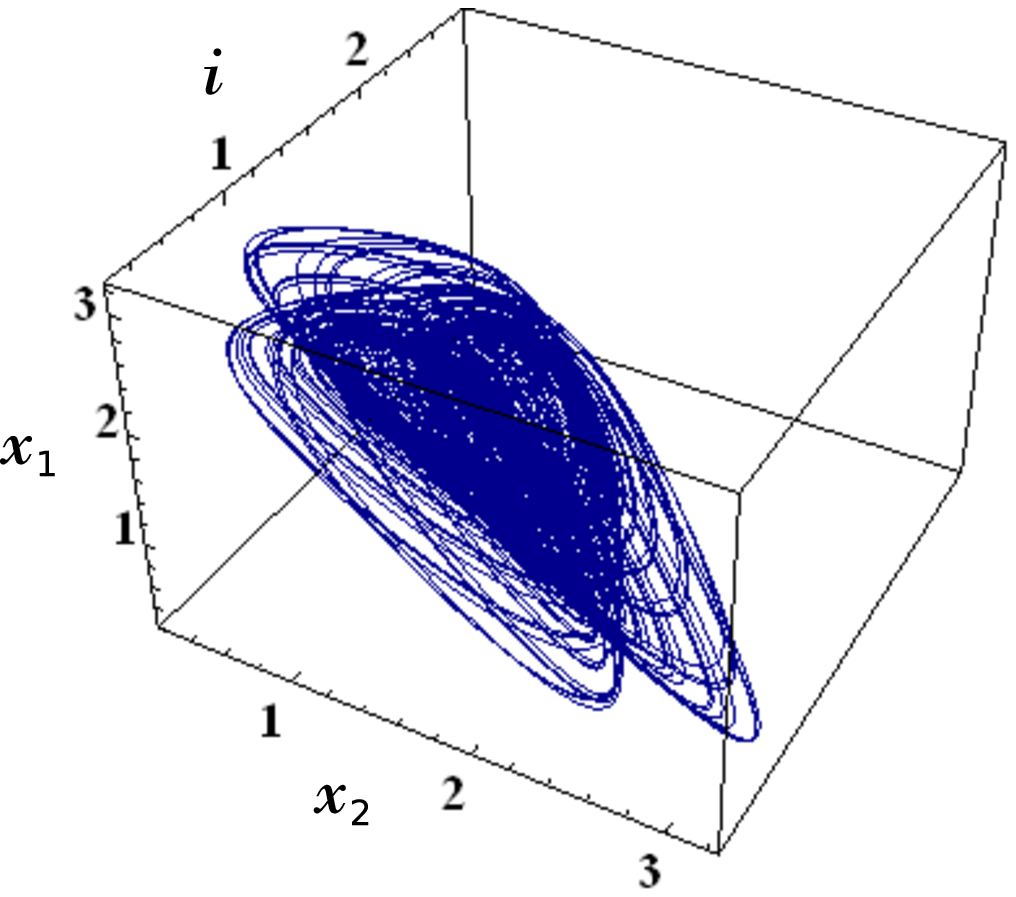, height=5.5cm}  & \hspace{5mm} &
    \psfig{file=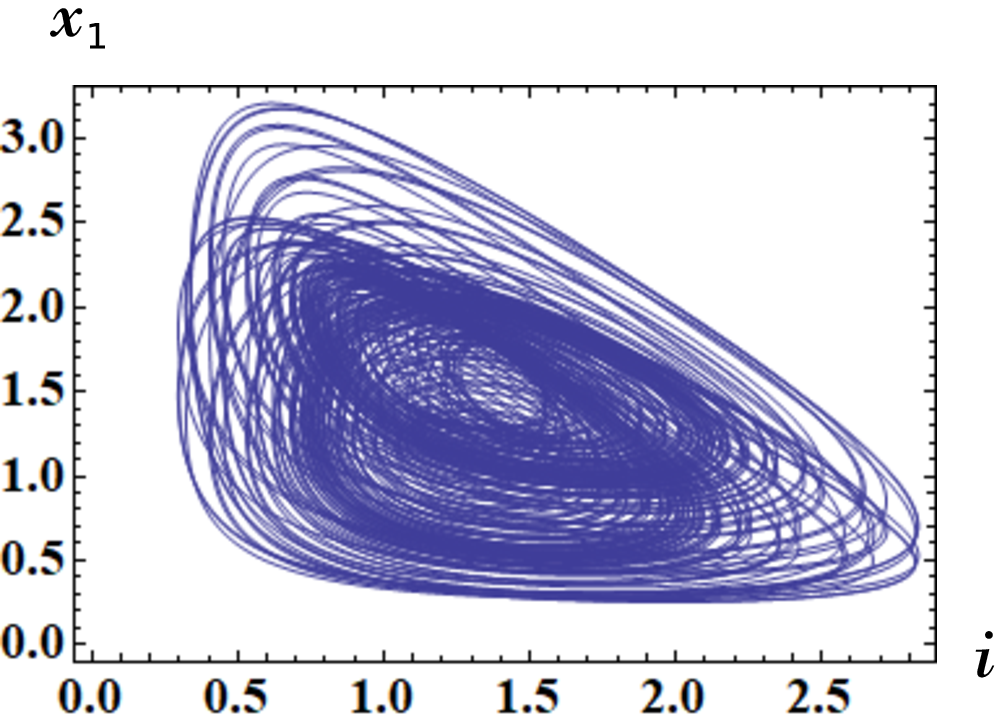, height=4.8cm} 
   \end{tabular} \\
    (a) non-periodic \vspace{10mm}\\
   \begin{tabular}{ccc}
    \psfig{file=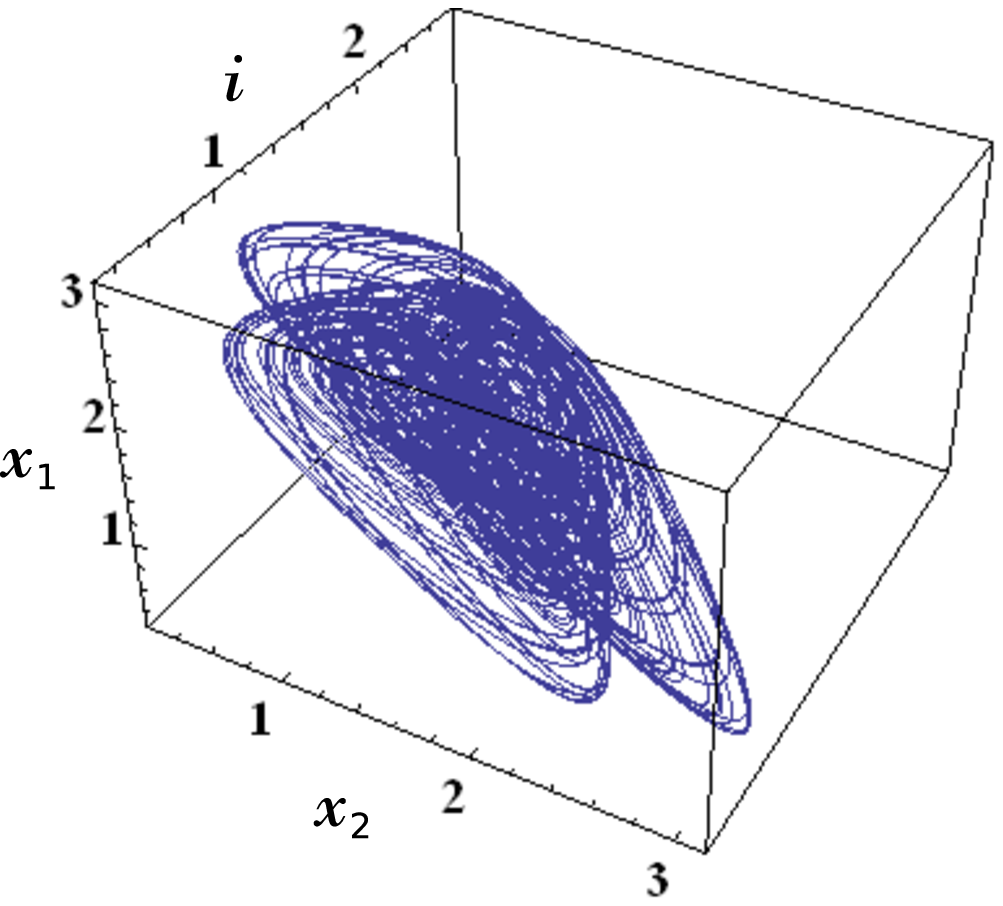, height=5.5cm}  & \hspace{5mm} &
    \psfig{file=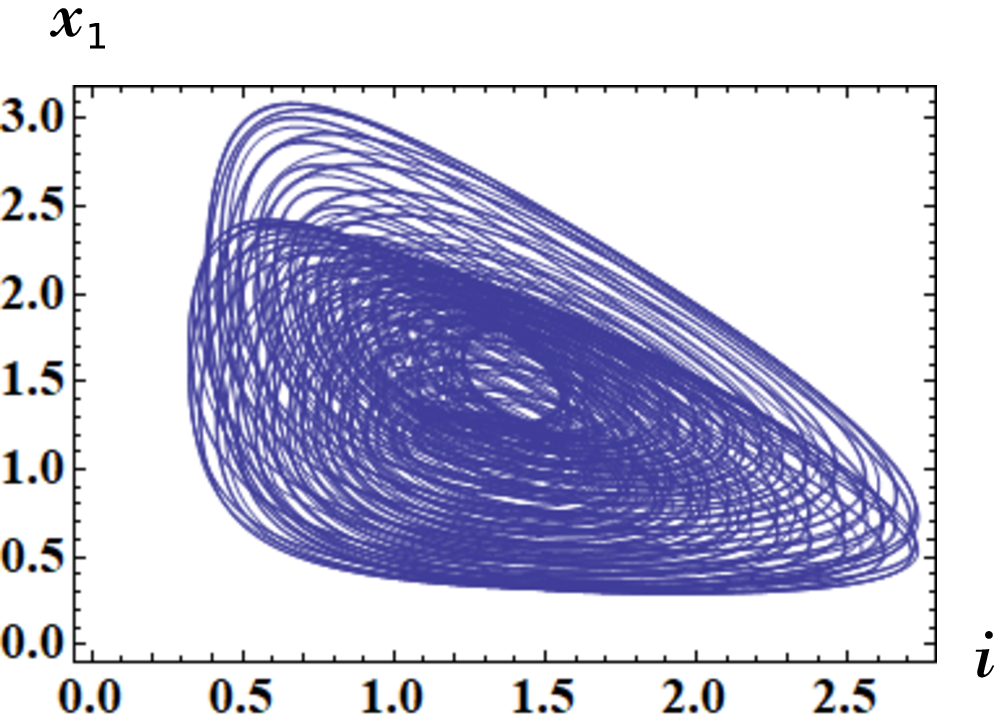, height=4.8cm}  \\
   \end{tabular} \\
    (b)  quasi-periodic  \vspace{5mm}\\   
  \caption{Non-periodic and quasi-periodic responses of 
   the forced $3$-dimensional memristor Lotka-Volterra equations (\ref{eqn: 3-Lotka-Volterra-5}). 
   Parameters:  $\ r = 0.5,  \ \omega = 1.1$.
   Initial conditions: (a) $i(0) = 1.121, \, x_{1}(0)= 1.2, \, x_{2}(0) =1.3$. \ \ (b) $i(0) = 1.09, \, x_{1}(0)= 1.2, \, x_{2}(0) =1.3$  }
  \label{fig:3-Lotka-attractor} 
\end{figure}
%
%

%---Fig. 56-------%
\begin{figure}[hpbt]
 \centering
   \begin{tabular}{c}
   \psfig{file=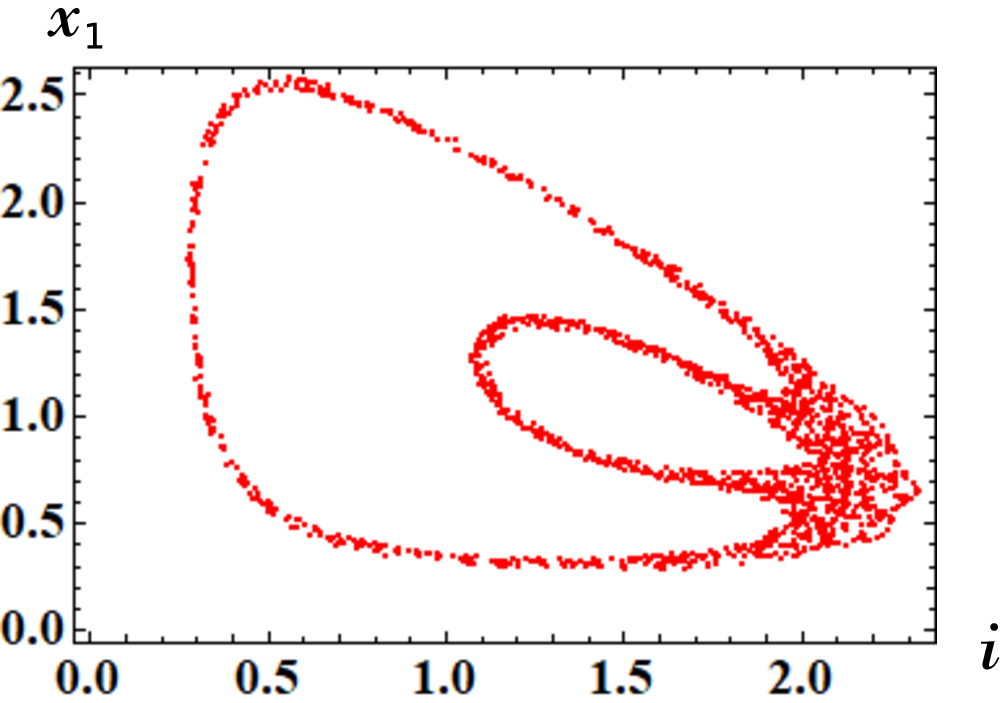, height=5.0cm} \\
   (a) non-periodic  \vspace{1mm} \\
   \psfig{file=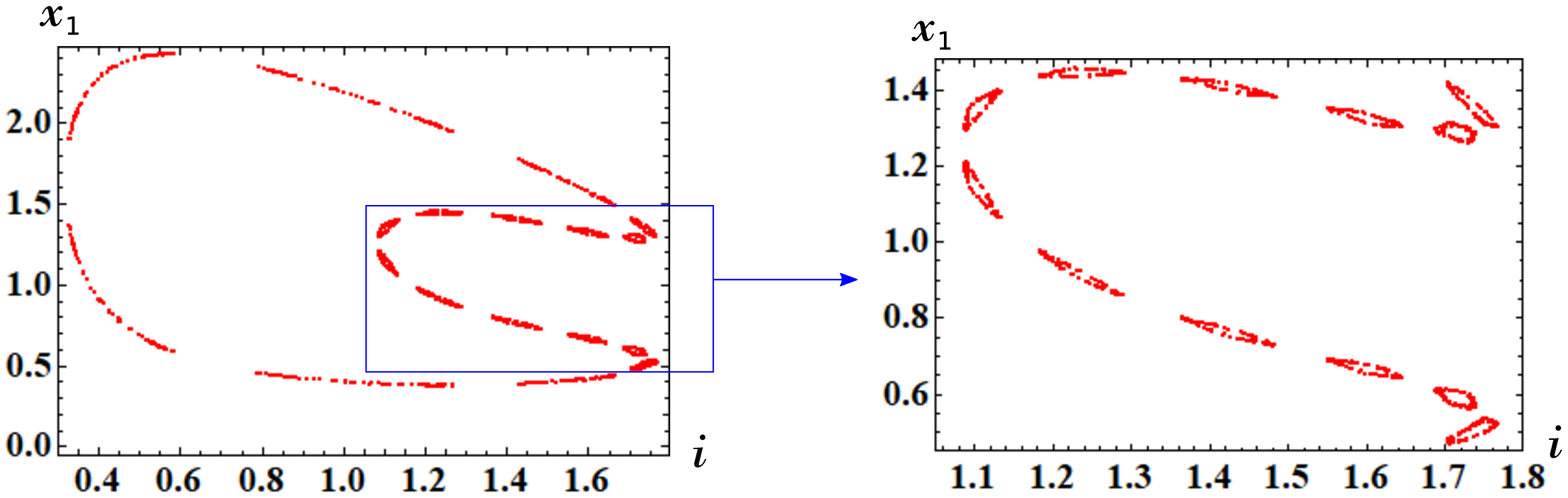, height=5.0cm}  \\
   (b) quasi-periodic   \vspace{1mm} \\
  \end{tabular}
  \caption{Poincar\'e maps of the forced $3$-dimensional memristor Lotka-Volterra equations (\ref{eqn: 3-Lotka-Volterra-5}).  
   Observe the islands of tori in Figure \ref{fig:3-Lotka-poincare}(b).  
   Parameters:  $\ r = 0.5,  \ \omega = 1.1$.
   Initial conditions: 
   (a) $i(0) = 1.121, \, x_{1}(0)= 1.2, \, x_{2}(0) =1.3$.  \ \ 
   (b) $i(0) = 1.09, \, x_{1}(0)= 1.2, \, x_{2}(0) =1.3$.}
  \label{fig:3-Lotka-poincare} 
\end{figure}
%
%

%
%---Fig. 57-------%
\begin{figure}[hpbt]
 \centering
   \begin{tabular}{cc}
    \psfig{file=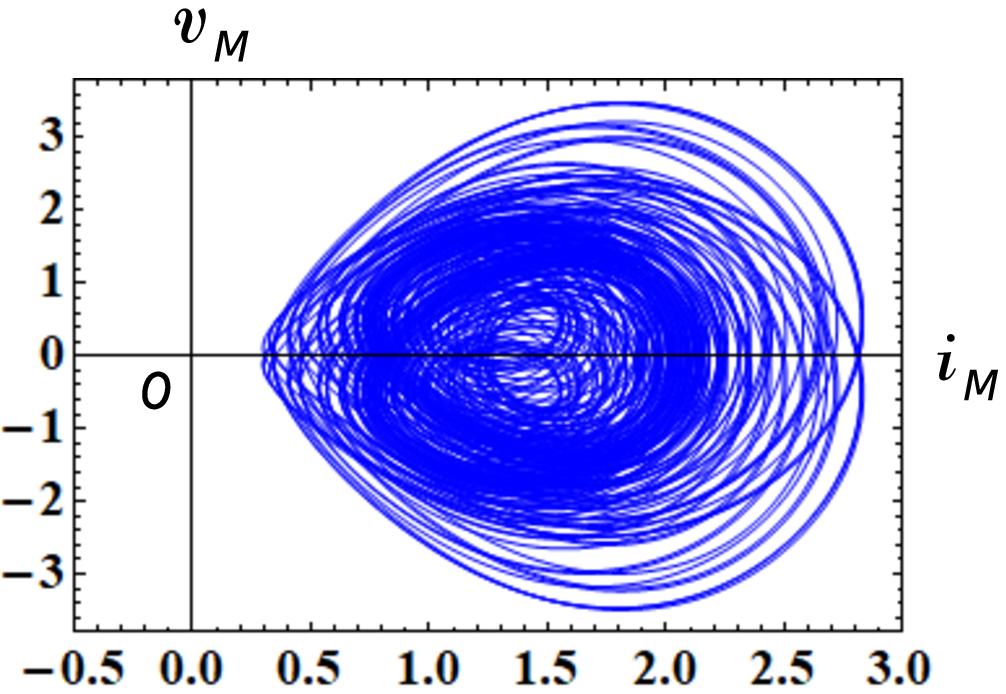, height=4.8cm}  & 
    \psfig{file=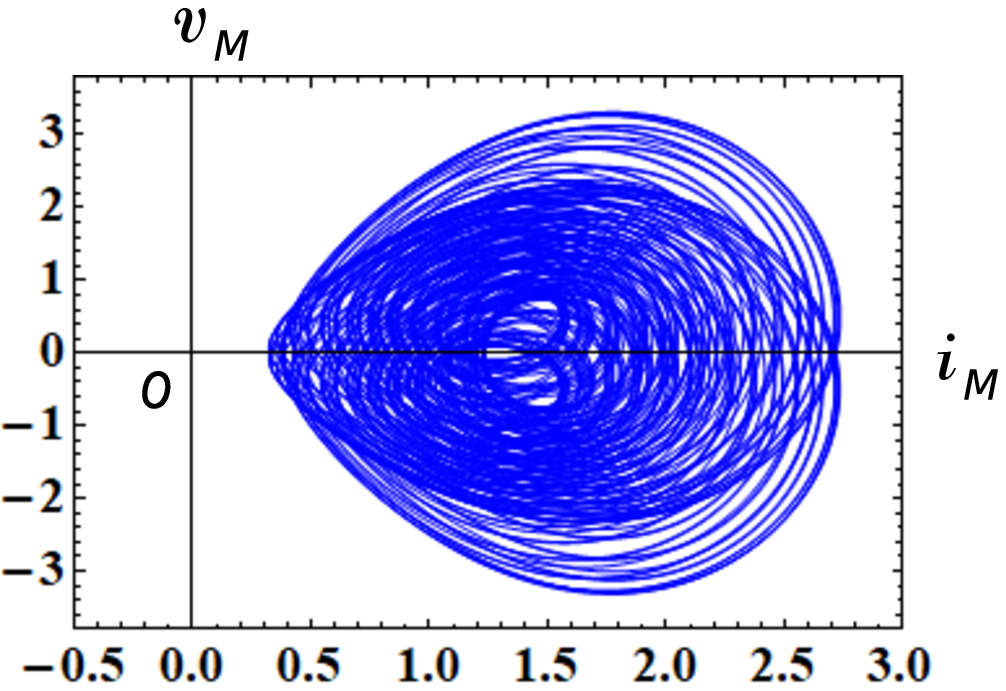, height=4.8cm}  \\
     (a) non-periodic & (b)  quasi-periodic 
    \end{tabular}
  \caption{ The $i_{M}-v_{M}$ loci of the forced $3$-dimensional memristor Lotka-Volterra equations (\ref{eqn: 3-Lotka-Volterra-5}). 
   Here, $v_{M}$ and  $i_{M}$ denote the terminal voltage and the terminal current of the current-controlled generic memristor.  
   Parameters:  $\ r = 0.5,  \ \omega = 1.1$.
   Initial conditions: 
   (a) $i(0) = 1.121, \, x_{1}(0)= 1.2, \, x_{2}(0) =1.3$.  \ \ 
   (b) $i(0) = 1.09, \, x_{1}(0)= 1.2, \, x_{2}(0) =1.3$.}
  \label{fig:3-Lotka-pinch} 
\end{figure}
%
%

%---Fig. 58-------%
\begin{figure}[hpbt]
 \centering
   \begin{tabular}{cc}
    \psfig{file=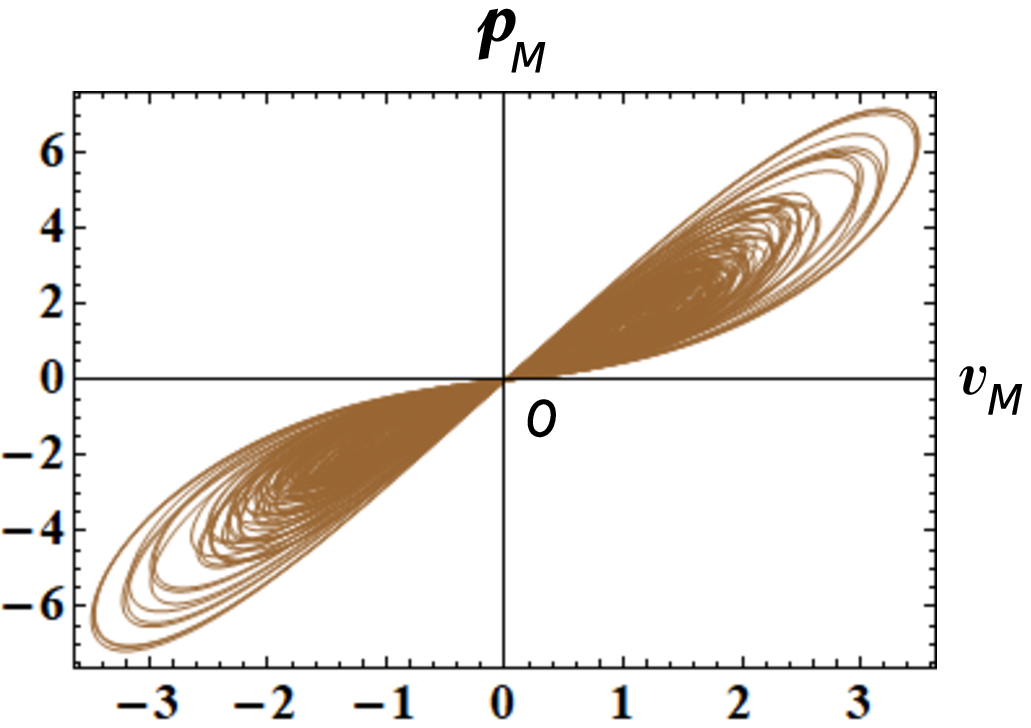, height=5cmm}  & 
    \psfig{file=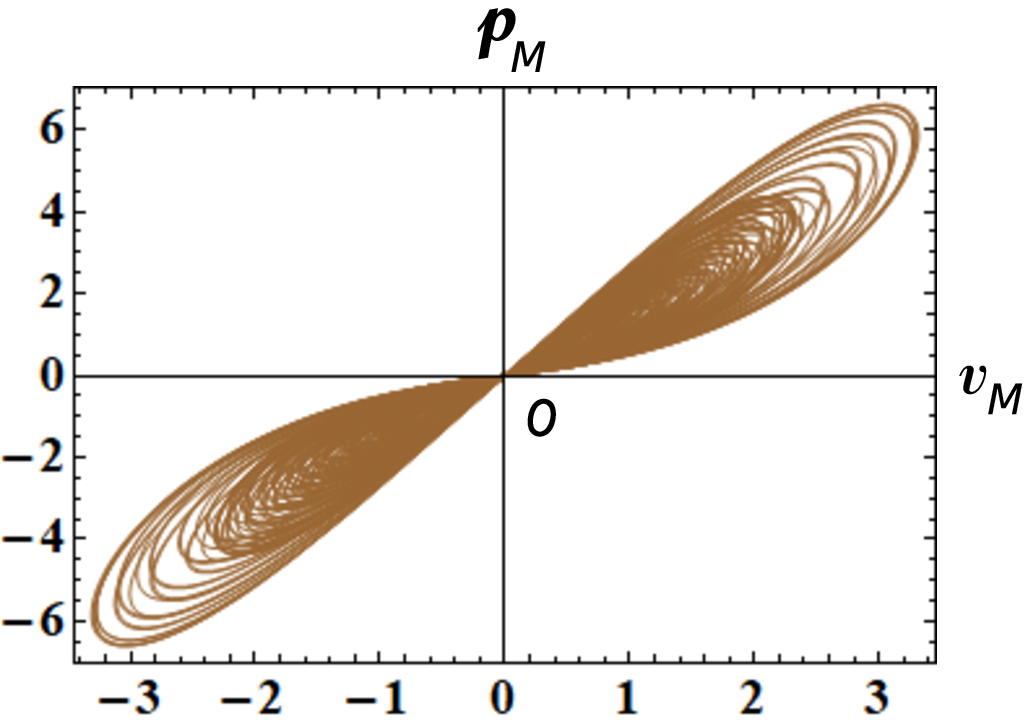, height=5cm}  \\
    (a) non-periodic & (b) quasi-periodic \\
   \end{tabular}
  \caption{ The $v_{M}-p_{M}$ locus of the forced $3$-dimensional memristor Lotka-Volterra equations (\ref{eqn: 3-Lotka-Volterra-5}). 
   Here, $p_{M}(t)$ is an instantaneous power defined by $p_{M}(t)=i_{M}(t)v_{M}(t)$, 
   and $v_{M}(t)$ and $i_{M}(t)$ denote the terminal voltage and the terminal current of the current-controlled generic memristor.  
   Observe that the $v_{M}-p_{M}$ locus is pinched at the origin, and the locus lies in the first and the third quadrants. 
   The memristor switches between passive and active modes of operation, depending on its terminal voltage $v_{M}(t)$.
   Parameters:  $\ r = 0.5,  \ \omega = 1.1$.
   Initial conditions: 
   (a) $i(0) = 1.121, \, x_{1}(0)= 1.2, \, x_{2}(0) =1.3$.  \ \ 
   (b) $i(0) = 1.09, \, x_{1}(0)= 1.2, \, x_{2}(0) =1.3$.}
  \label{fig:3-Lotka-power} 
\end{figure}
%
%

%---Fig. 59-------%
\begin{figure}[ht]
 \centering
   \begin{tabular}{c}
   \psfig{file=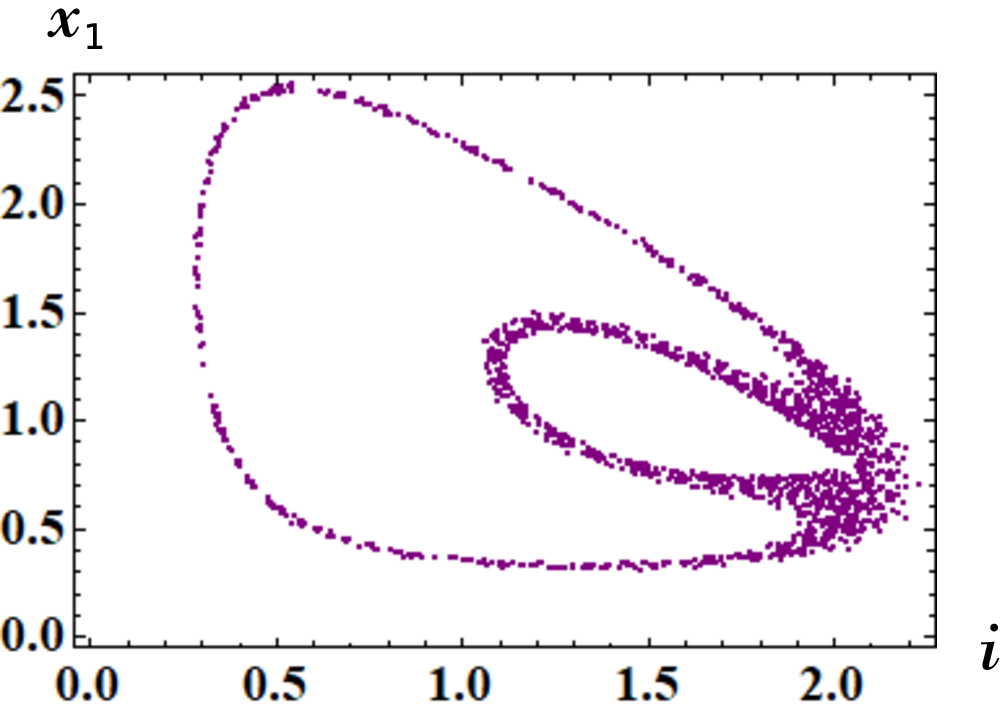, height=4.5cm} \\
   (a) non-periodic  \vspace{1mm} \\
   \psfig{file=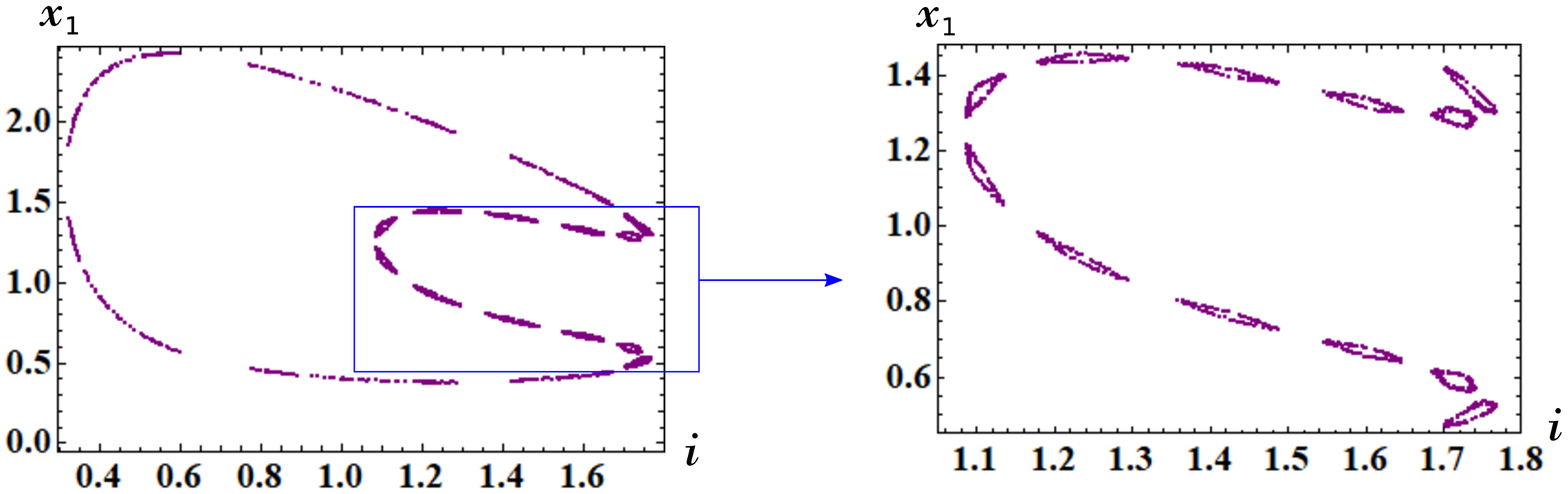,height=4.5cm} \\
   (b) quasi-periodic  \vspace{1mm} \\
  \end{tabular}
  \caption{Poincar\'e maps of the non-autonomous equations (\ref{eqn: 3-Lotka-Volterra-6}). 
   Observe the islands of tori in Figure \ref{fig:3-Lotka-poincare-10}(b).  
   Parameters:  $\ r = 0.5,  \ \omega = 1.1$.
   Initial conditions: 
   (a) $i(0) = 1.121, \, x_{1}(0)= 1.2, \, x_{2}(0) =1.3, \, k = i(0)+ x_1(0) + x_{2}(0) + r/\omega \approx 4.0756 $.  \ \ 
   (b) $i(0) = 1.09, \, x_{1}(0)= 1.2, \, x_{2}(0) =1.3, \, k = i(0)+ x_1(0) + x_{2}(0)  + r/\omega \approx 4.0446 $. }
  \label{fig:3-Lotka-poincare-10} 
\end{figure}
%
%

%---Fig. 60-------%
\begin{figure}[hpbt]
 \centering
   \begin{tabular}{cc}
   \psfig{file=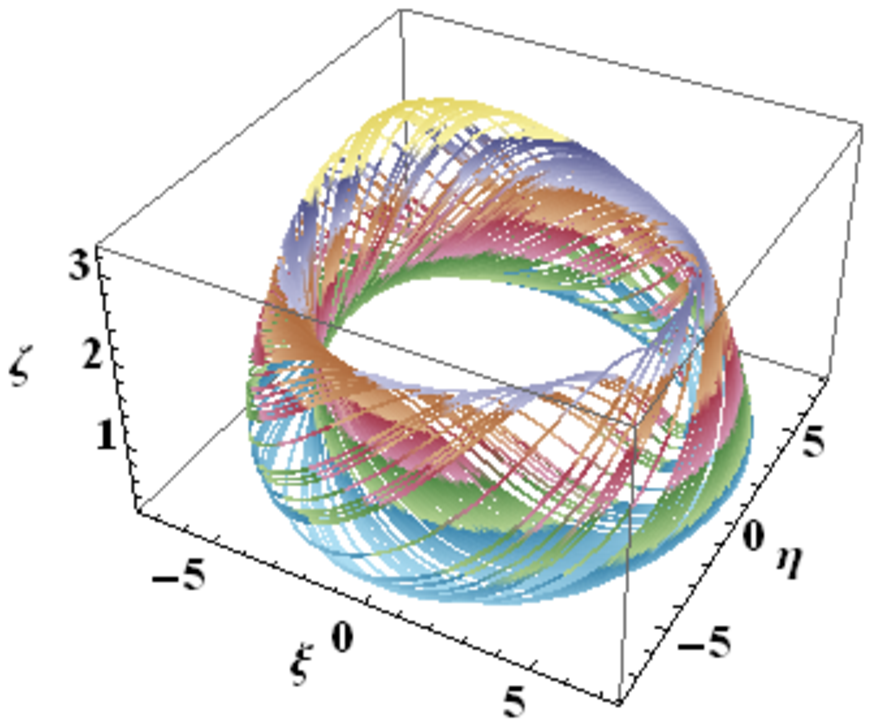, width=7.5cm} & 
   \psfig{file=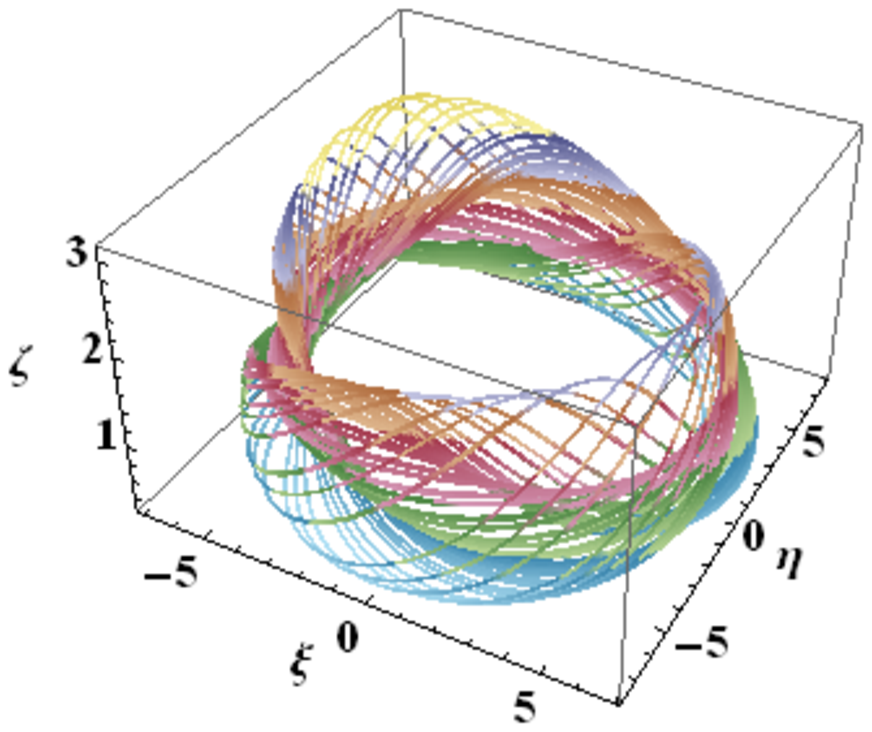, width=7.5cm} \vspace{1mm} \\
   (a) non-periodic & (b) quasi-periodic \\
  \end{tabular}
  \caption{The two trajectories of the forced $3$-dimensional memristor Lotka-Volterra equations (\ref{eqn: 3-Lotka-Volterra-5}), 
   which are projected into the $( \xi, \, \eta, \, \zeta )$-space via the coordinate transformation (\ref{eqn: 3-LV-projection}).  
   Observe that the trajectory in Figure \ref{fig:3-Lotka-torus}(b) is less dense than that in Figure \ref{fig:3-Lotka-torus}(a).    
   The trajectories are colored with the \emph{DarkBands} color code in Mathematica. 
   Parameters:  $\ r = 0.5,  \ \omega = 1.1$.
   Initial conditions: 
   (a) $i(0) = 1.121, \, x_{1}(0)= 1.2, \, x_{2}(0) =1.3$.  \ \ 
   (b) $i(0) = 1.09, \, x_{1}(0)= 1.2, \, x_{2}(0) =1.3$.}
  \label{fig:3-Lotka-torus} 
\end{figure}
%
%

%---Fig. 61-------%
\begin{figure}[hpbt]
 \centering
   \begin{tabular}{cc}
    \psfig{file=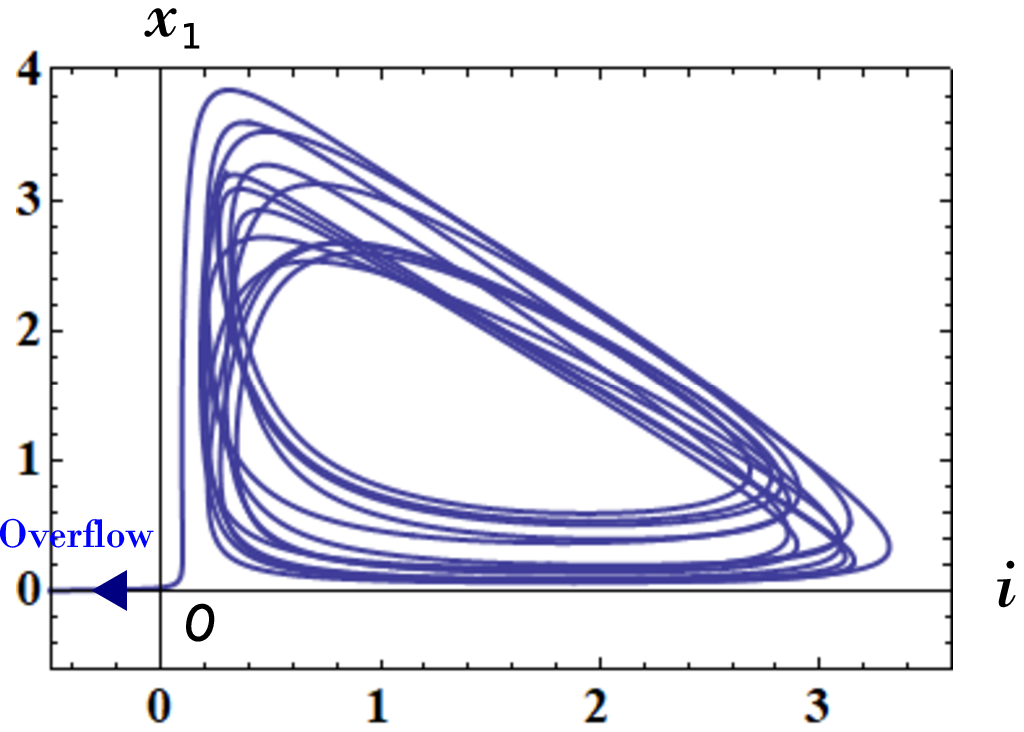, height=5cm}  & 
    \psfig{file=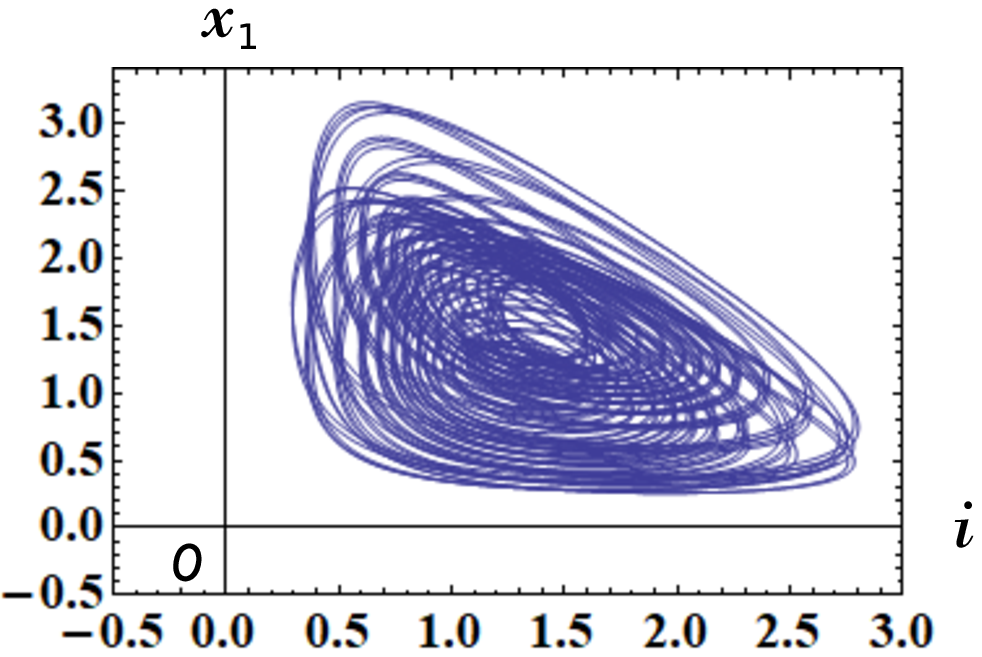, height=5cm}  \vspace{1mm} \\
   (a) $h=0.005$ & (b) $h=0.002$  \\ 
   \end{tabular}
  \caption{Behavior of the second-order non-autonomous differential equations (\ref{eqn: 3-Lotka-Volterra-6}).  
  If we choose $h=0.005$, then $i(t)$ rapidly decreases for $t \ge 6848$, 
  and an overflow occurs as shown in Figure \ref{fig:4-Lotka-trajectory}(a). 
  However, if we choose $h=0.002$, then the trajectory stays in the first-quadrant of the $(i, \ x_{1})$-plane 
  as shown in Figure \ref{fig:3-Lotka-trajectory}(b). 
  Here, $h$ denotes the maximum step size of the numerical integration.  
  Parameters:  $r = 0.5, \ \  \omega =1.1$.
  Initial conditions: $i(0) =1.2, \, x_{1}(0)= 1.121$.}
  \label{fig:3-Lotka-trajectory} 
\end{figure}
%
%

%---Fig. 62-------%
\begin{figure}[hpbt]
 \centering
   \begin{tabular}{ccc}
   \psfig{file=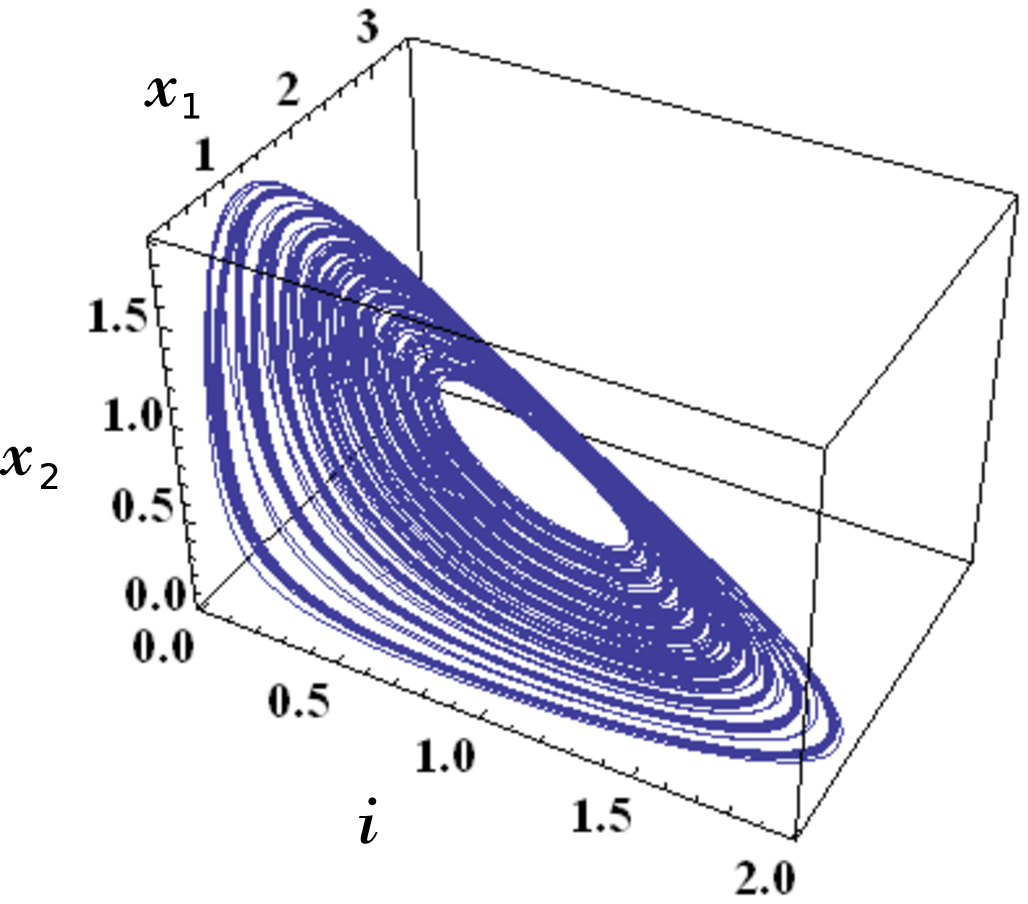, height=6.0cm} & \hspace{5mm} &
   \psfig{file=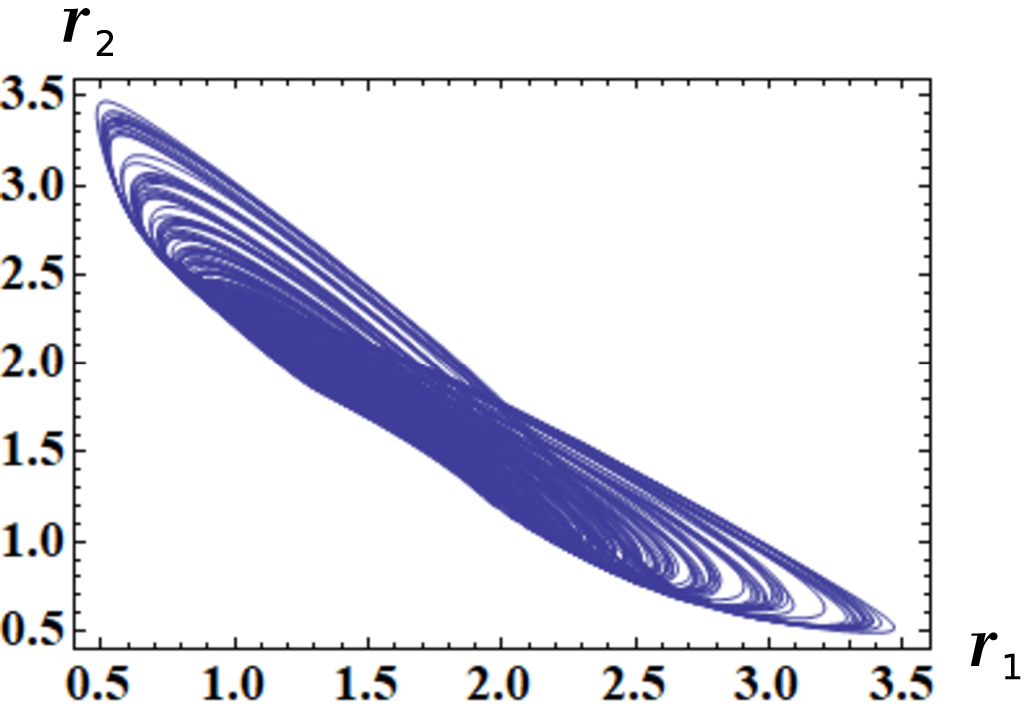, height=5.5cm} \\
   (a) $(i, \ x_{1}, \ x_{2})$-space & & (b) $(r_{1},\ r_{2})$-plane  \vspace{1mm} \\
  \end{tabular}
  \caption{Non-periodic responses of the forced $4$-dimensional memristor Lotka-Volterra equations (\ref{eqn: 4-Lotka-Volterra-3}), 
   where $r_{1} =\sqrt{i^{2}+{x_{1}}^{2}}$ and $r_{2} =\sqrt{{x_{2}}^{2}+{x_{3}}^{2}}$
   Parameters:  $\ r = 0.1,  \ \omega = 2$.
   Initial conditions: $i(0) = 0.6072, \, x_{1}(0)= 1.2, \, x_{2}(0) =1.3, \, x_{3}(0) =1.3$.}
  \label{fig:4-Lotka-attractor} 
\end{figure}
%
%

%---Fig. 63-------%
\begin{figure}[hpbt]
 \centering
   \begin{tabular}{ccc}
   \psfig{file=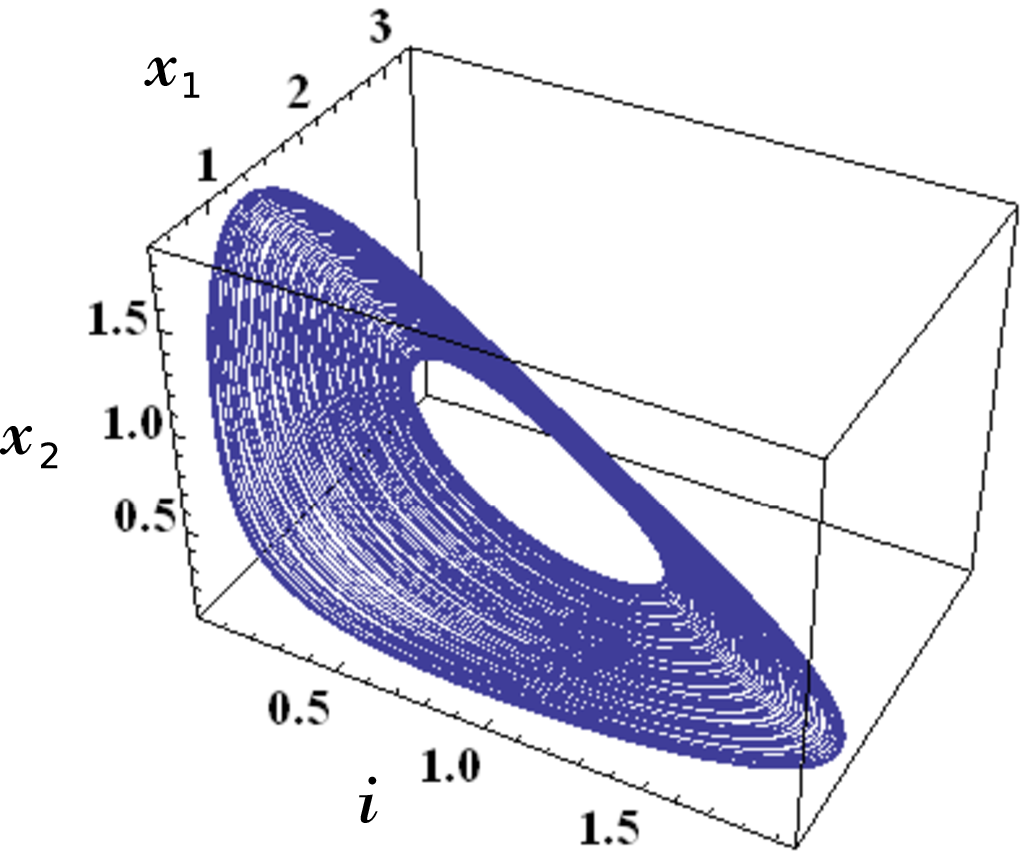, height=6.0cm} & \hspace{2mm} &
   \psfig{file=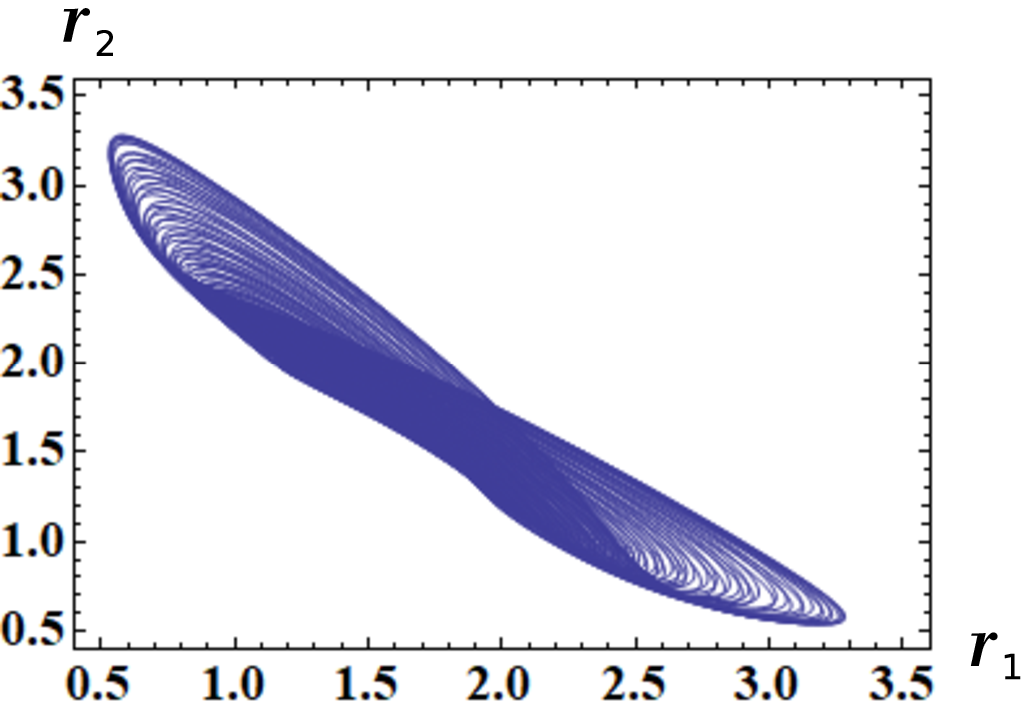, height=5.5cm} \\
   (a) $(i, \ x_{1}, \ x_{2})$-space & & (b) $(r_{1},\ r_{2})$-plane  \vspace{1mm} \\
  \end{tabular}
  \caption{Quasi-periodic responses of the forced $4$-dimensional memristor Lotka-Volterra equations (\ref{eqn: 4-Lotka-Volterra-3}), 
   where $r_{1} =\sqrt{i^{2}+{x_{1}}^{2}}$ and $r_{2} =\sqrt{{x_{2}}^{2}+{x_{3}}^{2}}$
   Parameters:  $\ r = 0.1,  \ \omega = 2$. \newline
   Initial conditions: $i(0) = 0.585, \, x_{1}(0)= 1.2, \, x_{2}(0) =1.3, \, x_{3}(0) =1.3$.}
  \label{fig:4-Lotka-attractor-2} 
\end{figure}
%
%

%---Fig. 64-------%
\begin{figure}[hpbt]
 \centering
   \begin{tabular}{cc}
   \psfig{file=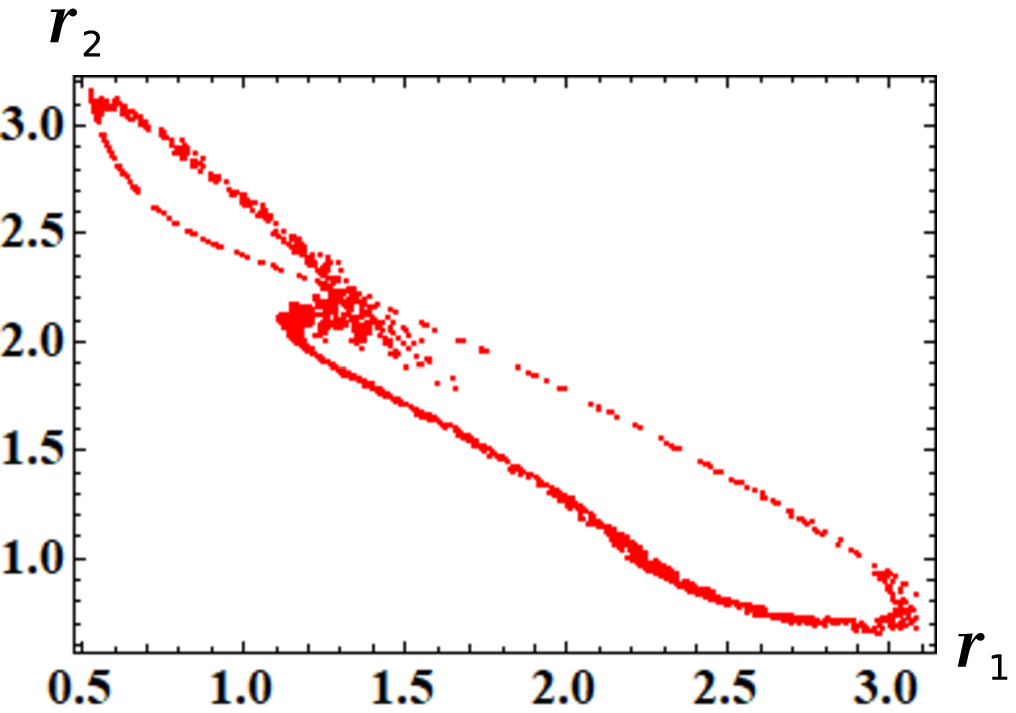, height=5.0cm} & 
   \psfig{file=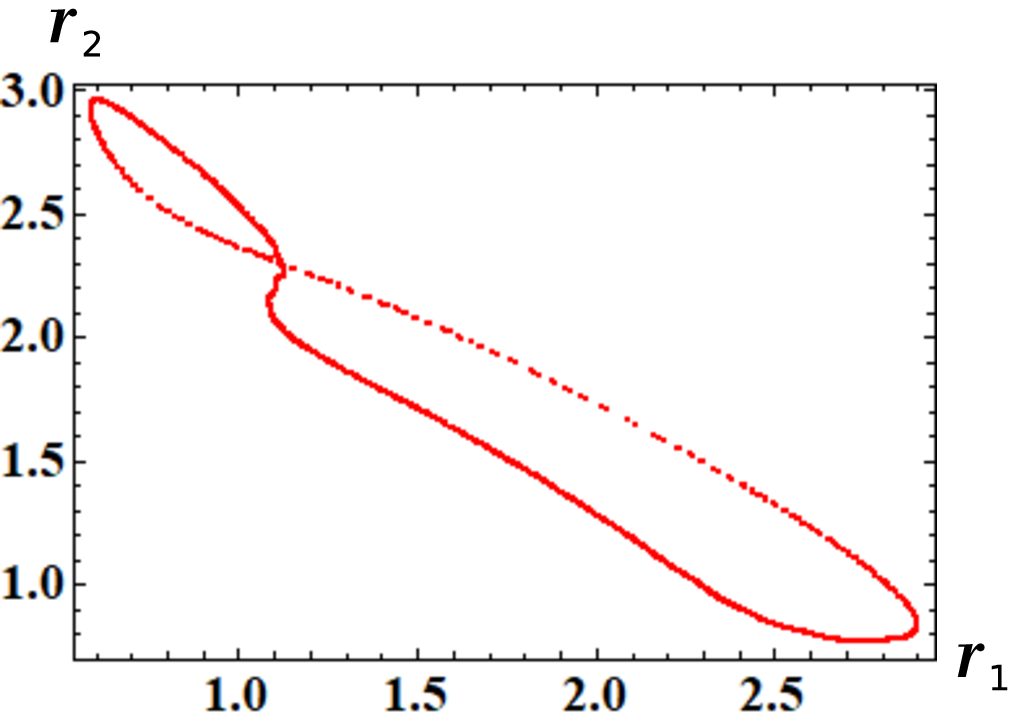, height=5.0cm} \vspace{1mm} \\
   (a) non-periodic & (b) quasi-periodic \\
  \end{tabular}
  \caption{Poincar\'e maps of the forced $4$-dimensional memristor Lotka-Volterra equations (\ref{eqn: 4-Lotka-Volterra-3}), 
   where $r_{1} =\sqrt{i^{2}+{x_{1}}^{2}}$ and $r_{2} =\sqrt{{x_{2}}^{2}+{x_{3}}^{2}}$.
   Parameters:  $\ r = 0.1,  \ \omega = 2$.
   Initial conditions: 
   (a) $i(0) = 0.6072, \, x_{1}(0)= 1.2, \, x_{2}(0) =1.3, \, x_{3}(0) =1.3$.  \ \ 
   (b) $i(0) = 0.585, \, x_{1}(0)= 1.2, \, x_{2}(0) =1.3, \, x_{3}(0) =1.3$.}
  \label{fig:4-Lotka-poincare} 
\end{figure}
%
%

%---Fig. 65-------%
\begin{figure}[hpbt]
 \centering
   \begin{tabular}{cc}
   \psfig{file=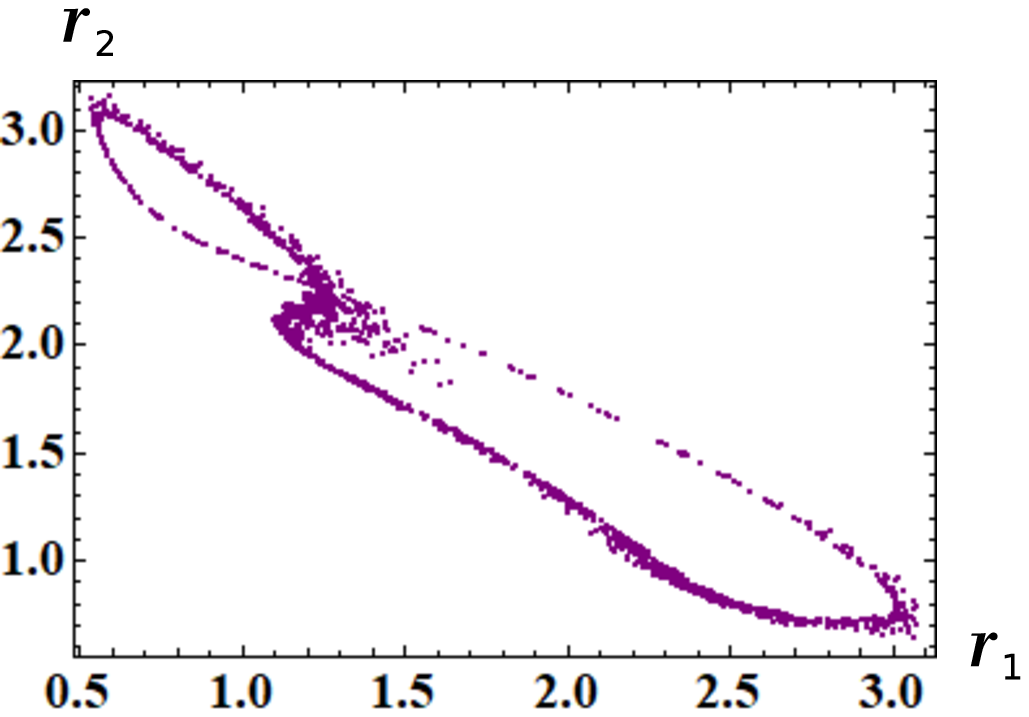, height=5.0cm} & 
   \psfig{file=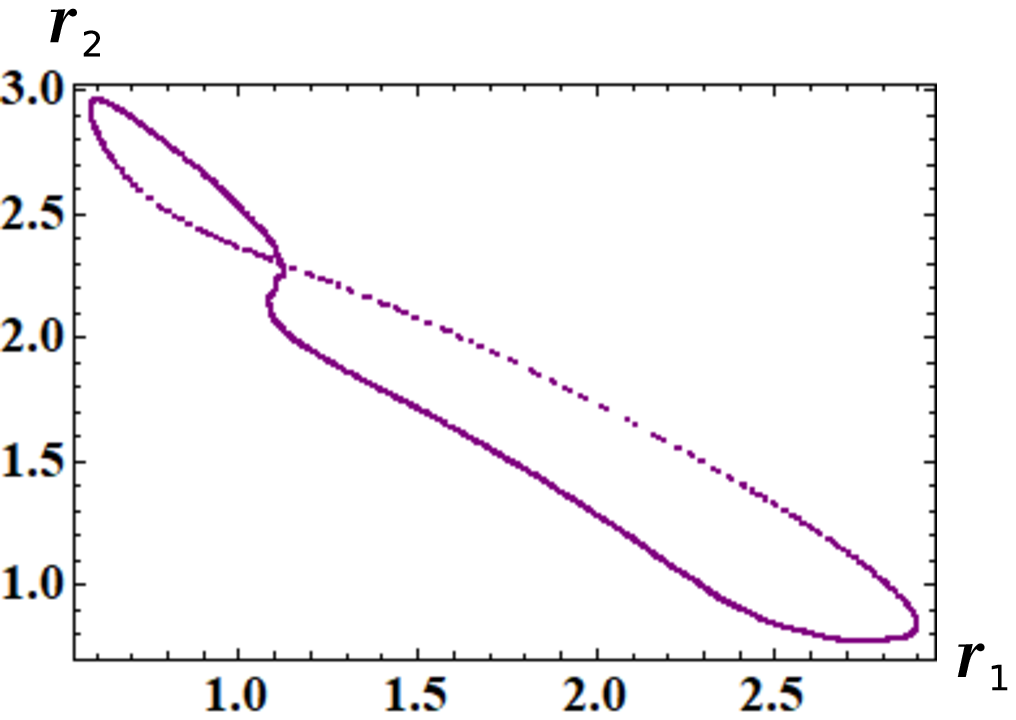, height=5.0cm} \vspace{1mm} \\
   (a) non-periodic & (b) quasi-periodic \\
  \end{tabular}
  \caption{Poincar\'e maps of the non-autonomous equations (\ref{eqn: 4-Lotka-Volterra-10}), 
   where $r_{1} =\sqrt{i^{2}+{x_{1}}^{2}}$, $r_{2} =\sqrt{{x_{2}}^{2}+{x_{3}}^{2}}$, 
   and  $x_{3} = K - i - x_{1} - x_{2}(t)- \frac{r}{\omega } \cos ( \omega t)$.   
   Parameters:  $\ r = 0.5,  \ \omega = 2$.
   Initial conditions: 
   (a) $i(0) = 0.6072, \, x_{1}(0)= 1.2, \, x_{2}(0) = 1.3, \, x_{3}(0) = 1.3, \, 
       k = i(0)+ x_1(0) + x_{2}(0) + x_{3}(0) + r/\omega \approx 4.0756 $.  \ \ 
   (b) $i(0) = 0.585, \, x_{1}(0)= 1.2, \, x_{2}(0) = 1.3, \, x_{3}(0) = 1.3, \, 
       k = i(0)+ x_1(0) + x_{2}(0)  + x_{3}(0) + r/\omega \approx 4.0446 $. }
  \label{fig:4-Lotka-poincare-10} 
\end{figure}
%
%

%\clearpage
%---Fig. 66-------%
\begin{figure}[hpbt]
 \centering
   \begin{tabular}{cc}
   \psfig{file=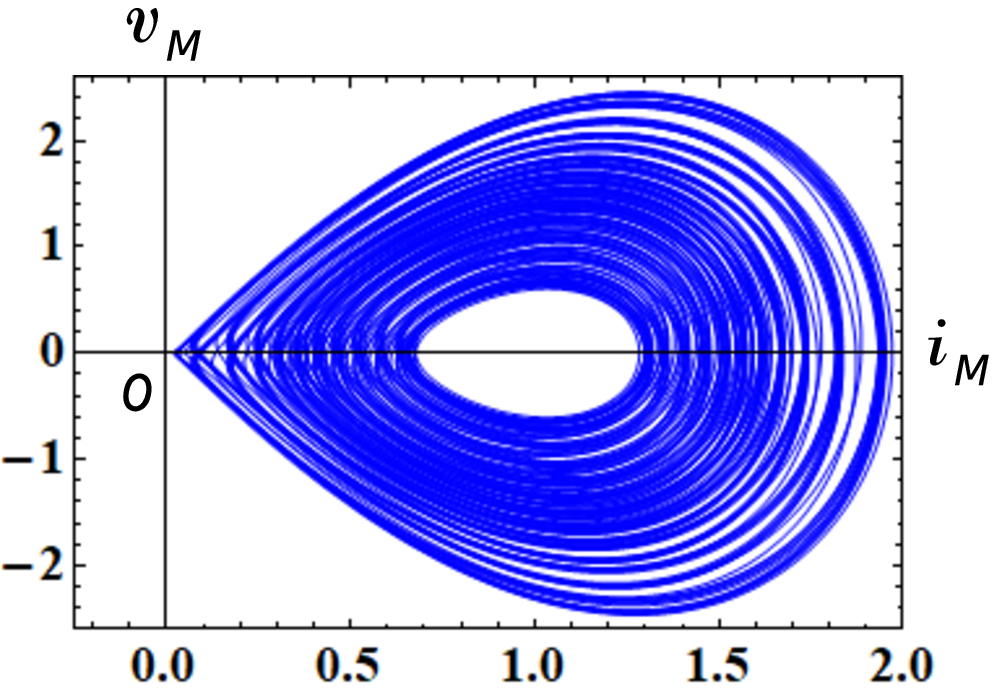, height=5.0cm} & 
   \psfig{file=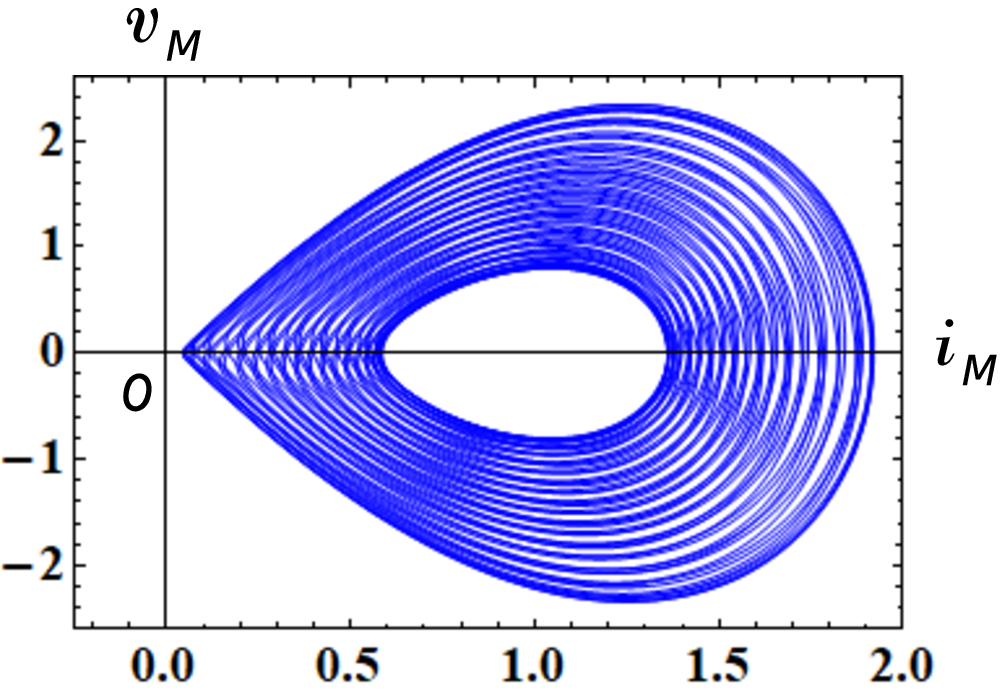, height=5.0cm} \vspace{1mm} \\
   (a) non-periodic & (b) quasi-periodic \\
  \end{tabular}
  \caption{ The $i_{M}-v_{M}$ loci of the forced $4$-dimensional memristor Lotka-Volterra equations (\ref{eqn: 4-Lotka-Volterra-3}).  
   Here, $v_{M}$ and $i_{M}$ denote the terminal voltage and the terminal current of the current-controlled generic memristor.  
   Parameters:  $\ r = 0.1,  \ \omega = 2$.
   Initial conditions: 
   (a) $i(0) = 0.6072, \, x_{1}(0)= 1.2, \, x_{2}(0) =1.3, \, x_{3}(0) =1.3$.  \ \ 
   (b) $i(0) = 0.585, \, x_{1}(0)= 1.2, \, x_{2}(0) =1.3, \, x_{3}(0) =1.3$.}
  \label{fig:4-Lotka-pinch} 
\end{figure}
%
%

%---Fig. 67-------%
\begin{figure}[hpbt]
 \centering
   \begin{tabular}{cc}
    \psfig{file=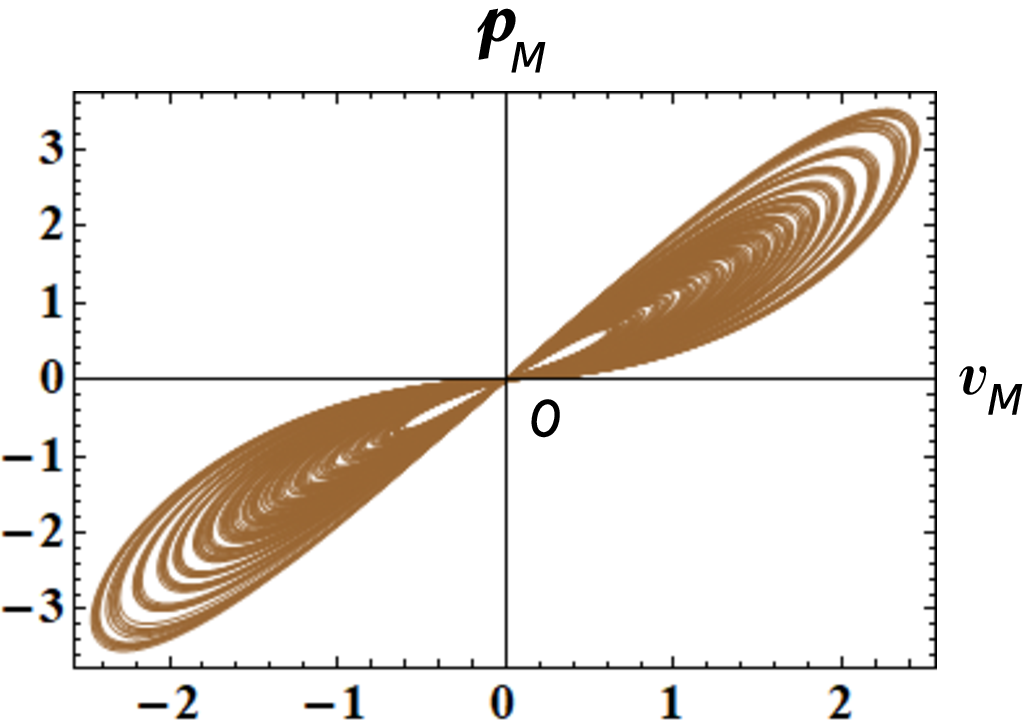, height=5cm}  & 
    \psfig{file=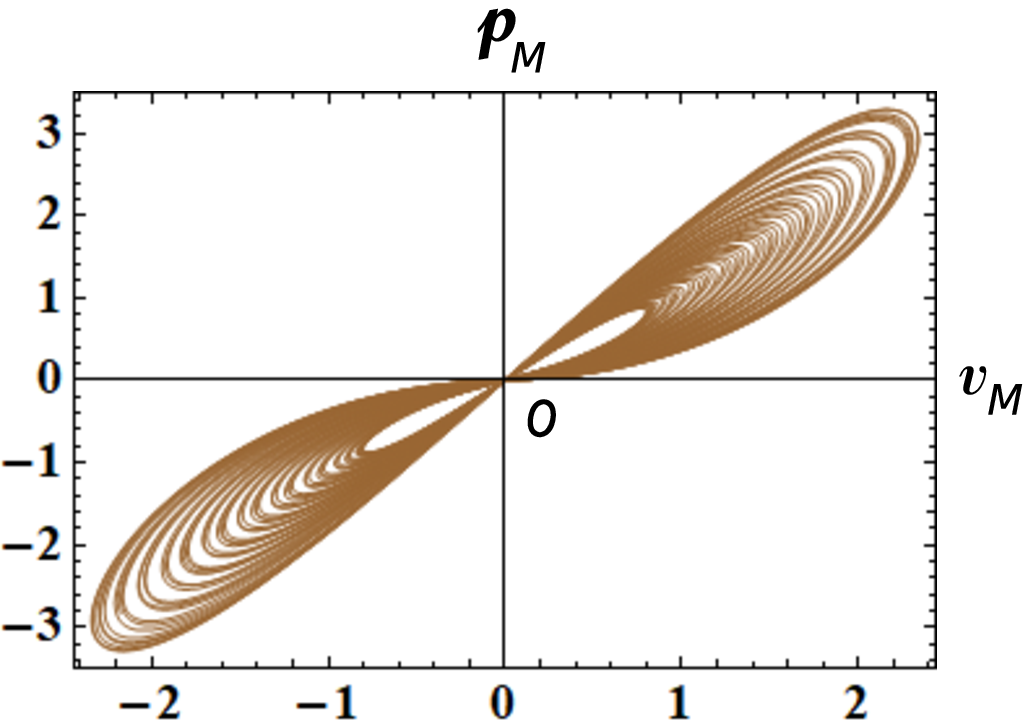, height=5cm}  \\
    (a) non-periodic & (b) quasi-periodic 
   \end{tabular}
  \caption{The $v_{M}-p_{M}$ locus of  the forced $4$-dimensional memristor Lotka-Volterra equations (\ref{eqn: 4-Lotka-Volterra-3}). 
   Here, $p_{M}(t)$ is an instantaneous power defined by $p_{M}(t)=i_{M}(t)v_{M}(t)$, 
   and $v_{M}(t)$ and $i_{M}(t)$ denote the terminal voltage and the terminal current of the current-controlled generic memristor.   
   Observe that the $v_{M}-p_{M}$ locus is pinched at the origin, and the locus lies in the first and the third quadrants. 
   The memristor switches between passive and active modes of operation, depending on its terminal voltage $v_{M}(t)$.
   Parameters:  $\ r = 0.1,  \ \omega = 2$.
   Initial conditions: 
   (a) $i(0) = 0.6072, \, x_{1}(0)= 1.2, \, x_{2}(0) =1.3, \, x_{3}(0) =1.3$.  \ \ 
   (b) $i(0) = 0.585, \, x_{1}(0)= 1.2, \, x_{2}(0) =1.3, \, x_{3}(0) =1.3$.}
  \label{fig:4-Lotka-power} 
\end{figure}
%
%

%---Fig. 68-------%
\begin{figure}[hpbt]
 \centering
   \begin{tabular}{cc}
   \psfig{file=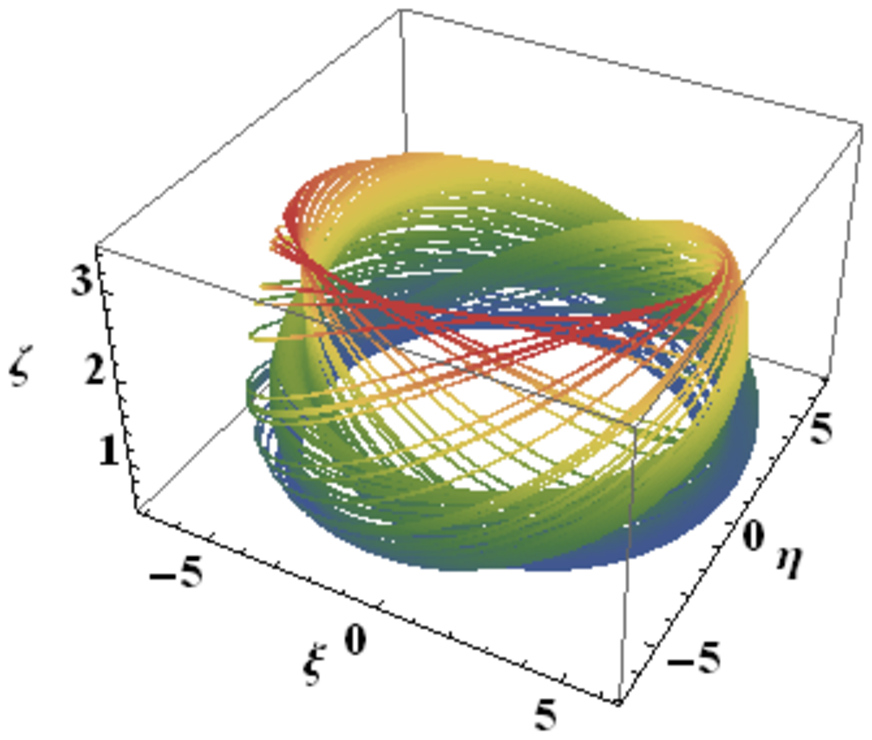, height=6cm} & 
   \psfig{file=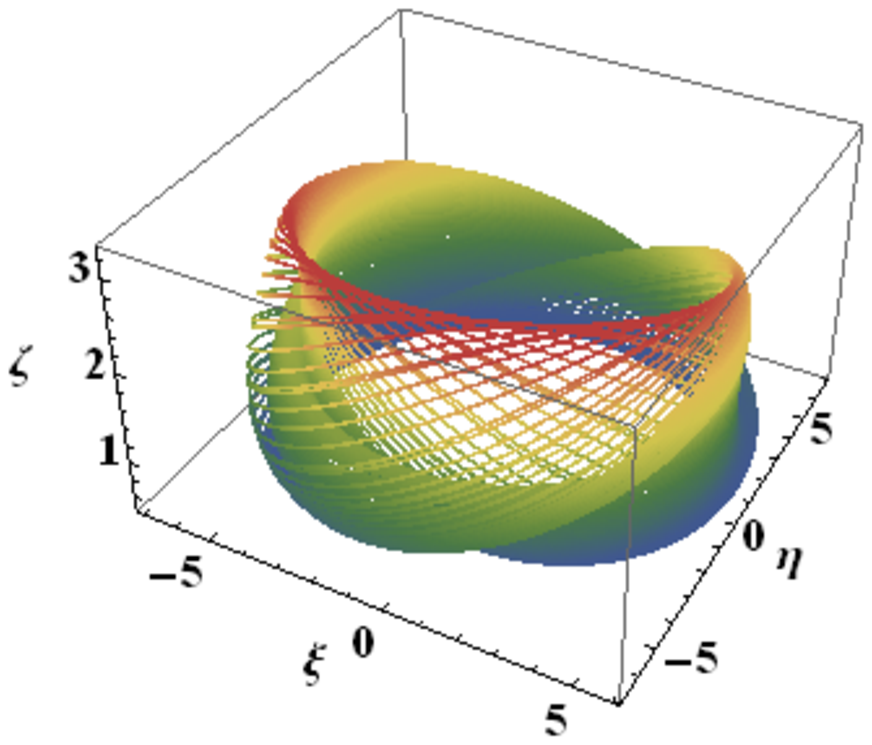, height=6cm} \\
   (a) non-periodic & (b) quasi-periodic 
  \end{tabular}
  \caption{The two trajectories of the forced $4$-dimensional memristor Lotka-Volterra equations (\ref{eqn: 4-Lotka-Volterra-3}), 
   which are projected into the $( \xi, \, \eta, \, \zeta )$-space via the coordinate transformation (\ref{eqn: 4-LV-projection}).  
   Observe that the trajectory in Figure \ref{fig:4-Lotka-torus}(a) is less dense than Figure \ref{fig:4-Lotka-torus}(b).   
   The trajectories are colored with the \emph{DarkRainbow} color code in Mathematica.
   Parameters:  $\ r = 0.1,  \ \omega = 2$.
   Initial conditions: 
   (a) $i(0) = 0.6072, \, x_{1}(0)= 1.2, \, x_{2}(0) =1.3, \, x_{3}(0) =1.3$.  \ \ 
   (b) $i(0) = 0.585, \, x_{1}(0)= 1.2, \, x_{2}(0) =1.3, \, x_{3}(0) =1.3$.}
  \label{fig:4-Lotka-torus} 
\end{figure}
%
%

%---Fig. 69-------%
\begin{figure}[hpbt]
 \centering
   \begin{tabular}{cc}
    \psfig{file=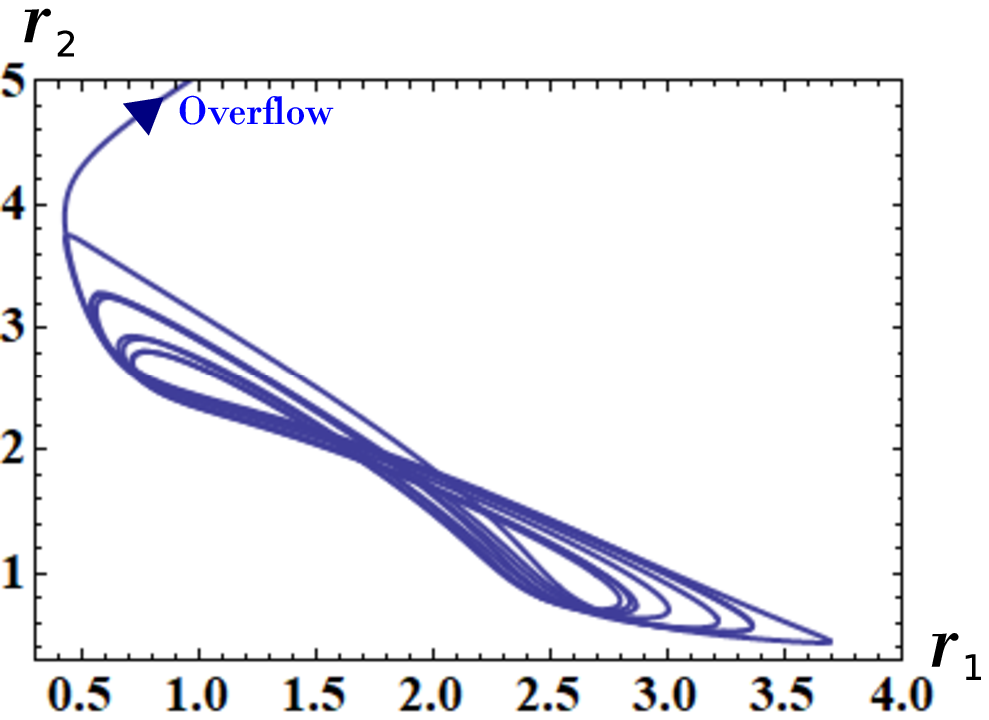, height=5cm}  & 
    \psfig{file=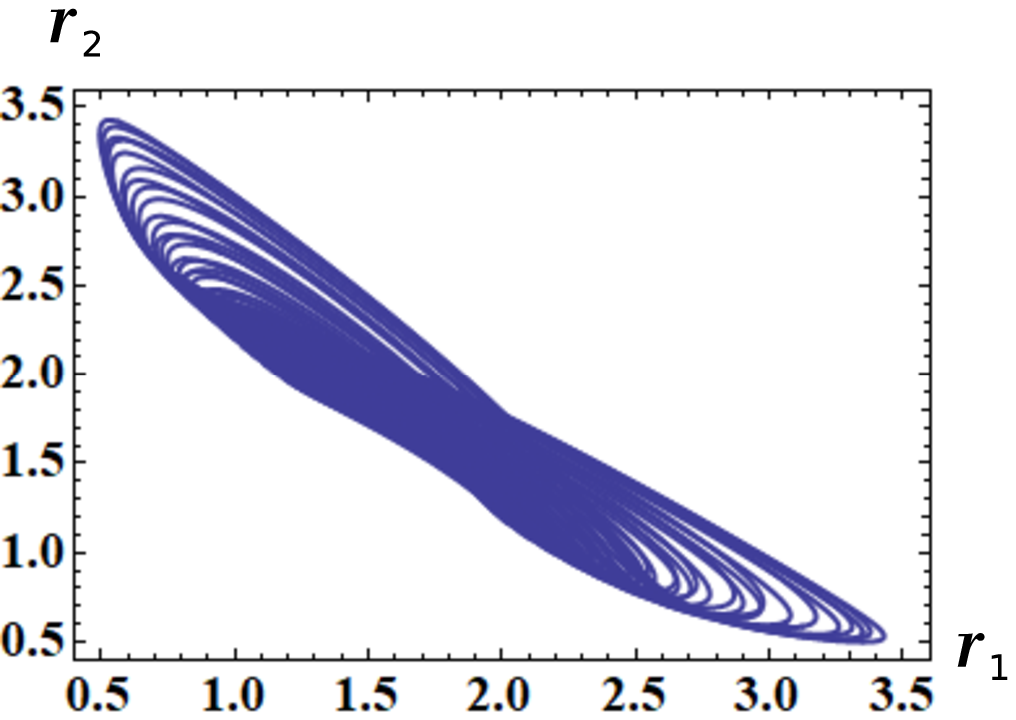, height=5cm}  \\
   (a) $h=0.002$ & (b) $h=0.001$  
   \end{tabular}
  \caption{Behavior of the forced memristor Lotka-Volterra equations (\ref{eqn: 4-Lotka-Volterra-3}). 
   If $h=0.002$, then the trajectory rapidly grows for $t \ge 1443$, and an overflow occurs as shown in Figure \ref{fig:Lotka-trajectory}(a). 
   However, if $h=0.001$, then the trajectory stays in a finite region of the $(r_{1}, \, r_{2})$-plane 
   as shown in Figure \ref{fig:Lotka-trajectory}(b).   
   Here, $h$ denotes the maximum step size of the numerical integration, 
   and $r_{1} =\sqrt{i^{2}+{x_{1}}^{2}}$, $r_{2} =\sqrt{{x_{2}}^{2}+{x_{3}}^{2}}$
   Parameters:  $r = 0.107, \ \  \omega =2$.
   Initial conditions: $i(0) = 0.608, \, x_{1}(0)= 1.2, \, x_{2}(0) =1.3, \, x_{3}(0) =1.3$.}
  \label{fig:4-Lotka-trajectory} 
\end{figure}
\clearpage

%-------------------------------------%
\subsubsection{$4$-dimensional ecological predator-prey model}
%-------------------------------------%
%
The $4$-dimensional ecological predator-prey model is given by \cite{Hirota(1985)} 
\begin{center}
\begin{minipage}{9.5cm}
\begin{shadebox}
\underline{\emph{$4$-dimensional ecological predator-prey model equations}}
\begin{equation}
\left. 
\begin{array}{ccl}
 \displaystyle \frac{d M_{1}}{dt} &=& (M_{4} - M_{2}) {M_{1}}^{2},  \vspace{2mm} \\
 \displaystyle \frac{d M_{2}}{dt} &=& (M_{1} - M_{3}) {M_{2}}^{2},  \vspace{2mm} \\
 \displaystyle \frac{d M_{3}}{dt} &=& (M_{2} - M_{4}) {M_{3}}^{2},  \vspace{2mm} \\
 \displaystyle \frac{d M_{4}}{dt} &=& (M_{3} - M_{1}) {M_{4}}^{2}. 
\end{array}
\right \}
\label{eqn: ecological-5}
\end{equation}
\end{shadebox}
\end{minipage}
\end{center}
If we change the variables  
\begin{equation}
   M_{1} = i_{1}, \ M_{2} = x_{1}, \ M_{3}  = i_{2} , \  M_{4} =  x_{2},     
\end{equation}
Eq. (\ref{eqn: ecological-5}) can be recast into the form 
\begin{center}
\begin{minipage}{11cm}
\begin{shadebox}
\underline{\emph{$4$-dimensional memristor ecological predator-prey model equations}}
\begin{equation}
\left. 
\begin{array}{ccc}
 \displaystyle \frac{d i_{1}}{dt} &=& (x_{2} - x_{1}) \, {i_{1}}^{2},  \vspace{2mm} \\
 \displaystyle \frac{d x_{1}}{dt} &=& (i_{1} - i_{2}) \, {x_{1}}^{2},  \vspace{2mm} \\
 \displaystyle \frac{d i_{2}}{dt} &=& (x_{1} - x_{2}) \, {i_{2}}^{2},  \vspace{2mm} \\
 \displaystyle \frac{d x_{2}}{dt} &=& (i_{2} - i_{1}) \, {x_{2}}^{2}. 
\end{array}
\right \}
\label{eqn: ecological-6}
\end{equation}
\end{shadebox}
\end{minipage}
\end{center}
Here, $i_{1}$ and $i_{2}$ denote the currents of two current-controlled extended memristors in Figure \ref{fig:memristor-inductor-N}.  
In this case, the small-signal \emph{memristances} of the extended memristors is defined by   
\begin{equation}
 \left.
 \begin{array}{cll}
  \hat{R}_{1}(x_{1}, \, x_{2}, \, i_{1})  &=& - (x_{2} - x_{1}) \, i_{1} \vspace{1mm} \\
  \hat{R}_{2}(x_{1}, \, x_{2}, \, i_{2})  &=& - (x_{1} - x_{2}) \, i_{2}.
 \end{array} 
\right \} 
\end{equation}
The terminal voltages $v_{n}$ and the terminal current $i_{n}$ of these extended memristors are given by ($n=1, \, 2$)
\begin{center}
\begin{minipage}{8.8cm}
\begin{shadebox}
\underline{\emph{V-I characteristics of the $2$ extended memristors}}
\begin{equation}
\begin{array}{c}
\left.
\begin{array}{cll}
  v_{1} &=& \hat{R}_{1}(x_{1}, \, x_{2}, \, i_{1}) \, i_{1} = - (x_{2} - x_{1}) \, {i_{1}}^{2}, 
  \vspace{1mm} \\
  \displaystyle \frac{d x_{1}}{dt} 
   &=& \tilde{f}_{1}(x_{1}, \, i_{1}, \, i_{2}) = (i_{1} - i_{2}) \, {x_{1}}^{2} 
\end{array}
\right \} 
\vspace{4mm} \\
\left. 
\begin{array}{cll}
  v_{2} &=& \hat{R}_{2}(x_{1}, \, x_{2}, \, i_{2}) \, i_{2}  = - (x_{1} - x_{2}) \, {i_{2}}^{2}, 
  \vspace{1mm} \\
  \displaystyle \frac{d x_{2}}{dt} 
   &=& \tilde{f}_{2}(x_{2}, \, i_{1}, \, i_{2}) = (i_{2} - i_{1}) \, {x_{2}}^{2}. 
\end{array}
\right \}
\end{array}
\label{eqn: extended-301}
\end{equation}
\end{shadebox}
\end{minipage}
\end{center}
Thus, Eq. (\ref{eqn: ecological-6}) can be realized by  the $4$-element memristor circuit in Figure \ref{fig:memristor-inductor-N}.  
The forced $4$-dimensional memristor ecological predator-prey model is given by 
\begin{center}
\begin{minipage}{12.5cm}
\begin{shadebox}
\underline{\emph{Forced $4$-dimensional memristor ecological predator-prey model equations}}
\begin{equation}
\left. 
\begin{array}{ccl}
 \displaystyle \frac{d i_{1}}{dt} &=& (x_{2} - x_{1}) \, {i_{1}}^{2} + r \sin ( \omega t),  \vspace{2mm} \\
 \displaystyle \frac{d x_{1}}{dt} &=& (i_{1} - i_{2}) \, {x_{1}}^{2},  \vspace{2mm} \\
 \displaystyle \frac{d i_{2}}{dt} &=& (x_{1} - x_{2}) \, {i_{2}}^{2},  \vspace{2mm} \\
 \displaystyle \frac{d x_{2}}{dt} &=& (i_{2} - i_{1}) \, {x_{2}}^{2}. 
\end{array}
\right \}
\label{eqn: ecological-7}
\end{equation}
where $r$ and $\omega$ are constants.  
\end{shadebox}
\end{minipage}
\end{center}

The memristor circuit equations (\ref{eqn: ecological-3}) and (\ref{eqn: ecological-7}) exhibit periodic behavior.  
If an external source is added as shown in Figure \ref{fig:memristor-inductor-source-N},
then the forced $4$-dimensional memristor ecological predator-prey model (\ref{eqn: ecological-7}) can exhibit quasi-periodic and non-periodic responses.  
We show the non-periodic response, quasi-periodic response, Poincar\'e maps, and $i_{j}-v_{j}$ loci of Eq. (\ref{eqn: ecological-7}) in Figures \ref{fig:Ecological-attractor}, \ref{fig:Ecological-attractor-2}, \ref{fig:Ecological-poincare}, and \ref{fig:Ecological-pinch}, respectively $(j=1, \, 2)$.  
The $i_{j}-v_{j}$ loci in Figure \ref{fig:Ecological-pinch} lie in the first and the fourth quadrants. 
Thus, the corresponding extended memristor is an active element. 
We show the $v_{j}-p_{j}$ locus in Figure \ref{fig:Ecological-power}, 
where $p_{j}(t)$ is an instantaneous power defined by $p_{j}(t)=i_{j}(t)v_{j}(t)$ $(j=1, \, 2)$.  
Observe that the $v_{j}-p_{j}$ locus is pinched at the origin, and the locus lies in the first and the third quadrants. 
Thus, the memristor switches between passive and active modes of operation, depending on its terminal voltage. 
We conclude as follow: \\
\begin{center}
\begin{minipage}{.7\textwidth}
\begin{itembox}[l]{Switching behavior of the memristor}
Assume that Eq. (\ref{eqn: ecological-7}) exhibits non-periodic or quasi-periodic oscillation.  
Then the extended memristor defined by Eq. (\ref{eqn: extended-301}) can switch between ``passive'' and ``active'' modes of operation, depending on its terminal voltage.  
\end{itembox} 
\end{minipage}
\end{center}
In order to view the Poincar\'e maps in Figure \ref{fig:Ecological-poincare} from a different perspective, 
let us project the trajectories into the $( \xi, \, \eta, \, \zeta )$-space via the transformation 
\begin{equation}
 \begin{array}{lll}
   \xi (\tau)   &=& (r_{1}( \tau) + 5 ) \cos \,( \omega \tau ), \vspace{2mm} \\ 
   \eta (\tau)  &=& (r_{1}( \tau) + 5 ) \sin \,( \omega \tau ), \vspace{2mm} \\ 
   \zeta (\tau) &=& r_{2}( \tau ),
 \end{array}  
\label{eqn: Ecological-projection}
\end{equation}
where $r_{1} =\sqrt{i_{1}^{2}+{x_{1}}^{2}}$ and $r_{2} =\sqrt{{i_{2}}^{2}+{x_{2}}^{2}}$.  
Observe that the irregular trajectory exists in Figure \ref{fig:Ecological-torus}(a).  
The following parameters are used in our computer simulations:
\begin{equation}
  r = 0.5,  \ \omega = 2.
\end{equation}

Note that in order to generate a non-periodic response, we have to choose the initial conditions carefully.  
We show its example in Figure \ref{fig:Ecological-trajectory}.  
Suppose that Eq. (\ref{eqn: ecological-7}) has the following parameters and initial conditions:  
\begin{equation}
 \begin{array}{ll}
  \text{Parameters: }         & r = 0.5, \ \  \omega = 2, \vspace{2mm} \\ 
  \text{Initial conditions: } &  i(0) = 0.098, \, x_{1}(0) = 0.5, \vspace{2mm} \\ 
                              & x_{2}(0)=1.1,  \, x_{3}(0)=1.3. 
 \end{array}  
\end{equation}
If we choose $h=0.01$, then the trajectory rapidly grows for $t \ge 5327$, and an overflow occurs as shown in Figure \ref{fig:Ecological-trajectory}(a). 
However, if we choose $h=0.005$, then the trajectory stays in a finite region as shown in Figure \ref{fig:Ecological-trajectory}(b).
The maximum step size of the numerical integration greatly affects the behavior of Eq. (\ref{eqn: ecological-7}).   
In this case, noise may considerably affect the behavior of the physical memristor circuit.

%---Fig. 70-------%
\begin{figure}[hpbt]
 \centering
   \begin{tabular}{ccc}
   \psfig{file=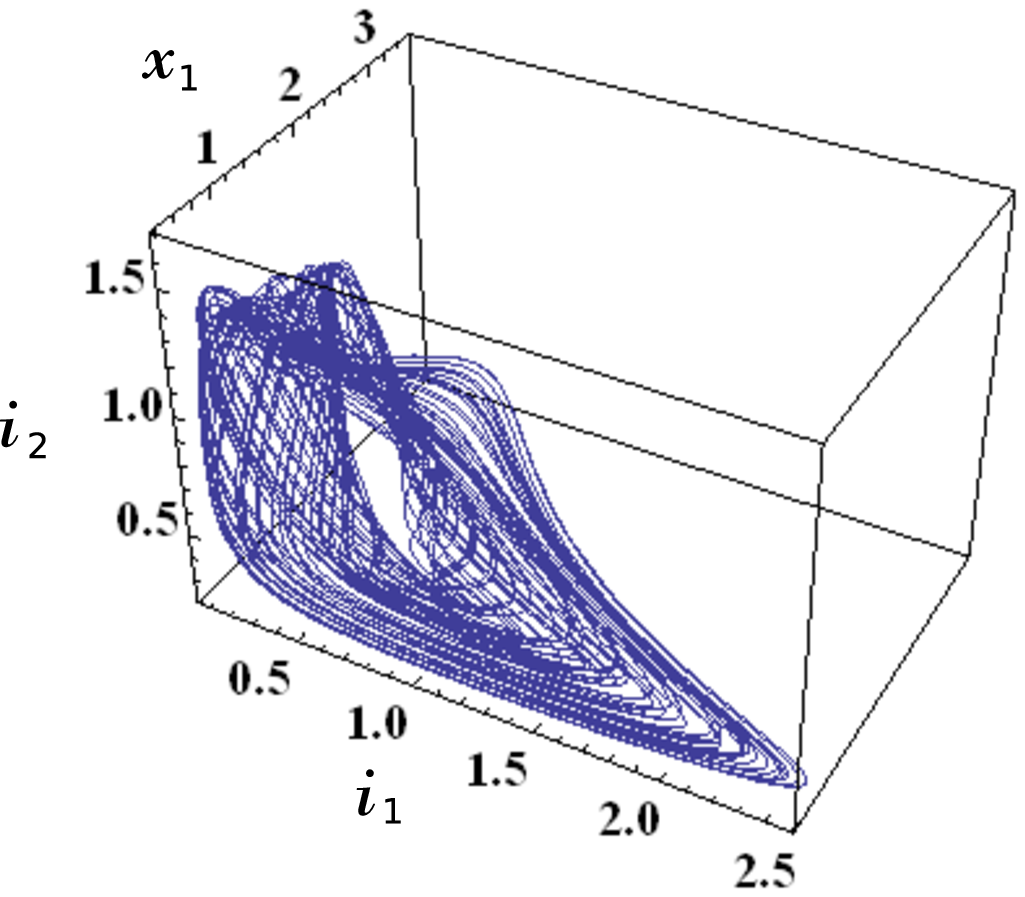, height=6.0cm} & \hspace{2mm} &
   \psfig{file=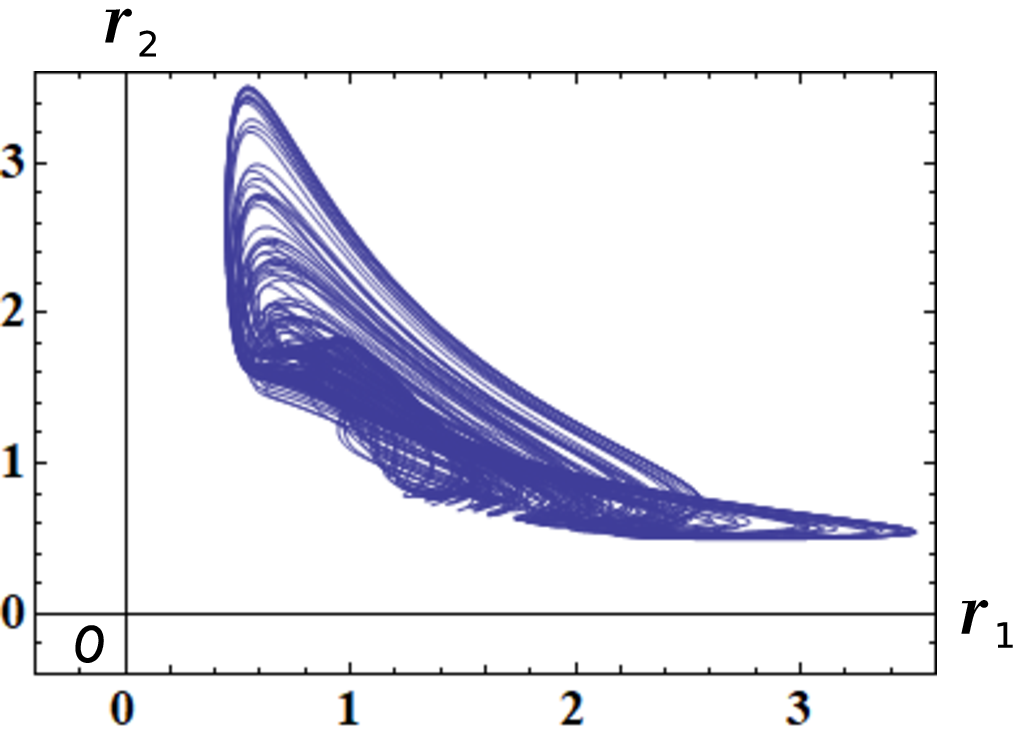, height=5.5cm} \\
   (a) $(i_{1}, \ x_{1}, \ i_{2})$-space & & (b) $(r_{1},\ r_{2})$-plane  \vspace{1mm} \\
  \end{tabular}
  \caption{Non-periodic response of the forced $4$-dimensional memristor ecological predator-prey equations (\ref{eqn: ecological-7}), 
   where $r_{1} =\sqrt{{i_{1}}^{2}+{x_{1}}^{2}}$ and $r_{2} =\sqrt{{i_{2}}^{2}+{x_{2}}^{2}}$.   
   Parameters:  $\ r = 0.5,  \ \omega = 2$.
   Initial conditions: $i(0) = 0.1, \, x_{1}(0)= 0.5, \, x_{2}(0) =1.1, \, x_{3}(0) =1.3$.}
  \label{fig:Ecological-attractor} 
\end{figure}
%
%

%---Fig. 71-------%
\begin{figure}[hpbt]
 \centering
   \begin{tabular}{ccc}
   \psfig{file=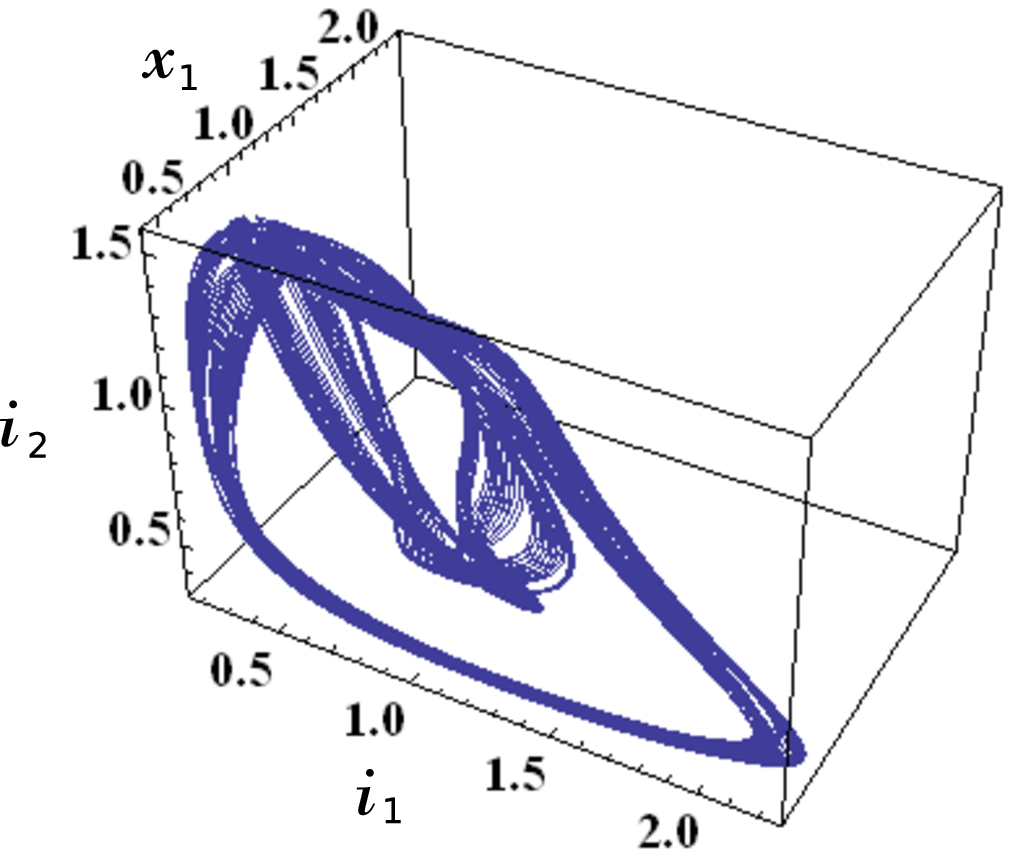, height=6.0cm} & \hspace{2mm} &
   \psfig{file=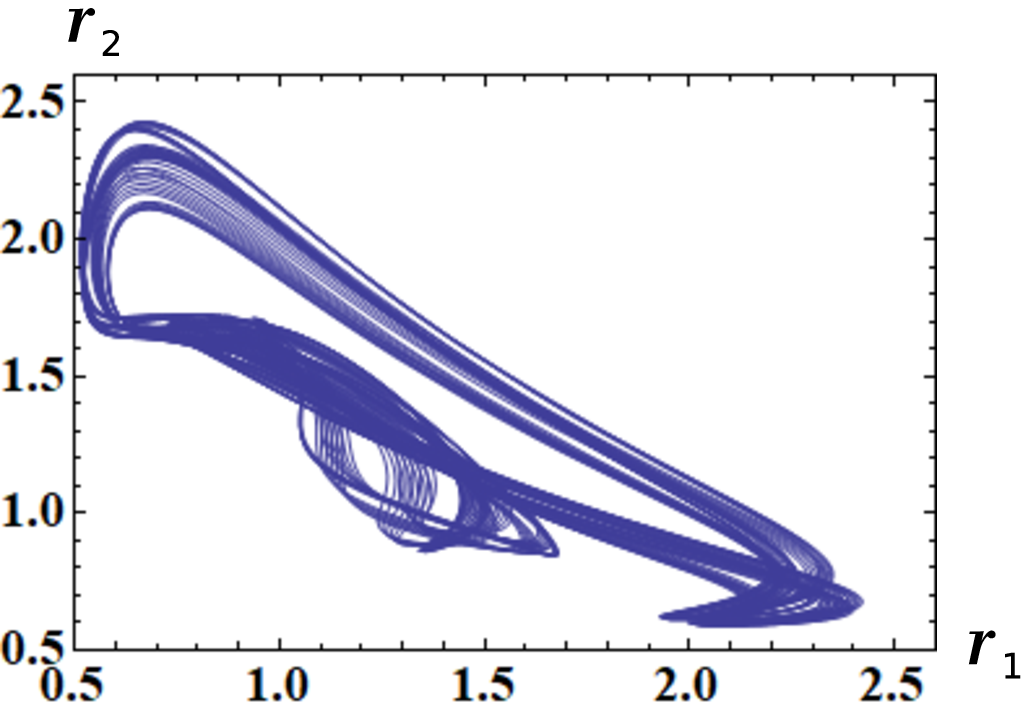, height=5.5cm} \\
   (a) $(i_{1}, \ x_{1}, \ i_{2})$-space & & (b) $(r_{1},\ r_{2})$-plane  \vspace{1mm} \\
  \end{tabular}
  \caption{Quasi-periodic response of the forced $4$-dimensional memristor ecological predator-prey equations (\ref{eqn: ecological-7}), 
   where $r_{1} =\sqrt{{i_{1}}^{2}+{x_{1}}^{2}}$ and $r_{2} =\sqrt{{i_{2}}^{2}+{x_{2}}^{2}}$.   
   Parameters:  $\ r = 0.5,  \ \omega = 2$.
   Initial conditions: $i_{1}(0) = 0.2, \, x_{1}(0)= 0.5, \, i_{2}(0) =1.1, \, x_{2}(0) =1.3$.}
  \label{fig:Ecological-attractor-2} 
\end{figure}
%
%

%\clearpage
%---Fig. 72-------%
\begin{figure}[hpbt]
 \centering
   \begin{tabular}{cc}
   \psfig{file=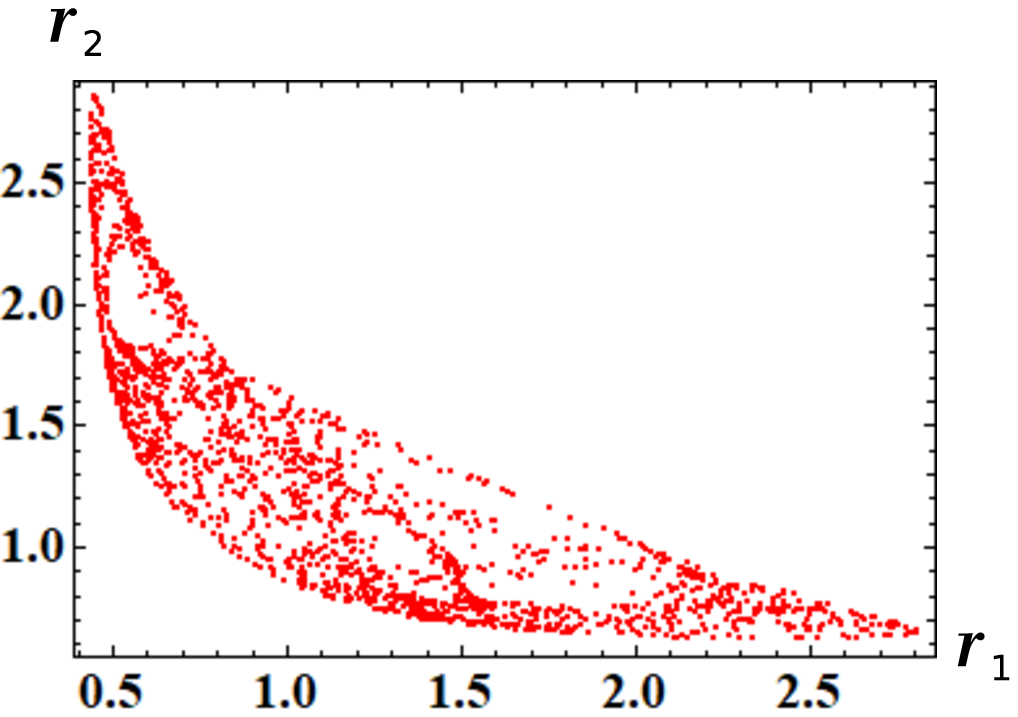, height=5.0cm} & 
   \psfig{file=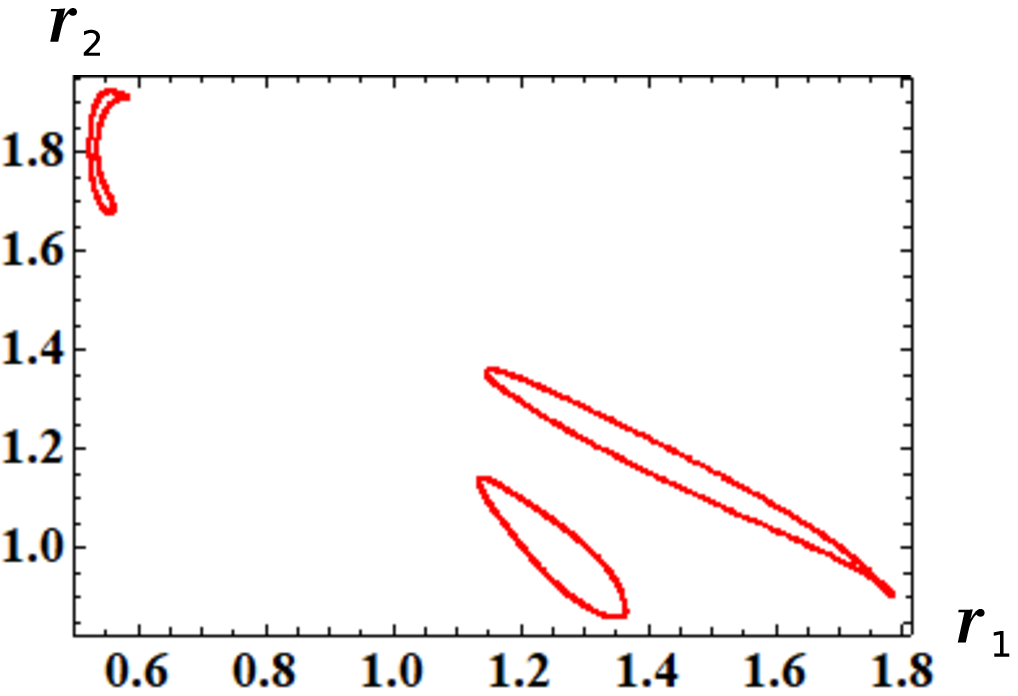, height=5.0cm} \vspace{1mm} \\
   (a) non-periodic & (b) quasi-periodic \\
  \end{tabular}
  \caption{Poincar\'e maps of the forced $4$-dimensional memristor ecological predator-prey equations (\ref{eqn: ecological-7}), 
   where $r_{1} =\sqrt{{i_{1}}^{2}+{x_{1}}^{2}}$ and $r_{2} =\sqrt{{i_{2}}^{2}+{x_{2}}^{2}}$.   
   Parameters:  $\ r = 0.5,  \ \omega = 2$.
   Initial conditions: 
   (a) $i_{1}(0) = 0.1, \, x_{1}(0)= 0.5, \, i_{2}(0) =1.1, \, x_{2}(0) =1.3$. \  \
   (b) $i_{1}(0) = 0.2, \, x_{1}(0)= 0.5, \, i_{2}(0) =1.1, \, x_{2}(0) =1.3$.}
  \label{fig:Ecological-poincare} 
\end{figure}
%
%

%---Fig. 73-------%
\begin{figure}[hpbt]
 \centering
   \begin{tabular}{cc}
   \psfig{file=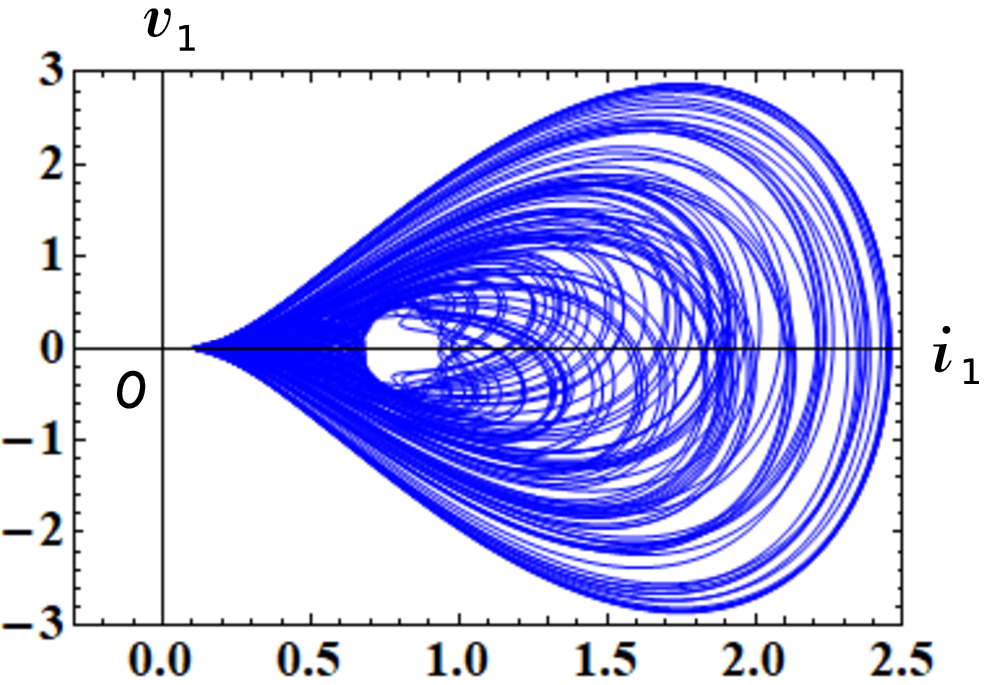, height=5.0cm} & 
   \psfig{file=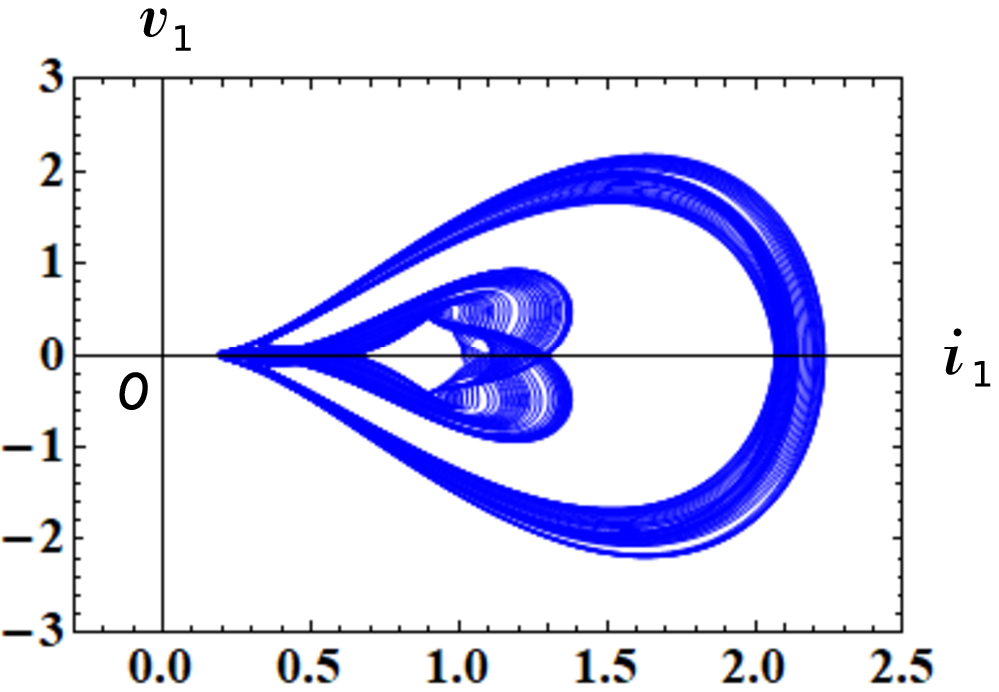, height=5.0cm} \vspace{1mm} \\
   (a) non-periodic $(i_{1}(0) = 0.1)$ & (c) quasi-periodic $(i_{1}(0) = 0.2)$ \vspace{5mm} \\
   \psfig{file=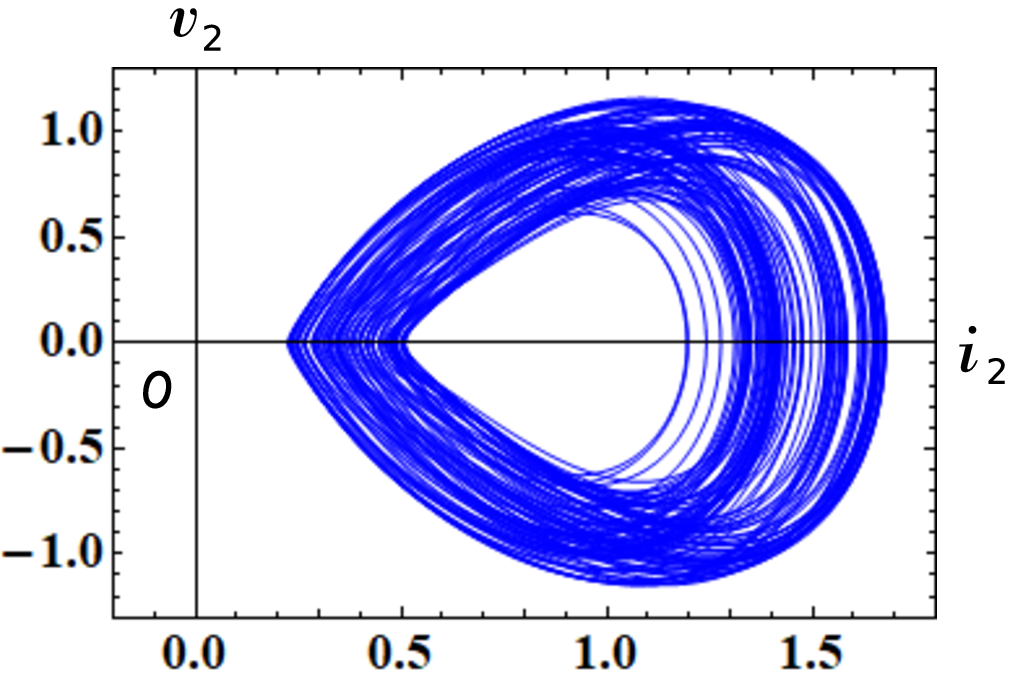, height=5.0cm} & 
   \psfig{file=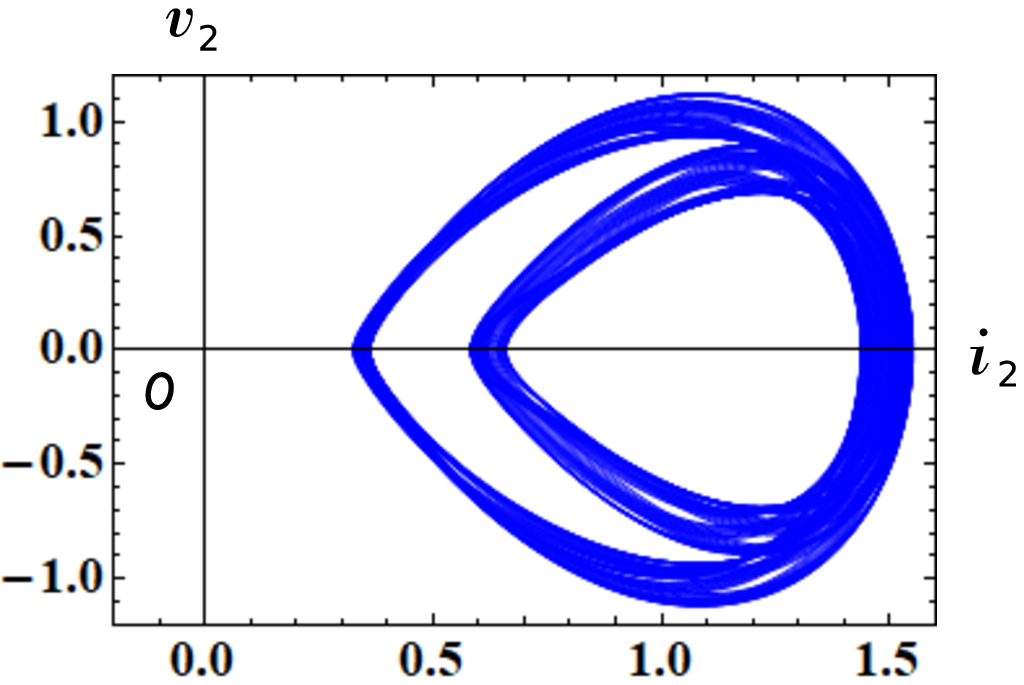, height=5.0cm} \vspace{1mm} \\
   (b) non-periodic  $(i_{1}(0) = 0.1)$ & (d) quasi-periodic  $(i_{1}(0) = 0.2)$\\
  \end{tabular}
  \caption{ The $i_{j}-v_{j}$ loci of the forced $4$-dimensional memristor ecological predator-prey equations (\ref{eqn: ecological-7}). 
   Here, $i_{j}$ and $v_{j}$ denote the terminal current and the voltage of the extended memristor, respectively $(j=1, \, 2)$.  
   Parameters:  $\ r = 0.5,  \ \omega = 2$.
   (a), (b) $i_{1}(0) = 0.1, \, x_{1}(0)= 0.5, \, i_{2}(0) =1.1, \, x_{2}(0) =1.3$. \  \
   (c), (d) $i_{1}(0) = 0.2, \, x_{1}(0)= 0.5, \, i_{2}(0) =1.1, \, x_{2}(0) =1.3$.}
  \label{fig:Ecological-pinch} 
\end{figure}
%
%

%\clearpage
%---Fig. 74-------%
\begin{figure}[hpbt]
 \centering
   \begin{tabular}{cc}
    \psfig{file=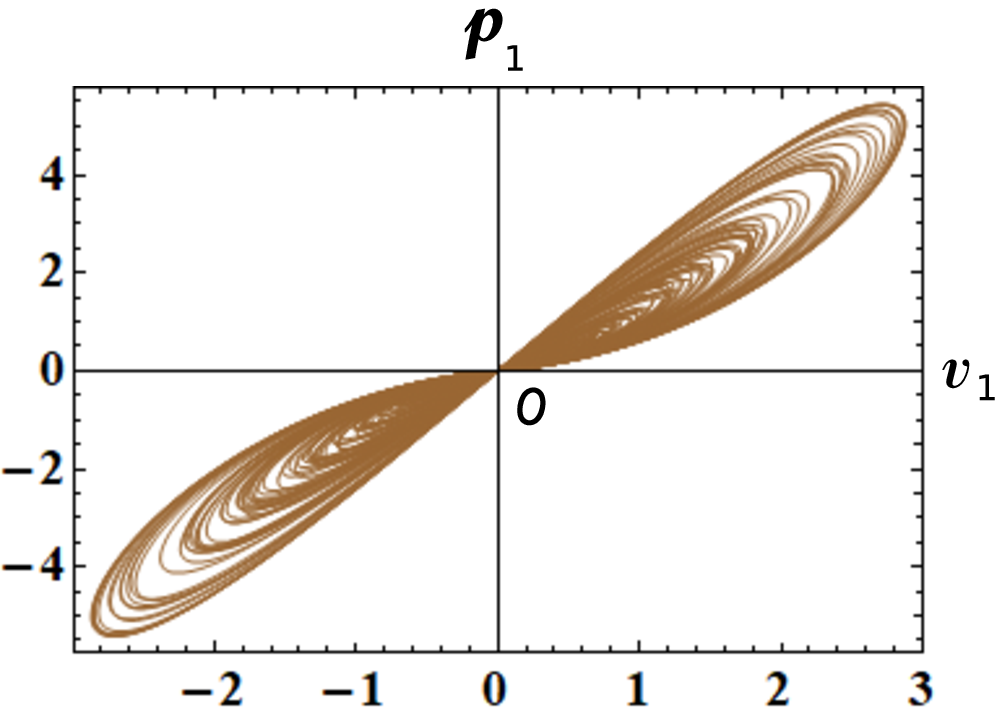, height=5.0cm}  & 
    \psfig{file=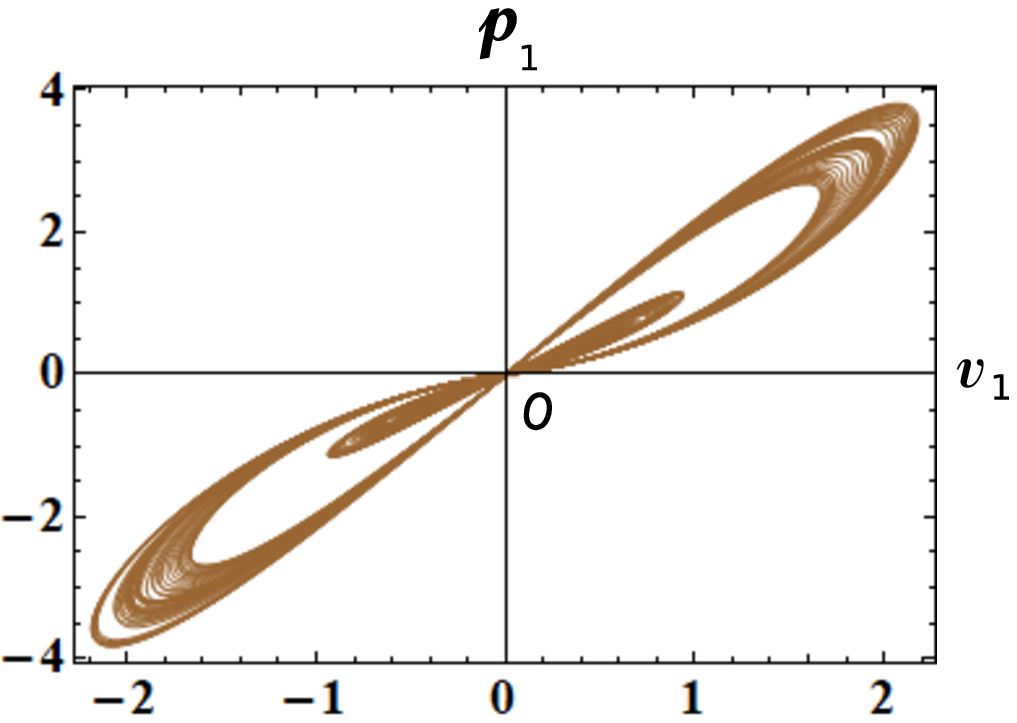, height=5.0cm}  \\
    (a) non-periodic $(i_{1}(0) = 0.1)$ & (c) quasi-periodic $(i_{1}(0) = 0.2)$ \vspace{5mm} \\
    \psfig{file=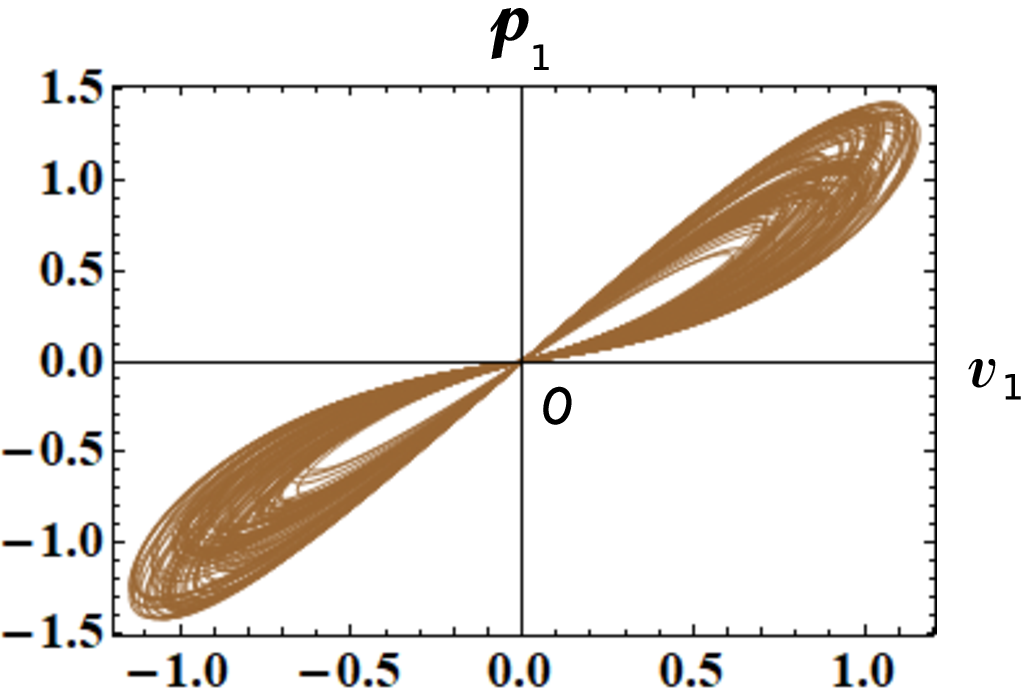, height=5.0cm}  & 
    \psfig{file=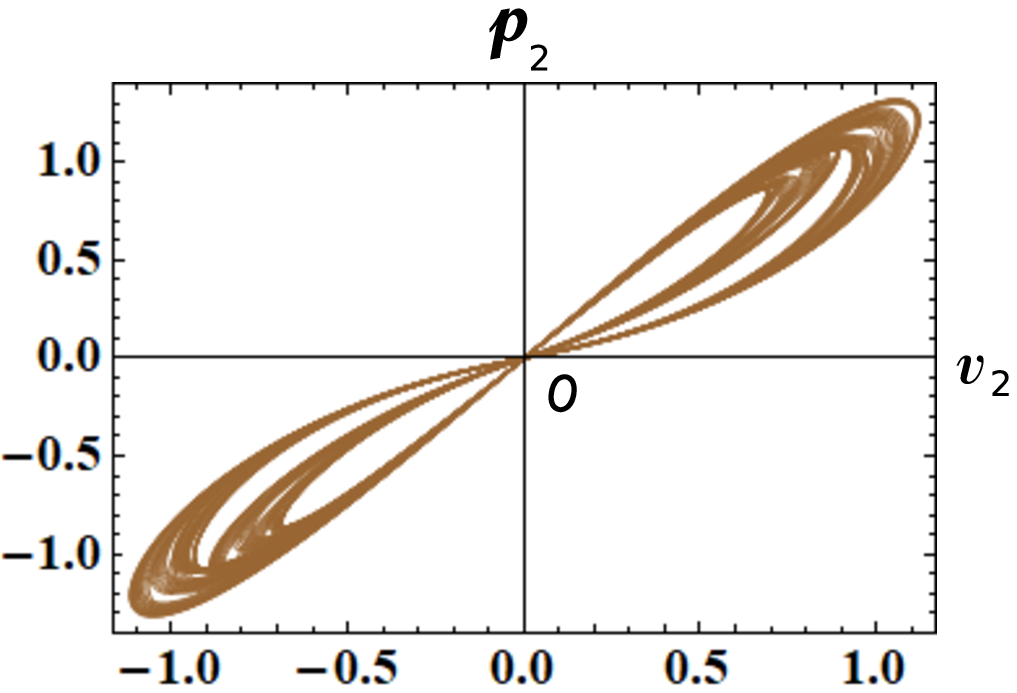, height=5.0cm}  \\
   (b) non-periodic  $(i_{1}(0) = 0.1)$ & (d) quasi-periodic  $(i_{1}(0) = 0.2)$\\
   \end{tabular}
  \caption{ The $v_{j}-p_{j}$ loci of the forced $4$-dimensional memristor ecological predator-prey equations (\ref{eqn: ecological-7}).   
   Here, $p_{j}(t)$ is an instantaneous power defined by $p_{j}(t)=i_{j}(t)v_{j}(t)$,  
   and $v_{j}(t)$ and $i_{j}(t)$ denote the terminal voltage and the terminal current of the $j$-th generic memristor, respectively 
   $(j=1, \, 2)$.
   Observe that the $v_{j}-p_{j}$ loci are pinched at the origin, and the locus lies in the first and the third quadrants. 
   The memristor switches between passive and active modes of operation, depending on its terminal voltage $v_{j}(t)$. 
   Parameters:  $\ r = 0.1,  \ \omega = 2$.
   Initial conditions: 
   (a), (b) $i_{1}(0) = 0.1, \, x_{1}(0)= 0.5, \, i_{2}(0) =1.1, \, x_{2}(0) =1.3$. \  \
   (c), (d) $i_{1}(0) = 0.2, \, x_{1}(0)= 0.5, \, i_{2}(0) =1.1, \, x_{2}(0) =1.3$.}
  \label{fig:Ecological-power} 
\end{figure}
%
%

%---Fig. 75-------%
\begin{figure}[hpbt]
 \centering
   \begin{tabular}{cc}
   \psfig{file=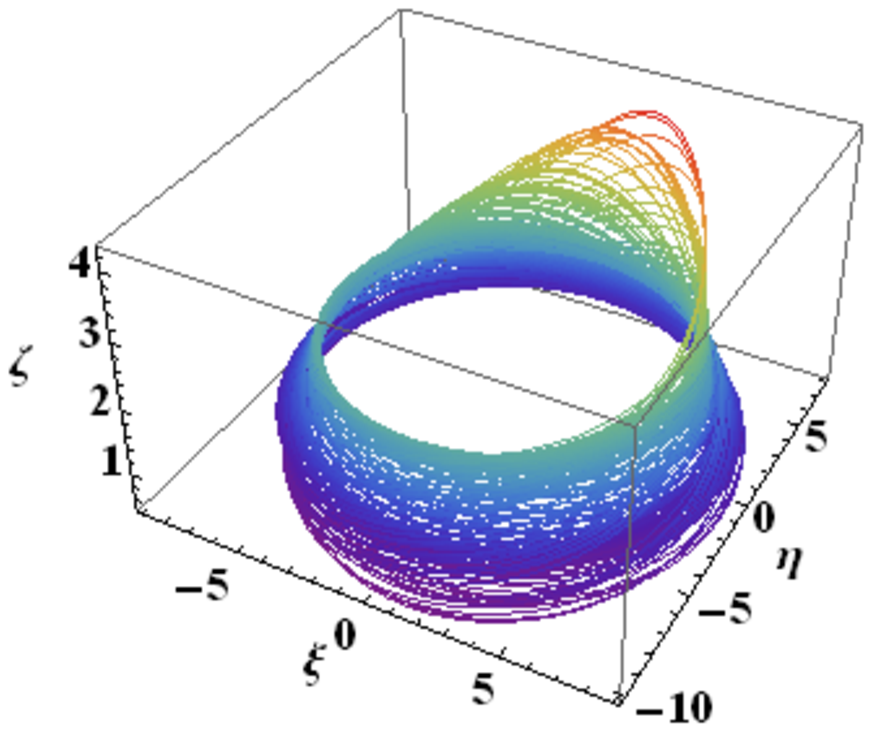, width=7.0cm} & 
   \psfig{file=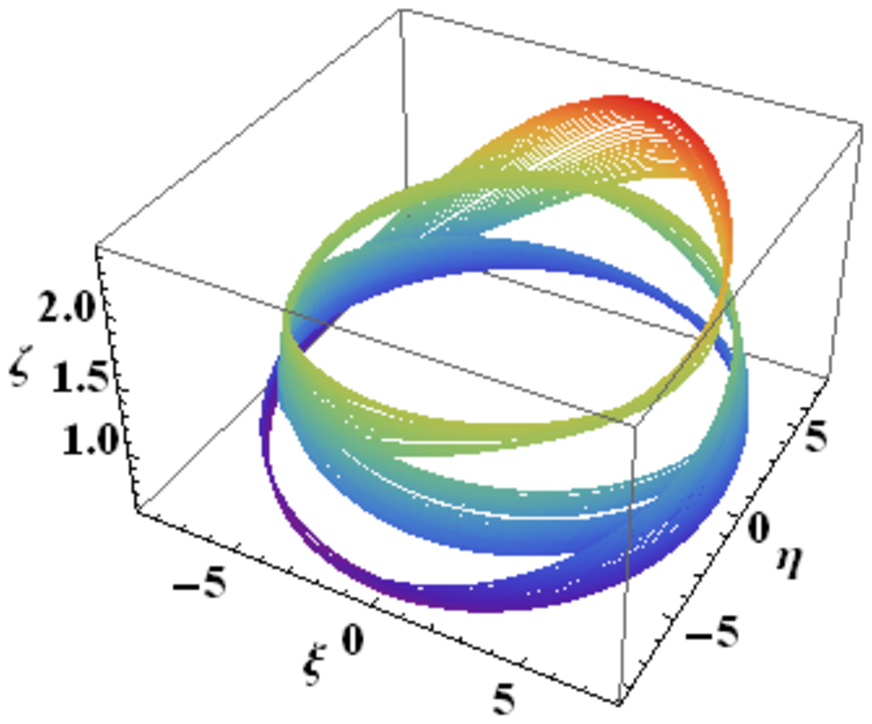, width=7.0cm} \vspace{1mm} \\
   (a) non-periodic & (b) quasi-periodic \\
  \end{tabular}
  \caption{The two trajectories of the forced $4$-dimensional memristor ecological predator-prey equations (\ref{eqn: ecological-7}), 
   which are projected into the $( \xi, \, \eta, \, \zeta )$-space via the coordinate transformation (\ref{eqn: Ecological-projection}).  
   We can observe a gap in Figure \ref{fig:Ecological-torus}(b).  
   Compare the trajectories in Figure \ref{fig:Ecological-torus} with the Poincar\'e maps in Figure \ref{fig:Ecological-poincare}.  
   The trajectories are colored with the \emph{Rainbow} color code in Mathematica.
   Parameters:  $\ r = 0.1,  \ \omega = 2$.
   Initial conditions: 
   (a) $i_{1}(0) = 0.1, \, x_{1}(0)= 0.5, \, i_{2}(0) =1.1, \, x_{2}(0) =1.3$. \  \
   (b) $i_{1}(0) = 0.2, \, x_{1}(0)= 0.5, \, i_{2}(0) =1.1, \, x_{2}(0) =1.3$.}
  \label{fig:Ecological-torus} 
\end{figure}
%
%

%\clearpage
%---Fig. 76-------%
\begin{figure}[hpbt]
 \centering
   \begin{tabular}{cc}
    \psfig{file=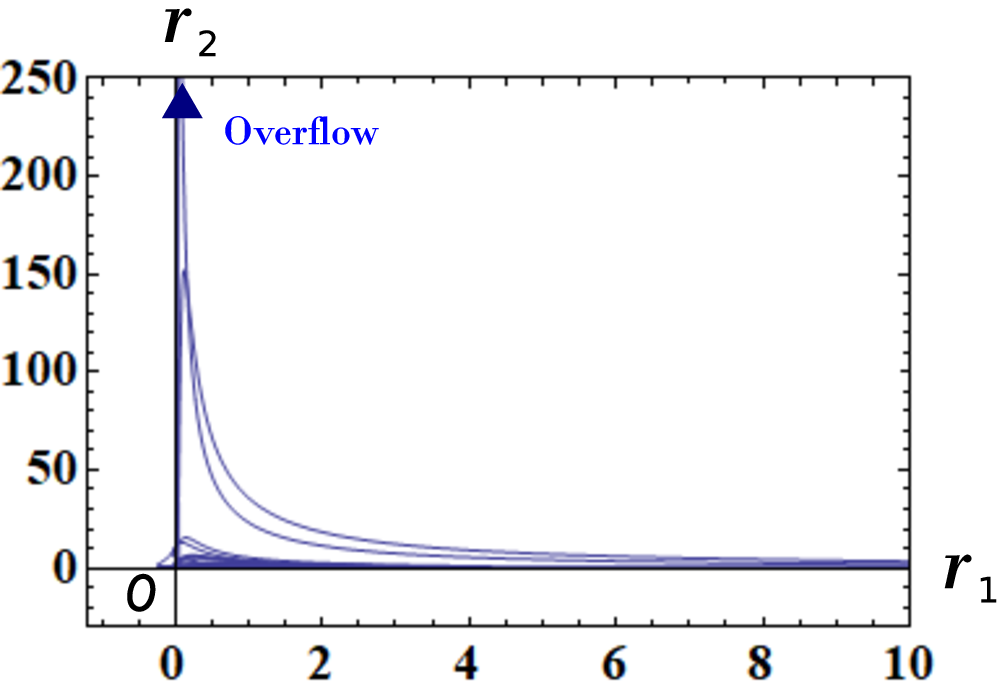, height=5cm}  & 
    \psfig{file=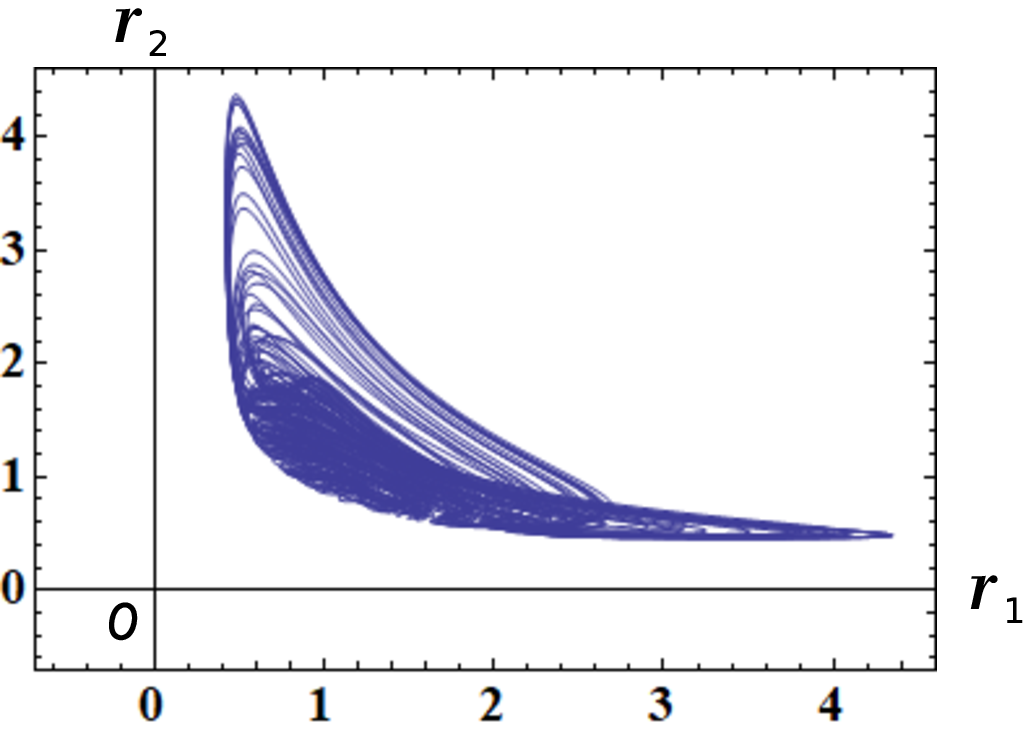, height=5cm}  \vspace{1mm} \\
   (a) $h=0.005$ & (b) $h=0.002$  \\ 
   \end{tabular}
  \caption{Behavior of the forced $4$-dimensional memristor ecological predator-prey model equations (\ref{eqn: ecological-7}). 
   If we choose $h=0.01$, then the trajectory rapidly grows for $t \ge 5327$, 
   and an overflow occurs as shown in Figure \ref{fig:Ecological-trajectory}(a). 
   However, if we choose $h=0.005$, then the trajectory stays in a finite region as shown in Figure \ref{fig:Ecological-trajectory}(b).
   Here, $h$ denotes the maximum step size of the numerical integration.  
   Parameters:  $r = 0.5, \ \  \omega =2$.  
   Initial conditions: $i_{1}(0) = 0.098, \, x_{1}(0) = 0.5, \, i_{2}(0)=1.1,  \, x_{2}(0)=1.3$.}
  \label{fig:Ecological-trajectory} 
\end{figure}
%
%

%\clearpage
%
%
%%%%%%%%%%%%%%%%%%%%%%%%%%%%%%%%%%%%%%%%%%%%%%%%%%%%%%%%%%%%%%%%%%%%%%%
\section{Exponential Coordinate Transformation}
%%%%%%%%%%%%%%%%%%%%%%%%%%%%%%%%%%%%%%%%%%%%%%%%%%%%%%%%%%%%%%%%%%%%%%%
%
%
In this section, we show that the dynamics of an n-dimensional autonomous system can be transformed into the dynamics of a three-element memristor circuit by using the exponential coordinate transformation \cite{Itoh(2011), Itoh(2014)}.

%-------------------------------------%
\subsection{Tennis racket equations}
\label{sec: tennis racket}
%-------------------------------------%

The components of the angular momentum of a tennis racket about its
center of mass are governed by the following equations \cite{Toda(1995)}:
\begin{center}
\begin{minipage}{8.7cm}
\begin{shadebox}
\underline{\emph{Tennis racket equations}}
\begin{equation}
\left. 
 \begin{array}{ccc}
  \displaystyle \frac{ d\omega_{1} }{dt} &=& - \omega_{2} \, \omega_{3}, \vspace{2mm} \\
  \displaystyle \frac{ d\omega_{2} }{dt} &=&   \omega_{3} \, \omega_{1}, \vspace{2mm} \\
  \displaystyle \frac{ d\omega_{3} }{dt} &=& - \omega_{1} \, \omega_{2},
 \end{array}
\right \}
\label{eqn: tennis-racket-1}
\end{equation}
where $\omega _{1}$, $\omega _{2}$, $\omega _{3}$ are the angular velocities about the object's three principal axes. 
\end{shadebox}
\end{minipage}
\end{center}
Equation (\ref{eqn: tennis-racket-1}) has the two integrals, since the solution satisfies 
\begin{center}
\begin{minipage}{.9\textwidth}
\begin{itembox}[l]{Integrals}
\begin{equation}
\left. 
 \begin{array}{c}
   \displaystyle \frac{d}{dt} \left ( {\omega_{1}}^{2} + {\omega_{2}}^{2} \right ) 
   = 2\omega_{1} \left( \frac{d\omega_{1}}{dt} \right ) +  2\omega_{2} \left ( \frac{d\omega_{2}}{dt} \right )
   =  2\omega_{1} (- \omega_{2} \, \omega_{3}) +  2\omega_{2}(\omega_{3} \, \omega_{1}) =  0, \vspace{2mm} \\
   \displaystyle \frac{d}{dt} \left ( {\omega_{2}}^{2} + {\omega_{3}}^{2} \right ) 
   = 2\omega_{2} \left ( \frac{d\omega_{2}}{dt} \right ) +  2\omega_{3} \left ( \frac{d\omega_{3}}{dt} \right )
   =  2\omega_{2} (\omega_{3} \, \omega_{1}) +  2\omega_{3}(- \omega_{1} \, \omega_{2}) = 0. \vspace{2mm} \\  
 \end{array}
\right \}
\end{equation}  
\end{itembox}
\end{minipage}
\end{center}

Substituting 
\begin{equation}
  \omega _{1} = \ln {|\, i \,|}, \ \omega_{2} = x_{1}, \ \omega_{3} = x_{2},  
\label{eqn: ln-1}
\end{equation}
into Eq. (\ref{eqn: tennis-racket-1}), 
we obtain 
\begin{center}
\begin{minipage}{8.7cm}
\begin{shadebox}
\underline{\emph{Memristor tennis racket equations}}
\begin{equation}
\left. 
\begin{array}{ccc}
  \displaystyle \frac{ di }{dt} &=& - x_{1} \, x_{2} \, i, \vspace{2mm} \\
  \displaystyle \frac{ dx_{1} }{dt} &=&  x_{2} \, \ln {|\, i \,|}, \vspace{2mm} \\
  \displaystyle \frac{ dx_{2} }{dt} &=& - x_{1} \, \ln {|\, i \,|}.
\end{array}
\right \}
\label{eqn: tennis-racket-2}
\end{equation}
\end{shadebox}
\end{minipage}
\end{center}
Since $|\,i_{1} \,| = e^{x_{1}}$, Eq. (\ref{eqn: ln-1}) indicates an exponential coordinate transformation \cite{Itoh(2014)}. 

Consider the three-element memristor circuit in Figure \ref{fig:memristor-inductor-battery}.  
The dynamics of this circuit given by Eq. (\ref{eqn: dynamics-1}).  
Assume that  Eq. (\ref{eqn: dynamics-1}) satisfies 
\begin{equation}
\left.
\begin{array}{ccc}
  L =1, && E=0,   \vspace{2mm} \\
  \hat{R}( x_{1}, \, x_{2},  \,  i_{M} ) &=&  x_{1} \, x_{2}, \vspace{2mm} \\
     \displaystyle  \tilde{f}_{1}(x_{1}, \ x_{2}, \ i) &=&  x_{2} \, \ln {|\, i_{M} \,|},
      \vspace{2mm} \\
     \displaystyle  \tilde{f}_{2}(x_{1}, \ x_{2}, \ i) &=& - x_{1} \, \ln {|\, i_{M} \,|}.     
\end{array}
\right \}
\end{equation}
Then Eq. (\ref{eqn: dynamics-1}) can be recast into Eq. (\ref{eqn: tennis-racket-2}).  
In this case, the extended memristor in Figure \ref{fig:memristor-inductor-battery} can be replaced by the generic memristor.  
That is, 
\begin{equation}
  \hat{R}( x_{1}, \, x_{2},  \,  i_{M} ) = \tilde{R}( x_{1}, \, x_{2} ) =  x_{1} \, x_{2}.
\end{equation}
The terminal voltage $v_{M}$ and the terminal current $i_{M}$ of the current-controlled generic memristor are described by
\begin{center}
\begin{minipage}{8.7cm}
\begin{shadebox}
\underline{\emph{V-I characteristics of the generic memristor}}
\begin{equation}
\begin{array}{lll}
  v_{M} &=& \tilde{R}( x_{1}, \, x_{2}) \, i_{M} =  x_{1} \, x_{2} \, i_{M},  \vspace{3mm} \\
     \displaystyle \frac{d x_{1}}{dt} &=&  x_{2} \, \ln {|\, i_{M} \,|},
      \vspace{2mm} \\
     \displaystyle \frac{d x_{2}}{dt} &=& - x_{1} \, \ln {|\, i_{M} \,|}. 
\end{array}
\label{eqn: tennis-racket-3}
\end{equation}
where $\hat{R}( x_{1}, \, x_{2} ) =  x_{1} \, x_{2}$ and $i_{M} = i$.  \vspace{2mm}
\end{shadebox}
\end{minipage}
\end{center}
Equation (\ref{eqn: tennis-racket-2}) has the two integrals, since the solution satisfies \vspace{2mm}
\begin{center}
\begin{minipage}{\textwidth}
\begin{itembox}[l]{Integrals}
\begin{equation}
\left. 
 \begin{array}{lll}
   \displaystyle \frac{d}{dt} \left ( {\ln {|\, i \,|}}^{2} + {x_{1}}^{2} \right ) &=& 
   \displaystyle 2 \, {\ln {|\, i \,|}} \left ( \frac{ d \ln {|\, i \,|}}{dt} \right ) + 2 {x_{1}} \left ( \frac{d x_{1}}{dt} \right )
   = 2 \, {\ln {|\, i \,|}} \left ( \frac{\frac{di}{dt}}{i} \right ) +  2 {x_{1}} \, ( x_{2} \, \ln {|\, i \,|} )  \\
   &=& 2 \, {\ln {|\, i \,|}} \, (- x_{1} \, x_{2}) +  2 {x_{1}} x_{2} \, \ln {|\, i \,|} = 0,
    \vspace{2mm} \\
   \displaystyle \frac{d}{dt} \left ( {x_{1}}^{2} + {x_{2}}^{2} \right ) 
   &=& \displaystyle 2{x_{1}} \left ( \frac{dx_{1}}{dt} \right ) + 2{x_{2}} \left ( \frac{dx_{2}}{dt} \right ) 
       =   2{x_{1}} \, \bigl ( x_{2} \, \ln {|\, i \,|} \bigr )  +  2{x_{2}} \, \bigl ( - x_{1} \, \ln {|\, i \,|} \bigr ) = 0,
 \end{array}
\right \}
\end{equation}  
where $i \ne 0$. 
\end{itembox}
\end{minipage}
\end{center}
It can exhibit periodic behavior.  
When an external source is added as shown in Figure \ref{fig:memristive-inductor-battery-source}, 
the forced memristor tennis racket equations can exhibit a non-periodic response.  
The dynamics of this circuit is given by 
\begin{center}
\begin{minipage}{8.9cm}
\begin{shadebox}
\underline{\emph{Forced memristor tennis racket equations}}
\begin{equation}
\left. 
\begin{array}{ccl}
  \displaystyle \frac{ di }{dt} &=& - x_{1} \, x_{2} \, i + r \sin ( \omega t),  \vspace{2mm} \\
  \displaystyle \frac{ dx_{1} }{dt} &=&  x_{2} \, \ln {|\, i \,|}, \vspace{2mm} \\
  \displaystyle \frac{ dx_{2} }{dt} &=& - x_{1} \, \ln {|\, i \,|}, 
\end{array}
\right \}
\label{eqn: tennis-racket-4}
\end{equation}
where $r$ and $\omega$ are constants.  
\end{shadebox}
\end{minipage}
\end{center}
The solution of Eq. (\ref{eqn: tennis-racket-4}) satisfies
\begin{equation}
  {x_{1}(t)}^{2} + {x_{2}(t)}^{2} = K,   
\label{eqn: tennis-circle}
\end{equation}
where $K$ is a constant. 
We show the non-periodic response, quasi-periodic response, Poincar\'e maps, and $i_{M}-v_{M}$ loci of Eq. (\ref{eqn: tennis-racket-4}) in Figures \ref{fig:Tennis-attractor}, \ref{fig:Tennis-attractor-2}, \ref{fig:Tennis-poincare}, and \ref{fig:Tennis-pinch}, respectively.  
The trajectories projected into the $(x_{1}, \, x_{2})$-plane are shown in Figure \ref{fig:Tennis-attractor}(b) and Figure \ref{fig:Tennis-attractor-2}(b), which moves on the circle defined by Eq. (\ref{eqn: tennis-circle}).  
Compare the two Poincar\'e maps in Figure \ref{fig:Tennis-poincare}.  
The $i_{M}-v_{M}$ loci in Figure \ref{fig:Tennis-pinch} lie in the first and the fourth quadrants. 
Thus, the generic memristor defined by Eq. (\ref{eqn: tennis-racket-3}) is an active element. 
We show the $v_{M}-p_{M}$ locus in Figure \ref{fig:Tennis-power}, 
where $p_{M}(t)$ is an instantaneous power defined by $p_{M}(t)=i_{M}(t)v_{M}(t)$.  
Observe that the $v_{M}-p_{M}$ locus is pinched at the origin, and the locus lies in the first and the third quadrants. 
Thus, the memristor switches between passive and active modes of operation, depending on its terminal voltage. 
We conclude as follow: \\
\begin{center}
\begin{minipage}{.7\textwidth}
\begin{itembox}[l]{Switching behavior of the memristor}
Assume that Eq. (\ref{eqn: tennis-racket-4}) exhibits non-periodic or quasi-periodic oscillation.  
Then the generic memristor defined by Eq. (\ref{eqn: tennis-racket-3}) can switch between ``passive'' and ``active'' modes of operation, depending on its terminal voltage.  
\end{itembox}
\end{minipage}
\end{center}

In order to view the above Poincar\'e maps from a different perspective, 
let us project the trajectory into the $( \xi, \, \eta, \, \zeta )$-space via the transformation 
\begin{equation}
 \begin{array}{lll}
   \xi (\tau)   &=& (i(\tau) + 5) \cos \,( \omega \tau ), \vspace{2mm} \\ 
   \eta (\tau)  &=& (i(\tau) + 5) \sin \,( \omega \tau ), \vspace{2mm} \\ 
   \zeta (\tau) &=& x_{1}(\tau).
 \end{array}  
\label{eqn: tennis-racket-projection}
\end{equation}
We show the projected trajectories in Figure \ref{fig:Tennis-torus}.   
Observe that the trajectory in Figure \ref{fig:Tennis-torus}(a) is less dense than Figure \ref{fig:Tennis-torus}(b).

%---Fig. 77-------%
\begin{figure}[hpbt]
 \centering
   \begin{tabular}{ccc}
    \psfig{file=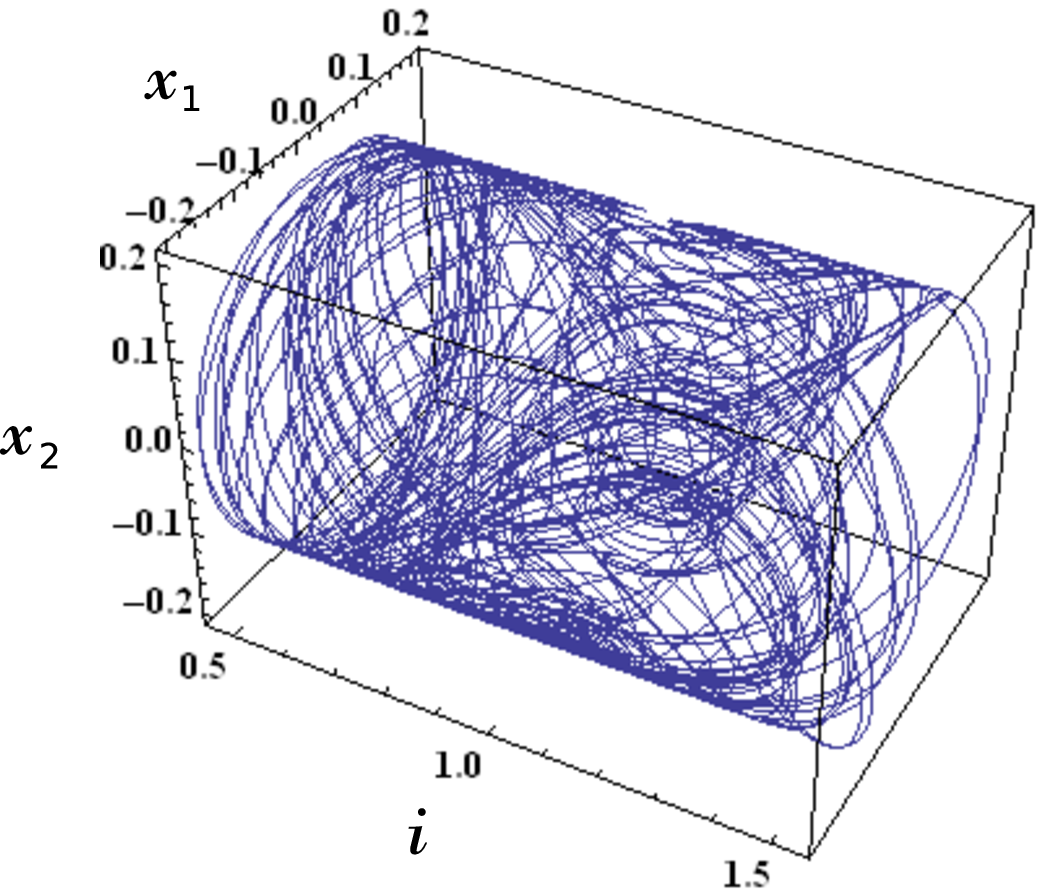, width=6.5cm}  & \hspace{5mm} &
    \psfig{file=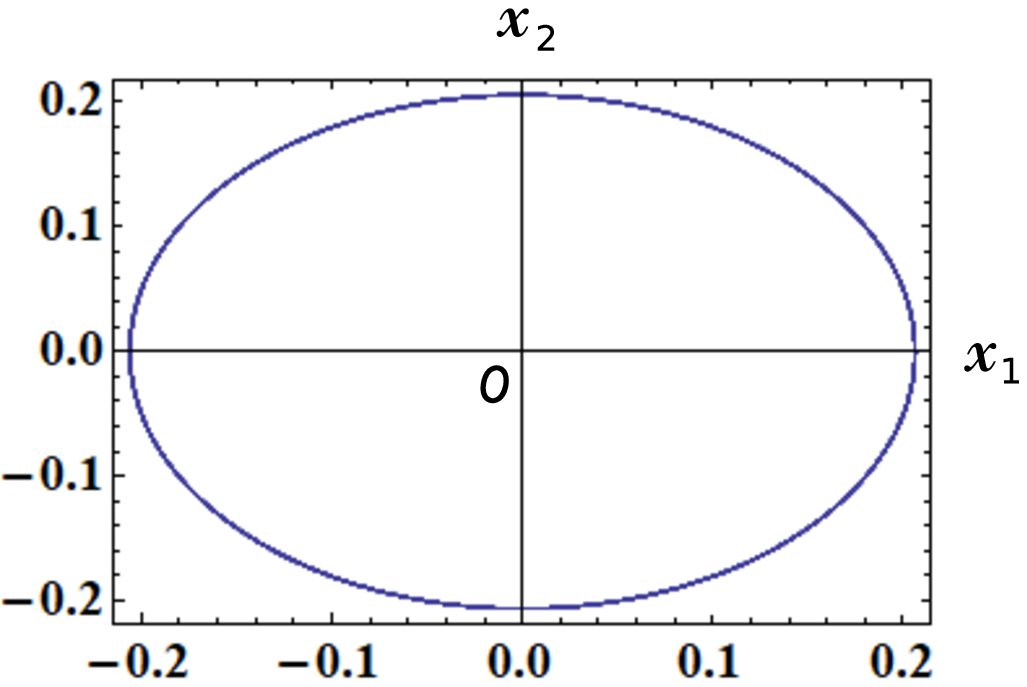, width=6cm}  \vspace{1mm} \\
   (a) $(i, \, x_{1}, \, x_{2})$-space & & (b) Projected into the $(x_{1}, \, x_{2})$-plane  \\ 
   \end{tabular}
  \caption{Non-periodic response of the forced memristor tennis racket equations (\ref{eqn: tennis-racket-4}). 
   (a) A non-periodic trajectory in the $(i, \, x_{1}, \, x_{2})$-space.  
   (b) A trajectory projected into the $(x_{1}, \, x_{2})$-plane.  
   It satisfies a circle equation: ${x_{1}(t)}^{2} + {x_{2}(t)}^{2} = 0.0425$.
   Parameters: $r = 0.087, \ \omega = 0.5$.  
   Initial conditions: $i(0)=e^{0.11}, \  x_{1}(0)=0.2, \ x_{2}(0)=0.05$.}
  \label{fig:Tennis-attractor} 
\end{figure}
%
%

%---Fig. 78-------%
\begin{figure}[hpbt]
 \centering
   \begin{tabular}{ccc}
    \psfig{file=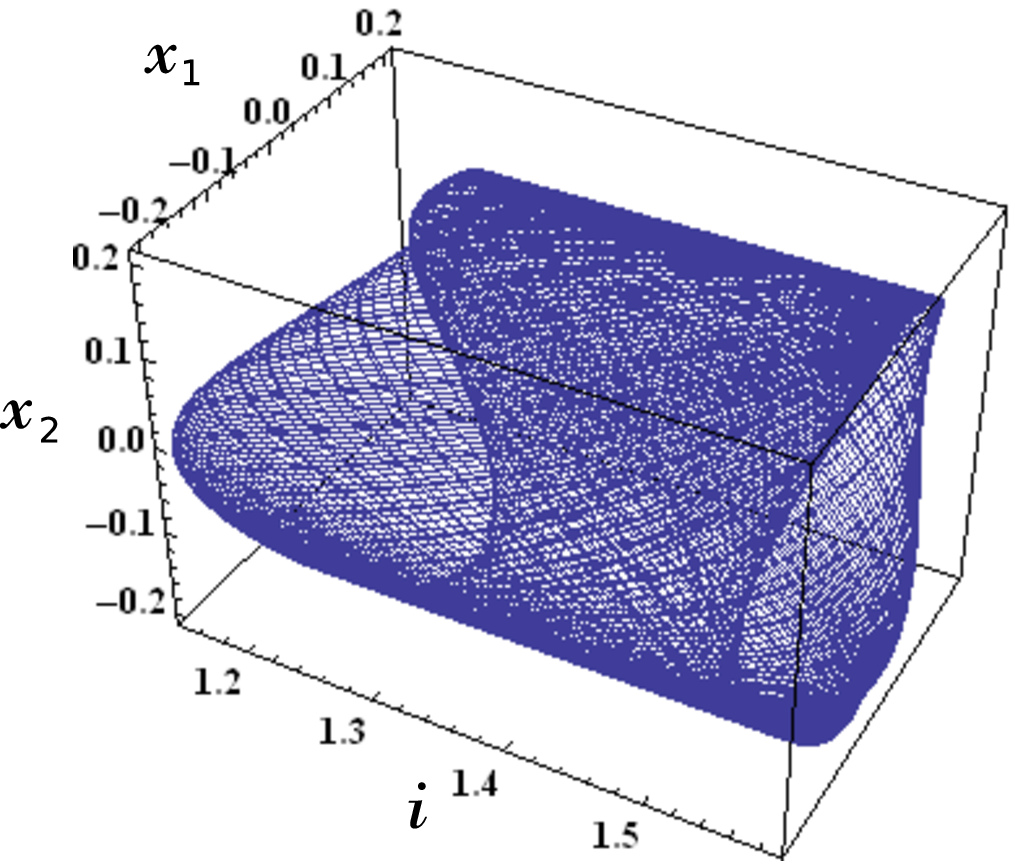, width=6.5cm}  & \hspace{5mm} &
    \psfig{file=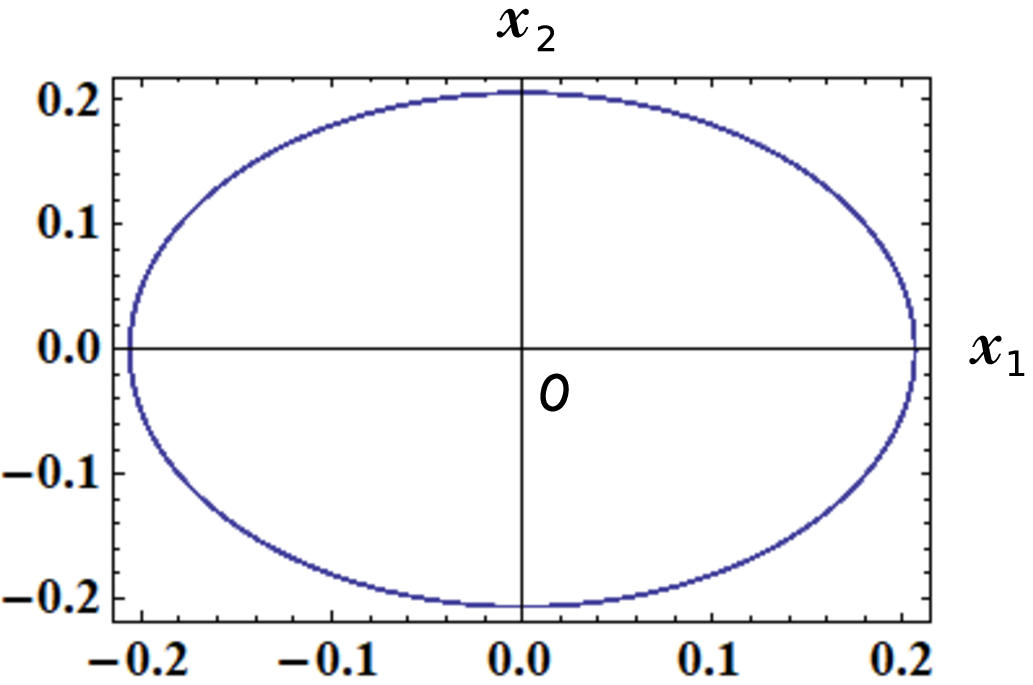, width=6cm}  \vspace{1mm} \\
   (a) $(i, \, x_{1}, \, x_{2})$-space & & (b) Projected into the $(x_{1}, \, x_{2})$-plane  \\ 
   \end{tabular}
  \caption{Quasi-periodic response of the forced memristor tennis racket equations (\ref{eqn: tennis-racket-4}). 
   (a) A quasi-periodic trajectory in the $(i, \, x_{1}, \, x_{2})$-space.  
   (b) A trajectory projected into the $(x_{1}, \, x_{2})$-plane.  
   It satisfies a circle equation: ${x_{1}(t)}^{2} + {x_{2}(t)}^{2} = 0.0425$.
   Parameters: $r = 0.087, \ \omega = 0.5$.  
   Initial conditions: $i(0)=e^{0.15}, \  x_{1}(0)=0.2, \ x_{2}(0)=0.05$.}
  \label{fig:Tennis-attractor-2} 
\end{figure}
%
%

%---Fig. 79-------%
\begin{figure}[hpbt]
 \centering
   \begin{tabular}{cc}
   \psfig{file=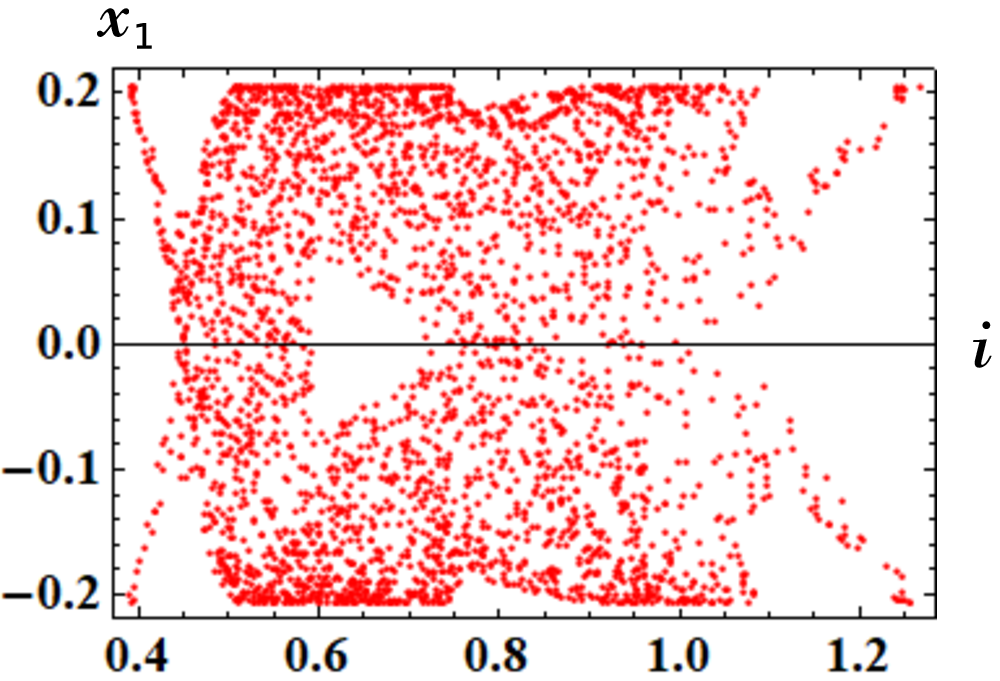, height=4.7cm} & 
   \psfig{file=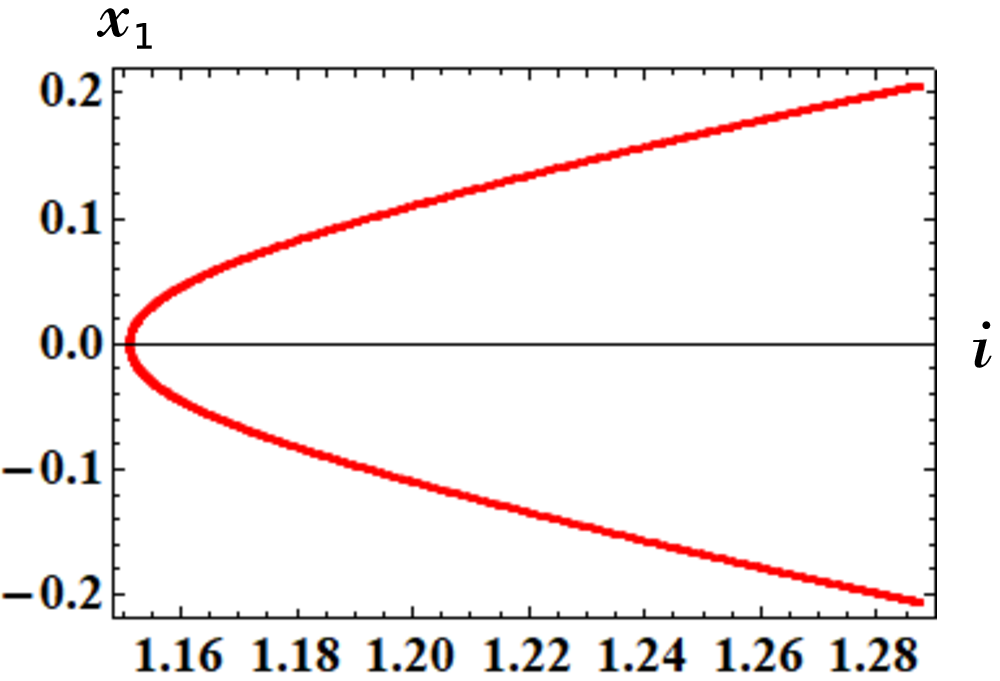, height=4.7cm} \vspace{1mm} \\
   (a) non-periodic & (b) quasi-periodic \\
  \end{tabular}
  \caption{Poincar\'e maps of the forced memristor tennis racket equations (\ref{eqn: tennis-racket-4}). 
   Parameters: $r = 0.087, \ \omega = 0.5$.  
   Initial conditions: (a) $i(0)=e^{0.11}, \  x_{1}(0)=0.2, \ x_{2}(0)=0.05$. \ \  
   (b) $i(0)=e^{0.15}, \  x_{1}(0)=0.2, \ x_{2}(0)=0.05$. }
  \label{fig:Tennis-poincare} 
\end{figure}
%
%

%---Fig. 80-------%
\begin{figure}[hpbt]
 \centering
   \begin{tabular}{cc}
   \psfig{file=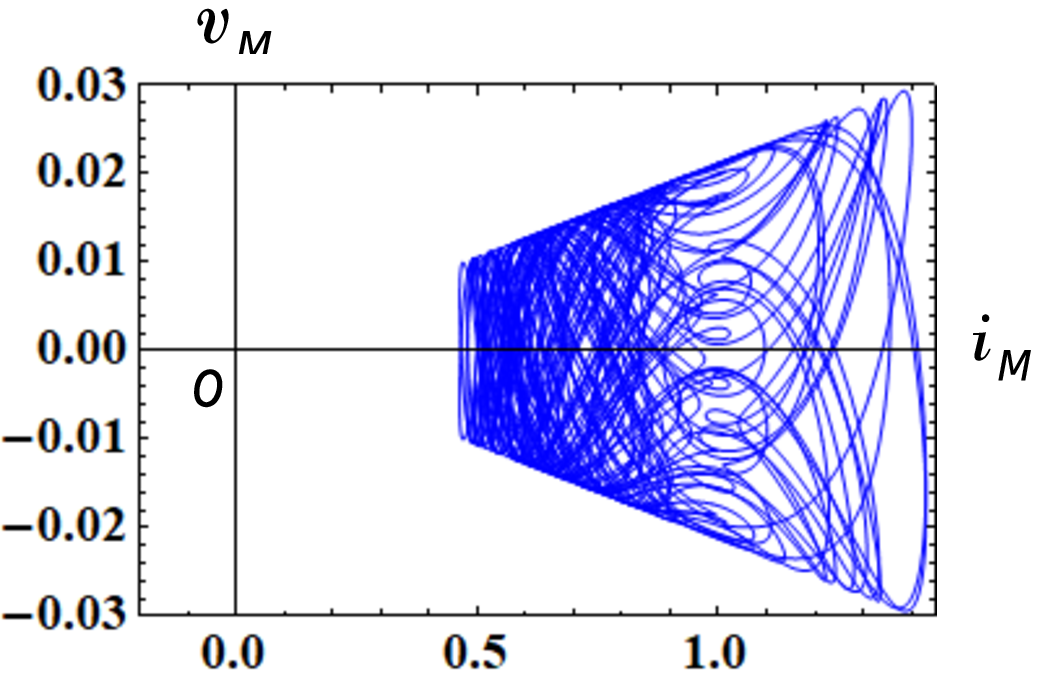, height=4.7cm} & 
   \psfig{file=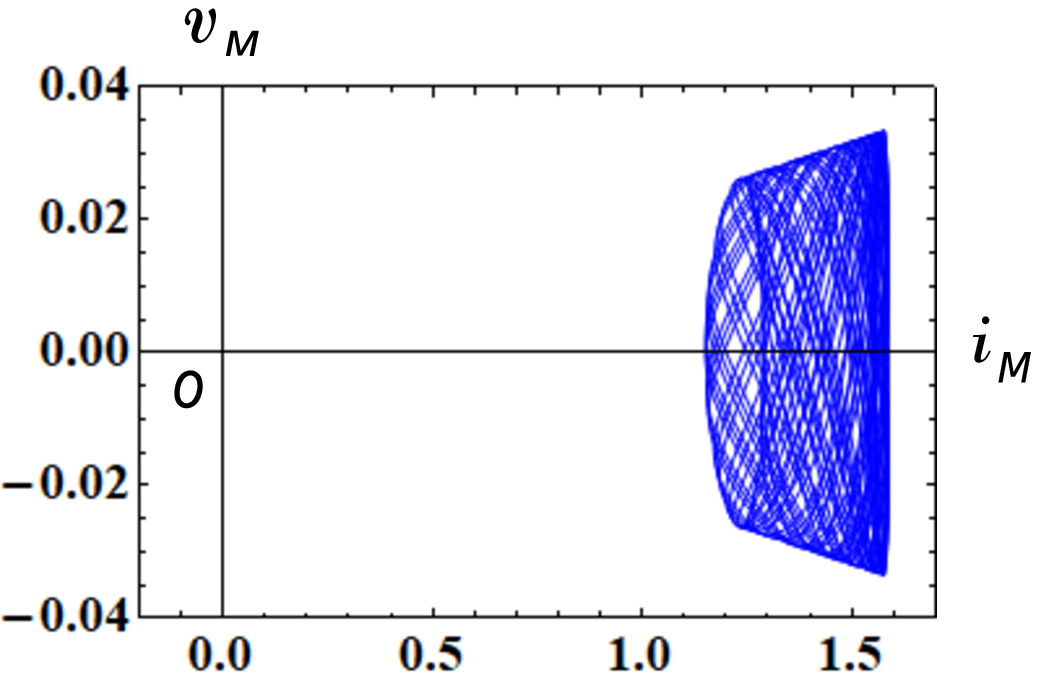,height=4.7cm} \vspace{1mm} \\
   (a) non-periodic & (b) quasi-periodic \\
  \end{tabular}
  \caption{The $i_{M}-v_{M}$ loci of the forced memristor tennis racket equations (\ref{eqn: tennis-racket-4}).  
   Here, $v_{M}$ and  $i_{M}$ denote the terminal voltage and the terminal current of the current-controlled generic memristor.  
   Parameters: $r = 0.087, \ \omega = 0.5$.  
   Initial conditions: (a) $i(0)=e^{0.11}, \  x_{1}(0)=0.2, \ x_{2}(0)=0.05$. \ \  
   (b) $i(0)=e^{0.15}, \  x_{1}(0)=0.2, \ x_{2}(0)=0.05$. }
  \label{fig:Tennis-pinch} 
\end{figure}
%
%

%---Fig. 81-------%
\begin{figure}[hpbt]
 \centering
   \begin{tabular}{cc}
    \psfig{file=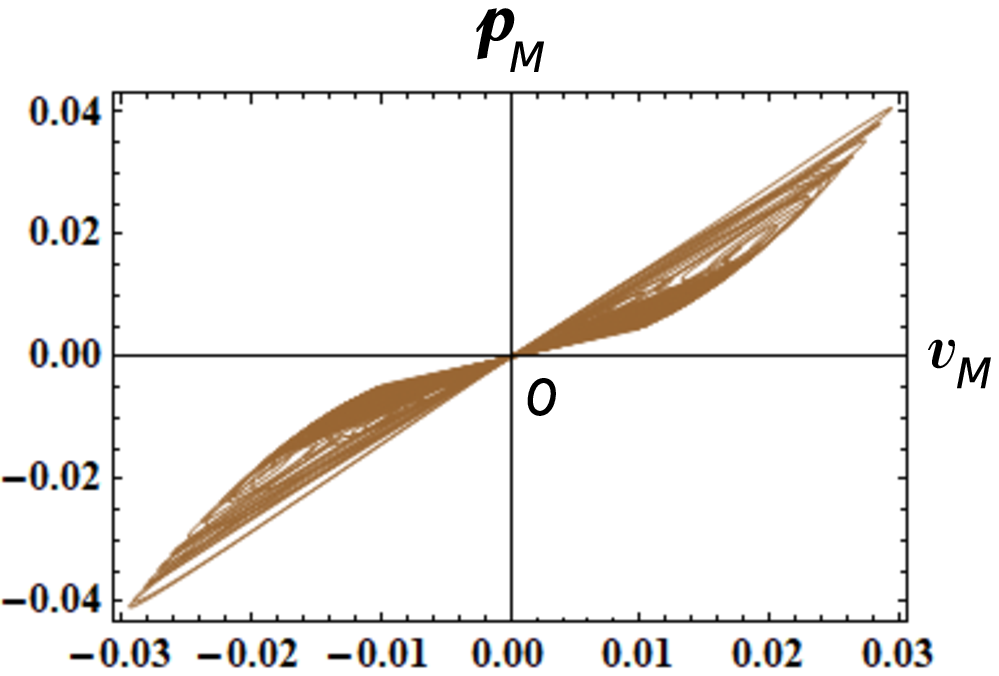, width=7cm}  & 
    \psfig{file=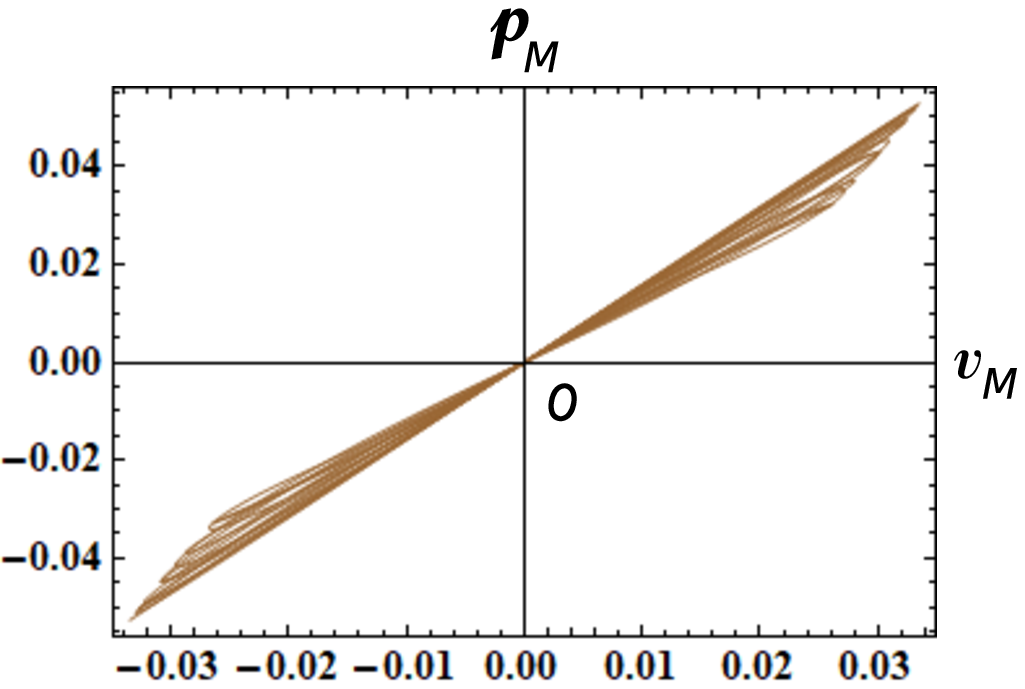, width=7cm}  \\
    (a) non-periodic & (b) quasi-periodic \\
   \end{tabular}
  \caption{ The $v_{M}-p_{M}$ loci of the forced memristor tennis racket equations (\ref{eqn: tennis-racket-4}). 
   Here, $p_{M}(t)$ is an instantaneous power defined by $p_{M}(t)=i_{M}(t)v_{M}(t)$,   
   and $v_{M}(t)$ and $i_{M}(t)$ denote the terminal voltage and the terminal current of the current-controlled generic memristor.  
   Observe that the $v_{M}-p_{M}$ locus is pinched at the origin, and the locus lies in the first and the third quadrants. 
   The memristor switches between passive and active modes of operation, depending on its terminal voltage $v_{M}(t)$.
   Parameters: $r = 0.087, \ \omega = 0.5$. 
   Initial conditions: 
   (a) $i(0)=e^{0.11}, \  x_{1}(0)=0.2, \ x_{2}(0)=0.05$. \ \  
   (b) $i(0)=e^{0.15}, \  x_{1}(0)=0.2, \ x_{2}(0)=0.05$.}
  \label{fig:Tennis-power} 
\end{figure}
%
%

%---Fig. 82-------%
\begin{figure}[hpbt]
 \centering
   \begin{tabular}{cc}
   \psfig{file=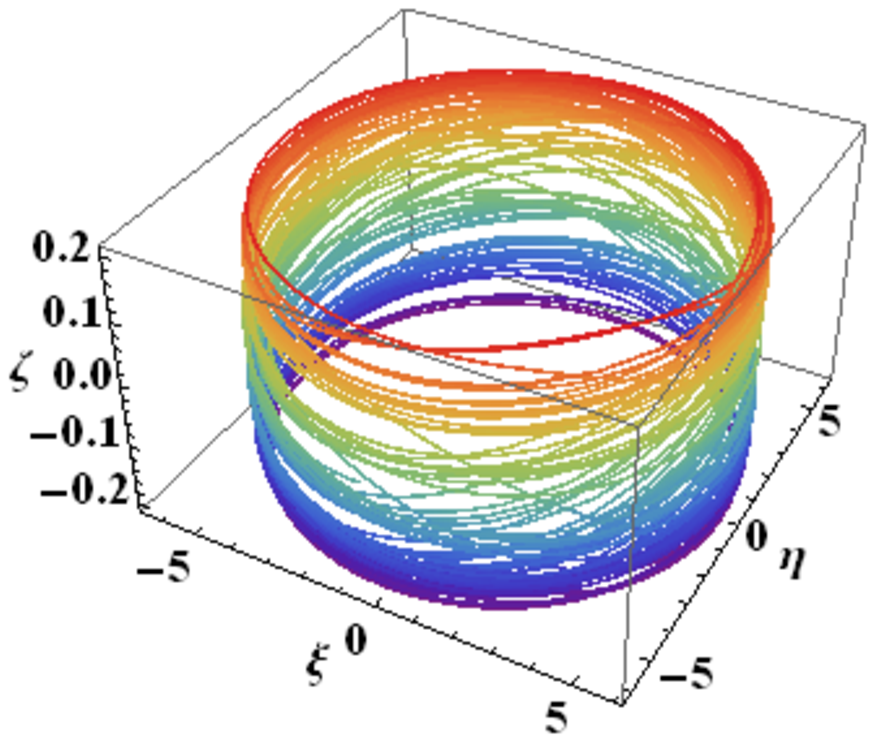, height=4.7cm} & 
   \psfig{file=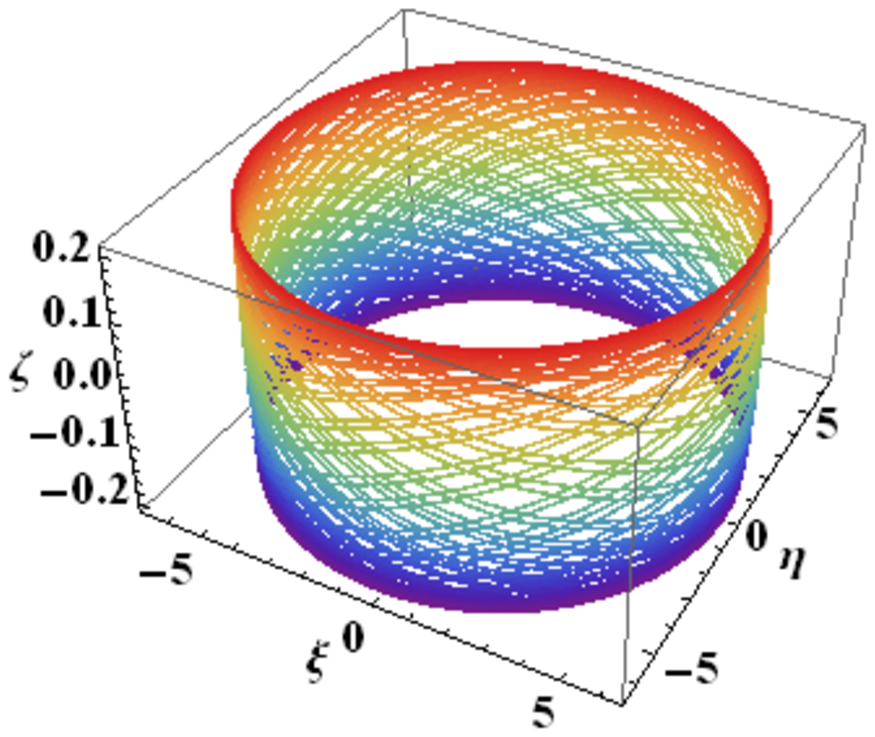, height=4.7cm} \vspace{1mm} \\
   (a) non-periodic & (b) quasi-periodic \\
  \end{tabular}
  \caption{The two trajectories of the forced memristor tennis racket equations (\ref{eqn: tennis-racket-4}, 
   which are projected into the $( \xi, \, \eta, \, \zeta )$-space via the coordinate transformation (\ref{eqn: tennis-racket-projection}).  
   Observe that the trajectory in Figure \ref{fig:Tennis-torus}(a) is less dense than Figure \ref{fig:Tennis-torus}(b). 
   Compare the trajectories in Figure \ref{fig:Tennis-torus}  with the Poincar\'e maps in Figure \ref{fig:Tennis-poincare}.  
   The trajectories are colored with the \emph{Rainbow} color code in Mathematica.
   Parameters: $r = 0.087, \ \omega = 0.5$. 
   Initial conditions: 
   (a) $i(0)=e^{0.11}, \  x_{1}(0)=0.2, \ x_{2}(0)=0.05$. \ \  
   (b) $i(0)=e^{0.15}, \  x_{1}(0)=0.2, \ x_{2}(0)=0.05$.}
  \label{fig:Tennis-torus} 
\end{figure}
%
%

%
%==============================================================%
\subsection{Pendulum equations}
%==============================================================%
%

The equation for a pendulum can be written as \cite{Andronov}
\begin{equation}
 \displaystyle \frac{d^{2} u}{dt^{2}} + \sin u = 0, 
\label{eqn: pendulum-1}
\end{equation}
where $u$ is the angle from the downward vertical.  
It is equivalent to the sine-Gordon equation in the absence of the diffusion term.  
Equation (\ref{eqn: pendulum-1}) can be recast into the form 
\begin{center}
\begin{minipage}{8.7cm}
\begin{shadebox}
\underline{\emph{Pendulum equations}}
\begin{equation}
\begin{array}{lll}
 \displaystyle \frac{d u}{dt} &=& v,   \vspace{2mm} \\
 \displaystyle \frac{d v}{dt} &=& - \sin u.  
 \end{array}
\label{eqn: pendulum-2}
\end{equation}
\end{shadebox}
\end{minipage}
\end{center}
Substituting $ v = \ln {|\, i \,|}$ and $u=x$ into Eq. (\ref{eqn: pendulum-1}), we obtain 
\begin{center}
\begin{minipage}{8.7cm}
\begin{shadebox}
\underline{\emph{Memristor pendulum equations}}
\begin{equation}
\begin{array}{lll}
 \displaystyle \frac{d i}{dt} &=& - i \sin x,  \vspace{2mm} \\
 \displaystyle \frac{d x}{dt} &=& \, \ln {|\, i \,|}.  
 \end{array}
\label{eqn: pendulum-3}
\end{equation}
\end{shadebox}
\end{minipage}
\end{center}

Consider the three-element memristor circuit in Figure \ref{fig:memristor-inductor-battery}.  
The dynamics of this circuit given by Eq. (\ref{eqn: dynamics-n-1}).  
Assume that Eq. (\ref{eqn: dynamics-n-1}) satisfies 
\begin{equation}
\left.
\begin{array}{ccc}
  E &=& 0,   \vspace{2mm} \\
  \hat{R}( x,  \,  i_{M} ) &=&  \sin x, \vspace{2mm} \\
  f_{1}(x, \, i) &=& \, \ln {|\, i \,|}.
\end{array}
\right \}
\end{equation}
Then, Eq. (\ref{eqn: dynamics-n-1}) can be recast into Eq. (\ref{eqn: pendulum-3}).  
In this case, the extended memristor in Figure \ref{fig:memristor-inductor-battery} can be replaced by by the generic memristor. 
Thus, 
\begin{equation}
  \hat{R}( x,  \,  i_{M} ) = \tilde{R}(x)= \sin x.
\end{equation}
The terminal voltage $v_{M}$ and the terminal current $i_{M}$ of the memristor are described by
\begin{center}
\begin{minipage}{8.7cm}
\begin{shadebox}
\underline{\emph{V-I characteristics of the generic memristor}}
\begin{equation}
\begin{array}{lll}
  v_{M} &=& \tilde{R}( x ) \, i_{M} =  ( \sin x ) \, i_{M},  \vspace{3mm} \\
     \displaystyle \frac{d x}{dt} &=& \, \ln {|\, i_{M} \,|},
\end{array}
\label{eqn: tennis pendulum-4}
\end{equation}
where $\hat{R}( x ) =  \sin x $ and $i_{M} = i$. 
\end{shadebox}
\end{minipage}
\end{center}
Equation (\ref{eqn: pendulum-3}) has the integral, since the solution satisfies 
\begin{center}
\begin{minipage}{.5\textwidth}
\begin{itembox}[l]{Integral}
\begin{equation}
  \frac{dt}{dt} \left \{ \frac{ { ( \, \ln {|\, i \,|}} \, )^{2} }{2}  +  \cos x  \right \} = 0.   
\end{equation}
\end{itembox}
\end{minipage}
\end{center}

The memristor pendulum equations (\ref{eqn: pendulum-3}) exhibit periodic behavior.  
If an external source is added as shown in Figure \ref{fig:memristive-inductor-battery-source}, then the forced memristor pendulum equations can exhibit non-periodic and quasi-periodic responses.  
The dynamics of this circuit is given by 
\begin{center}
\begin{minipage}{8.9cm}
\begin{shadebox}
\underline{\emph{Forced memristor pendulum equations}}
\begin{equation}
\begin{array}{lll}
 \displaystyle \frac{d i}{dt} &=& - i \sin x + r \sin ( \omega t),  \vspace{2mm} \\
 \displaystyle \frac{d x}{dt} &=& \, \ln {|\, i \,|},   
 \end{array}
\label{eqn: pendulum-5}
\end{equation}
where $r$ and $\omega$ are constants.  
\end{shadebox}
\end{minipage}
\end{center}
Equation (\ref{eqn: pendulum-5}) is invariant under the transformation $x \rightarrow x + 2 \pi$. 
We show their trajectories, Poincar\'e maps, and $i_{M}-v_{M}$ loci in Figures \ref{fig:Pendulum-attractor}, \ref{fig:Pendulum-poincare}, and \ref{fig:Pendulum-pinch}, respectively.  
The following parameters are used in our computer simulations:
\begin{equation}
 r = 0.04452, \ \omega = 1. 
\end{equation}

The $i_{M}-v_{M}$ loci in Figure \ref{fig:Pendulum-pinch} lie in the first and the fourth quadrant.  
Thus, the generic memristor defined by Eq. (\ref{eqn: tennis pendulum-4}) is an active element. 
We next show the $v_{M}-p_{M}$ locus in Figure \ref{fig:Pendulum-power}, 
where $p_{M}(t)$ is an instantaneous power defined by $p_{M}(t)=i_{M}(t)v_{M}(t)$.  
Observe that the $v_{M}-p_{M}$ locus is pinched at the origin, and the locus lies in the first and the third quadrants. 
Thus, the memristor switches between passive and active modes of operation, depending on its terminal voltage. 
We conclude as follow: \\
\begin{center}
\begin{minipage}{.7\textwidth}
\begin{itembox}[l]{Switching behavior of the memristor}
Assume that Eq. (\ref{eqn: pendulum-5}) exhibits non-periodic or quasi-periodic oscillation oscillation.  
Then the generic memristor defined by Eq. (\ref{eqn: tennis pendulum-4}) can switch between ``passive'' and ``active'' modes of operation, depending on its terminal voltage.  
\end{itembox}
\end{minipage}
\end{center}
%
%

%---Fig. 83-------%
\begin{figure}[hpbt]
 \centering
   \begin{tabular}{cc}
   \psfig{file=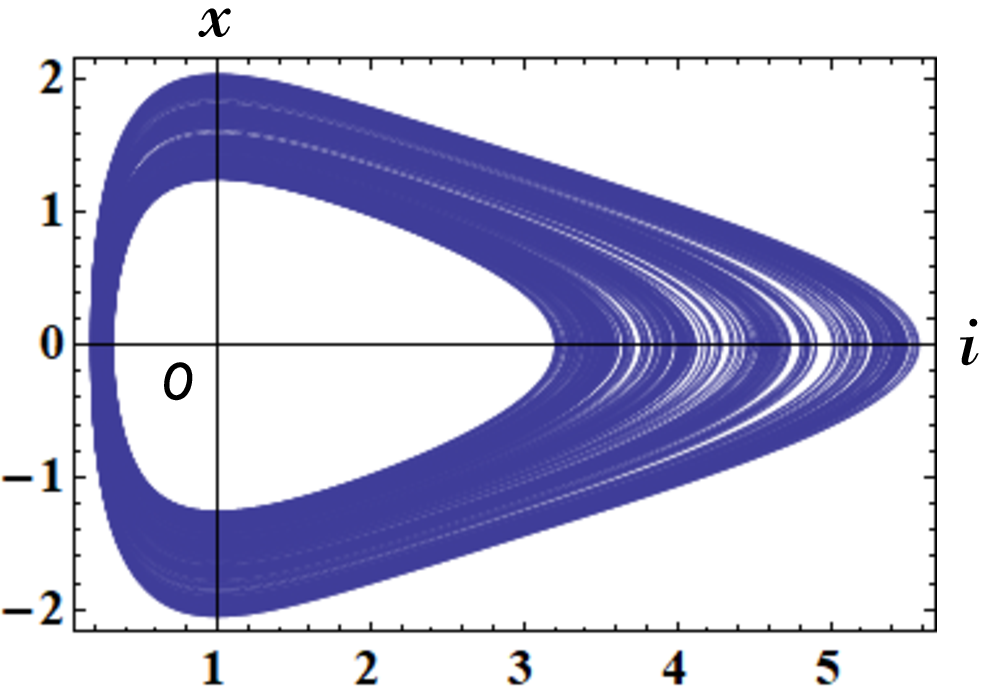, height=4.5cm} & 
   \psfig{file=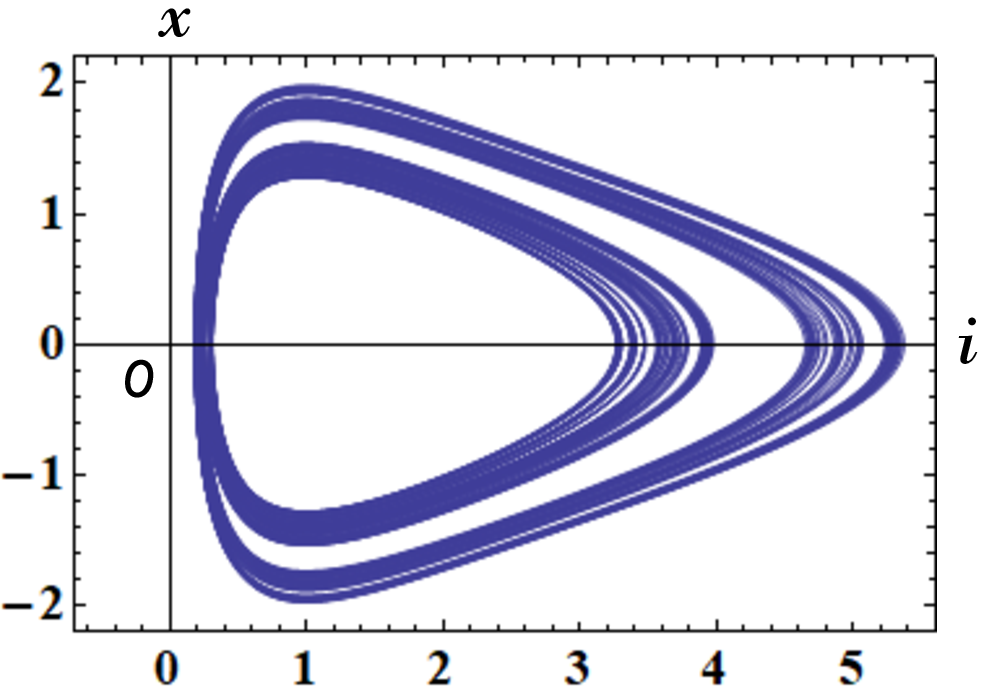, height=4.5cm} \\
   (a) non-periodic & (b) quasi-periodic 
  \end{tabular}
  \caption{Non-periodic and quasi-periodic responses of the forced memristor pendulum equations (\ref{eqn: pendulum-5}). 
   Parameters: $r = 0.04452, \ \omega = 1$.  
   Initial conditions: (a) $i(0)=0.187, \  x(0)=0.21$.  \ \ (b) $i(0)=0.187, \  x(0)=0.2$.}
  \label{fig:Pendulum-attractor} 
\end{figure}

In order to view the above Poincar\'e maps in Figure \ref{fig:Pendulum-poincare} from a different perspective, 
let us project the trajectory into the $( \xi, \, \eta, \, \zeta )$-space via the transformation 
\begin{equation}
 \begin{array}{lll}
   \xi (\tau)   &=& (i(\tau) + 5) \cos \,( \omega \tau ), \vspace{2mm} \\ 
   \eta (\tau)  &=& (i(\tau) + 5) \sin \,( \omega \tau ), \vspace{2mm} \\ 
   \zeta (\tau) &=& x_{1}(\tau).
 \end{array}  
\label{eqn: pendulum-projection}
\end{equation}
Compare the trajectories in Figure \ref{fig:Pendulum-torus} with the Poincar\'e maps in Figure \ref{fig:Pendulum-poincare}.   
We can observe a wide gap in Figure \ref{fig:Pendulum-torus}(b). 

Note that in order to obtain the Poincar\'e map in Figure \ref{fig:Pendulum-poincare}(a), we have to choose the parameters and initial conditions carefully.  
Furthermore, the maximum step size $h$ of the numerical integration must be sufficiently small ($h=0.003$) because of the numerical instability in long-time simulations.  
We show an interesting example in Figure \ref{fig:Pendulum-orbit}.  
Suppose that Eq. (\ref{eqn: pendulum-5}) has the following parameters and initial conditions:  
\begin{equation}
 \begin{array}{l}
  \text{Parameters: }  r = 0.0445, \ \  \omega = 1, \vspace{2mm} \\ 
  \text{Initial conditions: } i(0)=0.187, \  x(0)=0.187.
 \end{array}  
\end{equation}
If we choose $h=0.0005$, then $x(t)$ decreases gradually as time $t$ increases as shown in Figure \ref{fig:Pendulum-orbit}(a). 
However, if we choose $h=0.0003$, then the trajectory stays in a finite region of the $(i, \ x)$-plane as shown in Figure \ref{fig:Pendulum-orbit}(b).
The maximum step size of the numerical integration greatly affects the behavior of Eq. (\ref{eqn: pendulum-5}).

%---Fig. 84-------%
\begin{figure}[hpbt]
 \centering
   \begin{tabular}{c}
   \psfig{file=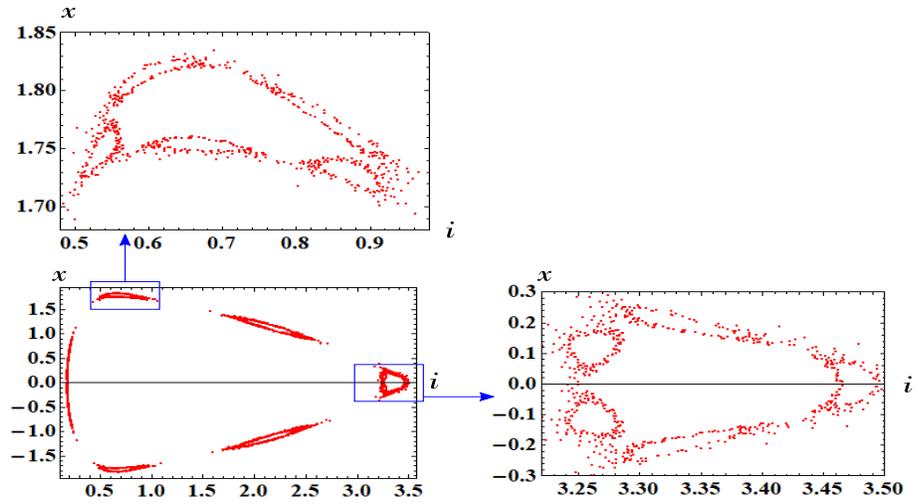, width=12cm, height=6.6cm} \\
   (a) non-periodic \vspace{1mm} \\
   \psfig{file=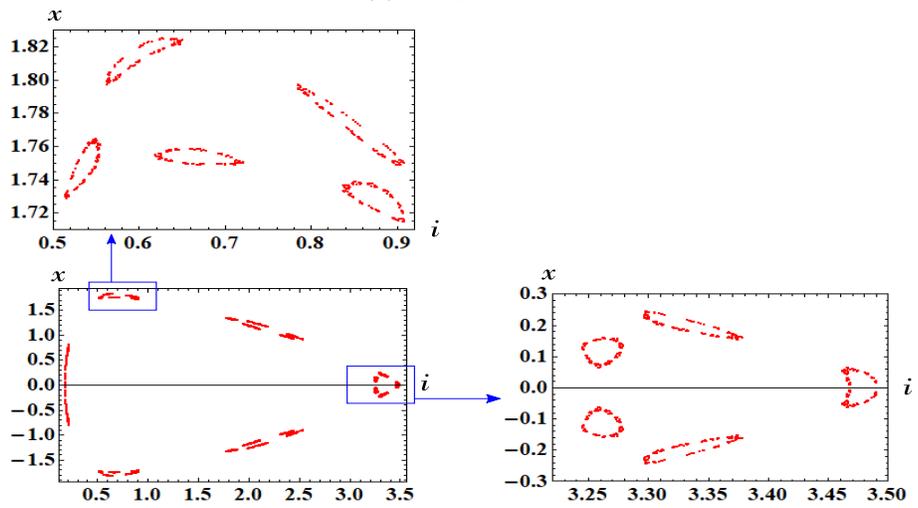, width=12cm, height=6.6cm} \\
   (b) quasi-periodic 
  \end{tabular}
  \caption{Poincar\'e maps of the forced memristor pendulum equations (\ref{eqn: pendulum-5}). 
   Parameters: $r = 0.04452, \ \omega = 1$.  
   Initial conditions: (a) $i(0)=0.187, \  x(0)=0.21$. \ \ (b) $i(0)=0.187, \  x(0)=0.2$.}
  \label{fig:Pendulum-poincare} 
\end{figure}

\clearpage 
%---Fig. 85-------%
\begin{figure}[hpbt]
 \centering
   \begin{tabular}{cc}
   \psfig{file=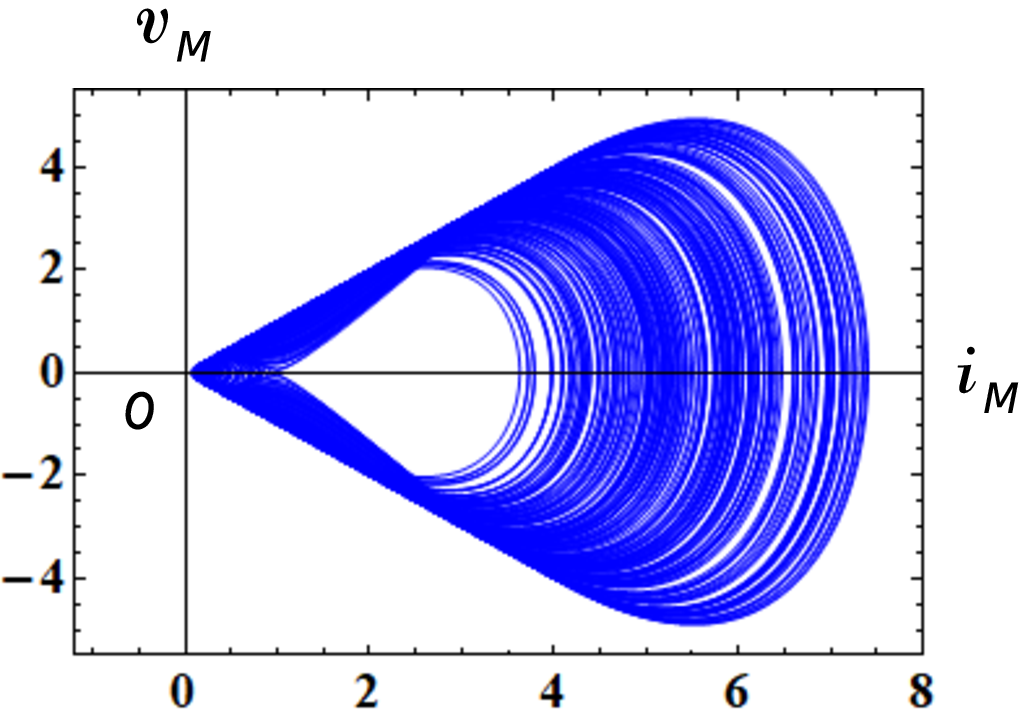, height=4.8cm} & 
   \psfig{file=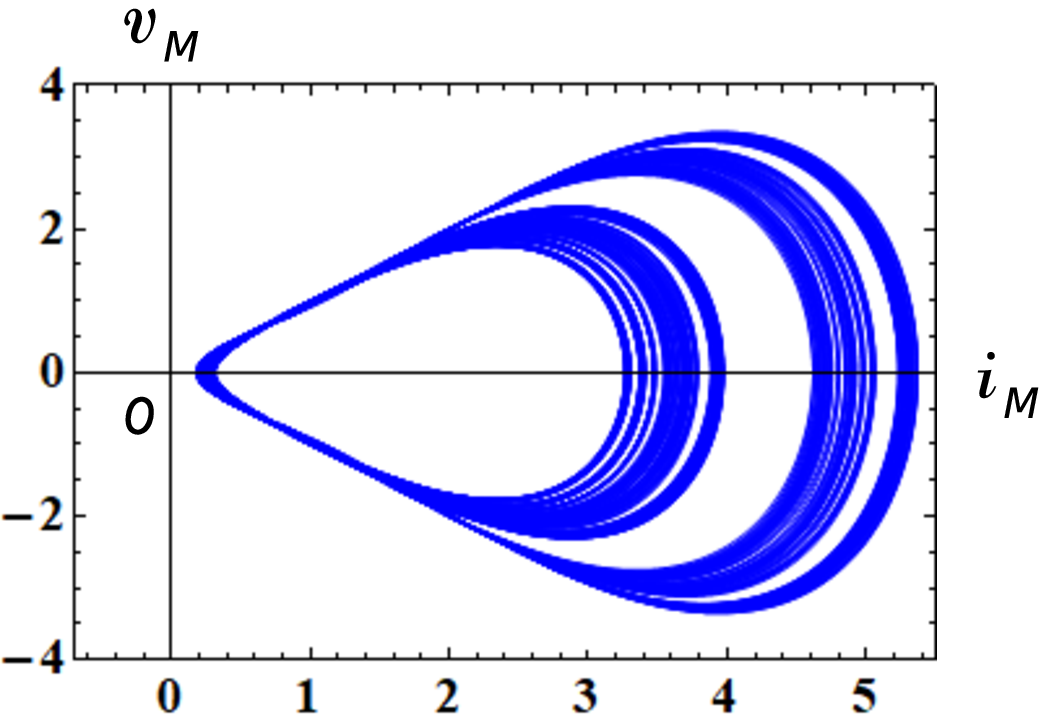, height=4.8cm} \vspace{1mm} \\
   (a) non-periodic & (b) quasi-periodic \\
  \end{tabular} 
  \caption{The $i_{M}-v_{M}$ loci of the forced memristor pendulum equations (\ref{eqn: pendulum-5}).  
   Here, $v_{M}$ and  $i_{M}$ denote the terminal voltage and the terminal current of the current-controlled generic memristor.  
   Parameters: $r = 0.04452, \ \omega = 1$.  
  Initial conditions: (a) $i(0)=0.187, \  x(0)=0.21$. \ \ (b) $i(0)=0.187, \  x(0)=0.2$.}
  \label{fig:Pendulum-pinch} 
\end{figure}
%
%

%---Fig. 86-------%
\begin{figure}[hpbt]
 \centering
   \begin{tabular}{cc}
    \psfig{file=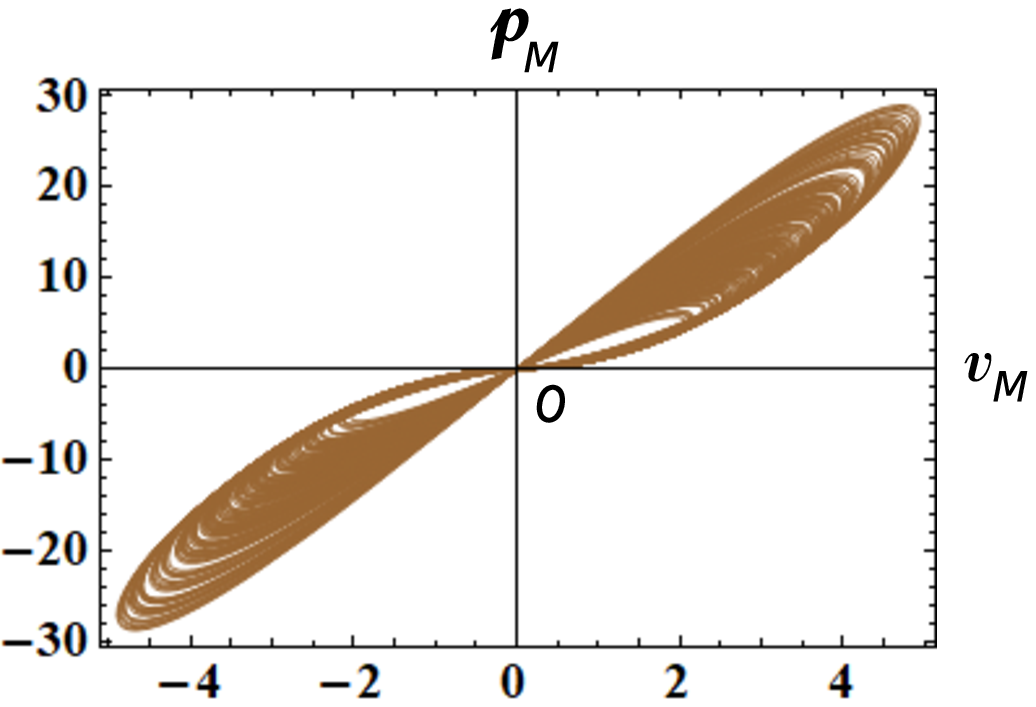, width=7cm}  & 
    \psfig{file=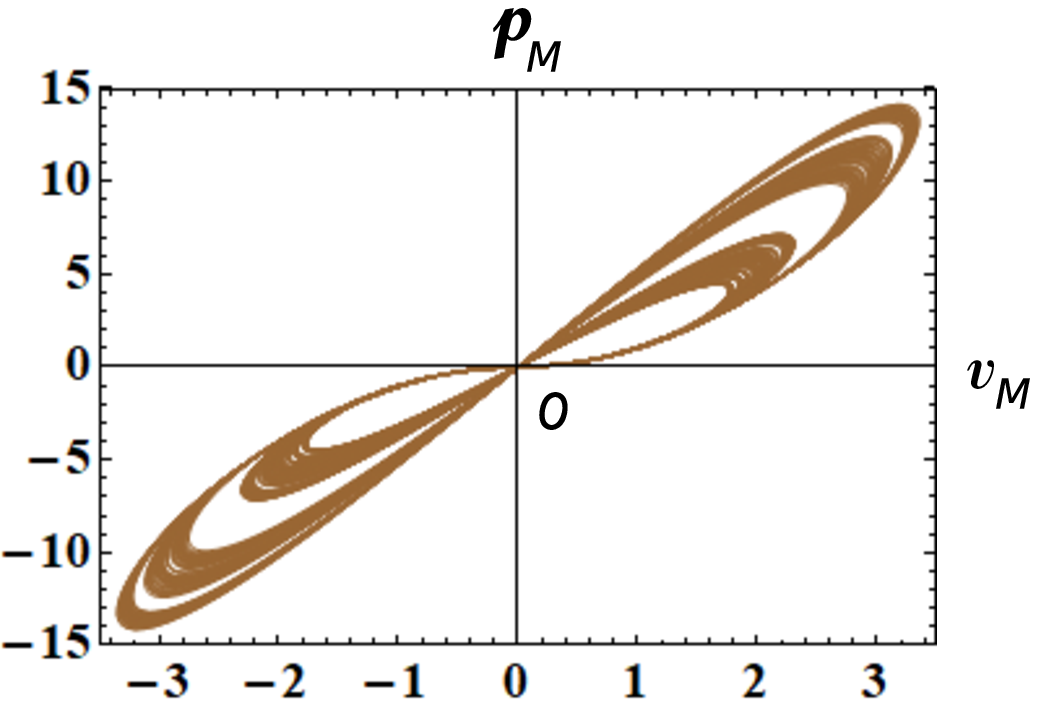, width=7cm}  \\
    (a) non-periodic & (b) quasi-periodic \\
   \end{tabular}
  \caption{ The $v_{M}-p_{M}$ locus of the forced memristor pendulum equations (\ref{eqn: pendulum-5}). 
   Here, $p_{M}(t)$ is an instantaneous power defined by $p_{M}(t)=i_{M}(t)v_{M}(t)$, 
   and $v_{M}(t)$ and $i_{M}(t)$ denote the terminal voltage and the terminal current of the current-controlled generic memristor.  
   Observe that the $v_{M}-p_{M}$ locus is pinched at the origin, and the locus lies in the first and the third quadrants. 
   The memristor switches between passive and active modes of operation, depending on its terminal voltage $v_{M}(t)$.
   Parameters: $r = 0.04452, \ \omega = 1$.  
   Initial conditions: (a) $i(0)=0.187, \  x(0)=0.21$. \ \ (b) $i(0)=0.187, \  x(0)=0.2$.}
  \label{fig:Pendulum-power} 
\end{figure}
%
%

%\clearpage
%---Fig. 87-------%
\begin{figure}[hpbt]
 \centering
   \begin{tabular}{cc}
   \psfig{file=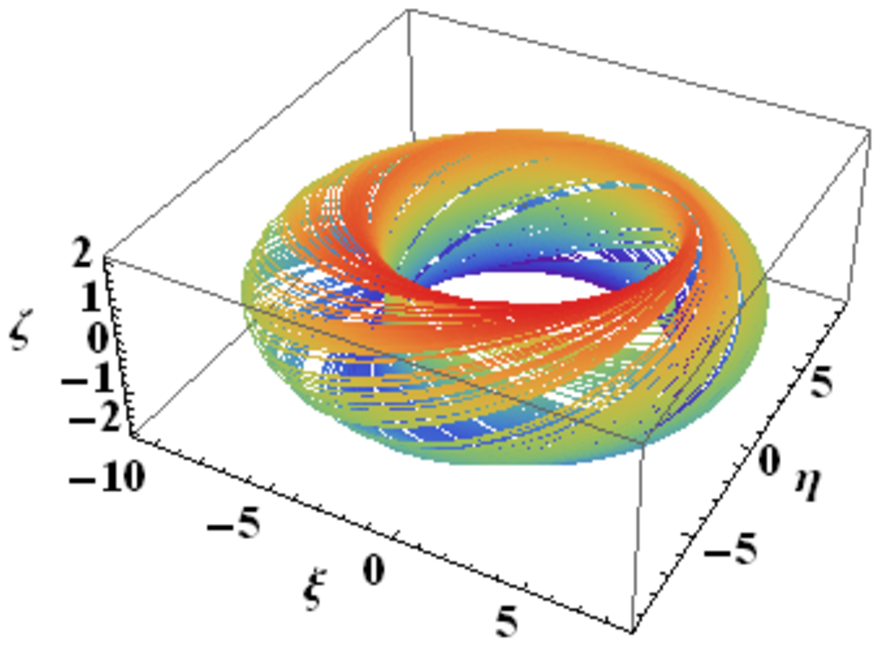, width=7.0cm} & 
   \psfig{file=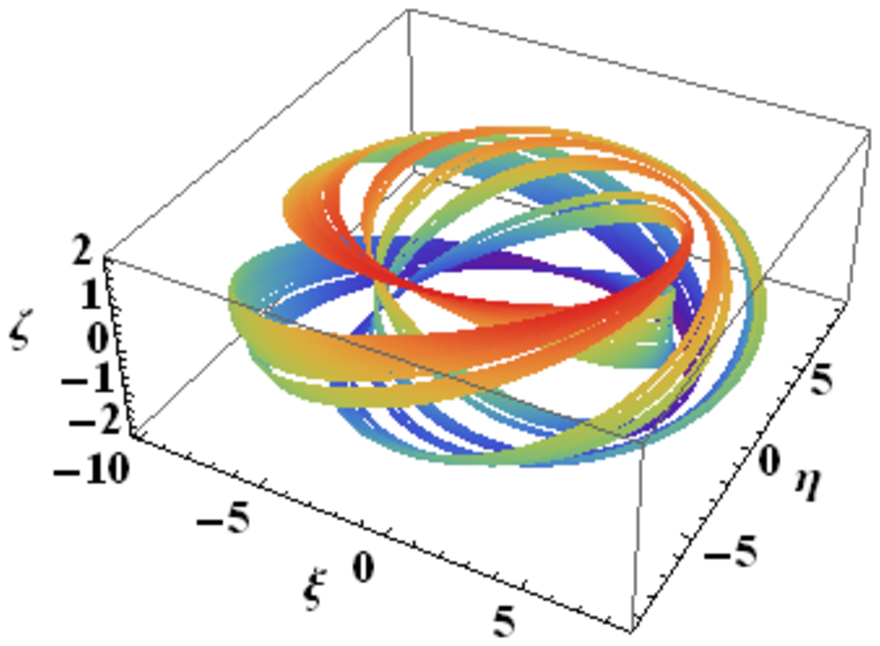, width=7.0cm} \vspace{1mm} \\
   (a) non-periodic & (b) quasi-periodic \\
  \end{tabular}
  \caption{The two trajectories of the forced memristor pendulum equations (\ref{eqn: pendulum-5}), 
  which are projected into the $( \xi, \, \eta, \, \zeta )$-space via the coordinate transformation (\ref{eqn: pendulum-projection}).  
  We can observe a wide gap in Figure \ref{fig:Pendulum-torus}(b). 
  Compare the trajectories in Figure \ref{fig:Pendulum-torus}  with the Poincar\'e maps in Figure \ref{fig:Pendulum-poincare}.  
  The trajectories are colored with the \emph{Rainbow} color code in Mathematica.
  Parameters: $r = 0.04452, \ \omega = 1$.  
   Initial conditions: (a) $i(0)=0.187, \  x(0)=0.21$. \ \ (b) $i(0)=0.187, \  x(0)=0.2$.}
  \label{fig:Pendulum-torus} 
\end{figure}
%
%

%---Fig. 88-------%
\begin{figure}[hpbt]
 \centering
   \begin{tabular}{cc}
    \psfig{file=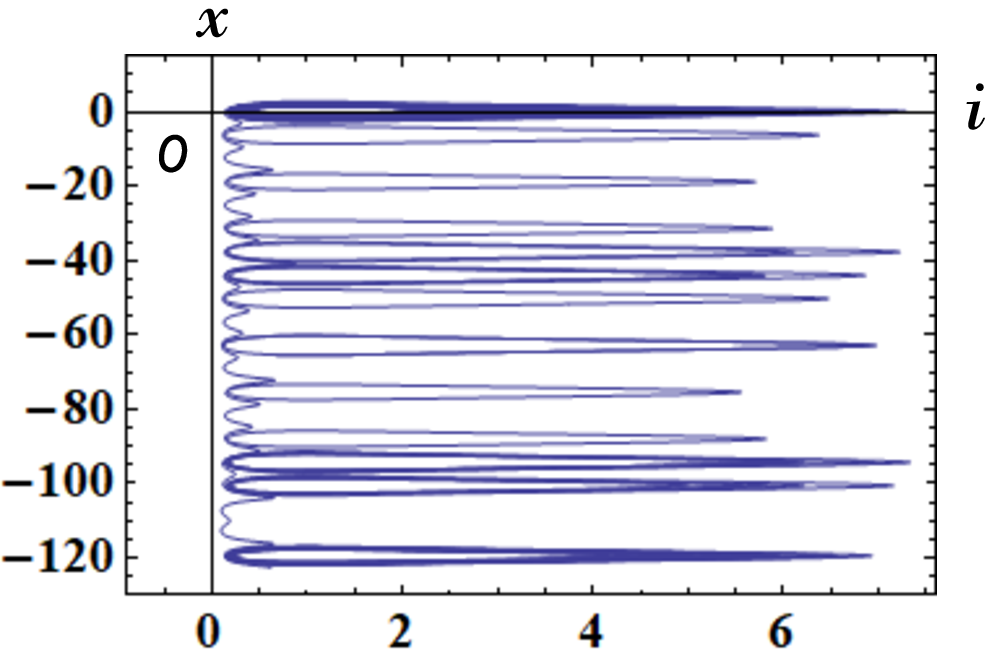, height=5cm}  & 
    \psfig{file=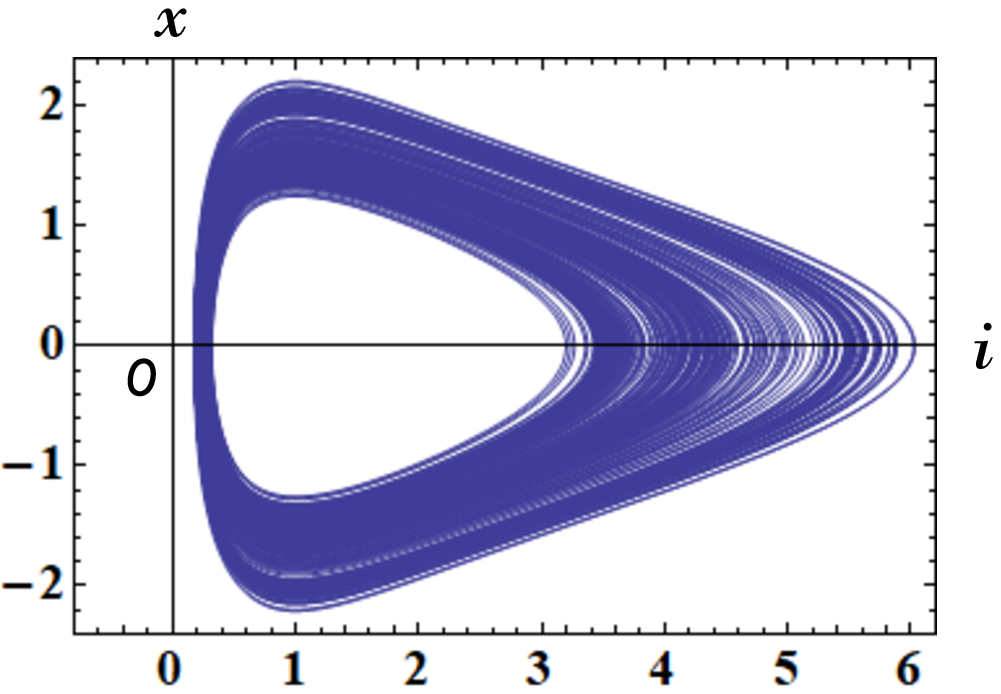, height=5cm}  \vspace{1mm} \\
   (a) $h=0.0005$ & (b) $h=0.0003$  \\ 
   \end{tabular}
  \caption{Behavior of the forced memristor pendulum equations (\ref{eqn: pendulum-5}) for $0 \le t \le 2000$. 
  Observe the difference between the two trajectories.  
  If we choose $h=0.0005$, then $x(t)$ decreases gradually as time $t$ increases 
  as shown in Figure \ref{fig:Pendulum-orbit}(a), where  $x(2000) \approx -122.7$. 
  However, if we choose $h=0.0003$,  then the trajectory stays in a finite region of the $(i, \ x)$-plane 
  as shown in Figure \ref{fig:Pendulum-orbit}(b), where  $x(2000) \approx -1.815$.    
  Here, $h$ denotes the maximum step size of the numerical integration.  
  Parameters: $r = 0.0445, \ \  \omega = 1$.  
  Initial conditions: $i(0)=0.187, \  x(0)=0.187$.}
  \label{fig:Pendulum-orbit} 
\end{figure}
\newpage

%
%==============================================================%
\subsection{Lorenz system}
%==============================================================%
%
The dynamics of the \emph{Lorenz system} \cite{Peitgen} is defined by a system of three ordinary differential equations: 
\begin{center}
\begin{minipage}{8.7cm}
\begin{shadebox}
\underline{\emph{Lorenz system}}
\begin{equation} 
 \left.
  \begin{array}{lll}
     \displaystyle \frac{d x}{dt} &=&   \sigma \, ( y - x ),
      \vspace{2mm} \\
     \displaystyle \frac{d y}{dt} &=&  x \, ( \rho - z ) - y,
      \vspace{2mm} \\    
     \displaystyle \frac{d z}{dt} &=&  x y - \beta \, z,    
  \end{array}
 \right \}
\label{eqn: lorenz-1}
\end{equation} 
where $\sigma = 10$, $\beta = \frac{8}{3}$, and $\rho = 28$.  
\end{shadebox}
\end{minipage}
\end{center}
The Lorenz system (\ref{eqn: lorenz-1}) is a simplified model of convection rolls in the atmosphere.  
When $\sigma = 10$, $\beta = \frac{8}{3}$, and $\rho = 28$, Eq. (\ref{eqn: lorenz-1}) has chaotic solutions, which resemble a butterfly or figure eight.

Substituting $x = \ln {|\, i \,|}$, $y=x_{1}$, and $z=x_{2}$ into Eq. (\ref{eqn: lorenz-1}), 
we obtain 
\begin{center}
\begin{minipage}{8.7cm}
\begin{shadebox}
\underline{\emph{Memristor Lorenz equations for Eq. (\ref{eqn: lorenz-1})}}
\begin{equation} 
 \left.
  \begin{array}{lll}
     \displaystyle \frac{d i}{dt} &=&   \sigma \, ( x_{2} -  \ln {|\, i \,|} \,) \, i,
      \vspace{2mm} \\
     \displaystyle \frac{d x_{1}}{dt} &=& ( \rho - x_{2} ) \, \ln {|\, i \,|}  - x_{1} ,
      \vspace{2mm} \\    
     \displaystyle \frac{d x_{2}}{dt} &=&  x_{1} - \beta \, x_{2},    
  \end{array}
 \right \}
\label{eqn: lorenz-2}
\end{equation} 
where $\sigma = 10$, $\beta = \frac{8}{3}$, and $\rho = 28$.  
\end{shadebox}
\end{minipage}
\end{center}
Consider the three-element memristor circuit 
in Figure \ref{fig:memristor-inductor-battery}.  
The dynamics of this circuit given by Eq. (\ref{eqn: dynamics-1}).  
Assume that Eq. (\ref{eqn: dynamics-1}) satisfies 
\begin{equation}
\left.
\begin{array}{ccc}
  L =1, && E=0,   \vspace{2mm} \\
  \hat{R}( x_{1}, \, x_{2},  \,  i_{M} ) &=&  - \sigma \, ( x_{2} -  \ln {|\, i_{M} \,|} \,) \vspace{2mm} \\
     \displaystyle  \tilde{f}_{1}(x_{1}, \ x_{2}, \ i) &=&  ( \rho - x_{2} ) \, \ln {|\, i_{M} \,|}  - x_{1},
      \vspace{2mm} \\
     \displaystyle  \tilde{f}_{2}(x_{1}, \ x_{2}, \ i) &=& x_{1} - \beta \, x_{2}.   
\end{array}
\right \}
\end{equation}
Then Eq. (\ref{eqn: dynamics-1}) can be recast into Eq. (\ref{eqn: lorenz-2}).  
Thus, the terminal voltage $v_{M}$ and the terminal current $i_{M}$ of the current-controlled extended memristor 
in Figure \ref{fig:memristor-inductor-battery} are described
by
\begin{center}
\begin{minipage}{8.7cm}
\begin{shadebox}
\underline{\emph{V-I characteristics of the extended memristor}}
\begin{equation}
\begin{array}{lll}
  v_{M} &=& \hat{R}( x_{1}, \, x_{2},  \,  i_{M} ) \, i_{M} \vspace{2mm} \\
        &=& - \sigma \, ( x_{2} -  \ln {|\, i_{M} \,|} \,) \, i_{M},   
  \vspace{1mm} \\
     \displaystyle \frac{d x_{1}}{dt} &=&  ( \rho - x_{2} ) \, \ln {|\, i_{M} \,|}  - x_{1},
      \vspace{2mm} \\
     \displaystyle \frac{d x_{2}}{dt} &=&  x_{1} - \beta \, x_{2}, 
\end{array}
\label{eqn: lorenz-3}
\end{equation}
where $\hat{R}( x_{1}, \, x_{2},  \,  i_{M} ) = - \sigma \, ( x_{2} -  \ln {|\, i_{M} \,|} \,) $. \vspace{2mm}
\end{shadebox}
\end{minipage}
\end{center}
Note that $i=0$ and $i_{M} = 0$ are not well-defined in Eq. (\ref{eqn: lorenz-2}) and Eq. (\ref{eqn: lorenz-3}), respectively.
Furthermore, the condition for the extended memristor, that is, $\hat{R}( x_{1}, \, x_{2}, \ 0) \ne \infty$ is not satisfied, 
since 
\begin{equation}
\scalebox{0.93}{$\displaystyle
 \displaystyle \lim_{i_{M} \to 0}| \hat{R}( x_{1}, \, x_{2}, \ i_{M})| \\
  = \lim_{i_{M} \to 0} \, \Bigl | - \sigma \, ( x_{2} -  \ln {|\, i_{M} \,|} \,) \Bigr |  \to \infty. 
$}
\end{equation}
However, if $i_{M} \to 0$, then $v_{M}$ satisfies 
\begin{equation}
  v_{M} = \hat{R}( x_{1}, \, x_{2},  \,  i_{M} ) \, i_{M} = - \sigma \, ( x_{2} -  \ln {|\, i_{M} \,|} \,) \, i_{M} \to 0.      
\end{equation}
Therefore, without loss of generality, we can regard this kind of memristor as the extended memristor.
For more details, see \cite{Itoh(2017)}. 

For $\sigma = 10$, $\beta = \frac{8}{3}$, and $\rho = 28$, the memristor Lorenz equations (\ref{eqn: lorenz-2}) also exhibit chaotic oscillation.   
Thus, an external periodic forcing is \emph{unnecessary} to generate chaotic oscillation. 
We show the chaotic attractor, Poincar\'e map, and $i_{M}-v_{M}$ locus in Figures \ref{fig:Lorenz-attractor}, \ref{fig:Lorenz-poincare}, and \ref{fig:Lorenz-pinch}, respectively.  
We can easily observe the folding action of the chaotic attractor as shown in Figure \ref{fig:Lorenz-poincare}.  
The $i_{M}-v_{M}$ locus in Figure \ref{fig:Lorenz-pinch} lies in the first and the fourth quadrants.   
Thus, the extended memristor defined by Eq. (\ref{eqn: lorenz-3}) is an active element. 
Let us show the $v_{M}-p_{M}$ locus in Figure \ref{fig:Lorenz-power}, where $p_{M}(t)$ is an instantaneous power defined by $p_{M}(t)=i_{M}(t)v_{M}(t)$.  
Observe that the $v_{M}-p_{M}$ locus is pinched at the origin, and the locus lies in the first and the third quadrants. 
Thus, when $v_{M}>0$, the instantaneous power $p_{M}$ delivered from the forced signal and the inductor is dissipated in the memristor.  
However, when $v_{M}<0$, the instantaneous power $p_{M}$ is not dissipated in the memristor. 
Hence, the memristor switches between passive and active modes of operation, depending on its terminal voltage. 
We conclude as follow: 
\begin{center}
\begin{minipage}{.7\textwidth}
\begin{itembox}[l]{Switching behavior of the memristor}
Assume that Eq. (\ref{eqn: lorenz-2}) exhibits chaotic oscillation.  
Then the extended memristor defined by Eq. (\ref{eqn: lorenz-3}) can switch between ``passive'' and ``active'' modes of operation, depending on its terminal voltage.  
\end{itembox}
\end{minipage}
\end{center}
%
%

%---Fig. 89-------%
\begin{figure}[hpbt]
 \centering
  \psfig{file=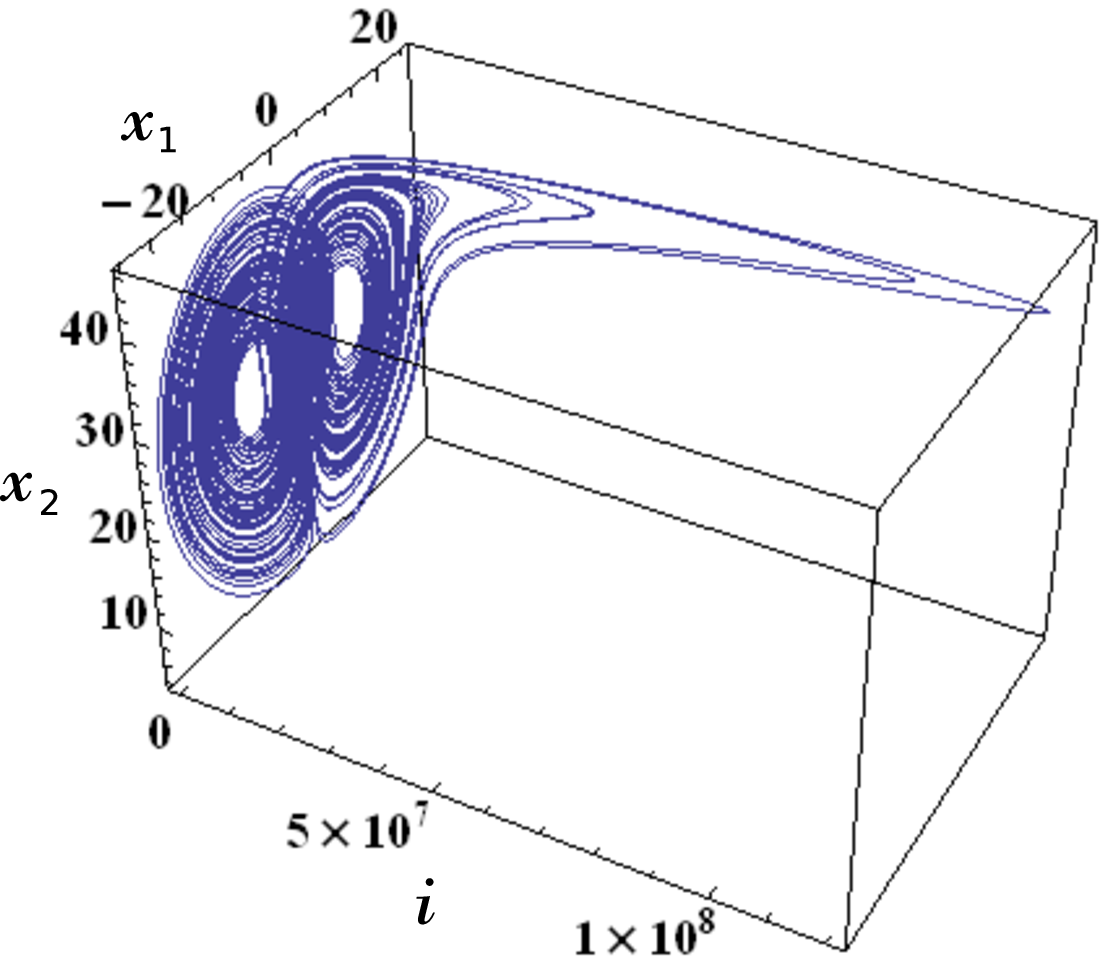, width=7.0cm}
  \caption{Chaotic attractor of the memristor Lorenz equations (\ref{eqn: lorenz-2}). 
  Parameters: $a = 0.2, \ b = 0.2, \ c = 5.7$.
  Initial conditions: $i(0)=0.1, \  x_{1}(0)=0.1, \  x_{2}(0)=0.1$.}
  \label{fig:Lorenz-attractor} 
\end{figure}
%
%

%---Fig. 90-------%
\begin{figure}[hpbt]
 \centering
  \psfig{file=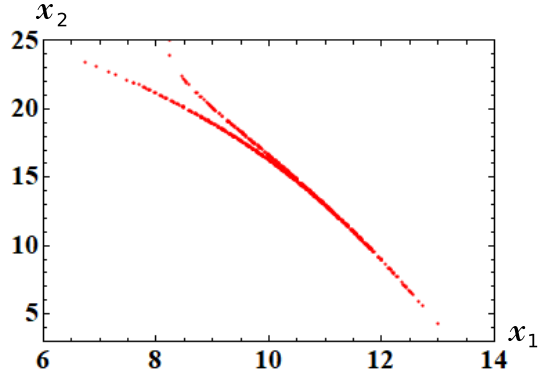, width=7.0cm}
  \caption{Poincar\'e map of the memristor Lorenz equations (\ref{eqn: lorenz-2}).     
   The Poincar\'e cross-section is defined by $\{(i, \, x_{1}, \, x_{2}) \in R^{3} \ | \ i= 500 \}$.  
   The chaotic trajectory of Eq. (\ref{eqn: lorenz-2}) crosses the above Poincar\'e cross-section (plane) many times.  
   Observe the folding action of the chaotic attractor.   
   Parameters: $a = 0.2, \ b = 0.2, \ c = 5.7$.
   Initial conditions: $i(0)=0.1, \  x_{1}(0)=0.1, \  x_{2}(0)=0.1$.}
  \label{fig:Lorenz-poincare} 
\end{figure}
%
%

%---Fig. 91-------%
\begin{figure}[hpbt]
  \centering
  \psfig{file=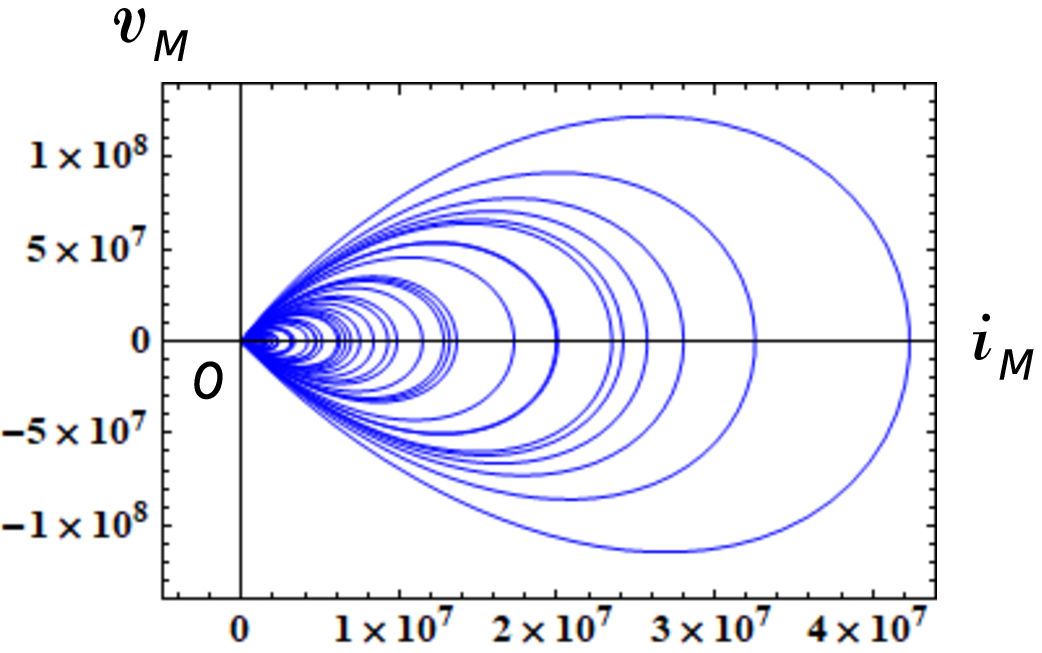, width=7.0cm}
  \caption{ The $i_{M}-v_{M}$ locus of the memristor Lorenz equations (\ref{eqn: lorenz-2}).  
   Here, $v_{M}$ and  $i_{M}$ denote the terminal voltage and the terminal current of the current-controlled extended memristor.  
   Parameters: $a = 0.2, \ b = 0.2, \ c = 5.7$.
   Initial conditions: $i(0)=0.1, \  x_{1}(0)=0.1, \  x_{2}(0)=0.1$.}
  \label{fig:Lorenz-pinch} 
\end{figure}
%
%

%---Fig. 92-------%
\begin{figure}[hpbt]
  \centering
  \psfig{file=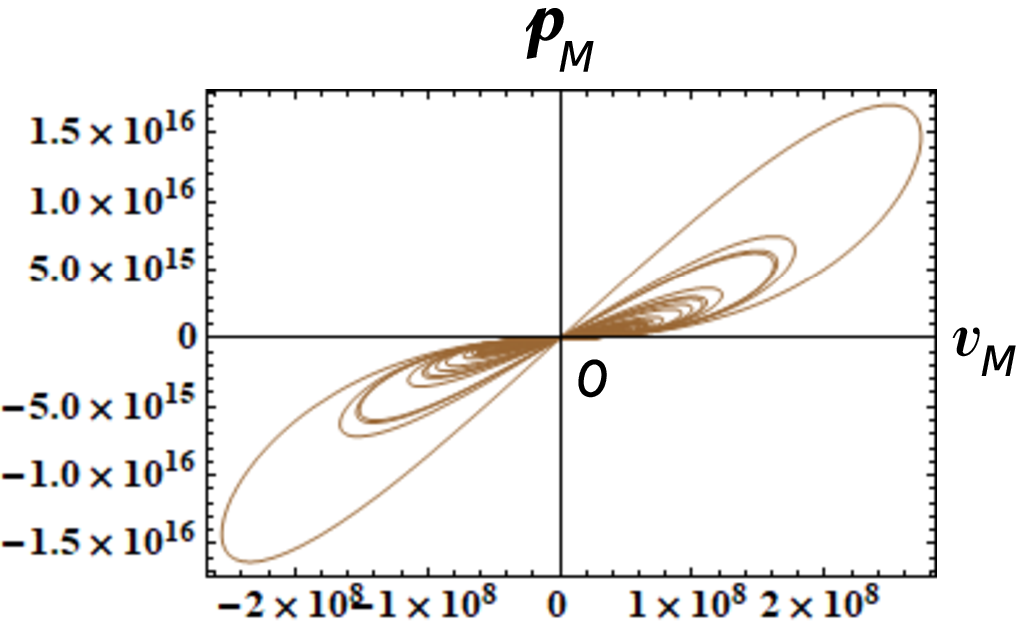, height=5cm}
  \caption{ The $v_{M}-p_{M}$ locus of the memristor Lorenz equations (\ref{eqn: lorenz-2}). 
   Here, $p_{M}(t)$ is an instantaneous power defined by $p_{M}(t)=i_{M}(t)v_{M}(t)$,   
   and $v_{M}(t)$ and $i_{M}(t)$ denote the terminal voltage and the terminal current of the current-controlled extended memristor. 
   Observe that the $v_{M}-p_{M}$ locus is pinched at the origin, and the locus lies in the first and the third quadrants.  
   The memristor switches between passive and active modes of operation, depending on its terminal voltage $v_{M}(t)$.
   Parameters: $a = 0.2, \ b = 0.2, \ c = 5.7$.
   Initial conditions: $i(0)=0.1, \  x_{1}(0)=0.1, \  x_{2}(0)=0.1$.}
  \label{fig:Lorenz-power} 
\end{figure}
%
%

%
%==============================================================%
\subsection{Two-variable Oregonator model}
%==============================================================%
%
Two-variable Oregonator model \cite{Zemskov(2011)} is defined by 
\begin{center}
\begin{minipage}{9.5cm}
\begin{shadebox}
\underline{\emph{Two-variable Oregonator model equations}}
\begin{equation} 
  \begin{array}{ccl}
     \displaystyle \frac{d u}{dt} &=&  \displaystyle \frac{1}{\epsilon }\left ( u - u^{2} - \frac{f v( u - q )}{u +q} \right ),
      \vspace{2mm} \\
     \displaystyle \frac{d v}{dt} &=&  u - v, 
  \end{array}
\label{eqn: Oregonator-1}
\end{equation} 
where $f = 1$, $q = 0.001$, and $\epsilon = 0.7$.  
\end{shadebox}
\end{minipage}
\end{center}

Substituting $u = \ln {|\, i \,|}$ and $v=x$, into Eq. (\ref{eqn: Oregonator-1}), 
we obtain 
\begin{center}
\begin{minipage}{9.5cm}
\begin{shadebox}
\underline{\emph{Memristor two-variable Oregonator model equations}}
\begin{equation} 
  \begin{array}{ccl}
     \displaystyle L \frac{d i}{dt} &=& 
      \displaystyle \left \{ \ln {|\, i \,|}  - ( \, \ln {|\, i \,|} \, )^{2} - \frac{f \, x \, ( \ln {|\, i \,|} - q )}{\ln {|\, i \,|} + q} \right \} i,
      \vspace{2mm} \\
     \displaystyle \frac{d x}{dt} &=&  \ln {|\, i \,|} - x, 
  \end{array}
\label{eqn: Oregonator-2}
\end{equation} 
where $f = 1$, $q = 0.001$, and $L = 0.7$.   
\end{shadebox}
\end{minipage}
\end{center}

Consider the three-element memristor circuit in Figure \ref{fig:memristor-inductor-battery}.  
The dynamics of this circuit given by Eq. (\ref{eqn: dynamics-n-1}).  
Assume that Eq. (\ref{eqn: dynamics-n-1}) satisfies 
\begin{equation}
\begin{array}{ccl}
  E &=& 0,   \vspace{2mm} \\
  \hat{R}( x,  \,  i_{M} ) &=&  \scalebox{0.85}{$\displaystyle 
          - \left \{ \ln {|\, i_{M} \,|}  - ( \, \ln {|\, i_{M} \,|} \, )^{2} - \frac{f \, x \, ( \ln {|\, i_{M} \,|} - q )}{\ln {|\, i_{M} \,|} + q} \right \} $}, \vspace{2mm} \\
  f_{1}(x, \, i) &=& \ln {|\, i_{M} \,|} - x.   
\end{array}
\end{equation}
Then Eq. (\ref{eqn: dynamics-n-1}) can be recast into Eq. (\ref{eqn: Oregonator-2}).  
The terminal voltage $v_{M}$ and the terminal current $i_{M}$ of the current-controlled extended memristor in Figure \ref{fig:memristor-inductor-battery} are described
by
\begin{center}
\begin{minipage}{10.0cm}
\begin{shadebox}
\underline{\emph{V-I characteristics of the extended memristor}}
\begin{equation}
\begin{array}{l}
  v_{M} = \hat{R}( x, \, i_{M} ) \, i_{M} \vspace{2mm} \\
        = \scalebox{0.9}{$\displaystyle 
          - \left \{ \ln {|\, i_{M} \,|}  - ( \, \ln {|\, i_{M} \,|} \, )^{2} - \frac{f \, x \, ( \ln {|\, i_{M} \,|} - q )}{\ln {|\, i_{M} \,|} + q} \right \} i_{M}$},
  \vspace{1mm} \\
     \displaystyle \frac{d x}{dt} = f_{1}(x, \, i) = \ln {|\, i_{M} \,|} - x, 
\end{array}
\label{eqn: Oregonator-3}
\end{equation}
where 
\[
\scalebox{0.82}{$\displaystyle \hat{R}( x,  i_{M} ) = - \left \{ \ln {|\, i_{M} \,|}  - ( \, \ln {|\, i_{M} \,|} \, )^{2} - \frac{f \, x \, ( \ln {|\, i_{M} \,|} - q )}{\ln {|\, i_{M} \,|} + q} \right \}. $} 
\]
%
%
%\vspace{2mm}
\end{shadebox}
\end{minipage}
\end{center}
Note that $\displaystyle \lim_{i_{M} \to 0}| \hat{R}( x, \, i_{M})| \to \infty$. 
Hence, the above memristor does not satisfy the condition of the extended memristor, that is, $\hat{R}( x, \, 0) \ne \infty$.      
Furthermore, $i=0$ and $i_{M} = 0$ are not well-defined in Eq. (\ref{eqn: Oregonator-2}) and Eq. (\ref{eqn: Oregonator-3}).
However, if $i_{M} \to 0$, then $v_{M} \to 0$.
Thus, without loss of generality, we can regard this kind of memristor as the extended memristor.  
For more details, see \cite{Itoh(2017)}.  

The memristor two-variable Oregonator model equations (\ref{eqn: Oregonator-2}) exhibit periodic oscillation.  
When an external source is added as shown in Figure \ref{fig:memristive-inductor-battery-source}, the forced two-variable Oregonator model equations can exhibit chaotic oscillation.  
The dynamics of this circuit is given by 
\begin{center}
\begin{minipage}{10.0cm}
\begin{shadebox}
\underline{\emph{Forced memristor two-variable Oregonator model equations}}
\begin{equation} 
  \begin{array}{ccl}
     \displaystyle L \frac{d i}{dt} &=& 
      \displaystyle \left \{ \ln {|\, i \,|}  - ( \, \ln {|\, i \,|} \, )^{2} 
                           - \frac{f \, x \, ( \ln {|\, i \,|} - q )}{\ln {|\, i \,|} + q}  
                     \right \} i \vspace{2mm} \\
       && + r \sin ( \omega t), 
      \vspace{2mm} \\
     \displaystyle \frac{d x}{dt} &=&  \ln {|\, i \,|} - x, 
  \end{array}
%\vspace{2mm}
\label{eqn: Oregonator-4}
\end{equation} 
where $r$ and $\omega$ are constants.  
\end{shadebox}
\end{minipage}
\end{center}
We show their chaotic attractor, Poincar\'e map, and $i_{M}-v_{M}$ locus in Figures \ref{fig:Oregonator-attractor}, \ref{fig:Oregonator-poincare}, and \ref{fig:Oregonator-pinch}, respectively.  
The following parameters are used in our computer simulations:
\begin{equation}
\left.
 \begin{array}{lll}
  f = 1, & q = 0.001, & L = 0.7.  \vspace{1mm} \\
  r = 0.01, & \omega = 0.79. &
 \end{array}
\right \}
\end{equation}
The $i_{M}-v_{M}$ locus in Figure \ref{fig:Oregonator-pinch} lies in the first and the fourth quadrants.  
Thus, the extended memristor defined by Eq. (\ref{eqn: Oregonator-3}) is an active element. 
Let us show the $v_{M}-p_{M}$ locus in Figure \ref{fig:Oregonator-power}, where $p_{M}(t)$ is an instantaneous power defined by $p_{M}(t)=i_{M}(t)v_{M}(t)$.  
Observe that the $v_{M}-p_{M}$ locus is pinched at the origin, and the locus lies in the first and the third quadrants. 
Thus, the memristor switches between passive and active modes of operation, depending on its terminal voltage. 
We conclude as follow: \\
\begin{center}
\begin{minipage}{.7\textwidth}
\begin{itembox}[l]{Switching behavior of the memristor}
Assume that Eq. (\ref{eqn: Oregonator-4}) exhibits chaotic oscillation.  
Then the extended memristor defined by Eq. (\ref{eqn: Oregonator-3}) can switch between ``passive'' and ``active'' modes of operation, depending on its terminal voltage.  
\end{itembox}
\end{minipage}
\end{center}
In order to obtain the above results, we have to choose the initial conditions carefully. 
It is due to the fact that a periodic response (drawn in magenta) coexists with a chaotic attractor (drawn in blue) 
as shown in Figure \ref{fig:Oregonator-attractor-coexistence}.   

%\newpage
%---Fig. 93-------%
\begin{figure}[hpbt]
 \centering
  \psfig{file=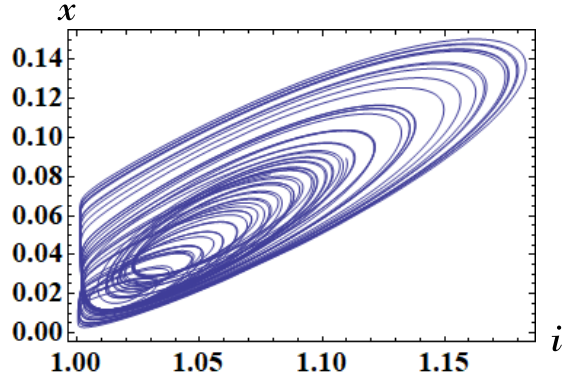, height=5cm}
  \caption{Chaotic attractor of the forced two-variable Oregonator model equations (\ref{eqn: Oregonator-4}). 
   Parameters: $f = 1, \ q = 0.001, \ L = 0.7, \ r = 0.01, \ \omega = 0.79$.  
   Initial conditions: $i(0)=e^{0.1}, \  x(0)=0.1$.}
  \label{fig:Oregonator-attractor} 
\end{figure}
%
%

%---Fig. 94-------%
\begin{figure}[hpbt]
 \centering
  \psfig{file=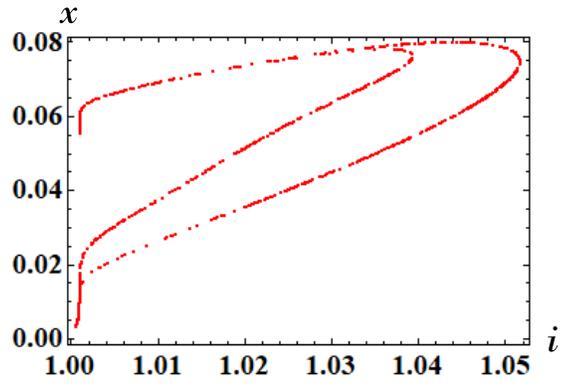, height=5cm}
  \caption{Poincar\'e map of the forced two-variable Oregonator model equations (\ref{eqn: Oregonator-4}). 
   Parameters: $f = 1, \ q = 0.001, \ L = 0.7, \ r = 0.01, \ \omega = 0.79$.  
   Initial conditions: $i(0)=e^{0.1}, \  x(0)=0.1$.}
  \label{fig:Oregonator-poincare} 
\end{figure}
%
%

%---Fig. 95-------%
\begin{figure}[hpbt]
  \centering
  \psfig{file=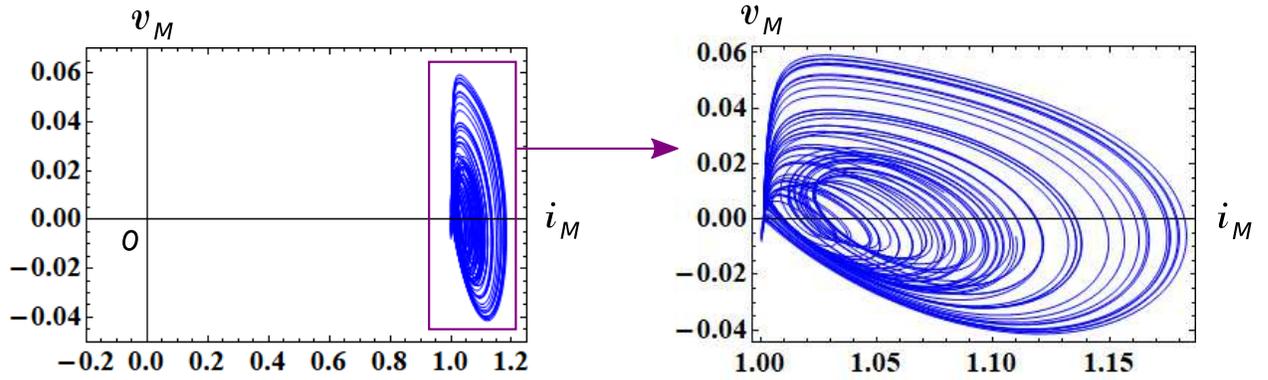, height=5cm}
  \caption{ The $i_{M}-v_{M}$ locus of the forced two-variable Oregonator model equations (\ref{eqn: Oregonator-4}). 
   Here, $v_{M}$ and  $i_{M}$ denote the terminal voltage and the terminal current of the current-controlled extended memristor.  
   Parameters: $f = 1, \ q = 0.001, \ L = 0.7, \ r = 0.01, \ \omega = 0.79$.  
   Initial conditions: $i(0)=e^{0.1}, \  x(0)=0.1$.}
  \label{fig:Oregonator-pinch} 
\end{figure}
%
%

%---Fig. 96-------%
\begin{figure}[hpbt]
  \centering
  \psfig{file=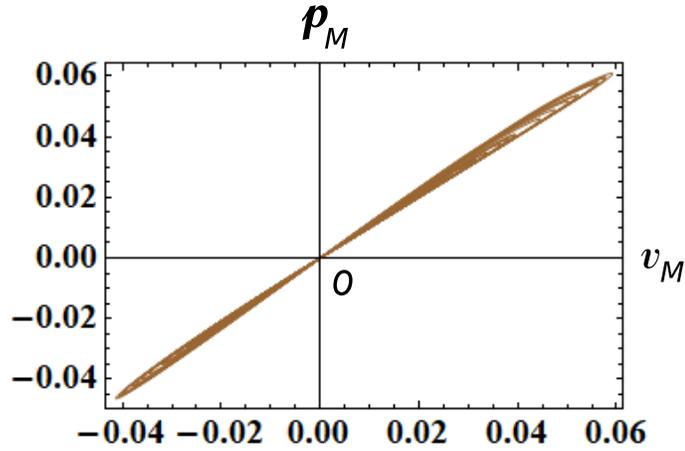, height=6cm}
  \caption{ The $v_{M}-p_{M}$ locus of the forced two-variable Oregonator model equations (\ref{eqn: Oregonator-4}). 
   Here, $p_{M}(t)$ is an instantaneous power defined by $p_{M}(t)=i_{M}(t)v_{M}(t)$,   
   and $v_{M}(t)$ and $i_{M}(t)$ denote the terminal voltage and the terminal current of the current-controlled extended memristor.    
   Observe that the $v_{M}-p_{M}$ locus is pinched at the origin, and the locus lies in the first and the third quadrants.  
   The memristor switches between passive and active modes of operation, depending on its terminal voltage $v_{M}(t)$.
   Parameters: $f = 1, \ q = 0.001, \ L = 0.7, \ r = 0.01, \ \omega = 0.79$. 
   Initial conditions: $i(0)=e^{0.1}, \  x(0)=0.1$.}
  \label{fig:Oregonator-power} 
\end{figure}
%
%

%---Fig. 97-------%
\begin{figure}[hpbt]
 \begin{center}
  \psfig{file=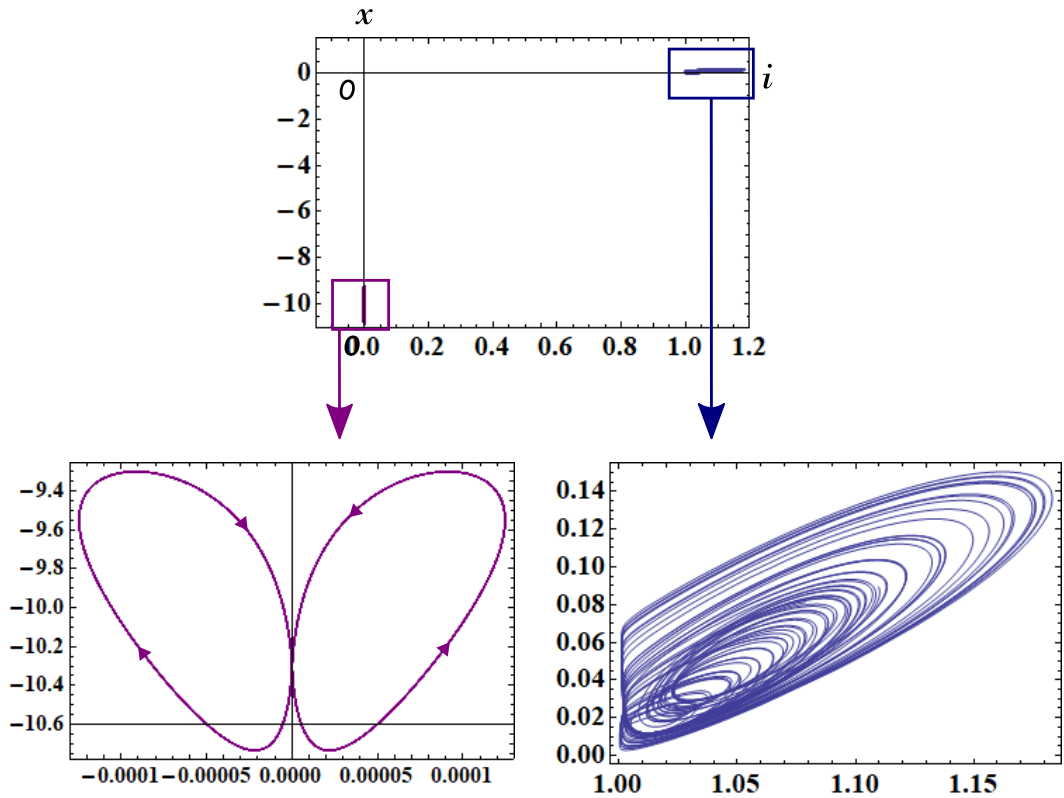, width=14cm} 
  \caption{A periodic response (purple) coexists with a chaotic attractor (blue).
  Parameters: $f = 1, \ q = 0.001, \ L = 0.7, \ r = 0.01, \ \omega = 0.79$.  
  Initial conditions for a periodic response: $i(0)=e^{-0.1}, \  x(0)=0.1$.   \ \
  Initial conditions for a chaotic attractor: $i(0)=e^{0.1}, \  x(0)=0.1$. }
  \label{fig:Oregonator-attractor-coexistence} 
 \end{center}
\end{figure}
%
%
%\clearpage

%\clearpage
%
%
%%%%%%%%%%%%%%%%%%%%%%%%%%%%%%%%%%%%%%%%%%%%%%%%%%%%%%%%%%%%%%%%%%%%%%%
\section{Ideal Memristor Circuit}
\label{sec: ideal}
%%%%%%%%%%%%%%%%%%%%%%%%%%%%%%%%%%%%%%%%%%%%%%%%%%%%%%%%%%%%%%%%%%%%%%%
%
%
%
In this section, we realize the dynamics of the systems by using \emph{ideal} memristors. 

%==============================================================%
\subsection{Van der Pol oscillator }
\label{sec: vdp}
%==============================================================%
%
The Van der Pol oscillator is defined by the second-order differential equations 
\begin{center}
\begin{minipage}{8.7cm}
\begin{shadebox}
\underline{\emph{Van der Pol equations}}
\begin{equation}
 \left.
  \begin{array}{lll}
   \displaystyle \frac{dx}{dt} &=& y - f(x), \vspace{2mm} \\
   \displaystyle \frac{dy}{dt} &=& - x,
  \end{array}
 \right \}
\label{eqn: vdp1}
\end{equation}
where $f(x)$ is a scalar function of a single variable $x$ defined by 
\begin{equation}
  f(x) = \frac{x^{3}}{3}-x. 
\end{equation}
\end{shadebox}
\end{minipage}
\end{center}
Equation (\ref{eqn: vdp1}) can be realized by the circuit in Figure \ref{fig:vdp-m} \cite{Itoh(2013)}.  
The circuit equations are given by 
\begin{center}
\begin{minipage}{8.7cm}
\begin{shadebox}
\underline{\emph{Memristor Van der Pol equations}}
\begin{equation}
 \left.
  \begin{array}{lll}
   \displaystyle L\frac{dq}{dt} &=& \varphi - f(q), \vspace{2mm} \\
   \displaystyle C\frac{d\varphi}{dt} &=& -q.
  \end{array}
 \right \}
\label{eqn: vdp2}
\end{equation}
\end{shadebox}
\end{minipage}
\end{center}
Here, $L=C=1$,  $q$ and $\varphi$ denote the charge of the inductor $L$ and the flux of the capacitors $C$, respectively, that is,
\begin{equation}
 \left.
  \begin{array}{cll} 
   q(t) 
    &\stackrel{\triangle}{=}
     & \displaystyle \int_{-\infty}^{t} i(t) dt, \vspace{1mm} \\
   \varphi (t)    
    & \stackrel{\triangle}{=} 
     & \displaystyle  \int_{-\infty}^{t} v(t) dt,
  \end{array}
 \right \}
\end{equation}
and the $q-\varphi$ curve of the charge-controlled memristor is given by
\begin{center}
\begin{minipage}{8.7cm}
\begin{shadebox}
\underline{\emph{$q-\varphi$ curve of the charge-controlled memristor}}
\begin{equation}
  \varphi = f(q) = \frac{q^{3}}{3}-q.   
\label{eqn: vdp2-curve}
\end{equation}
\end{shadebox}
\end{minipage}
\end{center}
Differentiating Eq. (\ref{eqn: vdp2}) with respect to time $t$, we obtain a set of differential equations
\begin{center}
\begin{minipage}{8cm}
\begin{itembox}[l]{Derivative of Eq. (\ref{eqn: vdp2})}
\begin{equation}
 \left.
  \begin{array}{rll}
   \displaystyle L\frac{di}{dt} &=& v - M(q)i, \vspace{2mm} \\
   \displaystyle C\frac{dv}{dt} &=& -i,        \vspace{2mm} \\
   \displaystyle  \frac{dq}{dt} &=& i.     
  \end{array}
 \right \}
\label{eqn: vdp2-derivative}
\end{equation}
\end{itembox}
\end{minipage} 
\end{center}
Here, $L=C=1$, $i$ and $v$ denote the current of the inductor $L$ and the voltage of the capacitor $C$, respectively, and $ M(q)$ is the small-signal memristance defined by 
\begin{equation}
   M(q) \stackrel{\triangle}{=} \displaystyle \frac{d f(q)}{dq} = q^{2} - 1.
\end{equation}
The terminal voltage $v_{M}$ and the terminal current $i_{M}$ of the ideal memristor 
in Figure \ref{fig:vdp-m} are given by
\begin{center}
\begin{minipage}{8.7cm}
\begin{shadebox}
\underline{\emph{V-I characteristics of the ideal memristor}}
\begin{equation}
    v_{M} = M(q) \, i_{M} =  (q^{2} - 1)\, i_{M}. 
\label{eqn: vdp2-vm-im}
\end{equation}
\vspace{0.5mm}
\end{shadebox}
\end{minipage}
\end{center}
Equations (\ref{eqn: vdp2}) and (\ref{eqn: vdp2-derivative}) exhibit periodic oscillation (limit cycle). 
If an external source is added as shown in Figure \ref{fig:vdp-m-source}, 
then the forced memristor Van der Pol equations can exhibit quasi-periodic oscillation \cite{Thompson}.    
The dynamics of this circuit is given by 
\begin{center}
\begin{minipage}{8.7cm}
\begin{shadebox}
\underline{\emph{Forced memristor Van der Pol equations}}
\begin{equation}
 \left.
  \begin{array}{rll}
   \displaystyle L\frac{di}{dt} &=& v - M(q)i + r \sin ( \omega t), \vspace{2mm} \\
   \displaystyle C\frac{dv}{dt} &=& -i,        \vspace{2mm} \\
   \displaystyle  \frac{dq}{dt} &=& i.    
  \end{array}
 \right \}
\label{eqn: vdp2-forced}
\end{equation} 
\end{shadebox}
\end{minipage}
\end{center}
We show the quasi-periodic attractor, Poincar\'e map, and $i_{M}-v_{M}$ locus of Eq. (\ref{eqn: vdp2-forced}) in Figures \ref{fig:vdp2-attractor}, \ref{fig:vdp2-poincare}, and \ref{fig:vdp2-pinch}, respectively.  
The $i_{M}-v_{M}$ locus in Figure \ref{fig:vdp2-pinch} lies in the all quadrants.   
Thus, the ideal memristor defined by Eq. (\ref{eqn: vdp2-curve}) or Eq. (\ref{eqn: vdp2-vm-im}) is an active element. 

Let us next show the $v_{M}-p_{M}$ locus in Figure \ref{fig:vdp2-power}, where $p_{M}(t)$ is an instantaneous power defined by $p_{M}(t)=i_{M}(t)v_{M}(t)$.  
Observe that the $v_{M}-p_{M}$ locus is pinched at the origin, and the locus lies in all quadrants, which is similar to the locus in Figure \ref{fig:Toda-C-power}(a).  
The memristor switches between four modes of operation:   
\begin{equation}
  (v_{M}, \, p_{M}) = ( +, \, + ), \ ( +, \, -), \  ( -, \, +), \  ( -, \, -). 
\label{eqn: vdp2-modes}
\end{equation}
Here, $(v_{M}, \, p_{M}) = (+, \, +)$ is read as $v_{M}>0$ and $p_{M}>0$,
$(v_{M}, \, p_{M}) = (+, \, -)$  is read as $v_{M}>0$ and $p_{M}<0$, and 
we excluded the special case where $(v_{M}, \, p_{M}) = (0, \, 0)$. 
Thus, the operation of the memristor has the high complexity.    
The operation modes (\ref{eqn: vdp2-modes}) can be coded by two bits: 
\[
  ( 0, \, 0 ), \ ( 0, \, 1), \  ( 1, \, 0), \  (1, \, 1),  
\]
where $+$ is coded to a binary number $0$ and $-$ to $1$.

%---Fig. 98-------%
\begin{figure}[hpbt]
 \centering
  \psfig{file=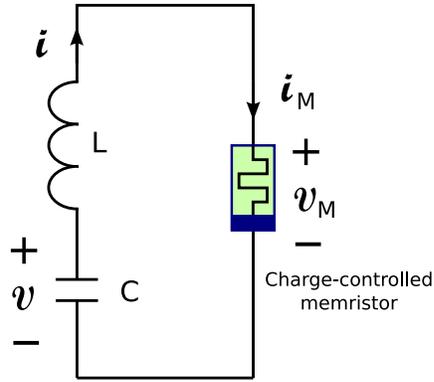,height=5cm}
  \caption{Memristor Van der Pol oscillator.   
   The $q-\varphi$ curve of the charge-controlled memristor is given by
   $\varphi = f(q) = \frac{q^3}{3}-q$.  
   Parameters: $L=C=1$.}
 \label{fig:vdp-m} 
\end{figure}
%
%

%---Fig. 99-------%
\begin{figure}[hpbt]
 \centering
  \psfig{file=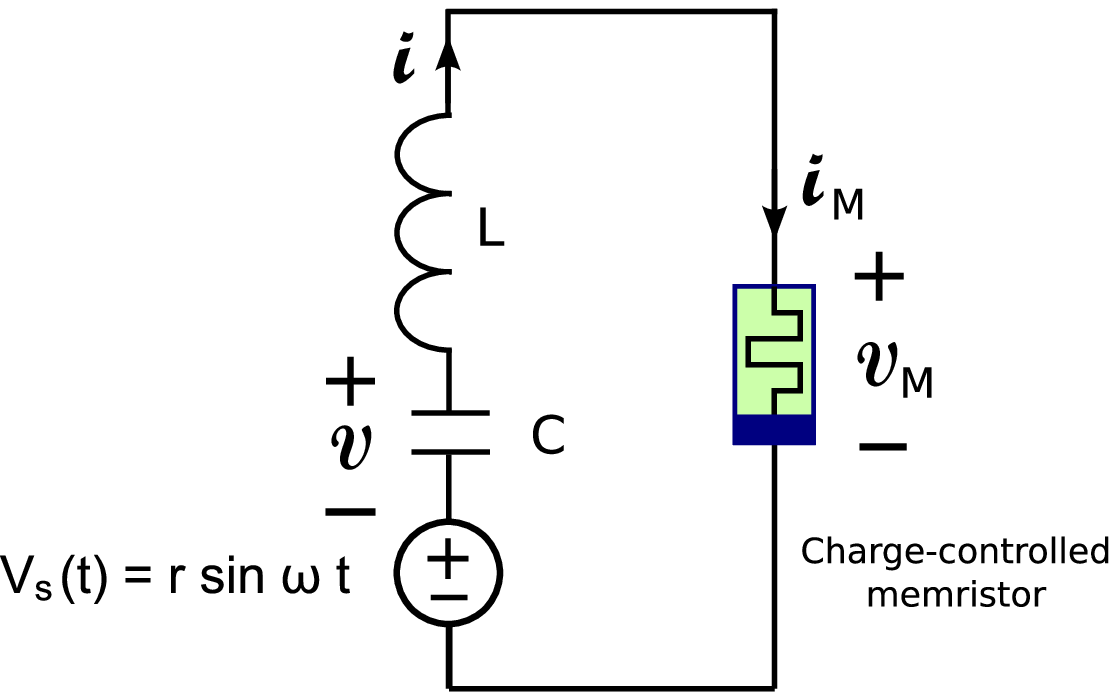,height=5cm}
  \caption{Memristor circuit driven by a periodic voltage source $v_{s}(t) = r \sin ( \omega t)$, where $r$ and $\omega$ are constants.}
  \label{fig:vdp-m-source} 
\end{figure}
%
%

%\newpage
%---Fig. 100-------%
\begin{figure}[hpbt]
\centering
  \psfig{file=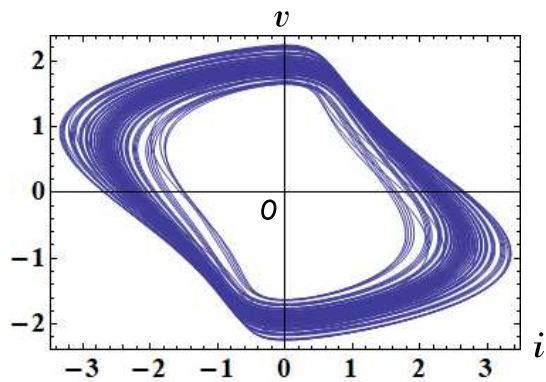, height=5.0cm} 
  \caption{Quasi-periodic attractor of the forced memristor Van der Pol equations (\ref{eqn: vdp2-forced}). 
  Parameters: $r = 0.59,  \ \omega = 1.1$.    
  Initial conditions: $i(0)=0.5, \  x(0)=0$.}
  \label{fig:vdp2-attractor} 
\end{figure}
%
%

%---Fig. 101-------%
\begin{figure}[hpbt]
\centering
  \psfig{file=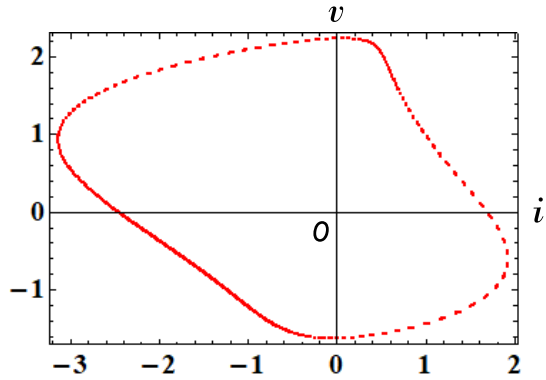, height=5.0cm}
  \caption{Poincar\'e map of the forced memristor Van der Pol equations (\ref{eqn: vdp2-forced}).    
  Parameters: $r = 0.59,  \ \omega = 1.1$.    
  Initial conditions: $i(0)=0.5, \  x(0)=0$.}
  \label{fig:vdp2-poincare} 
\end{figure}
%
%

%---Fig. 102-------%
\begin{figure}[hpbt]
\centering
  \psfig{file=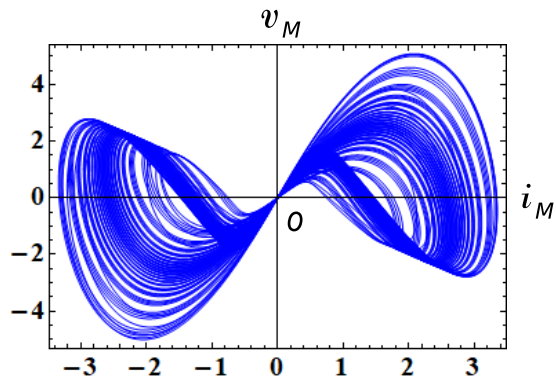, height=5.0cm}
  \caption{ The $i_{M}-v_{M}$ locus of the forced memristor Van der Pol equations (\ref{eqn: vdp2-forced}).
   Here, $v_{M}$ and  $i_{M}$ denote the terminal voltage and the terminal current of the charge-controlled memristor.  
   Observe that the ideal memristor defined by Eq. (\ref{eqn: vdp2-curve}) or Eq. (\ref{eqn: vdp2-vm-im})  is an active element.  
   Parameters: $r = 0.59,  \ \omega = 1.1$.    
   Initial conditions: $i(0)=0.5, \  x(0)=0$.}
  \label{fig:vdp2-pinch} 
\end{figure}
%
%

%---Fig. 103-------%
\begin{figure}[hpbt]
\centering
  \psfig{file=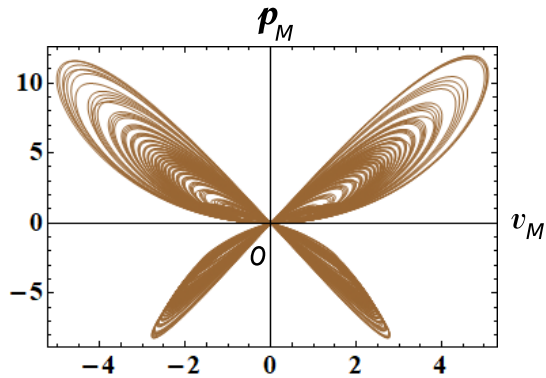, height=5.0cm}
  \caption{ The $v_{M}-p_{M}$ locus of the forced memristor Van der Pol equations (\ref{eqn: vdp2-forced}).  
   Here, $p_{M}(t)$ is an instantaneous power defined by $p_{M}(t)=i_{M}(t)v_{M}(t)$.  
   and $v_{M}(t)$ and $i_{M}(t)$ denote the terminal voltage and the terminal current of the charge-controlled memristor.   
   Observe that the $v_{M}-p_{M}$ locus is pinched at the origin, and the locus lies in all quadrants.  
   Parameters: $r = 0.59,  \ \omega = 1.1$.    
   Initial conditions: $i(0)=0.5, \  x(0)=0$.}
  \label{fig:vdp2-power} 
\end{figure}
\newpage

%
%
%%%%%%%%%%%%%%%%%%%%%%%%%%%%%%%%%%%%%%%%%%%%%%%%%%%%%%%%%%%%%%%%%%%%%%%
\subsection{Chua Circuit}
\label{sec: chua}
%%%%%%%%%%%%%%%%%%%%%%%%%%%%%%%%%%%%%%%%%%%%%%%%%%%%%%%%%%%%%%%%%%%%%%%
%
%
%
The dynamics of the Chua circuit \cite{{Hirsch(2003)}, {Madan}} is defined by 
\begin{center}
\begin{minipage}{8.7cm}
\begin{shadebox}
\underline{\emph{Chua circuit equations}}
\begin{equation} 
 \left.
  \begin{array}{lll}
     \displaystyle \frac{d x_{1}}{dt} &=& \alpha \bigl( x_{2} - x_{1} - g(x_{1}) \bigr ),
       \vspace{2mm} \\
     \displaystyle \frac{d x_{2}}{dt} &=&  x_{1} - x_{2} + x_{3},
       \vspace{2mm} \\    
     \displaystyle \frac{d x_{3}}{dt} &=&  - \beta x_{2}.    
  \end{array}
 \right \}
\label{eqn: chua-1}
\end{equation} 
\end{shadebox}
\end{minipage}
\end{center}
Here, $\alpha$ and $\beta$ are parameters, and $g(x_{1})$ is a scalar function of a single variable $x_{1}$ defined by 
\begin{equation} 
  g(x_{1}) = \frac{1}{16}x_{1}^3 - \frac{7}{6} x_{1}, 
\end{equation} 
which is a generalization from a continuous piecewise-linear function to a smooth function \cite{Hirsch(2003)}.
The original Chua circuit equations possess a piecewise-linear nonlinearity \cite{Madan}.  
That is, $g(x_{1})$ is a piecewise-linear function with discontinuous derivatives at the breakpoints.  
Equation (\ref{eqn: chua-1}) has a chaotic attractor similar to a double scroll attractor for certain values of the parameters $\alpha$ and $\beta$ \cite{{Hirsch(2003), {Madan}}}.  
This equation can also have a stable closed orbit outside of a chaotic attractor.  

Equation (\ref{eqn: chua-1}) can be realized by the circuit in Figure \ref{fig:chua-m} \cite{Itoh(2013)}.  
\begin{center}
\begin{minipage}{8.7cm}
\begin{shadebox}
\underline{\emph{Memristor Chua circuit equations}}
\begin{equation}
 \left.
  \begin{array}{rll}
   \displaystyle C_{1}\frac{d\varphi_{1}}{dt}   
    &=& \displaystyle \frac{\varphi_{2}-\varphi_{1}}{R} - f(\varphi_{1}),  \vspace{2mm} \\ 
   \displaystyle C_{2} \frac{d\varphi_{2}}{dt}  
    &=& \displaystyle q_{3} - \frac{\varphi_{2}-\varphi_{1}}{R},          \vspace{2mm} \\  
   \displaystyle L\frac{dq_{3}}{dt}       
    &=& - \varphi_{2}.           
  \end{array}
 \right \}
\label{eqn: chua-m}
\end{equation}
\end{shadebox}
\end{minipage}
\end{center}
Here, $\varphi_{1}$, $\varphi_{2}$, and $q_{3}$ denote the flux of the capacitor $C_{1}$, the flux of the capacitor $C_{2}$, and the charge of the inductor $L$, respectively, and the $\varphi-q$ curve of the flux-controlled memristor is given by
\begin{center}
\begin{minipage}{8.7cm}
\begin{shadebox}
\underline{\emph{$\varphi-q$ curve of the ideal memristor}}
\begin{equation}
   q = g(\varphi) = \frac{{\varphi}^{3}}{16} - \frac{7 \varphi}{6}.
\label{eqn: varphi-q-curve}
\end{equation}
\end{shadebox}
\end{minipage}
\end{center}

Differentiating Eq. (\ref{eqn: chua-1}) with respect to time $t$, we obtain 
\begin{center}
\begin{minipage}{8.5cm}
\begin{itembox}[l]{Derivative of Eq. (\ref{eqn: chua-1})}
\begin{equation}
 \left.
  \begin{array}{rll}
   \displaystyle C_{1}\frac{dv_{1}}{dt}   
    &=& \displaystyle \frac{v_{2}-v_{1}}{R} 
        - W (\varphi_{1} ) v_{1},  \vspace{2mm} \\ 
   \displaystyle C_{2} \frac{dv_{2}}{dt}  
    &=& \displaystyle i_{3} - \frac{v_{2}-v_{1}}{R},    \vspace{2mm} \\           
   \displaystyle L\frac{di_{3}}{dt}       
    &=& - v_{2}.                             
  \end{array}
 \right \}
\label{eqn: chua-2}
\end{equation}
\end{itembox}
\end{minipage}
\end{center}
Here, $v_{1}$, $v_{2}$, and $i_{3}$ denote the voltage across the capacitor $C_{1}$, the voltage across the capacitor $C_{2}$, and the current through the inductor $L$, respectively, and $W (\varphi_{1} )$ is the small-signal \emph{memductance} defined by 
\begin{equation}
   W (\varphi_{1} ) \stackrel{\triangle}{=} \displaystyle \frac{d g( \varphi_{1} )}{d \varphi_{1}} =  \frac{3}{16} {\varphi}^{2}- \frac{7}{6}.
\end{equation}
The terminal voltage $v_{M}$ and the terminal current $i_{M}$ of the ideal memristor 
in Figure \ref{fig:chua-m} are given by
\begin{center}
\begin{minipage}{8.7cm}
\begin{shadebox}
\underline{\emph{V-I characteristics of the ideal memristor}}
\begin{equation}
    i_{M} = W (\varphi_{1} )  \, v_{M} =  \left ( \frac{3}{16}{\varphi}^{2} - \frac{7}{6} \right ) \, v_{M}. 
\label{eqn: chua-im-vm}
\end{equation}
\vspace{0.5mm}
\end{shadebox}
\end{minipage}
\end{center}

We show the chaotic attractor, Poincar\'e map, and $v_{M}-i_{M}$ locus in Figures \ref{fig:chua-attractor}, \ref{fig:chua-poincare}, and \ref{fig:chua-pinch}, respectively.  
The following parameters are used in our computer simulations:
\begin{equation}
\left.
 \begin{array}{cccc}
  \displaystyle C_{1}  = \frac{1}{\alpha}, & \displaystyle C_{2}=1, & \displaystyle L = \frac{1}{\beta}, & R=1.  \vspace{2mm}\\ 
  \alpha = 10, & \beta =14. 
 \end{array}
 \right \}
\label{eqn: parameter-chua-100}
\end{equation}
Observe the folding action of the chaotic attractor in Figure \ref{fig:chua-poincare}.
The $v_{M}-i_{M}$ locus in Figure \ref{fig:chua-pinch} lies in the second and the fourth quadrants.   
Thus, the ideal memristor defined by Eq. (\ref{eqn: varphi-q-curve}) or Eq. (\ref{eqn: chua-im-vm}) is an active element. 

Let us next show the $i_{M}-p_{M}$ locus in Figure \ref{fig:vdp2-power}, where $p_{M}(t)$ is an instantaneous power defined by $p_{M}(t)=i_{M}(t)v_{M}(t)$.  
Observe that the $i_{M}-p_{M}$ locus is pinched at the origin, and the locus lies in the third and the fourth quadrants.  
The memristor switches between two operation modes:   
\begin{equation}
  (i_{M}, \, p_{M}) = ( +, \, - ), \ ( -, \, -). 
\label{eqn: chua-modes}
\end{equation}
Here, $(i_{M}, \, p_{M}) = (+, \, -)$ is read as $i_{M}>0$ and $p_{M}<0$,
$(i_{M}, \, p_{M}) = (-, \, -)$  is read as $i_{M}<0$ and $p_{M}<0$, and 
we excluded the special case where $(i_{M}, \, p_{M}) = (0, \, 0)$. 
These memristor's operation modes are quite different from those of the other systems.  
The operation modes (\ref{eqn: chua-modes}) can be coded by two bits: 
\[
  ( 0, \, 1 ), \ ( 1, \, 1), 
\]
where $+$ is coded to a binary number $0$ and $-$ to $1$.\footnote{Similarly, the forced memristor Brusselator equations (\ref{eqn: Brusselator-4}) have the different operation modes, which are given by  $(v_{M}, \, p_{M}) = ( +, \, + )$.  
It can be coded by $( 0, \, 0 )$.  It is due to the reason that the extended memristor is passive.  
See the $v_{M}-p_{M}$ locus shown in Figure \ref{fig:Brusselator-power} and Appendix B for more details.}  

%---Fig. 104-------%
\begin{figure}[hpbt]
 \centering
  \psfig{file=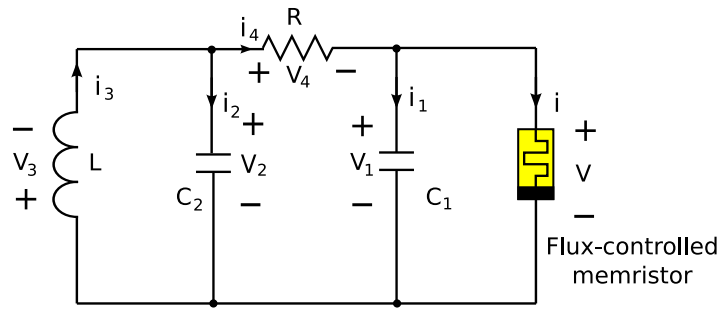,height=4.0cm}
  \caption{Memristor Chua circuit.  
   It contains five circuit elements: two passive capacitors, one passive inductor, one passive resistor, and one flux-controlled memristor.}
  \label{fig:chua-m}
\end{figure}
%
%

%---Fig. 105-------%
\begin{figure}[hpbt]
 \centering
   \begin{tabular}{cc}
    \psfig{file=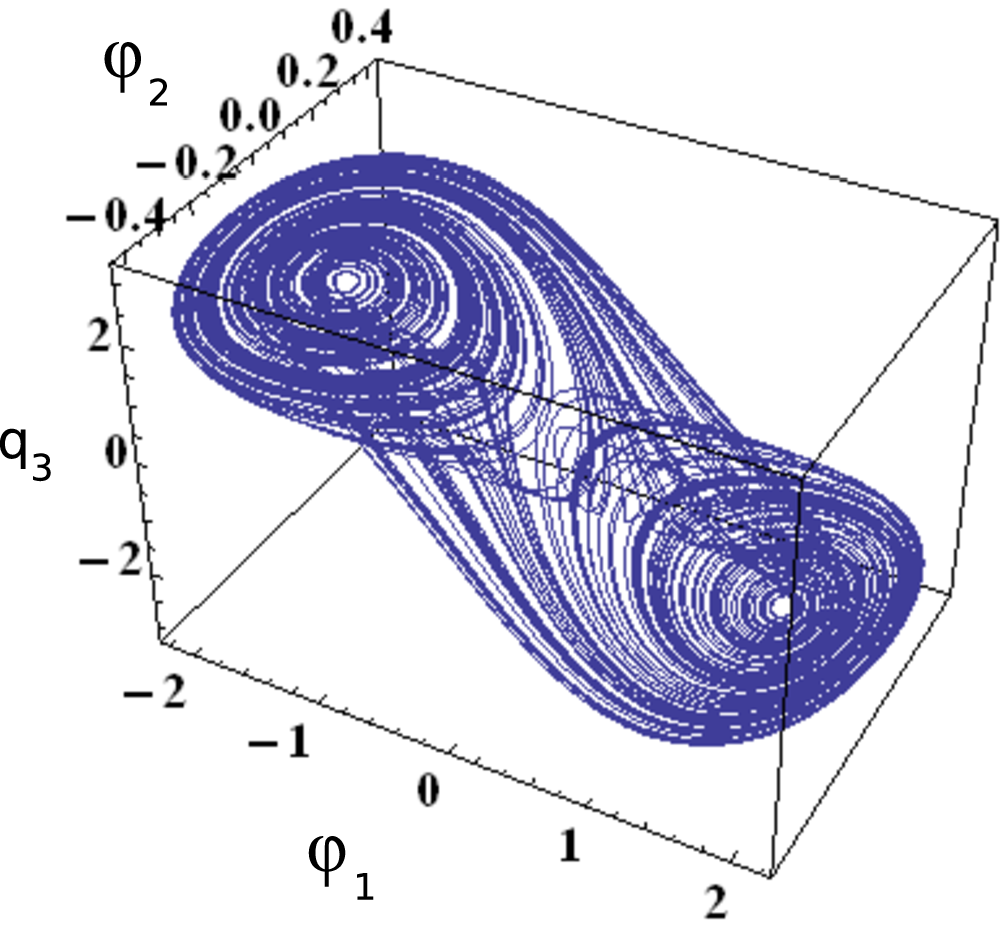, height=5.5cm}  & 
    \psfig{file=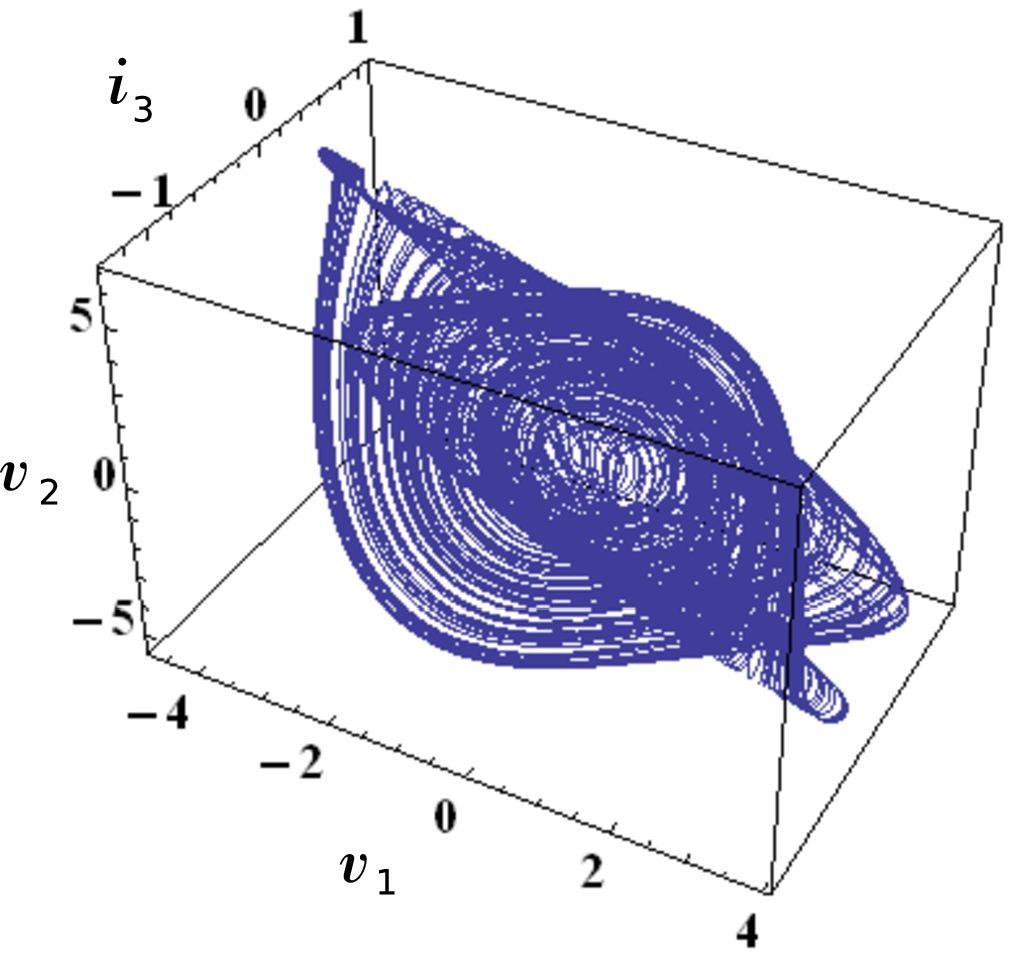, height=5.5cm}  \\
   (a) $(\varphi_{1}, \, \varphi_{1}, \, q_{3})$-space  & (b) $(v_{1}, \, v_{2}, \, i_{3})$-space 
   \end{tabular}
  \caption{Chaotic attractors of the memristor Chua circuit equations. 
   (a) Chaotic attractor of Eq. (\ref{eqn: chua-m}).    (b) Chaotic attractor of Eq. (\ref{eqn: chua-2}).  
   Parameters: $\displaystyle C_{1}  = \frac{1}{10}, \,  C_{2}=1, \, \displaystyle L = \frac{1}{14}, \, R=1$. 
   Initial conditions:  $(v_{1}(0), \ v_{2}(0),  \ i_{3}(0), \ \varphi_{1}(0))=(0.1, \ 0.1, \ 0.1, \ 0)$.}
  \label{fig:chua-attractor} 
\end{figure}
%
%

%---Fig. 106-------%
\begin{figure}[hpbt]
 \begin{center}
  \psfig{file=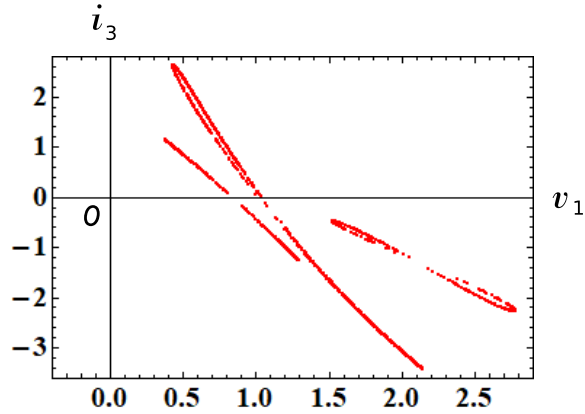, height=5.5cm}
  \caption{Poincar\'e map of the chaotic attractor for the memristor Chua circuit equations (\ref{eqn: chua-2}).   
   The Poincar\'e cross-section is defined by $\{(v_{1}, \, v_{2}, \, i_{3}) \in R^{3} \ | \ v_{2} = 0.8 \}$.  
   The chaotic trajectory of Eq. (\ref{eqn: chua-2}) crosses the above Poincar\'e cross-section (plane) many times.   
   Observe the folding action of the chaotic attractor. 
   Parameters: $\displaystyle C_{1}  = \frac{1}{10}, \, C_{2}=1, \, \displaystyle L = \frac{1}{14}, \, R=1$. 
   Initial conditions:  $(v_{1}(0), \ v_{2}(0),  \ i_{3}(0), \ \varphi_{1}(0))=(0.1, \ 0.1, \ 0.1, \ 0)$.}
  \label{fig:chua-poincare} 
 \end{center}
\end{figure}
%
%

%---Fig. 107-------%
\begin{figure}[hpbt]
\centering
  \psfig{file=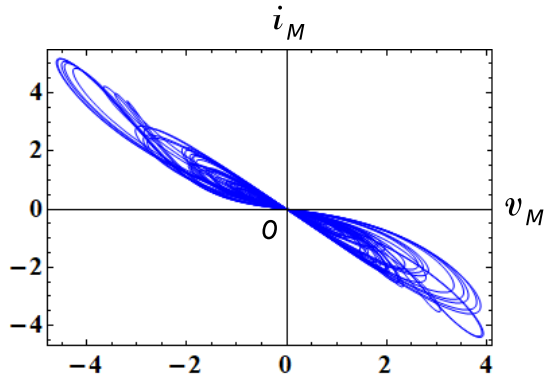, height=5.0cm}
  \caption{ The $v_{M}-i_{M}$ locus of the memristor Chua circuit equations (\ref{eqn: chua-2}).  
   Here, $v_{M}$ and  $i_{M}$ denote the terminal voltage and the terminal current of the flux-controlled memristor.  
   Observe that the ideal memristor defined by Eq. (\ref{eqn: varphi-q-curve}) or Eq. (\ref{eqn: chua-im-vm}) is an active element,
   since the locus lies in the second and the fourth quadrants.  
   Parameters: $\displaystyle C_{1}  = \frac{1}{10}, \, C_{2}=1, \, \displaystyle L = \frac{1}{14}, \, R=1$. 
   Initial conditions:  $(v_{1}(0), \ v_{2}(0),  \ i_{3}(0), \ \varphi_{1}(0))=(0.1, \ 0.1, \ 0.1, \ 0)$.}
  \label{fig:chua-pinch} 
\end{figure}
%
%

%---Fig. 108-------%
\begin{figure}[hpbt]
\centering
  \psfig{file=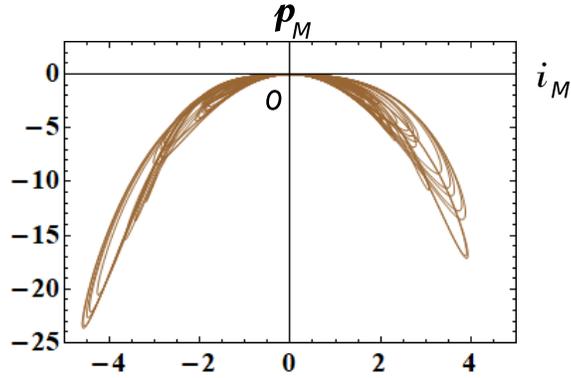, height=5.0cm}
  \caption{ The $i_{M}-p_{M}$ locus of the memristor Chua circuit equations (\ref{eqn: chua-2}).  
   Here, $p_{M}(t)$ is an instantaneous power defined by $p_{M}(t)=i_{M}(t)v_{M}(t)$, 
   and $v_{M}(t)$ and $i_{M}(t)$ denote the terminal voltage and the terminal current of the flux-controlled memristor.
   Observe that the $i_{M}-p_{M}$ locus is pinched at the origin, and the locus lies in the third and the fourh quadrants.  
   Parameters: $\displaystyle C_{1}  = \frac{1}{10}, \, C_{2}=1, \, \displaystyle L = \frac{1}{14}, \, R=1$. 
   Initial conditions:  $(v_{1}(0), \ v_{2}(0),  \ i_{3}(0), \ \varphi_{1}(0))=(0.1, \ 0.1, \ 0.1, \ 0)$.}
  \label{fig:chua-power} 
\end{figure}
\newpage

%\clearpage
%
%
%%%%%%%%%%%%%%%%%%%%%%%%%%%%%%%%%%%%%%%%%%%%%%%%%%%%%%%%%%%%%%%%%%%%%%%
\section{Conclusion}
%%%%%%%%%%%%%%%%%%%%%%%%%%%%%%%%%%%%%%%%%%%%%%%%%%%%%%%%%%%%%%%%%%%%%%%
%
%
%
We have shown that the dynamics of a wide variety of nonlinear systems such as engineering, physical, chemical, biological, and ecological systems, can be simulated or modeled by the dynamics of memristor circuits.  
The resulting memristor circuits can exhibit quasi-periodic, non-periodic, or chaotic behavior by supplying the external source.  
If they have an integral invariant, their behavior greatly depends on the initial conditions, the circuit parameters, and the maximum step size of the numerical integration.   
Furthermore, an overflow (outside the range of data) is likely to occur due to the numerical instability in long-time simulations.      
We have also shown that we can reconstruct chaotic attractors by using the terminal voltage and current of the memristor.   
Furthermore, in many circuits, the active memristor switches between passive and active modes of operation, depending on its terminal voltage.  
However, we found that the memristor's operation modes exhibit the higher complexity in the forced memristor Toda lattice equations.    
We note that almost all results in this paper were obtained by using ``NDSolve'' in Mathematica ($32$ bit version).  
If we use other softwares to solve differential equations, we might obtain slightly different results. 
%\clearpage

%\clearpage
%
%\section*{Appendices} 
%==============================================================%
%==============================================================%

%==============================================================%
%---Appendix A-------------%
%--------------------------%
\subsection*{Appendix A \ \it{Classification of Memristors}}
\label{sec:memristors}
\emph{Memristor} is a $2$-terminal electronic device, which was postulated in \cite{{Chua(2012)}, {Chua(1971)}, {Chua(1976)}}.  
An \emph{ideal} memristor can be described by a 
constitutive relation between the charge $q$ and the flux $\varphi$, 
\begin{center}
\begin{minipage}{8cm}
\begin{shadebox}
\vspace{-1mm}
\begin{equation}
  q = g( \varphi ) \ \ \text{or} \ \ \varphi = f(q),
\vspace{1mm}
\label{eqn: mem-q-g}
\end{equation}
\end{shadebox}
\end{minipage}
\end{center}
where $g(\cdot)$ and $f(\cdot)$ are differentiable scalar-valued functions. 
Its terminal voltage $v$ and terminal current $i$ are described by 
\begin{center}
\begin{minipage}{8cm}
\begin{shadebox}
\begin{equation}
  i = G( \varphi ) v \ \ \text{or} \ \  v = R(q) i,
\label{eqn: mem-i-v}
\end{equation}
\end{shadebox}
\end{minipage}
\end{center}
where 
\begin{equation}
 v = \frac{d\varphi}{dt} \ \ \text{and} \ \  i = \frac{dq}{dt},
\label{eqn: v-i}
\end{equation}
which represent \emph{Faraday's induction law} and its dual law, respectively.  
The nonlinear functions $G( \varphi )$ and $R( \varphi )$, called the small-signal \emph{memductance} and small-signal \emph{memristance}, respectively, are defined by, is defined by
\begin{equation}
  G(\varphi) \stackrel{\triangle}{=} \frac{dg(\varphi)}{d\varphi},
\end{equation}
and 
\begin{equation}
  R(q) \stackrel{\triangle}{=} \frac{df(q)}{dq},
\end{equation}
representing the \emph{slope} of the scalar function $q = g( \varphi )$, and $\varphi = f(q)$, respectively, called the \emph{memristor constitutive relation}.

All voltage-controlled memristors can be classified into four classes \cite{Chua(2015)}:  
\begin{itemize}
\item voltage-controlled ideal memristor
\begin{center}
\begin{minipage}{6cm}
\begin{shadebox}
\begin{equation}
\begin{array}{lll}
  i &=& G( \varphi )v,  
  \vspace{1mm} \\
  \displaystyle \frac{d\varphi}{dt}  &=& v.
\end{array}
\label{eqn: mem-ideal}
\end{equation}
\end{shadebox}
\end{minipage}
\vspace{2mm}
\end{center}

\item voltage-controlled ideal generic memristor
\begin{center}
\begin{minipage}{6cm}
\begin{shadebox}
\begin{equation}
\begin{array}{lll}
  i &=& G ( x )v,  
  \vspace{1mm} \\
  \displaystyle \displaystyle \frac{dx}{dt} &=& \hat{g}(x)v.
\end{array}
\label{eqn: generic}
\end{equation}
\end{shadebox}
\end{minipage}
\vspace{2mm}
\end{center}

\item voltage-controlled generic memristor
\begin{center}
\begin{minipage}{6cm}
\begin{shadebox}
\begin{equation}
\begin{array}{lll}
  i &=& \tilde{G}( \bd{x} )v,  
  \vspace{1mm} \\
  \displaystyle \frac{d\bd{x}}{dt} &=& \tilde{\bd{g}}(\bd{x}, \ v).
\end{array}
\label{eqn: generic2}
\end{equation}
\end{shadebox}
\end{minipage}
\vspace{2mm}
\end{center}

\item voltage-controlled extended memristor
\begin{center}
\begin{minipage}{6cm}
\begin{shadebox}
\begin{equation}
\begin{array}{lll}
  i &=& \hat{G}( \bd{x}, \ v )v,  \\ 
    &&  \hat{G}( \bd{x}, \ 0) \ne \infty,
  \vspace{1mm} \\
  \displaystyle \frac{d\bd{x}}{dt} &=& \tilde{\bd{g}}(\bd{x}, \ v).
\end{array}
\label{eqn: extended}
\end{equation}
\end{shadebox}
\end{minipage}
\end{center}
Here, $G(\cdot)$, $\tilde{G}(\cdot)$, $\hat{G}(\cdot)$, and $\hat{g}(\cdot)$ are continuous scalar-valued functions, $\bd{x} = (x_{1}, \, x_{2}, \, \cdots, \, x_{n}) \in \mathbb{R}^{n}$, and $\tilde{\bd{g}} = (\tilde{g}_{1}, \, \tilde{g}_{2}, \, \cdots, \, \tilde{g}_{n}): \mathbb{R}^{n} \rightarrow \mathbb{R}^{n}$.  
\end{itemize}

Similarly, all current-controlled memristors can be classified into four classes \cite{Chua(2015)}:  
\begin{itemize}
\item current-controlled ideal memristor
\begin{center}
\begin{minipage}{6cm}
\begin{shadebox}
\begin{equation}
\begin{array}{lll}
  v &=& R( q )i,  
  \vspace{1mm} \\
  \displaystyle \frac{dq}{dt}  &=& i.
\end{array}
\label{eqn: mem-ideal-2}
\end{equation}
\end{shadebox}
\end{minipage}
\vspace{2mm}
\end{center}

\item current-controlled ideal generic memristor
\begin{center}
\begin{minipage}{6cm}
\begin{shadebox}
\begin{equation}
\begin{array}{lll}
  v &=& R( x )i,  
  \vspace{1mm} \\
  \displaystyle \displaystyle \frac{dx}{dt} &=& \hat{f}(x)i.
\end{array}
\label{eqn: generic-2}
\end{equation}
\end{shadebox}
\end{minipage}
\vspace{2mm}
\end{center}

\item current-controlled generic memristor
\begin{center}
\begin{minipage}{6cm}
\begin{shadebox}
\begin{equation}
\begin{array}{lll}
  v &=& \tilde{R}( \bd{x} )i,  
  \vspace{1mm} \\
  \displaystyle \frac{d\bd{x}}{dt} &=& \tilde{\bd{f}}(\bd{x}, \ i).
\end{array}
\label{eqn: generic2-2}
\end{equation}
\end{shadebox}
\end{minipage}
\vspace{2mm}
\end{center}

\item current-controlled extended memristor
\begin{center}
\begin{minipage}{6cm}
\begin{shadebox}
\begin{equation}
\begin{array}{lll}
  v &=& \hat{R}( \bd{x}, \ i )i,  \\ 
    && \hat{R}( \bd{x}, \ 0) \ne \infty,
  \vspace{1mm} \\
  \displaystyle \frac{d\bd{x}}{dt} &=& \tilde{\bd{f}}(\bd{x}, \ i).
\end{array}
\label{eqn: extended-2}
\end{equation}
\end{shadebox}
\end{minipage}
\end{center}
Here, $R(\cdot)$, $\tilde{R}(\cdot)$, $\hat{R}(\cdot)$, and $\hat{f}(\cdot)$ are continuous scalar-valued functions, $\bd{x} = (x_{1}, \, x_{2}, \, \cdots, \, x_{n}) \in \mathbb{R}^{n}$, and $\tilde{\bd{f}} = (\tilde{f}_{1}, \, \tilde{f}_{2}, \, \cdots, \, \tilde{f}_{n}): \mathbb{R}^{n} \rightarrow \mathbb{R}^{n}$.  
\end{itemize}

%\clearpage
%==============================================================%
%---Appendix B-------------%
%--------------------------%
\subsection*{Appendix B \ \it{$v_{M}-p_{M}$ locus of the forced memristor Brusselator equations }}
\label{sec:Brusselator-appendix}

Consider the forced memristor Brusselator equations defined by Eq. (\ref{eqn: Brusselator-4}), that is,
\begin{equation}
\left.
\begin{array}{lll}
 \displaystyle L \frac{d i}{dt} &=& A + \bigl \{ ix - (B+1) \bigr \} i + r \sin ( \omega t), 
 \vspace{1mm} \\
 \displaystyle \frac{d x}{dt} &=& B \, i - i^{2}\, x,
\end{array}
\right \}
\label{eqn: Brusselator-104}
\end{equation}
where   
\begin{equation}
  A = 0.4, \ B = 1.2, \ r = 0.05, \ \omega = 0.81.  
\end{equation}
Assume that the terminal voltage $v_{M}$ and the terminal current $i_{M}$ of the extended memristor are given by Eq. (\ref{eqn: Brusselator-3}), 
that is,  
\begin{equation}
\left. 
\begin{array}{lll}
  v_{M} &=& \hat{R}(x, \, i_{M}) \, i_{M},  
  \vspace{1mm} \\
 \displaystyle \frac{d x}{dt} &=& B \, i_{M} - {i_{M}}^{2}\, x,
\end{array}
\right \}
\label{eqn: Brusselator-3-ap}
\end{equation}
where $\hat{R}(x, \, i_{M}) =  - \bigl \{ i_{M} \, x - (B+1) \bigr \}$ and $i_{M}=i$.  
Then the forced Brusselator equations (\ref{eqn: Brusselator-104}) can be realized by the circuit in Figure \ref{fig:memristive-inductor-battery-source}, 
where $L=1$ and $E=A=0.4$.  
 
As stated in Sec. \ref{sec: Brusselator}, the $i_{M}-v_{M}$ locus moves in the first quadrant only, that is, it moves in the \emph{passive} region (see Figure \ref{fig:Brusselator-pinch}(a)).  
Consider next the instantaneous power of the extended memristor, which is defined by 
\begin{equation}
  p_{M}(t)=i_{M}(t)v_{M}(t).  
\end{equation}
Then the $v_{M}-p_{M}$ locus of Eq. (\ref{eqn: Brusselator-104}) is not pinched at the origin, and the locus lies in the first quadrant only as shown in Figure \ref{fig:Brusselator-power}(a).    
Thus, the memristor's operation mode is given by 
\begin{equation}
  (v_{M}, \, p_{M}) = ( +, \, + ),   
\label{eqn: Brusselator-modes}
\end{equation}
where $(v_{M}, \, p_{M}) = (+, \, +)$ is read as $v_{M}>0$ and $p_{M}>0$.   
The operation mode (\ref{eqn: Brusselator-modes}) can be coded by 
\begin{equation}
  ( 0, \, 0 ),
\label{eqn: Brusselator-binary-modes}
\end{equation}
where $+$ is coded to a binary number $0$ and $-$ to $1$.  
The binary mode (\ref{eqn: Brusselator-binary-modes}) is equivalent to the one-bit coding defined by $( 0 )$. 

Define next the instantaneous power of the two elements, that is, the instantaneous power of the extended memristor and the battery by 
\begin{equation}
  p_{ME}(t)=i_{M}(t)\, v_{ME}(t), 
\end{equation}
where $E$ denotes the voltage of the battery and $v_{ME}(t) = v_{M}(t)-E$.   
That is, $v_{ME}(t)$ denotes the voltage across the extended memristor and the battery.  
We show the $v_{ME}-p_{ME}$ locus in Figure \ref{fig:Brusselator-power}(b).  
Observe that the locus is pinched at the origin, and it lies in the first and the third quadrants.  
Thus, the instantaneous power $p_{ME}$ delivered from the forced signal and the inductor is dissipated when $v_{M}(t) - E > 0$.   
However, the instantaneous power $p_{ME}$ is \emph{not} dissipated when $v_{M}(t) - E <0 $.
Thus, the operation modes of the two elements is given by 
\begin{equation}
  (v_{ME}, \ p_{ME}) = ( +, \, + ), \ ( -, \, - ).  
\label{eqn: Brusselator-modes-2}
\end{equation}
They are coded by 
\begin{equation}
 ( 0, \, 0 ), \ ( 1, \,1),    
\end{equation}
which are equivalent to the one bit coding defined by 
\begin{equation}
 ( 0 ), \  ( 1 ).    
\end{equation}
%
%

%---Fig. 109-------%
\begin{figure}[ht]
 \centering
   \begin{tabular}{cc}
   \psfig{file=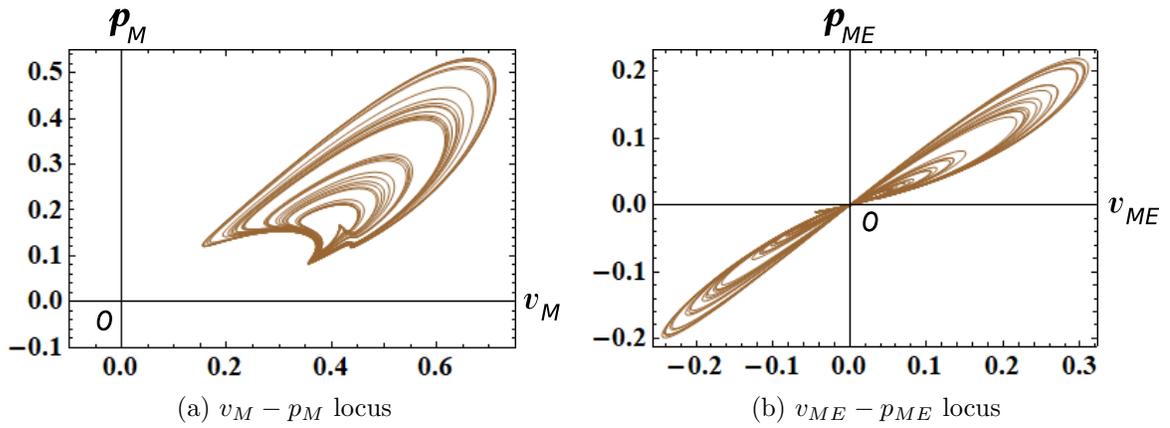, height=5.0cm}  & 
   \psfig{file=./figure/Brusselator-power-2, height=5.0cm} \\
   (a) $v_{M}-p_{M}$ locus  & (b) $v_{ME}-p_{ME}$ locus
   \end{tabular}  
  \caption{ The $v_{M}-p_{M}$ and $v_{ME}-p_{ME}$ loci of the forced memristor Brusselator equations (\ref{eqn: Brusselator-4}). 
   Here, $p_{M}(t)$ and $p_{ME}(t)$ are the instantaneous powers defined by 
   $p_{M}(t)=i_{M}(t) \, v_{M}(t)$ and $p_{ME}(t)=i_{M}(t)\, v_{ME}(t)$, respectively, 
   $v_{M}$ and  $i_{M}$ denote the terminal voltage and the terminal current of the current-controlled extended memristor, 
   $v_{ME}(t) = v_{M}(t)-E$, and $E$ denotes the voltage of the battery.   
   Observe that the $v_{M}-i_{M}$ locus is \emph{not} pinched at the origin, and the locus lies in the first quadrant only.  
   However, the $v_{ME}-p_{ME}$ locus is pinched at the origin, and it lies in the first and the third quadrants.  
   Parameters: $A = 0.4, \ B = 1.2, \ r = 0.05, \ \omega = 0.81$. 
   Initial conditions: $i(0)=1.1, \  x(0)=1.1$.}
  \label{fig:Brusselator-power} 
\end{figure}

\end{document}